\documentclass[10pt,openright,twoside,a4]{book}

\usepackage{amsmath,amsfonts,amssymb} 
\usepackage{bm} 
\usepackage{theorem}

\usepackage{sfmath}

\usepackage{makeidx} 
\makeindex 

\usepackage{abbrevs}

\usepackage{nomencl} 
\renewcommand{\nomgroup}[1]{%
\ifthenelse{\equal{#1}{S}}{\vspace*{\baselineskip}\item[\textbf{Symbols}]\vspace*{\baselineskip}}{%
\ifthenelse{\equal{#1}{A}}{\vspace*{\baselineskip}\item[\textbf{Abbrivations}]\vspace*{\baselineskip}}{}}} 

\usepackage{subfigure}
\usepackage{graphics}
\usepackage{epsfig}
\usepackage{graphicx,epic,eepic}
\usepackage{graphics}
\usepackage[dvips]{color}
\usepackage{multicol}        		

\usepackage{sectsty}

\usepackage{units}		
\usepackage{mathrsfs}	

\usepackage{fancyheadings}
\usepackage{titlesec}
\usepackage{titletoc}

\usepackage{appendix}
\usepackage{helvet}
\renewcommand{\familydefault}{\sfdefault} 
\renewcommand{\rmdefault}{\sfdefault}
 
\sectionfont{\sffamily \mdseries}
\subsectionfont{\sffamily \mdseries}
\subsubsectionfont{\sffamily }

\chapternumberfont{ \sffamily \mdseries}

\makeatletter 
\def\@seccntformat#1{\protect\makebox[0pt][r]{\sf  \csname
 the#1\endcsname $\,|$ }} 
\makeatother

\titleformat{\chapter}[display] 
{\bfseries \Large} 
{\filleft\MakeUppercase{\chaptertitlename} \Huge \thechapter} 
{4ex} 
{\titlerule 
\vspace{2ex}%
\filright} 
[\vspace{2ex}%
\titlerule
]

\renewcommand{\chaptermark}[1]{\markboth{Chapter \textbf{\thechapter}.\ {#1}}{}}
\renewcommand{\sectionmark}[1]{\markright{\thesection\ \boldmath {#1}\unboldmath}}
\def\A{{\mathrm A}} 

\def\B{{\mathrm B}} 

\def\d{{d}} 
\def\diag{{\mathrm{diag}}}	 
\def\det{{\mathrm{det}}} 

\def\eg{{\emph{e.g.}}} 
\def\e{\mathrm{e}} 

\def\G{{\mathrm{\Gamma}}} 
\definecolor{Gruen}{rgb}{.322,.537,.035}

\def\half{\textstyle{\frac{1}{2}}} 

\def\ie{{\emph{i.e.}}} 
\def\implies{\Rightarrow} 
\def\ipd{{\mathrm{\Theta}}} 

\def\K{{\mathrm{K}}} 

\def\lint{\hbox{\Large $\displaystyle\int$}} 

\def\Pl{{\mathrm{Planck}}} 

\def\Rb{{\mathrm{Rb}}} 
\def\r{{\mathbf{r}}} 

\def\tr{{\mathrm{tr}}} 
\def\T{{\mathrm{T}}} 

\newcommand{\aop}[1]{{\hat{a}_{k}^{\reg{#1}}}} 
\newcommand{\aopneg}[1]{{\hat{a}_{-k}^{\reg{#1}}}} 
\newcommand{\aopdag}[1]{{\hat{a}_{k}^{\reg{#1}^{{\small  \dag}}}}} 
\newcommand{\aopdagneg}[1]{{\hat{a}_{-k}^{\reg{#1}^{{\small  \dag}}}}} 
\newcommand{\abs}[1]{\left\vert #1 \right\vert} 

\newcommand{\betrag}[1]{\left\vert#1\right\vert}
\newcommand{\Ba}[2]{\alpha_{k}^{\reg{#1}\reg{#2}}} 
\newcommand{\Bac}[2]{{\alpha_{k}^{\reg{#1}\reg{#2}}}^{*}} 
\newcommand{\Bb}[2]{\beta_{k}^{\reg{#1}\reg{#2}}} 
\newcommand{\Bbc}[2]{{\beta_{k}^{\reg{#1}\reg{#2}}}^{*}} 
\newcommand{\Blue}[1]{\color{blue} #1}

\newcommand{\comBig}[2]{\Big[ #1,#2 \Big]} 

\newcommand{\Cyan}[1]{\color{cyan} #1}

\newcommand{\Green}[1]{\color{Gruen} #1}

\newcommand{\mf}[1]{u^{\reg{#1}}_{k} } 
\newcommand{\mfc}[1]{ {u^{\reg{#1}}_{k}}^{*} } 
\newcommand{\mfE}[1]{v^{\reg{#1}}_{k} } 
\newcommand{\mfEc}[1]{ {v^{\reg{#1}}_{k}}^{*} } 
\newcommand{\mfqp}[1]{\tilde{u}^{\reg{#1}}_{k} } 
\newcommand{\Mf}[1]{\vec{U}^{\reg{#1}}_{k}} 
\newcommand{\MfE}[1]{\vec{V}^{\reg{#1}}_{k}} 

\newcommand{\MatrixB}[4]{\left[ \begin{array}{cc} #1 & #2 \\ #3 & #4 \end{array}\right]}

\newcommand{\op}[1]{\hat{#1}} 

\newcommand{\refb}[1]{(\ref{#1})}
\newcommand{\reg}[1]{\,\hbox{\b{\tiny  #1\,}}}
\newcommand{\Reg}[1]{\hbox{\b{  #1 } }}
\newcommand{\regreg}[3]{\, \hbox{\b{\tiny #1\,} {\tiny #2} \b{\tiny #3 \,}}}
\newcommand{\Red}[1]{\color{red} #1}

\newcommand{\sound}[1]{c_{0}^{\reg{#1}} } 

\newcommand{\silkychaptermark}[1]{\renewcommand{\chaptermark}[1]{\markboth{ #1}{}}}

\def\tx{(t,\mathbf{x})} 
\def\txp{(t,\mathbf{x}')} 

\newcommand{\vstate}[1]{\vert 0^{\reg{#1}} \rangle} 
\newcommand{\vstatebra}[1]{\langle 0^{\reg{#1}} \vert} 

\input{abbreviations} 
\begin{document}
%

\begin{titlepage}
\center

\vspace*{1cm}


%


\includegraphics[width=5in]{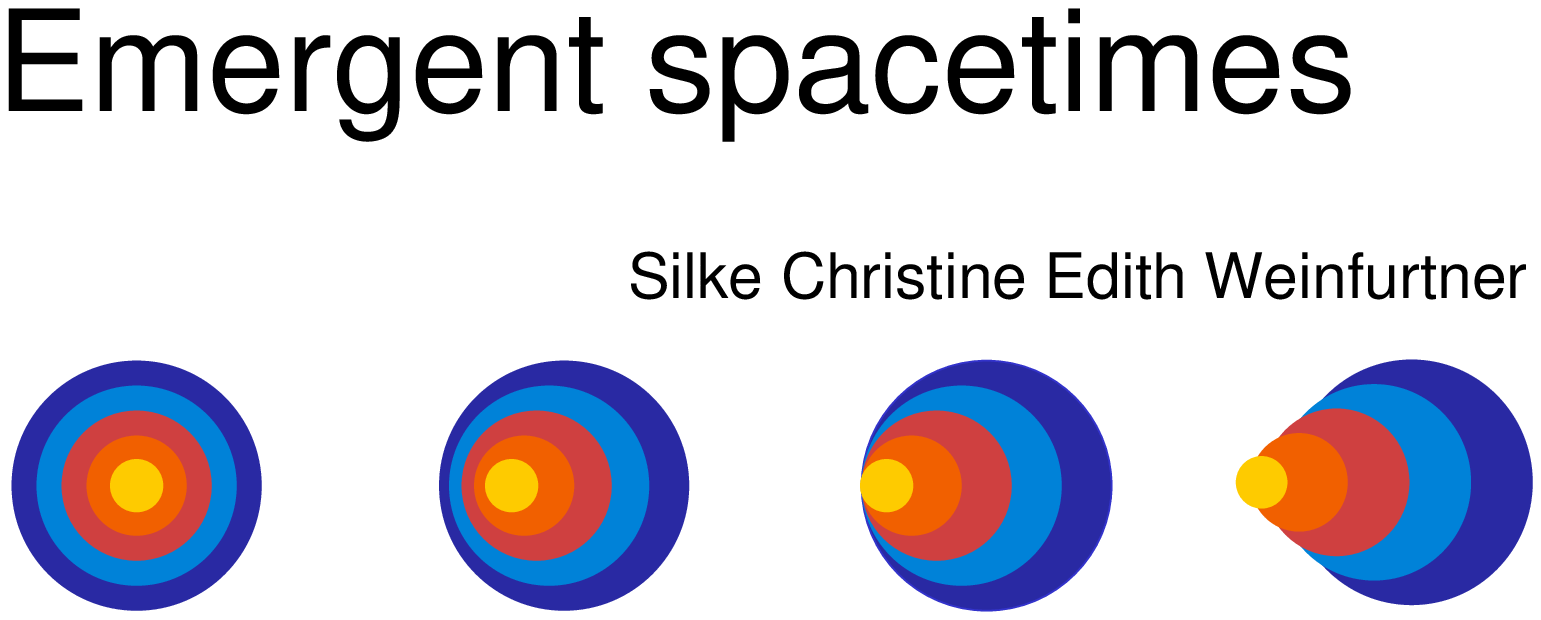}


{\small \noindent
Department of Mathematics, Statistics and Computer Science |
Te Kura Tatau\\[1cm]}

\vspace*{4.5cm}

\noindent
\includegraphics[width=2.0in]{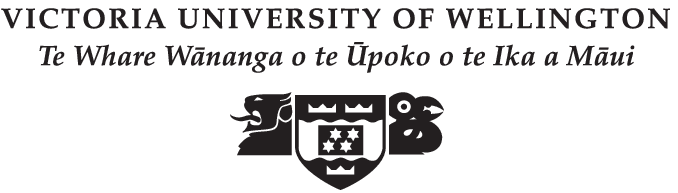}

\vspace*{1.5cm}

{\small \noindent
Thesis submitted for the degree of Doctor of Philosophy\\
at the Victoria University of Wellington.} 
%
%


\end{titlepage}

\newpage \thispagestyle{empty}
$\,$
\newpage \thispagestyle{empty}
$\,$
\vspace{5cm}
\begin{verse}
\begin{center}
$\;\;\;$ In Memory of \\
Johann Weinfurtner
\end{center}
\end{verse}
\newpage \thispagestyle{empty}
\frontmatter
\addcontentsline{toc}{chapter}{Abstract}
\renewcommand{\chaptermark}[1]{\markboth{ ABSTRACT}{}}
\silkychaptermark{ABSTRACT}
\chapter*{Abstract}

In this thesis we discuss the possibility that spacetime geometry may be an emergent phenomenon. This idea has been motivated by the Analogue Gravity programme. 
An ``effective gravitational field'' dominates the kinematics of small perturbations in an Analogue Model.
In these models there is no \emph{obvious} connection between the ``gravitational'' field tensor and the Einstein equations, as the emergent spacetime geometry arises as a consequence of linearising around some classical field.
After a brief survey of the most relevant literature on this topic, we present our contributions to the field. \\

First, we show that the spacetime geometry on the equatorial slice through a rotating Kerr black hole is formally equivalent to the geometry felt by phonons entrained in a rotating fluid vortex.
The most general acoustic geometry is compatible with the fluid dynamic equations in a collapsing/ expanding perfect-fluid line vortex. We demonstrate that there is a suitable choice of coordinates on the equatorial slice through a Kerr black hole that puts it into this vortex form; though it is not possible to put the entire Kerr spacetime into perfect-fluid ``acoustic'' form.  

We then discuss an analogue spacetime based on the propagation of excitations in a 2-component Bose--Einstein condensate. This analogue spacetime has a very rich and complex structure, which permits us to provide a mass-generating mechanism for the quasi-particle excitations. 
Additionally, we show that the analogue spacetime based on 2-component Bose--Einstein condensates provides models not just for standard general relativistic spacetimes, but also for the more general bi-metric, and even more general pseudo--Finsler spacetimes.

Furthermore, at short distances, where microscopic corrections due to the substructure (\ie, the fundamental Bosons) can no longer be neglected, and even in the mono-metric regime, one begins to see deviations from ``Lorentz invariance'' --- these deviations are qualitatively of the type encountered in ``quantum gravity phenomenology'', with the interesting property that the Lorentz violating physics is naturally suppressed by powers of the quasi-particle mass divided by the mass of the fundamental bosons that form the condensate.

A completely different issue can be probed in a single component BEC. This system naturally exhibits a microscopic mechanism allowing us to perform controlled signature change between Lorentzian and Riemannian geometries. We calculate the number of particles produced from a finite-duration Euclidean signature event, focussing on its impact on particle production in the ultraviolet regime, and the possibility of using the proposed signature change event as an amplifier for pre-existing fluctuations in condensed matter experiments.

Last but not least, we investigate cosmological particle production in a Bose--Einstein condensate with tunable microscopic interaction strength. Here Lorentz invariance emerges in the infrared limit, but is explicitly broken in the ultraviolet regime. Thus these  models are similar to many (but not all) models of quantum gravity, where a breakdown of Lorentz invariance is expected for ultraviolet physics around the Planck/ string scale. Motivated by previous studies on spacetimes emerging from a microscopic substrate, we show how these modifications naturally lead to momentum-dependent rainbow metrics.
In detail we investigate the robustness of the particle production process against the model-specific modifications, and also encounter cosmological particle production in ``rainbow inflation''. \\

We conclude with a brief discussion of lessons learned from these emergent spacetime models.
 
\newpage 
\renewcommand{\chaptermark}[1]{\markboth{ {#1}}{}}
\addcontentsline{toc}{chapter}{Contents}
\tableofcontents 
\newpage 
\addcontentsline{toc}{chapter}{List of Figures}
\listoffigures 
\newpage 
\addcontentsline{toc}{chapter}{List of Tables}
\listoftables
\newpage 
\silkychaptermark{ACKNOWLEDGMENTS}
\chapter{Acknowledgments}
Dear Matt, I cannot thank you enough for your tremendous support during the last three years. In endless conversations, discussions and explanations you have helped me to find my way through analogue models for gravity. With your infectious passion for science, you have motivated and encouraged me to pursue my scientific career. Rather than imposing your ideas, you helped me find my own way, and supported a (guided) scientific independence. You kept on pushing me to another level, the $(3+1)$-sentence ``Finish the paper, now!'', will always be indelibly branded in my mind.
Also on Martin's behalf, I thank you for your friendliness, many interesting conversations and numerous excursions around New Zealand. I am looking forward to hosting you in Vancouver, to continue our lunch time conversations and collaborations.  \\

Next I would like to thank Dr.~Stefano Liberati, with his infectious passion for the analogue model programme. Over the last three years you were a source of knowledge and inspiration, and this thesis would not be the same without you.
Someone whose opinion I regard highly once told me, that the transformation from an undergraduate student to a Postdoctoral fellow is similar to the metamorphosis that is involved turning a caterpillar into a butterfly. As a matter of fact, you never treated me like a graduate student, more as a collaborator and friend. I always felt very welcome when I stayed with you and your wife Paola. I would like to thank you both, for your support and friendship. \\

A couple of years ago Dr.~Susan Scott and Dr.~Craig Savage suggested a collaboration between their PhD student Angela White and myself, to work on issues related to analogue models in Bose--Einstein-condensates. I would like to thank Susan and Craig for their scientific and financial support. In both visits I felt warmly welcomed and enjoyed working at the Australian National University (ANU). Dear Angela, thank you for your hospitality and your friendship.  \\

I also would like to thank my New Zealand collaborators Prof.~Crispin Gardiner and his PhD student Piyush Jain. In many stimulating and fruitful conversations I had the opportunity to gain a deeper and different insight into condensed matter physics, and I would like to thank you both for that. Dear Piyush, thank you for your friendship which made the initial time in New Zealand so much easier. \\

I also would like to thank my colleagues Petarpa Boonserm and Celine Catto{\"{e}}n for sharing an office with me, and especially their support during the last few weeks. I wish you both good luck with your future careers, and I am looking forward to seeing you again. Dear Petarpa, whenever you feel a longing to revisit the Bavarian Forest, my mother and I will always welcome you as our special guest. \\

I shall not forget to thank the staff at the Department for Mathematics, Statistics and Computer Science for providing an ideal working environment. I really enjoyed studying for my PhD at Victoria University, and generally the time I spent in Aotearoa (the land of the long white cloud).

\newpage 
%

\mainmatter
\renewcommand{\chaptermark}[1]{\markboth{Chapter \textbf{\thechapter}.\ {#1}}{}}
\renewcommand{\sectionmark}[1]{\markright{\thesection\ \boldmath {#1}\unboldmath}}
\lhead[\fancyplain{}{\thepage}]         	{\fancyplain{}{\rightmark}}
\chead[\fancyplain{}{}]                 	{\fancyplain{}{}}
\rhead[\fancyplain{}{\leftmark}]       	{\fancyplain{}{\thepage}}

%
%
\chapter{Introduction \label{Chap:Introduction}}
%
This chapter is an introduction to \emph{Analogue Models of (and possibly for) Gravity}. We shall review important issues of general relativity  and quantum aspects of gravity, that are necessary to derive the notion of emergent spacetime geometry, without \emph{obvious} connection with the Einstein field equations. 
%
%
\section{Spacetime geometry \label{Sec:Spacetime.Geometry}}
%
In 1915 Albert Einstein presented to the world a theory of gravitation where gravity is not a force but a consequence of \emph{spacetime geometry}~\cite{Einstein:1915ca,Einstein:1915by,Einstein:1916vd}. This theory is based on the unification of $d$ dimensional space \emph{and} time, such that time is playing the role of an ``extra'' dimension in a $n=(d+1)$ dimensional \emph{spacetime}. The geometry of spacetime --- in $3$ spatial dimensions --- is characterized by the metric tensor
\begin{equation}
\label{Eq:Covariant.Metric.Tensor}
g_{ab} = 
\left[
\begin{array}{cccc}
g_{00}  & g_{01}  & g_{02} & g_{03}  \\
g_{01}  & g_{11}  & g_{12} & g_{13}  \\
g_{02}  & g_{12}  & g_{22} & g_{23}  \\
g_{03}  & g_{13}  & g_{23} & g_{33}  
\end{array}
\right] \, ,
\end{equation}
a symmetric covariant rank $2$ tensor. Without any further specification, \eg, the Einstein equations, coordinate conditions, or spacetime symmetries, the \emph{metric tensor}~\refb{Eq:Covariant.Metric.Tensor} has $n(n+1)/2$ independent --- here $10$ --- components. Each component may be a function of time and space $g_{ab}=g_{ab}(t,\mathbf r)$.

In general relativity \emph{free} particles are \emph{freely falling} particles; they are free from any external force but remain under the influence of the spacetime geometry. Under the assumption that the mass of the object is small enough not to interfere with the gravitational field $g_{ab}$, the \emph{kinematic} equations of motion are obtained by the principle of least action. The world line for a \emph{free test particle} connecting two points $A$ and $B$ in the $n$ dimensional spacetime extremizes the ``distance''
\begin{equation}
s_{AB} = \int_{A}^{B} \, ds = \int_{A}^{B} \left[  -g_{ab} \; dx^{a} \, dx^{b}  \right]^{1/2} \, ,
\end{equation}
between them. The world lines are called geodesics; we are dealing with a curved spacetime geometry, as opposed to Newtonian gravity where the three dimensional space is considered to be \emph{Euclidean} and flat. 
[$\blacktriangleright$ The \emph{signature} of flat or curved $n$ dimensional spacetime differs from that of a flat or curved $n$ dimensional space. We study this topic in greater detail in Chapter~\ref{Chap:Early.Quantum.Cosmology}.] \\

This geometrical interpretation of the gravitational field is only one aspect of Einstein's theory of general relativity.

\subsection{Einstein field equations \label{Sec:Einstein.Equation}}
The second aspect of general relativity is to identify --- in a coordinate covariant manner --- the density and flux of energy and momentum in the $n$ dimensional spacetime as the source of the gravitational field $g_{ab}$, in a manner similar to the way that mass is the source for the gravitational force in Newtonian gravity. 

The \emph{Einstein field equations},
\begin{equation}
\label{Eq:Einstein.Equations}
G_{ab} = 8 \pi G_{\mathrm{N}} \, T_{ab} \, ,
\end{equation}
relate (part of) the local curvature in the gravitational field $G_{ab}$ to the matter energy density $T_{ab}$. The current value --- provided by the Committee on Data for Science and Technology (CODATA) --- for the universal \emph{Newtonian constant of gravitation} is $G_{\mathrm{N}}=6.674 28(67) \times 10^{-11} \; m^{3} \, kg^{-1} \, s^{-2}$.

The covariant tensor $G_{ab}$, referred to as the  Einstein tensor, 
\begin{equation}
G_{ab} = R_{ab} - \half R \, g_{ab} \, ,
\end{equation}
depends on second derivatives of the gravitational field, the Ricci tensor $R_{ab}=R^{c}{}_{acb}$, and Ricci scalar $R=R^{bc}{}_{bc}$.~\footnote{In general the Riemann tensor
\begin{equation}
R^{a}{}_{bcd} = - \left( \partial_{d} \Gamma^{a}{}_{bc} - \partial_{b} \Gamma^{a}{}_{bd} + \Gamma^{a}{}_{md} \Gamma^{m}{}_{bc} - \Gamma^{a}{}_{mc} \Gamma^{m}{}_{bd} \right) \, ,
\end{equation}
is given in terms of first derivatives and quadratic combinations of \emph{Christoffel symbols},
$\Gamma^{a}{}_{bc} = \half \, g^{ad} \left( \partial_{b} g_{cd} + \partial_{c} g_{bd} - \partial_{d} g_{bc} \right)$.
Note, that $\Gamma^{a}{}_{bc}$ is \emph{not} a tensor. 
In Riemann normal coordinates, $\Gamma^{a}{}_{(bc,d)} \vert_{p} =0$ at the point $p$, the Ricci tensor simplifies to
$R_{ab} = \half \, g^{cd} \, \left( g_{cd,b,a} - g_{db,c,a} + g_{ab,c,d} - g_{ac,b,d}  \right) $.
}
On the right hand side of Eq.~\refb{Eq:Einstein.Equations} we have the stress-energy tensor, defined as
\begin{equation}
T^{ab} = 
\left[
\begin{array}{c|c} 
\mbox{energy density} & \mbox{energy fluxes}\\
\hline
\mbox{momentum densities} & \mbox{stress tensor} 
\end{array}\right] \, .
\end{equation}
The stress-energy tensor is a covariantly conserved entity, such that $\nabla_{b}T^{ab}=0$.~\footnote{As pointed out in the beginning of this section, the tensorial structure of the Einstein equations is maintained in any coordinatization. This has been achieved by the replacement of partial derivatives $\partial$ with \emph{covariant} derivatives $\nabla$. The covariant derivative $\nabla_{a}$ acting on an arbitrary tensor field $T^{b_{1} \cdots b_{k}}{}_{c_{1} \cdots c_{l}}$ is 
\begin{equation}
\nabla_{a}  T^{b_{1} \cdots b_{k}}{}_{c_{1} \cdots c_{l}} = 
\partial_{a} T^{b_{1} \cdots b_{k}}{}_{c_{1} \cdots c_{l}}
+ \sum_{\alpha=1}^{k} \Gamma^{b_{\alpha}}{}_{\sigma a} T^{b_{1} \cdots \sigma \cdots b_{k}}{}_{c_{1} \cdots c_{l}} 
+ \sum_{\beta=1}^{l} \Gamma^{\sigma}{}_{c_{\beta} a} T^{b_{1} \cdots b_{k}}{}_{c_{1} \cdots \sigma \cdots c_{l}} \, ,
\end{equation}
a combination of partial derivatives and Christoffel symbols.} \\

Automatically, the Einstein equations are such that they fulfill the contracted Bianchi identities
\begin{equation}
\nabla_{b} G^{ab} =0 \, .
\end{equation} 
The contracted Bianchi identities reflect the coordinate freedom and impose $n$ conditions on the $n$-dimensional gravitational field. (So in $3+1$ dimensions the metric effectively has $6$ free components.) \\

There are three different approaches to the Einstein field equations: (1) For any given metric tensor we can calculate the stress-energy density (and then check to see if it is ``physically reasonable''), (2) for any given stress-energy density we can calculate the corresponding gravitational field (possibly up to a few integration constants), and (3) the Einstein field equations can be viewed as \emph{dynamical} equations for the gravitational field, correlating changes in the stress-energy density (the cause of gravitation) with changes in the local spacetime curvature (the effect of gravitation).

[$\blacktriangleright$ In~\cite{Boonserm:2005qr} we study exact solutions for perfect fluid spheres. Based on known solutions we show how to generate new solutions without explicitly solving the Einstein equation. For the abstract of this article see Appendix~\ref{Sec:Perfect.Fluid.Spheres}. In the subsequent paper~\cite{Boonserm:2006ve} we connected the differences in the gravitational field between ``old'' and ``new'' solutions with changes in the stress-energy density; for abstract see Appendix~\ref{Sec:TOV.Equations}] \\

Thus at first glance it seems that the notion of spacetime geometry is inevitably associated with the Einstein field equations. However, in the following we present a model for an effective $(d+1)$ dimensional spacetime geometry, where the metric tensor $g_{ab}$ is \emph{not} obtained by solving the Einstein equations. In the literature this phenomenon is referred to as an \emph{analogue model for gravity} or an \emph{analogue  spacetime}.

\subsection{Analogue spacetime geometry \label{Sec:Analogue.Models.For.Gravity}}
The first modern paper on analogue spacetime geometry was published in 1981 by \emph{Bill Unruh}~\cite{Unruh:1981bi}, followed by \emph{Matt Visser}~\cite{Visser:1993tk} in 1993. The authors demonstrated that the effective propagation of \emph{sound waves} in a fluid flow is given in terms of geodesics in an \emph{acoustic} ``spacetime geometry''. Since then an increasing number of articles have been published on this subject, now commonly referred to as ``Analogue Gravity''~\cite{Barcelo:2005ln}.
In the following sections we briefly summarize aspects of analogue gravity relevant for this thesis, and refer the reader interested in more details to the review article ``Analogue Gravity'', freely available online through \emph{Living Reviews in Relativity}~\cite{Barcelo:2005ln}.

\subsubsection{A simple example: Sound waves in a fluid flow \label{Sec:Sound.Waves.Convergent.Fluid.Flow}}
The \emph{kinematic} equations for small --- classical or quantum --- perturbations (\ie, sound waves) in a barotropic, inviscid and irrotational fluid are given by
\begin{equation}
\label{Eq:Intro.Klein.Gordon.Equation}
\frac{1}{\sqrt{- g }} \, \partial_{a} \left( \sqrt{- g } g^{ab} \partial_{b} \phi \right)  = 0 \, , 
\end{equation} 
which represents a massless minimally coupled scalar field in an \emph{effective} curved spacetime. The sound waves experience an acoustic metric,
\begin{equation} 
\label{Eq:Intro.Geometric.Acoustics}
g_{ab} \propto \left[ \begin{array}{cc} -(c^{2}-v^{2}) & -v_{j} \\ -v_{i} & \delta_{ij} \end{array} \right] \; ,
\end{equation}
whose entries include the purely collective variables $c=c(t,\mathbf{x})$, the sound speed, and $\mathbf{v}=\mathbf{v}(t,\mathbf{x})$, the background velocity. Here the indices run from $1$ to $d$, where $d$ is the spatial dimension of the fluid. For a detailed derivation of this result see Refs.~\cite{Unruh:1981bi,Visser:1993tk,Matt.-Visser:2002ot,Barcelo:2005ln}. 

Therefore it is possible to read off the metric components from the ``covariant'' classical $\phi$ or quantum $\hat \phi$ field equation~\refb{Eq:Intro.Klein.Gordon.Equation}. In a broader sense we are dealing effectively with a semi-classical model, where the matter fields are quantized $\hat \phi$, but the entries in the \emph{acoustic} metric remain classical. We will elaborate on this in Section~\ref{Sec:Intro.SemiClassical.Quantum.Gravity}.\\

\paragraph{Vorticity.}
The macroscopic fluid parameters can easily be chosen to design emergent non-rotating black holes~\cite{Unruh:1981bi,Jacobson:1991sd,Visser:1993tk}. [$\blacktriangleright$ In Chapter~\ref{Chap:Acoustic.Black.Holes} we analyzed the possibility of developing acoustic rotating Kerr black hole geometries, where we analyze the most general vortex flow in fluids.]
Since many ``real'' fluids exhibit vortices, it is --- \emph{not only} from a gravity point of view, see Chapter~\ref{Chap:Acoustic.Black.Holes} --- desirable to extend the concept of acoustic metrics to rotational fluids. First attempts to derive the acoustic metric in a (barotropic and inviscid) fluid with vorticity have been carried out in~\cite{Perez-Bergliaffa:2001lq}.

\paragraph{Substantive derivative.}
The \emph{substantive} derivative (also referred to as \emph{fluid following}, or \emph{material}, or \emph{advective} derivative) in a fluid is defined as
\begin{equation}
\frac{D \, \star}{Dt} = \frac{\partial \, \star}{\partial t} + \vec{v} \cdot \nabla \, \star \, . 
\end{equation}
[$\blacktriangleright$ The material derivative plays an important role in quantum field theory in effective curved spacetimes, as we show explicitly in~\ref{subsec:A-spacetime} and in~\ref{Sec:Commutation.Relations}, where we relate the material derivative to the conjugate momentum on the emergent spacetime geometry; see also \cite{Unruh:2003ln}.] 

\paragraph{Geometrical and physical acoustics.}
In Eq.~\refb{Eq:Intro.Geometric.Acoustics} we give an expression for the acoustic metric up to an overall conformal factor. This representation is referred to as the \emph{geometrical acoustics approximation}, which is good enough to describe how sound rays propagate in the fluid. Since the overall conformal factor can in general be a function of space and time, and also explicitly depending on the spatial dimensions, we require --- for the full kinematic equations --- the ``complete'' metric, this is referred to as the \emph{physical acoustics approximation}. [$\blacktriangleright$ In Chapter~\ref{Chap:Acoustic.Black.Holes} we discuss geometrical and physical acoustics in detail, and outline some of the consequences of the overall conformal factor.]

\paragraph{Zero background velocity}
For $\vec{v}=\vec{0}$ we obtain
\begin{equation} 
\label{Eq:Intro.Geometric.Acoustics.Minkowski}
g_{ab} \to \eta_{ab} \propto \left[ \begin{array}{cc} -c^{2} & 0 \\ 0& \delta_{ij} \end{array} \right] \; ,
\end{equation}
an acoustic Lorentzian flat spacetime (or acoustic Minkowski) geometry, as long as $c^{2}>0$. In general we notice that the acoustic metric for $2$ and $3$ spatial dimensions describes a Lorentzian (pseudo-Riemannian) geometry. [$\blacktriangleright$ In Chapter~\ref{Chap:Early.Quantum.Cosmology} we present a specific analogue model, that exhibits a natural mechanism to change between Lorentzian $(-,+++)$, and Riemannian $(+,+++)$ \emph{effective} spacetime geometries.] \\

\subsubsection{A catalogue of models \label{Sec:Intro.A.Catalogue.Of.Models}}
Over the last 25 years the basic concept of analogue models has been transferred to many different media. There is a broad class of systems that possess an emergent spacetime in the low-energy regime. We would like to list some of the different models that have been studied in the past. More background information and current developments can be found in~\cite{Matt.-Visser:2002ot,Barcelo:2005ln}. 

\begin{description}
\item[Sound in moving fluids]{The fluid can be any barotropic, inviscid and irrotational fluid. As mentioned above, the first (modern) papers on this subject were written by \emph{Bill Unruh}~\cite{Unruh:1976aa,Unruh:1981bi,Unruh:1995aa} and \emph{Matt Visser}~\cite{Visser:1993tk,Visser:1998ra,Visser:1998gn,Visser:1998gf}. (For more details see above, or in Refs.~~\cite{Matt.-Visser:2002ot,Barcelo:2005ln}. We will revisit these models below when we discuss the quantum aspects of analogue spacetime geometry.)}
\item[Shallow water waves (gravity waves)]{The natural occurrence of an acoustic metric governing the propagation of ripples in a basin filled with liquid has been pointed out by \emph{Ralf Sch{\"{u}}tzhold} and \emph{Bill Unruh} in 2002~\cite{Schutzhold:2002di}. The acoustic line element is given in terms of the background velocity parallel to the basin $\vec{v}_{B}^{II}$, as
\begin{equation}
ds^{2} = \frac{1}{c^{2}} \, \left[ -(c^{2}-(v_{B}^{II})^{2}) \, dt^{2} - 2 \, \vec{v}_{B}^{II} \cdot d\vec{x} \, dt + d\vec{x}^{T} \cdot d \vec{x} \right] \, .
\end{equation}
Noticeable in this model is the extreme ease with which the velocity of the surface waves,
\begin{equation}
c^{2} = g \, h_{B} \, ,
\end{equation}
can be adjusted by changing the depth $h_{B}$ of the basin. (The variations in the gravity waves at the surface of the fluid in the basin are caused by the presence of Earth's gravity.)}
\item[Classical refractive index]{The Maxwell equations inside a dielectric material in flat $4$ dimensional Minkowski spacetime can be written in a very condensed way,
\begin{equation}
\partial_{b} \, \left( Z^{abcd} \, F_{cd} \right) = 0.
\end{equation}
As usual $F_{cd}=- 2 \, A_{[c,d]}$ is the electromagnetic tensor, and $Z^{abcd}$ is a rank $4$ tensor whose components are functions of the components of the $3\times 3$ permittivity tensor $\mathbf{\epsilon}$, and the $3\times 3$ permeability tensor $\mathbf{\mu}$ of the dielectric. 

The dielectric can be adjusted such that the equations for the electric field (in flat spacetime inside the dielectric) are similar to the equation for an electric field in an \emph{effective} curved spacetime inside vacua. [$\blacktriangleright$ In~\cite{Boonserm:2005wu} we have shown the reverse analogy, how weak-field gravity can be interpreted in terms of an \emph{effective} refractive index. For abstract see~\ref{Sec:Effective.Refactive.Index}.] }
\item[Electromagnetic wave-guide]{In 2006 \emph{Ralf Sch{\"{u}}tzhold} and \emph{Bill Unruh}~\cite{Schutzhold:2005aa} showed the possibility of manipulating electromagnetic waves in a wave-guide so that they experience an effective curved spacetime (in the form of a $(2+1)$ dimensional Painl\'eve--Gullstrand--Lema\^itre geometry).

The wave-guide consists of a ladder circuit with a time-dependent capacitance $C_{n}(t)$ at each rung of the ladder, while the coil in each loop has inductance $L_{n}=L=\mbox{constant}$. The rungs are spatially separated by equal distances $\Delta x$. The current in each circuit is given by $I_{n}$, and we can define an effective potential $A_{n}$, such that $I_{n}=A_{n+1}-A_{n}$. For wave-length with $\lambda \gg \Delta x$ the discreteness of the $x$-axis is negligibly small, $A_{n} \to A(x)$, and in the continuum limit the wave equation is given by
\begin{equation}
\left[ \left( \partial_{t} + v \, \partial_{x} \right) \, 1/c^{2} \, \left( \partial_{t} + v \, \partial_{x} \right) - \partial_{x}^{2} \right] A(x) = 0 \, .
\end{equation}
Here the spacetime dependent propagation velocity can be arranged to be
\begin{equation}
c_{n}(t) = \Delta x / \sqrt{L \, C_{n}(t)} \to c(x+v \, t) \, ,
\end{equation}
a function of time and space, and the constant velocity $v$. (For smaller wavelengths the sound speed will be momentum-dependent, in a manner similar to Lattice field theory. We will turn back to this issue shortly, when we discuss the quantum aspects of analogue spacetime geometries, and also in the section about quantum gravity phenomenology.)}
%
%
\item[Condensed matter systems]{
To the best of our knowledge the first to point out the analogy between curved spacetimes and condensed matter systems such as superfluid Helium, was \emph{Grisha Volovik} in 1995~\cite{Volovik:1995za}. This was followed (2000/01) by \emph{Luis Garay et al.}~\cite{Garay:2000ez,Garay:2001in}, and \emph{Carlos Barcel\'o} and \emph{Matt Visser}~\cite{Barcelo:2001gt} focussing on analogue models arising from Bose--Einstein condensates. Both systems are prime models for superfluids (zero viscosity), and yield similar kinematics to sound waves in a flowing fluid --- as explained above.

In Chapers~\ref{Chap:Geometries.Masses}-\ref{Chap:Late.Quantum.Cosmology} we have focused on the latter, on spacetimes emerging from Bose gases. A derivation of the effective spacetime geometry starting from the many-body Hamiltonian for an ultra-cold weakly interacting gas of Bosons can be found in those chapters. Before we jump to the main part of this thesis, we would like to briefly summarize the (for this work) relevant applications and properties of analogue spacetime geometries. (While many of the presented topics are valid far beyond the specific model we discuss, the main focus of this thesis is on \emph{emergent spacetime geometries} from Bose gases.)}
\end{description}

We would like to comment on the \emph{incompleteness} of the list we have just presented. To do justice to all existing models would go far beyond the scope of this thesis, to give a brief introduction to the field. We refer the interested reader to~\cite{Matt.-Visser:2002ot,Barcelo:2005ln} for more detail, for example to see a study on analogue models in quantum dielectrics~\cite{Leonhardt:2000aa,Unruh:2003ln}.

%
\section{Quantum aspects of analogue gravity \label{Sec:Intro.Quanatum.Aspects.Gravity}}
%
After more than 80 years of devoted research, the quest to unify quantum mechanics and gravity (general relativity) continues. While remarkable progress has been achieved in the unification of three out of the four forces of nature --- the Weak force (Electroweak theory), the Electromagnetic force (Quantum Electrodynamics), and the Strong force (Quantum Chromodynamcis) --- the Gravitational force (General Relativity) remains excluded. All direct attempts to quantize general relativity have been in vain, for example see~\cite{Isham:1975aa,Isham:1979aa}. (However, limited progress has been made in other quantum gravity candidates, \eg, string models.)

(From an emergent spacetime point of view any attempt to quantize the ``emergent gravitational field'' cannot be successful in reproducing the overall quantum theory of the analogue model. A quantization of the macroscopic fluid variables (\eg, the Bose--Einstein condensate) will \emph{not} reveal the microscopic degrees of freedom of the underlying substructure (\eg, the ultra-cold weakly interacting gas of Bosons); for example see~\cite{Barcelo:2001lu,Hu:2005wu,Volovik:2003jn}.)

From a general standpoint the question has to be asked if there is anything at all we can say about the occurrence of quantum phenomena in gravitational fields. Especially important for the analogue model programme is the demand for quantum effects where the (fully interacting) dynamical behavior in the gravitational field can be approximated by kinematical equations (no back-reaction). Only for those cases we can guarantee the \emph{obvious} validity of the analogy, and employ the analogue for mimicking quantum field theory phenomena in \emph{effective} curved spacetimes.

\subsection{Semi-classical quantum gravity \label{Sec:Intro.SemiClassical.Quantum.Gravity}}
One particular approximation that circumvents the problem of quantizing the gravitational field itself is that called \emph{semi-classical} quantum gravity. Here the gravitational field is treated as being a classical field, while  the matter fields are treated as being quantum fields.

The stress-enery tensor of a matter field is given in terms of
\begin{equation}
T_{ab}(\mathbf x) = \frac{2}{\sqrt{-g}} \, \frac{\delta S^{\mathrm{matter}}[g_{ab},\phi] }{\delta g^{ab}(x)} \, ,
\end{equation}
the functional derivative of the matter action $S^{\mathrm{matter}}$ with respect to metric tensor. Therefore the quantization of the right hand side of the Einstein field equations~\refb{Eq:Einstein.Equations} is relatively straightforward. It can be shown that the coupling of the matter fields to the gravitational field yields
\begin{equation}
G_{ab} = 8 \pi G_{\mathrm{N}} \, \langle \hat T_{ab} \rangle \, .
\end{equation}
These are \emph{semi-classical Einstein equations}, where vacuum fluctuations in the matter fields contribute to the gravitational field via the expectation value for the quantized stress-energy tensor. (A
derivation of the semi-classical Einstein equations can be found \emph{inter alia} in~\cite{Birrell:1984aa,Fulling:1989aa,Mukhanov:2007aa}.) \\

Below we will focus on cases where back-reaction effects are negligibly small, and quantum field theory in curved spacetime is a sufficiently good approximation. In Chapters~\ref{Chap:Early.Quantum.Cosmology} and \ref{Chap:Late.Quantum.Cosmology} we study quantum effects in a covariant \emph{free} field, where no \emph{external} forces act on the field, and the ``gravitational force'' is encoded in the spacetime geometry, $\eta_{ab} \to g_{ab}$. Within this approximation (semi-classical quantum gravity for small quantum effects), which only involves the kinematics of the gravitational field, analogue models for gravity can be employed for mimicking quantum aspects of ``gravity''.

\subsubsection{Semi-classical quantum geometry \label{Sec:Intro.Semi.Classical.Quantum.Geometry}}
While it is relatively straightforward to derive semi-classical quantum gravity, the situation is much more difficult in the analogue gravity programme. Here the fundamental equations for the effective spacetime geometry are obtained from various media.

However, in~\cite{Barcelo:2001ay,Barcelo:2005ln} the authors showed that the phenomenon of quantum field theory in \emph{effective} curved spacetimes can be generalized in the following way. 
Small perturbations around some background solution $\phi_{0} (t,\mathbf x) $, 
\begin{equation}
\label{Eq:Intro.Linear.Field}
\phi(t,\mathbf x) = \phi_{0} (t,\mathbf x) + \epsilon \, \phi_{1} (t, \mathbf x) + \frac{\epsilon^{2}}{2} \, \phi_{2}(t, \mathbf x) \, ,
\end{equation}
in a generic Lagrangian $\mathcal{L}(\partial_{a} \phi, \phi)$, depending only on a single scalar field and its first derivatives yields an effective spacetime geometry,
\begin{equation}
g_{ab}(\phi_{0}) = \left[ - \det \left( \frac{\partial^{2}  \mathcal{L}}{\partial(\partial_{a} \phi) \, \partial (\partial_{b}  \phi) } \right)  \right]^{\frac{1}{d-1}} \Bigg\vert_{\phi_{0}} \;  \;
\left( \frac{\partial^{2}  \mathcal{L}}{\partial(\partial_{a} \phi) \, \partial (\partial_{b}  \phi) } \right)^{-1} \Bigg\vert_{\phi_{0}} \, ,
\end{equation}
for the classical\,/\,quantum fluctuations. 
The equation of motion for small perturbations around the background $\phi_{0}$
are then given by
\begin{equation}
\left( \Delta _{g(\phi_{0})} - V(\phi_{0}) \right) \phi_{1} = 0 \, ,
\end{equation}
where $\Delta_{g(\phi_{0})}$ is the d'Alembertian operator associated with the metric $g(\phi_{0})$ and $V(\phi_{0})$ is potential depending on the background.
That is, we have derived an \emph{effective} curved-spacetime quantum field theory description of the linearization process, under the premise that back-reaction effects are negligibly small. [$\blacktriangleright$ In Chapter~\ref{Chap:Geometries.Masses} we show that multi-scalar fields yield interesting extensions of this framework for the analogue gravity programme. There we show that a $2$-component Bose--Einstein condensate can be used for mimicking massive scalar fields in mono- and bi-metric Lorentzian, or even Finslerian spacetimes. A more general study on this topic has been carried out in~\cite{Barcelo:2002dp,Visser:2002vp} by \emph{Carlos Barcel\'o}, \emph{Matt Visser} and \emph{Stefano Liberati}.] \\

Before we address specific quantum aspects of gravity, we wish to add a few comments regarding the quantum aspects of emergent spacetimes arising from condensed matter systems. The entries in the emergent spacetime metric are collective variables associated with the expectation value of the macroscopic mean-field description of the superfluid. In momentum space the condensate itself represents the (large) occupation number of the lowest energy state in the system. In some sense the Bose--Einstein condensate itself is quantum fluctuation where $k=0$. Any change in the occupation number depletes the condensate, and therefore yields changes in the emergent gravitational field. So strictly speaking we are dealing with a \emph{semi-classical quantum geometry}. \\

First attempts to include the back-reaction effects (here due to the Hawking effect, see below) of quantum fluctuations into the analogue gravity programme have been made 2004 by \emph{Roberto Balbinot et al.}, where the authors assumed the validity of the Einstein equations for the acoustic spacetime geometry, see~\cite{Balbinot:2005aa,Balbinot:2005ab,Balbinot:2004aa}. (While there is no \emph{obvious} connection between the analogue models and the Einstein equations, there are some tentative indications in favor of emergent Einstein field equations, see Section~\ref{Sec:Inspirations.Quantum.Gravity}.)

A completely different point of view on this topic has been presented 2005 by \emph{Ralf Sch{\"{u}}tzhold et al.}~\cite{Schutzhold:2005wt}, where the back-reaction effects of quantum fluctuations (here Hawking radiation, see below) have been derived at the level of the fundamental Hamiltonian for the Bose gas. In this approach the quantum fluctuations are not necessarily expected to be small; so that significant corrections to the linearization process can be obtained.

[$\blacktriangleright$ In~\cite{Jain:2006ki} we have taken an alternative route. We have employed \emph{Classical Phase Space Methods}~\cite{Gardiner:2000aa} to numerically simulate quantum effects (here Cosmological Particle Production, see below) in a  ``realistic'' Bose--Einstein condensate, where (besides other deviations) back-reaction effects are automatically included. From our simulation it can be seen that indeed (as expected) back-reaction effects are negligibly small in the semi-classical quantum geometry picture.] \\

After a brief discussion of the validity of both theories, ``usual'' quantum field theory and \emph{emergent} quantum field theory, we present three different cases where non-trivial gravitational\,/\,effective gravitational effects in quantum fields are expected.

\subsubsection{Quantum aspects of black holes \label{Sec:Intro.Qunatum.Black.Holes}}
We expect serious disturbances for quantum fields in the vicinity of a black hole, for field modes whose wavelengths are comparable or larger than the black hole radius. The first semi-classical treatment of quantum field theory in a black hole geometry has been carried out by \emph{Stephen Hawking} in 1974~\cite{Hawking:1974rv,Carr:1974nx}. Hawking showed that a Schwarzschild black hole emits a thermal radiation at a temperature $T$ proportional to the surface gravity $\kappa$ at the horizon,\footnote{Here we use natural units, where $G=c=\hbar=1$.}
\begin{equation}
T = \frac{\kappa}{2 \pi } \, .
\end{equation}
Unfortunately, the surface temperature for example of a solar mass black hole, $\sim 10^{-7}$ Kelvin,\footnote{
The temperature of a black hole is inverse proportional to its mass $m$,
\begin{equation}
T=\frac{\hbar \, c^{3}}{8\pi \, G_{\mathrm{N}} \, k_{\mathrm B} \, m} \, .
\end{equation}
Here $k_{B}$ is the Boltzmann constant, and $\hbar$ the Planck constant.} when compared to the $2.7$ Kelvin of the surrounding cosmic background radiation is far too small for experimental verification. (The only hope is the existence of  ``primordial'' black holes, relics of the early universe, that are expected to have masses of order $10^{15}$ gm.) \\

On this account analogue black hole geometries are of great interest, since they might be the only chance for experimentally measuring \emph{effective} Hawking radiation. Indeed, the first modern analogue model, by \emph{Bill Unruh}, explicitly discussed the possibility of ``Experimental black hole evaporation''~\cite{Unruh:1981bi}.

\begin{figure}[t]
 \begin{center}
 \input{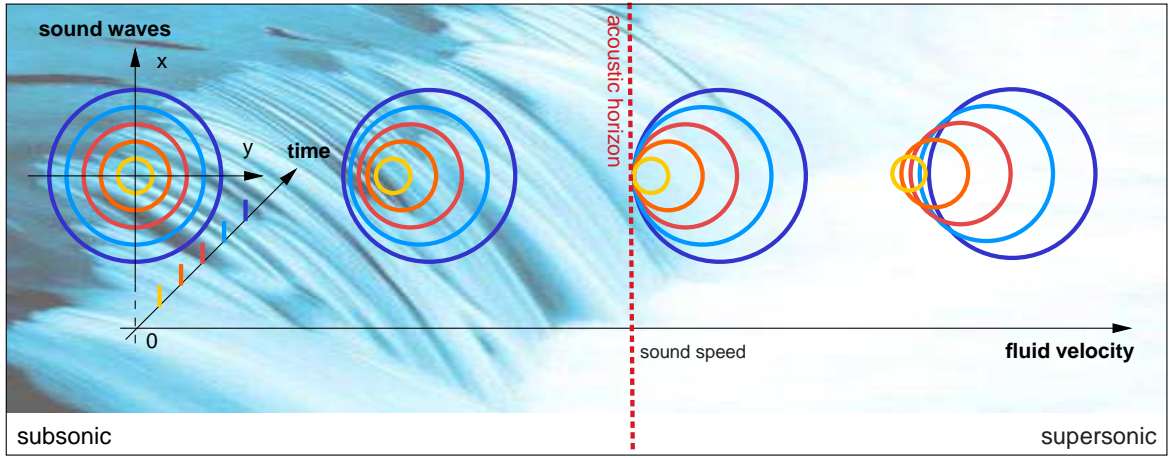}
 \caption[Sound waves in acoustic black hole spacetime.]  {\label{Fig.Acoustic.Black.Hole}
 We illustrate the propagation of sound waves in convergent fluid flow.}
 \end{center}
\end{figure}

The acoustic metric for sound waves in a fluid, as given in Eq.~\refb{Eq:Intro.Geometric.Acoustics}, shows that we expect an \emph{apparent horizon} --- a locally definable horizon --- at $c(t,\mathbf x)=\vert \vec v(t,\mathbf x) \vert$, where $g_{00}$ vanishes. In the region denoted by $c(t,\mathbf x)>\vert \vec v(t,\mathbf x) \vert$ the sound waves are propagating faster than the background velocity, and the fluid is said to be \emph{subsonic}. On the other hand, for $c(t,\mathbf x)<\vert \vec v(t,\mathbf x) \vert$ the sound waves propagate with less speed than the background flow. Here the fluid flow dominates and the flow is said to be \emph{supersonic}. Thus  sound waves in the supersonic region are not free to propagate in any direction, they get dragged along with the fluid flow. In Figure~\ref{Fig.Acoustic.Black.Hole} we illustrated sound waves at different parts of a lake that flows into a river via a waterfall. Let us further assume that the waterfall splits the stream into a subsonic and a supersonic part. This is in analogy to a black hole, where the supersonic region corresponds to the region bounded by the \emph{apparent} horizon. Here sound rays behave like light rays inside black holes, they are trapped. Sound waves in our gedanken-experiment can only cross the horizon from one direction, from subsonic to supersonic.\\
%
%

As pointed out above, Hawking radiation is a semi-classical quantum gravity effect, that does not directly involve the validity of the Einstein equations. This is in contrast to the notion of \emph{Bekenstein entropy}, a geometrical entropy formulated in 1973~\cite{Bekenstein:1972tm,Bekenstein:1973ur}. The derivation for Bekenstein entropy \emph{directly} involves the Einstein equations. In the context of the analogue model programme the issue of the ``Essential and inessential features of Hawking radiation'' has been summarized by \emph{Matt Visser}~\cite{Visser:2003gc}.

Hawking radiation requires --- besides the existence of a Lorentzian metric --- (1) an apparent horizon (at least asymptotically), (2) non-zero ``surface gravity'',
\begin{equation}
\label{Eq:Hawking.Temperature}
\kappa(t)  \Big\vert_{\mathrm{Horizon}}  = c_{\mathrm{sound}} \, \frac{d \left[ c(t, \mathbf x) - \vert \vec v(t, \mathbf x) \vert \right]}{d\mathbf x} \bigg\vert_{\mathrm{Horizon}} \, ,
\end{equation}
and (3) a relatively slow evolution of the geometry,
\begin{equation}
 \frac{d \left[ c(t, \mathbf x) - \vert \vec v(t, \mathbf x) \vert \right]}{d\mathbf x} \gg \frac{\dot c_{\mathrm{sound}}}{c_{\mathrm{sound}}}  \bigg\vert_{\mathrm{Horizon}} \, .
\end{equation}
The latter is important to guarantee the dominance of spatial gradients over temporal gradients. The slower the geometry evolves, the closer the \emph{apparent} horizon sits to the \emph{event} horizon (absolute horizon), and the closer the actual spectrum is to a quasi-thermal spectrum~\cite{Visser:2003gc}. (More recently, in~\cite{Barcelo:2006vs,Barcelo:2003ke} the authors argued that the presence of a pre-existing apparent horizon can be further relaxed.)

\paragraph{Acoustic Hawking radiation from analogue models.}
Quantum aspects of analogue black holes have been studied in various media: For example in fluids~\cite{Unruh:1994zw,Unruh:1981bi,Schutzhold:2002di}, slow light~\cite{Unruh:2003ln,Leonhardt:2000aa}, and in electro-magnetic wave-guides~\cite{Unruh:2003ln}, in Bose--Einstein condensates~\cite{Garay:2000ez,Garay:2001in,Barcelo:2001ay,Barcelo:2006vs,Barcelo:2003ke}, in superfluid Helium~\cite{Jacobson:2002ab,Jacobson:1998ni,Volovik:1995za}.

\paragraph{Robustness of Hawking radiation.}
Besides all the reasons mentioned above, there is another property of the Hawking process that makes the analogue model programme both useful and interesting. That is the fact that the Hawking effect is largely independent of the choice for the high-frequency cutoff.  This is important for any analogue model --- as it has per definition nothing to do with ``real'' gravity --- its ``true'' nature reveals itself in terms of a high-frequency cutoff. The cutoff is related to the microscopic physics of the media. [$\blacktriangleright$ We show this explicitly in Chapter~\ref{Chap:QGP}, where we study Lorentz symmetry breaking in spacetimes emerging from a Bose gas.] \\

The first author to study the \emph{robustness} of Hawking radiation against the introduction of a high-frequency cutoff was \emph{Ted Jacobson} in 1991~ \cite{Jacobson:1991sd,Jacobson:1993ab}. Subsequently this topic has been further investigated in~\cite{Unruh:1994zw,Unruh:2005aa,Brout:1995aa}. [$\blacktriangleright$ In Chapter~\ref{Chap:Late.Quantum.Cosmology} we study the robustness of cosmological particle production for inflation in a Bose gas.] 

\paragraph{Rotating black holes.}
The quantum aspects of rotating black holes are interesting in two regards. Firstly, we also expect Hawking radiation related to the outer event horizon, and secondly the \emph{ergoregion} (see Chapter~\ref{Sec:Ergosphere}) yields  \emph{superradiant scattering}~\cite{Penrose:1969pc,Williams:1995aa,Winstanley:2001aa,Glampedakis:2001aa,Williams:2004aa,Lepe:2005aa,Gariel:2007aa,Jiang:2007aa}. [$\blacktriangleright$ In Chapter~\ref{Chap:Acoustic.Black.Holes} we study the possibility of setting up analogue models for Kerr black holes.] \\

The superradiance from analogue black holes has been studied in~\cite{Basak:2003aa,Basak:2003ab,Basak:2005aa,Slatyer:2005aa}. However, we are still lacking a complete derivation of the most general vortex geometry\,/\,collapsing vortex geometry, as it involves fluids with vorticity~\cite{Perez-Bergliaffa:2001lq}.

\subsubsection{Quantum aspects of cosmology \label{Sec:Intro.Quantum.Cosmology}}
Similarly, we also expect a time-dependent gravitational field to have a non-trivial effect on quantum field modes, if the time-scale $t_{s}$ over which the gravitational field varies is faster than the oscillation frequency of the mode, $t_{s}^{-1}\gtrsim \omega_{k}$. Pioneer work on the theory of cosmological particle production in time-dependent universes has been carried out in 1969 by \emph{Leonard Parker}~\cite{Parker:1968aa,Parker:1969aa,Parker:1969aa,Parker:1971aa}. \\

An expanding\,/\,collapsing universe is described by the Friedmann--Robertson--Walker line-element, representing a time-dependent conformally flat metric tensor. [$\blacktriangleright$ In Chapter~\ref{Chap:Late.Quantum.Cosmology} we study a special sub-case that is of particular interest for cosmology, the \emph{de Sitter} universe, where the scale factor is given by an exponentially expanding (or contracting) universe. In the relevant chapter we also explain why a rapidly exponentially changing universe is believed to account for the large-scale structure of our universe, as seen in our microwave background~\cite{Guth:1981aa,Linde:1990aa,Albrecht:1982aa,Hawking:1982aa,PhysRevD.22.1882,Turner:1993aa,Brandenberger:1983aa,Rothman:1993aa,Dodelson:2003aa}.] \\

The de Sitter universe contains so-called \emph{cosmological horizons}~\cite{Rothman:1993aa}, such that every point in a de Sitter geometry is surrounded by a boundary that can \emph{never} be crossed. In 1977~\cite{Gibbons:1977aa} \emph{Gibbons} and \emph{Hawking} were able to show that the thermal spectrum --- as expected from cosmological particle production in an infinitely-long lasting expanding de Sitter universe --- is equivalent to the temperature calculated from the surface gravity at the cosmological horizon, compare with Eq.~\refb{Eq:Hawking.Temperature}. An extensive treatment of both quantum effects, the Hawking radiation and cosmological particle production, can be found in~\cite{Birrell:1984aa,Fulling:1989aa,Mukhanov:2007aa}. [$\blacktriangleright$ In Chapter~\ref{Chap:Late.Quantum.Cosmology} we discuss the existence of cosmological horizons in emergent spacetimes.] \\

In Chapter~\ref{Sec:Acoustic.Spherically.Geometries} we show that is is always possible to cast the acoustic metric, see Eq.~\refb{Eq:Intro.Geometric.Acoustics} into Friedman--Robertson--Walker (FRW) form, and consequently reproduce cosmological particle production for any time-dependent FRW geometry.

\paragraph{Analogue FRW spacetimes and cosmological particle production.}
While the Hawking effect could in principle occur in many different media --- \eg, ordinary fluids, electromagnetic wave-guide and superfluids --- the only \emph{available} analogues for FRW geometries are superfluids. 

The first to point out the analogy in superfluid Helium was \emph{Grisha Volovik}~\cite{Volovik:1996ka,Volovik:2003jn} in 1996. A few years later the analogy was transfered to sound waves in \emph{freely expanding} Bose--Einstein condensates, by \emph{Luis Garay}~\cite{Garay:aa} and \emph{Uwe Fischer} and his student in 2003~\cite{Fischer:2003ns,Fedichev:2006mc,Fischer:2004hr,Fischer:2004iy,Fedichev:2004fi,Fedichev:2004on}. An alternative route of achieving FRW geometry in a Bose--Einstein condensate was suggested by \emph{Carlos Barcel\'o}, \emph{Stefano Liberati} and \emph{Matt Visser} shortly afterwards in~\cite{Barcelo:2003ia,Barcelo:2003yk,Barcelo:2005ln}. Here the authors showed that in a Bose--Einstein condensate with time-dependent atomic interaction strength the sound waves propagate with a time-dependent speed of sound, and consequently experience an \emph{effective} FRW universe. [$\blacktriangleright$ Our analysis in Chapter~\ref{Chap:Late.Quantum.Cosmology} is based on the latter approach.] In \cite{Uhlmann:2005rx} \emph{Michael Uhlmann}, \emph{Yan Xu}, and \emph{Ralf Sch{\"{u}}tzhold} focused on the approach suggested by \emph{Uwe Fischer}, and calculated the density-density fluctuations during \emph{effective} inflation based on a freely expanding condensate.

\paragraph{Trans-Planckian physics and cosmological particle production.}
In~\cite{Niemeyer:2001aa,Martin:2003rp,Mersini-Houghton:2001aa,Bastero-Gil:2003aa} the authors analyzed the dependence of the particle spectra on \emph{ad-hoc} artificially imposed \emph{Lorentz symmetry breaking} modifications to the dispersion relation. This question is particularly interesting for the analogue model programme, where the Lorentz symmetry only emerges for the low-energy perturbations in the system. [$\blacktriangleright$ In Chapter~\ref{Chap:Late.Quantum.Cosmology} we study the model-dependent high-energy deviations for the particle production during \emph{effective} inflation in a Bose--Einstein condensate. We show that the full picture, including non-perturbative corrections in the dispersion relation, relates to momentum-dependent rainbow geometries, see also~\cite{Jain:2006ki}.]

\subsubsection{Quantum aspects of signature changes \label{Sec:Intro.Quantum.Signature}}
A different problem that can be treated by similar mathematical methods is short sudden signature change --- from Lorentzian $(-,+++)$ to Euclidean $(+,+++)$ ---  in the spacetime geometry. The particle production process can be understood in terms of the drastic change in the behavior of the quantum field modes. In the Euclidean region the dispersion relation of the modes turns imaginary, and thus the modes are exponentially growing and decaying. This is similar to the behavior of ``frozen'' modes in the de Sitter universe, after the frequency of the mode drops below the scaling time of the surrounding universe, see also~\cite{Hu:2003,Calzetta1:2003xb,Calzetta:2005yk}. The quantum mode then gets ``dragged along'' with the spacetime fabric. [$\blacktriangleright$ We explain this effect in depth in Chapter~\ref{Chap:Late.Quantum.Cosmology}.] \\

A thorough treatment of signature change events in general relativity has been carried out by \emph{Tevian Dray et al.}, and can be found in~\cite{Dray:1991zz,Dray:1992sx,Dray:1993xu,Dray:1994kv,Dray:1995fb,Dray:1996cw,Dray:1996cx,Dray:1996dc,Dray:2000hb,Dray:2004aa}. Here the authors (in a slightly different context) not only studied the particle production from a finite Euclidean region, they also studied the behavior of the Einstein equations for Euclidean geometries, and how to patch geometries with different signature together. \\

A more physical motivation for signature change events has been suggested in 1983 by \emph{James Hartle} and \emph{Steven Hawking}~\cite{Hartle:1983aa} --- they consider a signature change at extremely early times, in the very early stages of the big bang. Consequently the existence of physical time, and hence the existence of our universe, is associated with a signature change event from Euclidean to Lorentzian geometry (the ``no boundary'' proposal).

\paragraph{Analogue signature change events from Bose gas}
In 2001 \emph{Donley et al.} carried out a Bose--Einstein condensate experiment that can be
viewed as the first analogue model experiment. In the experiment the atomic interactions
were driven --- for a finite time duration --- into a weakly attractive regime. This triggered a
controlled condensate collapse~\cite{Donley:2001aa,Roberts:2001aa}. Two years later, the theoretical
work of \emph{Esteban Calzetta} and \emph{Bei-Lok Hu}  \cite{Calzetta1:2003xb,Calzetta:2005yk}, connected the so-called Bose-nova phenomenon with the amplification, mixing of positive and negative modes, and squeezing of vacuum fluctuations due to a signature change event. [$\blacktriangleright$ In Chapter~\ref{Chap:Early.Quantum.Cosmology} we study the influence of the ``trans-Planckian'' cutoff in this analogue model for the quantum effects in the ``signature change event''.] \\

Next we would like to focus on the ``trans-Planckian'', or model-dependent modifications that \emph{must} arise in any analogue model --- that in some sense makes the analogue model programme vulnerable in in the ultraviolet regime. The key questions have to be asked: ``Is it possible to classify the type of Lorentz symmetry violation in a specific model?'' Can we find similar Lorentz symmetry breaking mechanisms motivated by various candidates for quantum gravity? 

\subsection{Quantum gravity phenomenology \label{Sec:Intro.QGP}}
So-called ``quantum gravity phenomenology'' summarizes all possible phenomenological consequences from quantum gravity. While different quantum gravity candidates may have completely distinct physical motivation, they can yield similar \emph{observable} consequences. One example is that in many quantum gravity models the fundamental symmetry of Lorentz invariance is \emph{not} an exact symmetry. The investigation of the possible consequences, and detectability of Lorentz symmetry breaking has been reviewed by \emph{David Mattingly} in~~\cite{Mattingly:2005aa}, (freely available online through \emph{Living Reviews in Relativity}). In the following we would like to summarize some relevant aspects of Lorentz symmetry breaking. \\

There are many different quantum gravity models that suggest the violation of Lorentz symmetry at ultraviolet scales, for example string theory~\cite{Kostelecky:1989aa,Ellis:1999aa}, wraped brane worlds~\cite{Burgess:2002aa}, and loop quantum gravity~\cite{Gambini:1999aa}. (For more references herein please consult~~\cite{Mattingly:2005aa}.)
If Lorentz symmetry is not an exact symmetry, we expect due to the tight experimental constraints that the violations have to take place around the Planck-scale $\approx 10^{19} GeV$ or higher. \\

There are two fundamentally different frameworks for Lorentz violation: (1) due to the presence of a preferred frame of observers or background field that decouples from the notion of spacetime metric, and (2) all frames remain equal, but transformation laws between the frames are modified (\eg, see ``Doubly special'' relativity (DSR)~\cite{Judes:2002bw,Liberati:2004ju}).  \\

It is possible to analyze kinematic effects for testing Lorentz symmetry~\cite{Amelino-Camelia:1998aa,Mansouri:1977aa,Lightman:1973aa,Robertson:1949aa} in a way similar to analyzing curved-spacetime quantum-field-theory --- where full (dynamical) quantum gravity is not \emph{necessarily} required. \\

\paragraph{Analogue Lorentz symmetry breaking}
It is easy to see that the symmetry breaking mechanism in different analogue models will lead to model-specific modifications in the dispersion relation. For example it is common knowledge that the ``full'' dispersion relation for perturbations in a Bose--Einstein condensate is given by
\begin{equation}
\omega_{k}^{2} = c_{0}^{2} \, k^{2} + \epsilon_{\mathrm{qp}}^{2} \, k^{4} \, ,
\end{equation}
the so-called Bogoliubov dispersion relation~\cite{Pethick:2001aa,Visser:1993tk}, while the dispersion relation for the specific ``lumped element'' electromagnetic wave-guide considered in~\cite{Schutzhold:2005aa} is given by
\begin{equation}
\omega_{k}^{2} = \frac{4}{L\,C} \, \sin^{2}\left( \frac{k \, \Delta x}{2} \right) \approx c^{2} \, k^{2} - \frac{\Delta x^{4} }{12 \, L \, C} \, k^{4} \, .
\end{equation}
Both models are capable of mimicking \emph{effective} black hole geometries within the infrared limit, but already start to deviate at the kinematic level, since the Bose--Einstein condensate yields a \emph{supersonic} dispersion relation, while the electromagnetic wave-guide exhibits a \emph{subsonic} dispersion relation. (However, the robustness of the Hawking effect~\cite{Jacobson:1991sd,Jacobson:1993ab,Unruh:1994zw,Unruh:2005aa,Brout:1995aa} assures us that there will be similar quantum effects.) It also demonstrates, that despite all the fundamental differences both models exhibit very similar Lorentz symmetry breaking mechanisms in terms of a quartic modification in $\Delta \omega_{k}^{2} \sim \pm k^{4}$. \\

A possible explanation might be related to the fact that \emph{any} emergent spacetimes based on analogue models \emph{per definition} have a preferred frame. This is the frame of the \emph{external observer}. At high energies the ``microscopic'' substructure starts to reveal itself, and all the fundamental particles the microscopic substructure are compared to the \emph{external} frame. On the other hand, low-energy perturbations experience an \emph{emergent Lorentz symmetry}.\footnote{A nice explanation for \emph{External and inner observers} can be found in~\cite{Volovik:2003jn}, in Chapter~(4.3.2).} Thus we conclude that analogue models of gravity are classified under type (1) Lorentz symmetry breaking, see discussion above. [$\blacktriangleright$ In Chapter~\ref{Chap:Geometries.Masses} we introduce a specific model for an emergent massive scalar field. In the subsequent Chapter~\ref{Chap:QGP} we employ this model to analyze the ``Lorentz symmetry breaking'' in the presence of ``massive particles''.] 

\subsection{Gravity as an emergent phenomenon?\label{Sec:Inspirations.Quantum.Gravity}}
In 1968 \emph{Andrei Sakharov}~\cite{Sakharov:1968dy} proposed the concept of ``induced gravity''  providing a natural framework for the emergence of ``effective geometrodynamics''. His ideas have been revisited by \emph{Matt Visser et al.} in~\cite{Barcelo:2001ay,Barcelo:2001lu,Visser:2002hf}, and below we would like to summarize the key points of ``induced gravity''.

In Section~\ref{Sec:Intro.Semi.Classical.Quantum.Geometry} we briefly pointed out that the notion of an ``effective metric'' generally arises for linear perturbations around some classical background field described by a generic Lagrangian $\mathcal{L}(\partial_{a} \phi, \phi)$ depending only on a single scalar field and its first derivatives. The emergent spacetime geometry therefore can be derived under the application of classical physics. 

Sakharov's ``induced gravity'' instead is based on the quantum nature of the system. Here we explicitly focus on small \emph{quantum fluctuations} $\phi_{1} \to \hat\phi_{1}$ (compare with Eq.~\refb{Eq:Intro.Linear.Field}) around the background field, employing Feynman path integral methods~\cite{Birrell:1984aa,Fulling:1989aa,Peskin:1995aa,Zee:2004aa,Mukhanov:2007aa} to calculate the one-loop effective  action. After regularization and renormalization, the one-loop effective action
\begin{eqnarray}
\nonumber
\Gamma[g(\phi_{0}),\phi_{0}] &=&  S[\phi_{0}] + \hbar \, \int \sqrt{-g} \, \kappa \, \left[ -2\Lambda + R(g) - 6V(\phi_{0}) \right] d^{d+1} x \\ &\;& + \hbar \, X[g(\phi_{0}),\phi_{0}] + \mathcal{O}(\hbar^{2}) \, ,
\end{eqnarray}
induces an Einstein--Hilbert term,
\footnote{Variations of the Einstein--Hilbert action with respect to the metric field tensor $g^{ab}$ yields the Einstein tensor, for an explicit derivation see~\cite{Poisson:2004aa}.} 
and the cosmological constant $\Lambda$.
\footnote{The cosmological constant has been introduced by \emph{Albert Einstein} as a generalization of the Einstein field equations. The cosmological constant can be viewed as an intrinsic energy density of the vacuum, and is considered to be related to expansion rate of the recent expanding universe, for example see~\cite{Dodelson:2003aa}.} 
The term $X[g(\phi_{0}),\phi_{0}]$ is a finite contribution to the renormalized one-loop effective action. For more details see~\cite{Barcelo:2001ay,Barcelo:2001lu,Visser:2002hf}.

Note that, explicitly through $\phi_{0}$ and implicitly through $g(\phi_{0})$, the emergent ``Einstein-Hilbert term'' depends on the background field. Since the ``effective metric'' only depends on the single scalar field $\phi_{0}$ it \emph{seems} not possible to reproduce full Einstein gravity.

Altogether we notice that Sakharov's ``induced gravity'' provides us with Einstein-\emph{like} equations. 
In this context it does not seem to be completely hopeless to expect that Einstein gravity itself may not a be ``fundamental theory'', instead it \emph{might} emerge from ``some other'' more fundamental theory. \\

For alternative viewpoints on emergent gravity consult~\cite{Hu:2005wu,Volovik:2003jn,Volovik:2006aa,Jacobson:1995eg}. 

\subsubsection{Summary}
We trust that this introduction has provided sufficient guidance to place the various topics to be discussed in this thesis into a wider perspective. Briefly, analogue spacetimes are a generic tool for probing the interface between gravity and quantum physics, and \emph{might} ultimately point to a deeper fundamental theory.

%
\chapter{Acoustic analogues of the Kerr black hole \label{Chap:Acoustic.Black.Holes}}
	 The spacetime geometry on the equatorial slice through a Kerr black
  hole is formally equivalent to the geometry felt by phonons
  entrained in a rotating fluid vortex. We analyse this example of
  ``analogue gravity'' in some detail: First, we find the most general
  acoustic geometry compatible with the dynamical equations for the fluid flow in a
  collapsing or expanding perfect-fluid line vortex. Secondly, we
  demonstrate that there is a suitable choice of coordinates on the
  equatorial slice through a Kerr black hole 
  that puts the spacetime geometry into this vortex form; though it is not possible to put the entire Kerr
  spacetime into perfect-fluid ``acoustic'' form.  Finally, we discuss
  the implications of this formal equivalence; both with respect to
  gaining insight into the Kerr spacetime and with respect to possible
  vortex-inspired experiments, and indicate ways in which more general
  analogue spacetimes might be constructed.

%
\section{Geometrical and physical acoustics \label{Sec:Geometrical.Physical.Acoustics}}
%
In the geometrical acoustics approximation, spacetime geometry emerges in a straightforward manner by considering the way the ``sound cones'' are dragged along by the fluid flow, thereby obtaining the conformal class
of metrics (see, for instance,~\cite{Matt.-Visser:2002ot}) given by,
\begin{equation}
g_{\mu\nu} \propto
\left[\begin{array}{c|c}{-(c^2-h^{mn}\;v_n\;v_n}) & -v_j\\
\hline
-v_i & h_{ij} 
\end{array}\right] \, .
\end{equation}
Here $c$ is the velocity of sound, $v$ is the velocity of the fluid, and $h_{ij}$ is the 3-metric of the ordinary Euclidean space of Newtonian mechanics (possibly in curvilinear coordinates).\\

Previously in Section \ref{Sec:Analogue.Models.For.Gravity} we have indicated how that the wave-equation for sound can be derived by linearizing and combining the Euler equation, the continuity equation, and a barotropic equation
of state~\cite{Unruh:1981bi,Unruh:1995aa,Visser:1993tk,Visser:1998gn}.  For irrotational flow
this process leads to the curved-space d'Alembertian equation (or, Klein--Gordon equation) and in
particular now fixes the overall conformal factor. The resulting acoustic metric in (3+1) dimensions is (see, for instance,~\cite{Unruh:1981bi,Unruh:1995aa,Visser:1993tk,Visser:1998gn,Matt.-Visser:2002ot,Barcelo:2004aa}, and references therein):
\begin{equation}
\label{Eq:Physical.Acoustic.Line.Element}
g_{\mu\nu} = \left({\rho\over c}\right)\;
\left[\begin{array}{c|c}{- (c^2-h^{mn}\;v_n\;v_n} ) & -v_j\\
\hline
-v_i & h_{ij} 
\end{array}\right].
\end{equation}
Here, in physical acoustics, the density of the fluid, $\rho$, is entering the emergent metric. Sound is then described by a massless minimally coupled scalar field propagating in this acoustic geometry.\\
In the presence of vorticity, a more complicated wave equation may still be derived~\cite{Pierce:1990aa,Perez-Bergliaffa:2001lq}. Then for frequencies large compared to the local vorticity, 
\begin{equation}
c\,k \gg \vert\vert\nabla\times \vec v \vert\vert
\end{equation}
this wave equation reduces to the d'Alembertian and the acoustic geometry can be identified in the usual manner. For more details see~\cite{Perez-Bergliaffa:2001lq}.\\

The overall conformal factor derived in the physical acoustics approximation~\refb{Eq:Physical.Acoustic.Line.Element} depends explicitly on the dimension $d$ of the spatial slice.
%
\subsection{The role of dimension\label{Sec:Role.Of.Dimension}}
The role of spacetime dimension in these acoustic geometries is sometimes a bit surprising and potentially confusing. 
There is a real physical distinction between truly two-dimensional systems and systems that are effectively two dimensional, such as three-dimensional systems with cylindrical symmetry.
We emphasize that in cartesian coordinates the wave equation is given by
\begin{equation} 
\label{wavef_eq}
\frac{\partial}{\partial {x^{\mu}}} 
\left( f^{\mu \nu} \frac{\partial}{\partial {x^{\nu}}}\; \psi \right)=0 \, ,
\end{equation}
where
\begin{equation}
f^{\mu \nu}=\left[
\begin{array}{ccc}
-{\rho}/{c^2}            &|& -{\rho}\;{v^j} /{c^2}\\
\hline
-{\rho}\; {v^i}/{c^2}  &|& 
\rho \;\{\delta^{ij} - v^i v^j /{c^2}\}
\end{array}
\right],
\end{equation}
holds \emph{independent} of the dimensionality of spacetime. It depends only on the Euler equation, the continuity equation, a barotropic equation of state, and the assumption of irrotational flow~\cite{Visser:1993tk}.

Introducing the acoustic metric $g^{\mu \nu}$, defined by
\begin{equation}
f^{\mu \nu}=\sqrt{-g} \, g^{\mu \nu}; 
\qquad
g=\frac{1}{\det(g^{\mu \nu})}
\end{equation} 
the wave Eq.~\refb{wavef_eq} corresponds to the massless
Klein--Gordon equation (d'Alembertian wave equation) in a curved
space-time with contravariant metric tensor:
\begin{equation}
\label{Eq:Acoustic.Line.Element.d.Cartesian.Cooradinates}
g^{\mu \nu}=\left({\rho\over c}\right)^{-2/(d-1)}
\left[
\begin{array}{c|c}
-1/c^2            & -\vec{v}^{\,T}/c^2 \\
\hline
\vphantom{\Big|}
- \vec{v}/c^2  & 
\mathbf{I}_{d\times d} - \vec{v}\otimes \vec{v}^{\,T}/c^2
\end{array}
\right] \, ,
\end{equation}
where $d$ is the dimension of \emph{space} (not spacetime).

The covariant acoustic metric is
\begin{equation}
g_{\mu \nu}=\left( \frac{\rho}{c} \right)^{2/(d-1)}
\left[
\begin{array}{ccc}
-\left( c^2-v^2 \right)       &|& -\vec{v}^{\,T} \\
\hline
-\vec{v}  &|& \mathbf{I}_{d\times d}
\end{array}
\right] \, . 
\end{equation}

\subsubsection{$d=3$\label{Sec:d.3}}
The acoustic line-element for three space and one time dimension reads
\begin{equation} 
g_{\mu \nu}=\left( \frac{\rho}{c} \right)
\left[
\begin{array}{ccc}
-\left( c^2-v^2 \right)       &|& -\vec{v}^{\,T} \\
\hline
-\vec{v}  &|& \mathbf{I}_{3\times 3}
\end{array}
\right] \, . 
\end{equation}
This is the primary case of interest in this chapter.
%
\subsubsection{$d=2$\label{Sec:d.2}}
The acoustic line-element for two space and one time dimension reads
\begin{equation} 
g_{\mu \nu}=\left( \frac{\rho}{c} \right)^2
\left[
\begin{array}{ccc}
-\left( c^2-v^2 \right)       &|&- \vec{v}^{\,T} \\
\hline
-\vec{v}  &|& \mathbf{I}_{2 \times 2}
\end{array}
\right] \, . 
\end{equation}
This situation would be appropriate when dealing with surface waves or excitations confined to a particular substrate, for example at the surface between two superfluids. An important physical point is that, due to the fact that one can always find a conformal transformation to make the two-dimensional
spatial slice flat~\cite{Wald:1984rg}, essentially all possible 2+1 metrics can in principle be reproduced by acoustic metrics~\cite{Volovik:2003jn}. (The only real restriction is the quite physical constraint that there be no closed timelike curves in the
spacetime.)

\subsubsection{$d=1$\label{Sec:d.1}}
The naive form of the acoustic metric in (1+1) dimensions is ill-defined, because the conformal factor is raised to a formally infinite power --- this is a side effect of the well-known conformal invariance of the Laplacian in 2 dimensions. The wave equation in terms of $f^{\mu\nu}$ continues to make good sense --- it is only the step from $f^{\mu\nu}$ to the effective metric that breaks down.

Note that this issue only presents a difficulty for physical systems that are intrinsically one-dimensional. A three-dimensional system with plane symmetry, or a two-dimensional system with line symmetry,
provides a perfectly well behaved model for (1+1) dimensions, as in the cases $d=3$ and $d=2$ 
above.\\

Before we go ahead with our programme, we present explicit expressions for the acoustic line-element in cylindrical polar coordinates and spherical polar coordinates.
%
\subsection{Various coordinate systems\label{Sec:Acoustic.In.Various.Coordinates}}
The acoustic line element for $d$ dimensions in cartesian coordinates, see Eq.~\refb{Eq:Acoustic.Line.Element.d.Cartesian.Cooradinates}, is defined by
\begin{equation}
ds^{2} = \left( \frac{\rho}{c} \right)^{2/(d-2)} \left[ -(c^{2}-v^{2})\, dt^{2} - 2 \, \vec v  \cdot d\vec r \, dt + dx^{2} + dy^{2} + dz^{2} \right] \, ,
\end{equation}
where $d\vec r^{T} = [ dx , \, dy , \, dz ]^{T} $ .
For computational convenience it is advantageous to change to cylindrical coordinates.
%
\subsubsection{Cylindrical polar coordinates\label{Sec:Acoustic.In.Cylindrical.Polar.Coordinates}}
The coordinate transformation from cartesian to cylindrical coordinates is specified by 
$x = r \, \cos \phi$, $y = r \, \sin \phi $, and $z = z$, where $\phi$ is the angle between the position vector $\mathbf r$ and the $x$-axis, and $r$ the length of the position vector. In acoustic cylindrical polar coordinates we get
\footnote{
Here we used the formal replacement,
\begin{equation}
\nonumber
d\vec r = \left( \begin{array}{c} \cos \phi \, dr - r\, \sin \phi \, d \phi \\ \sin\phi \, dr + r\, \cos\phi \, d\phi  \\ dz  \end{array} \right) 
\quad \to \quad 
\left( \begin{array}{r} dr \\ r\, d\phi \\ dz \end{array} \right) =
\left( \begin{array}{c} \cos\phi \, dx + \sin\phi \, dy \\ -  \sin\phi \, dx +  \cos\phi \, dy \\ dz  \end{array} \right) 
\, .
\end{equation}}
,
\begin{equation}
ds^{2} = \left( \frac{\rho}{c} \right)^{2/(d-2)} \left[ -(c^{2}-v^{2})\, dt^{2} - 2 \left( v_{\hat r} \, dr + v_{\hat \phi} \, r \, d\phi + v_{z} dz \right)\, dt 
+ dr^{2} + r^{2} \, d\phi^{2} + dz^{2} \right] \, .
\end{equation}
If we take the cylindrical polar coordinate system and simply drop the $z$ direction, setting $dz=0$, everything reduces to a $(2+1)$ dimensional problem. However, the acoustic metric depends on the total spacetime dimension so there is a physical difference between a $(3+1)$ dimensional problem that has cylindrical symmetry (\eg, a line vortex), and an intrinsically $(2+1)$ dimensional system. For surface phonons trapped on a two-dimensional substrate, and so living in $(2+1)$ dimensional spacetime we have to use the appropriate conformal factor resulting from $d=2$.
\subsubsection{Spherical polar coordinates\label{Sec:Acoustic.In.Spherical.Polar.Coordinates}}
In similar manner we obtain the acoustic line element in spherical polar coordinates. The mapping between cartesian to spherical polar coordinates is given by
$x = r \, \sin\theta \, \cos\phi$, $y = r \, \sin\theta \, \sin\phi$, and $z = r \, \cos\theta$, where $\phi$ represents the angle between the position vector $\vec r$ and the $x$-axis, and $\theta$ the angle between the position vector $\vec r$ and the $z$-axis. 
Altogether we get
\footnote{ Here we used the formal replacement,
\begin{eqnarray}
\nonumber
d\vec r = \left( 
\begin{array}{c} 
\sin\theta \, \cos\phi \, dr + r\, \cos\theta \, \cos\phi \, d\theta - r \, \sin\theta \, \sin\phi \, d\phi   \\ 
\sin\theta \, \sin\phi \, dr + r\, \cos\theta \, \sin\phi \, d\theta + r \, \sin\theta \, \cos\phi \, d\phi    \\ 
\cos\theta \, dt - r \, \sin\theta \, d\theta 
\end{array} \right) 
\quad \quad \quad \quad \quad \quad \quad \quad \quad
\\ 
\quad \to \quad \quad \quad  
\left( \begin{array}{r} dr \\ r\, \sin\theta \, d\phi \\ r \, d\theta \end{array} \right) =
\left( \begin{array}{c} 
\sin\theta \, \cos\phi \, dx + \sin\theta \, \sin\phi \, dy + \cos\theta \, dz  \\ 
- \sin\phi \, dx + \cos\phi \, dy \\ 
\cos\theta \, \cos\phi \, dx + \cos\theta \, \sin\phi \, dy - \sin\theta \, dz  
\end{array} \right)  \, .
\end{eqnarray}}
\begin{equation}
ds^{2} = \left( \frac{\rho}{c} \right)^{\frac{2}{d-2}} \left[ -(c^{2}-v^{2})\, dt^{2} - 2 \left( v_{\hat r} \, dr + v_{\hat \phi} \, r \, \sin\phi \, d\phi + v_{\hat \theta} \, r \, d\theta \right)\, dt 
+ dr^{2} + r^{2}\left(\sin^{2}(\theta) \, d\phi^{2} + d\theta^{2}\right) \right]  .
\end{equation}
Please note, that for $\theta = \pi/2$, that is if we are in the $r-\phi$-plane, the spherical polar coordinates are identical to the reduced, that is $z=0$, cylindrical polar coordinates. 

Now because the ordinary Euclidean space ($h_{ij}$) appearing in these perfect fluid acoustic geometries is Riemann flat, the 3-dimensional space (given by the constant-time slices) in any acoustic geometry is forced to be conformally flat, with 3-metric $g_{ij} = (\rho/c)\; h_{ij}$. This constraint places very strong restrictions on the class of (3+1)-dimensional geometries that can be cast into perfect fluid
acoustic form.  While many of the spacetime geometries of interest in general relativity can be put into this acoustic form, there are also important geometries that \emph{cannot} be put into this form; at
least for a literal interpretation in terms of flowing perfect fluid liquids and physical sound. In the following we briefly summarize the situation for non-rotating acoustic black holes in ordinary fluids.
%

%
\section{Acoustic non-rotating black holes \label{Sec:Non.Rotating.Black.Holes}}
%
In particular, the Schwarzschild geometry can, at the level of geometrical acoustics, certainly be cast into this perfect fluid acoustic form~\cite{Visser:1998gn}.  However, at the level of physical
acoustics (and working within the context of the Painl\'eve--Gullstrand coordinates) there is a technical difficulty in that the Euler and continuity equations applied to the background fluid flow yield a nontrivial and unwanted (3+1) conformal factor $\rho/c$~\cite{Visser:1998gn}. This overall conformal factor is annoying, but it does not affect the sound cones, and so does not affect the ``causal structure'' of the resulting analogue spacetime~\cite{Barcelo:2004aa}.  Furthermore any overall conformal factor will not affect the surface gravity~\cite{Jacobson:1993aa}, so it will not directly affect the acoustic analogue of Hawking radiation~\cite{Unruh:1981bi,Visser:1998gn}. So for most of the interesting questions one might ask, the Schwarzschild geometry can for all practical purposes be cast into acoustic form.

To begin with we illustrate in detail the problems involved with the ``acoustic Painl\'eve--Gullstrand coordinates''.
Alternatively, we demonstrate how the Schwarzschild geometry can also be put into acoustic form in different manner by using isotropic coordinates. 
%
\subsection{Geometrical acoustics and the Painl\'eve--Gullstrand geometry\label{Sec:Geometrical.Acoustic.Painleve.Gullstrand}}
We consider a time-independent fluid flow with a velocity profile as a function of $r$. The continuity equation~\refb{cy_eq} in this situation implies
\begin{equation}
\label{Eq:One.D.Velocity.Profile}
\rho(r) \, v(r) = \mbox{constant} \, ,
\end{equation}
thus we have a fixed relationship between the velocity and density profile.
The physical acoustic geometry for this case is given by
\begin{equation}
\label{Eq:Acoustic.Slab.Geometry}
ds^{2} = \left( \frac{\mbox{constant}}{v(z)\,c(z)} \right)^{\frac{2}{d-1}} \, 
\left[ -\left(c(r)^{2} -v(r)^{2} \right) \, dt^{2} +  2 \, v(r) \, dt \, dr  + dr^{2} + r^{2} \, \left( d\theta^{2} + \sin^{2}(\theta) \, d\phi^{2}  \right)  \right] \, .
\end{equation}

The Painl\'eve--Gullstrand, also often called the Lema\^itre, line element is 
\begin{equation}
ds^{2} = -dt^{2} + \left(dr \pm \sqrt{\frac{2m}{r}} \, dt \right)^{2} + r^{2} \, \left( d\theta^{2} + \sin^{2}(\theta) \, d\phi^{2} \right) \, ,
\end{equation}
where the $+ /  -$ represent ingoing\,/\,outgoing coordinates. 

Comparing the two line-elements, one might be tempted to simply adjust the density, $\rho(r)=\rho_{0}$ (\ie, an incompressible fluid), and sound speed, $c(r)=c_{0}$, to be position independent constants, and simultaneously set $v=2 m / r $. The crux is that this violates the continuity equation.

Alternatively, if we pick $c=1$, and $v=\frac{2\, m}{r}$, the continuity equation (in spherical coordinates), 
\begin{equation}
\nabla \, \left( \rho \cdot \vec{v} \right) = \frac{1}{r^{2}} \, \frac{1}{\partial r} \left( r^{2} \, \rho \, \vec{v} \right) = 0 \, ,
\end{equation}
can only be satisfied for
\begin{equation}
\rho(r) = \frac{\mbox{constant}}{\sqrt{2\,m}}\, r^{-3/2} \, ,
\end{equation}
a compressible fluid.

The resulting physical acoustic line-element,
\begin{equation}
ds^{2} =  \left(\frac{\mbox{constant}}{ \sqrt{2\,m}}\right)^{\frac{2}{1-d}} \, r^{-\frac{3}{1-d}} \,
\left[  -dt^{2} + \left(dr \pm \sqrt{\frac{2m}{r}} \, dt \right)^{2} + r^{2} \, \left( d\theta^{2} + \sin^{2}(\theta) \, d\phi \right)  \right] \, ,
\end{equation}
is \emph{up to a conformal factor} equivalent to the black hole geometry in  Painl\'eve--Gullstrand coordinates. Thus the analogy only holds at the level of geometrical acoustics, since the light-cone structure is unchanged by any overall conformal factor.
\subsection{Acoustic isotropic version of the Schwarzschild geometry\label{Sec:Acoustic.Isotropic.Schwarzschild}}
The isotropic version of the Schwarzschild geometry can be cast \emph{exactly} into acoustic form, without any extraneous conformal factors. Since this is not the standard way of viewing the acoustic analogue of the Schwarzschild geometry~\cite{Visser:1998gn}, and since the algebra is simple enough to be done in closed form, it is worthwhile taking a small detour.\\

In isotropic coordinates the Schwarzschild geometry reads
\begin{equation}
\d s^2 = - {(1-{m\over2r})^2\over(1+{m\over2r})^2}\;\d t^2 
+ \left(1+{m\over2r}\right)^4 
\left[ \d r^2 + r^2\left(\d\theta^2 +\sin^2\theta\;\d\phi^2\right) \right],
\end{equation}
and so in these coordinates the acoustic analogue in $(3+1)$ dimensions corresponds to
\begin{eqnarray}
v &=& 0;\\
\rho &=& \rho_\infty\;\left(1 - {m^2\over 4r^2} \right);\\
c^2 &=&   c^2_\infty\; {(1-{m\over2r})^2\over(1+{m\over2r})^6}.
\end{eqnarray}
The external force required to hold this configuration in place against the pressure gradient is
\begin{equation}
f_{\hat r} = {\d p\over\d r} = c^2 \; {\d\rho\over\d r} =  
\rho_\infty \; c^2_\infty \; 
{(1-{m\over2r})^2\over(1+{m\over2r})^6} \;  {m^2\over 2 r^3}.
\end{equation}
The pressure itself (normalizing $p_\infty\to0$) is then
\begin{equation}
p(r) =  -  \rho_\infty \; c^2_\infty \;  m^2 \; 
\left({ 120 r^3 - 20 r^2 m+10m^2r+m^3\over 480\;r^5\;(1+{m\over2r})^5} \right),
\end{equation}
and by eliminating $r$ in favour of $\rho$ using
\begin{equation}
r = {m\over2\sqrt{1-\rho/\rho_\infty}}
\end{equation}
we can deduce the equation of state
\begin{equation}
p(\rho) = - (\rho_\infty-\rho)\;   c^2_\infty \;
\left({ 
20-5\rho/\rho_\infty-4\sqrt{1-\rho/\rho_\infty}
-\rho/\rho_\infty\sqrt{1-\rho/\rho_\infty}
\over 
15\; (1+\sqrt{1-\rho/\rho_\infty})^5}
\right).
\end{equation}
At the horizon, which occurs at $r_H=m/2$ in these coordinates, both
$\rho=0$ and $c=0$ in the acoustic analogue, while
\begin{equation}
p_H =  -{ \rho_\infty \; c^2_\infty \over 30}
\end{equation}
is finite, so there is a finite pressure drop between asymptotic infinity and the horizon. Everything is now simple enough to be fully explicit, and although we now have an analogue model that reproduces the \emph{exterior} part of the Schwarzschild geometry \emph{exactly}, we can also clearly see the two forms in which fine tuning arises --- in specifying the external force, and in the equation of state.

More generally, any spherically symmetric geometry (static or otherwise) can be put into acoustic form, as we shall see now.  
%
\subsection{Acoustic spherically symmetric geometries \label{Sec:Acoustic.Spherically.Geometries}}
It is a standard result that \emph{any} spherically symmetric geometry can be put into isotropic form
\begin{equation}
\d s^2 = - \exp(-2\Phi)\;\d t^2 + \exp(-2\Psi)
\left[ \d r^2 + r^2\left(\d\theta^2 +\sin^2\theta\;\d\phi^2\right) \right].
\end{equation}
This can be put into acoustic form in a particularly simple manner by setting
\begin{eqnarray}
v(r) &=& 0; \\
\rho(r) &=& \rho_\infty\; \exp(-\Phi-\Psi); \\
c(r) &=&   c_\infty\;  \exp(\Psi-\Phi).
\end{eqnarray}
The pressure can now be formally evaluated as
\begin{equation}
p(r) =   \rho_\infty \; c_\infty^2 \int_r^\infty \exp(\Psi-3\Phi) \; \partial_r (\Phi+\Psi) \; \d r,
\end{equation}
and comparison with $\rho(r)$ can be used to construct a formal equation of state, $\rho(p)$. \\

This of course implies that the Reissner--Nordstr\"om geometry can be put into acoustic form. In particular, the FRW cosmologies can also be put into acoustic form, both at the level of physical acoustics and at the level of geometrical acoustics. In fact there are two rather different routes for doing so: Either by causing the fluid to explode or by adjusting the speed of sound~\cite{Barcelo:2003ia,Barcelo:2003yk,Fedichev:2004on,Fedichev:2006mc,Fedichev:2004fi,Lidsey:2004aa,Weinfurtner:2004wt,Weinfurtner:2005aa}.
%

\subsection{The problem with acoustic rotating black holes \label{Sec:Prob.Acousitc.Black.Holes}}
There is however a fundamental restriction preventing the Kerr geometry [and Kerr--Newman geometry] from being cast into perfect fluid acoustic form. It has recently been established~\cite{Garat:2000aa,Kroon:2003aa,Kroon:2004aa} that no possible time-slicing of the full Kerr geometry can ever lead to
conformally flat spatial 3-slices.  Faced with this fact, we ask a more modest question: Can we at least cast a \emph{subspace} of the Kerr geometry into perfect fluid acoustic form? Specifically, since we
know that the effective geometry of a generalized line vortex (a ``draining bathtub'' geometry) contains both horizons and ergosurfaces \cite{Visser:1998gn,Matt.-Visser:2002ot}, one is prompted to ask: If we look at the equatorial slice of the Kerr spacetime can we at least put that into
acoustic form? If so, then this opens the possibility of finding a physically reasonable analogue model based on a vortex geometry that might mimic this important aspect of the Kerr geometry.  Thus we have
three independent physics questions to answer:

\begin{itemize}
\item[Sec.~\ref{Sec:Vortex.Flow} $\vert$]{ What is the most general perfect fluid acoustic metric that can (even in principle) be constructed for the most general (translation invariant) line vortex geometry?}
\item[Sec.~\ref{Sec:Rotating.Black.Holes} $\vert$]{ Can the equatorial slice of Kerr then be put into this form? \\ (And if not, how close can one get?)}
\item[Sec.~\ref{Sec:Acoustic.Kerr.Equator} $\vert$]{By generalizing the analogue model to something more complicated than a perfect fluid, can we do any better?}
\end{itemize}
We shall now explore these three issues in some detail.
%

%
\section{General vortex flow in fluids \label{Sec:Vortex.Flow}}	
%
The Euler (zero-viscosity Navier--Stokes) equation for fluids sets tight constraints on the currents in moving fluids. In the following we derive the most general (translation invariant) line vortex geometry in a barotropic fluid.
%
\subsection{General framework\label{Sec:General.Framework}}
The background fluid flow (on which the sound waves are imposed) is governed by three key equations, the continuity equation, the Euler equation, and a barotropic equation of state;
\begin{eqnarray} 
\label{cy_eq}
\frac{\partial \rho}{\partial t}+\nabla\cdot (\rho \; \vec{v})=0 \, ; \\
\label{eu_eq}
\rho \left[  \frac{\partial \vec{v}}{\partial t} + (\vec{v} \cdot \nabla) \vec{v}  \right]=-\nabla p + \vec{f} \, ; \\
\label{Eq:Equation.Of.State}
p=p(\rho) \, .
\end{eqnarray}
Here we have included for generality an arbitrary external force $\vec f$, possibly magneto-hydrodynamic in origin, that we can in principle think of imposing on the fluid flow to shape it in some desired fashion. From an engineering perspective the Euler equation is best rearranged as
\begin{equation}
\label{euler-rearranged}
\vec{f} =  
\rho \left[
\frac{\partial \vec{v}}{\partial t} + (\vec{v} \cdot \nabla) \vec{v} \right]
+\nabla p,
\end{equation}
with the physical interpretation being that $\vec f$ is now telling you what external force you would need in order to set up a specified fluid flow.

\subsubsection{Magneto-hydrodynamic fluids\label{Sec:Magneto.Hydrodynamic.Fluids}}
Magneto-hydrodynamic (MHD) fluids are conductive fluids (\eg, plasmas, liquid metals, or salt water), where under the application of an external magnetic field controlled currents in the fluid are induced. MHD fluids are described by a combination of the Navier--Stokes equation of the fluid and Maxwell's equation of electromagnetism. 

MHD fluids allow a sophisticated manipulation of the fluid flow, and as we shall see shortly, are a necessity  to set up an acoustic analogue for  a rotating back hole.
%
\subsection{Zero radial flow\label{Sec:Zero.Radial.Flow}}
Assuming now a cylindrically symmetric time-independent fluid flow without any sinks or sources we have a line vortex aligned along the $z$ axis with fluid velocity $\vec{v}$:
\begin{equation} 
\label{2d_v}
\vec v(r) =   v_{\hat\theta}(r)\;\hat{\theta} \, .
\end{equation}
The continuity equation~\refb{cy_eq} for this geometry is trivially satisfied and we write the fluid acceleration $(\mathbf v \cdot \nabla) \, \mathbf v$ in cylindrical polar coordinates. The gradient vector is $\nabla^{T} = (\partial_{r},r^{-1} \partial_{\phi})^{T}$, and the unit vectors $\hat r$ and $\hat \theta$ are
\begin{eqnarray}
\hat r = d\vec{r} \,  \vert dr \vert^{-1} = \cos\theta \, \hat x + \sin\theta \, \hat y \, ; \\
\hat \phi = d\vec{\phi} \,  \vert d\phi \vert^{-1} = -\sin\theta \, \hat x + \cos\theta \, \hat y \, .
\end{eqnarray}
Further calculations show that
\begin{equation}
\partial_{t}\hat r = 0; \quad \partial_{\theta} \hat r = \hat \theta; \quad \partial_{r}\hat\theta = 0; \quad \partial_{\theta}\hat\theta = - \hat r\, ,
\end{equation}
and we get
\begin{equation}
\vec a = (\vec{v}\cdot \nabla) \vec{v} =
- \frac{v_{\hat \phi}(r)^2}{r} \; \hat{r} \, ,
\end{equation}
for the fluid acceleration. A substitution into the rearranged Euler equation\refb{euler-rearranged} yields
\begin{equation}
\vec{f} = f_{\hat{r}} \; \hat{r} = 
\left\{ -\rho(r) \; \frac{v_{\hat{\phi}}(r)^2}{r} 
+c^2(r) \; \partial_{r} \rho(r)  \right\} \hat{r} \, .
\end{equation} 
%

\subsubsection{Physical interpretation\label{Sec:Physical.Interpretation.Zero.Rad.Flow}}
%
\begin{figure}[ht]
\label{F:vortex-non-collapse}
\centerline{{\includegraphics[width=3in]{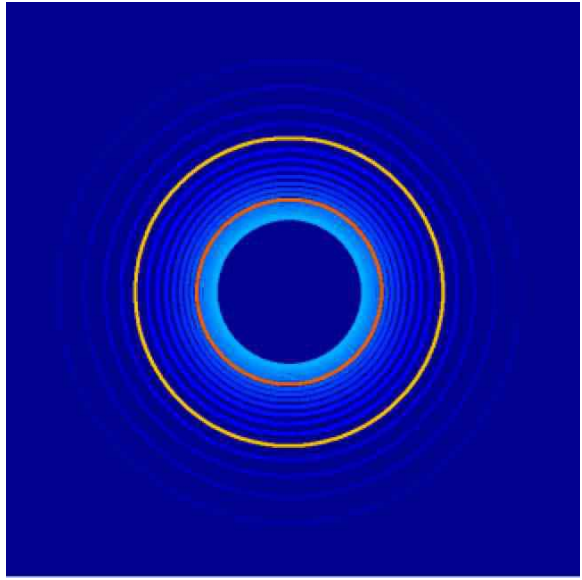}}}
\caption[Non-collapsing vortex geometry.]{ A simple non-collapsing vortex geometry: The orange colored circles denote the streamlines of the fluid flow. The brighter the circles the faster the angular velocity of the fluid. The outer golden circle represents the ergosurface, where the fluid velocity reaches the speed of sound. The inner orange circle (which need not exist in general) represents a possible (outer) horizon where the speed of sound goes to zero.}
\end{figure}
%

The external force $\vec{f}_{\hat{r}}$ must be chosen to precisely cancel against the combined effects of centripetal acceleration and pressure gradient. The angular-flow is not completely controlled by this external force, but is instead an independently specifiable quantity. (There is only \emph{one} relationship between $f_r(r)$, $\rho(r)$, $c(r)$, and $v_\phi(r)$, which leaves \emph{three} of these quantities as arbitrarily specifiable functions.)

We are now considering the equation of state to be an \emph{output} from the problem, rather than an \emph{input} to the problem. If for instance $\rho(r)$ and $c^2(r)$ are specified then the pressure can be
evaluated from
\begin{equation}
p(r) = \int c^2(r) \; {\d \rho\over\d r} \; \d r,
\end{equation}
and then by eliminating $r$ between $p(r)$ and $\rho(r)$, the EOS $p(\rho)$ can in principle be determined.

For zero external force (and no radial flow), which arguably is the most natural system to set up in the laboratory, we have
\begin{equation}
{v_{\phi}^2(r)\over c^2(r)} =  r \; \partial_r \ln\rho(r),
\end{equation}
which still has two arbitrarily specifiable functions.

In the geometric acoustics regime the acoustic line-element for this zero-source\,/\,zero-sink line vortex is
\begin{eqnarray}
\d s^2 &\propto& -\left( c^2-v_{\hat\phi}^2\right) \d t^2 - 2 v_{\hat\phi} \; r \; \d\phi \; \d t +\d r^2 + r^2 \d\phi^2 + \d z^2.
\end{eqnarray}
In the physical acoustics regime the acoustic line-element is
\begin{equation}
\d s^2=\left( \frac{\rho}{c} \right) \Big[ -\left( c^2-v_{\hat\phi}^2\right) \d t^2  - 2 v_{\hat\phi} \; r \d\phi \;\d t +\d r^2 + r^2 \d\phi^2 +\d z^2 \Big].
\end{equation}
The vortex quite naturally has an ergosurface where the speed of the fluid flow equals the speed of sound in the fluid. This vortex geometry may or may not have a horizon --- since $v_{\hat r}$ is identically zero the occurrence or otherwise of a horizon depends on whether or not the speed of sound $c(r)$ exhibits a
zero. \footnote{The vanishing of $c(r)$ at a horizon is exactly what happens for Schwarzschild black holes (or their analogues) in either Schwarzschild or isotropic coordinate systems. This is rather different from the behavior in Painleve--Gullstrand coordinates, but is a quite standard signal for the presence of a horizon.} We shall later see that this class of acoustic geometries is the most natural for building analogue models of the equatorial slice of the Kerr geometry.
%

\subsection{General analysis with radial flow}
For completeness we now consider the situation where the vortex contains a sink or source at the origin.  (A concrete example might be the ``draining bathtub'' geometry where fluid is systematically extracted from a drain located at the centre.)  Assuming now a cylindrically symmetric time-independent fluid flow with a line vortex aligned along the $z$ axis, the fluid velocity $\vec{v}$ is
\begin{equation} 
\label{2d_vb}
\vec{v}=v_{\hat r}(r)\;\hat{r}+v_{\hat\phi}(r)\;\hat{\phi}.
\end{equation}
Wherever the radial velocity $v_{\hat r}(r)$ is nonzero the entire vortex should be thought of as collapsing or expanding.  
%
\begin{figure}[htp]
\label{F:vortex}
\centerline{{\includegraphics[width=3in]{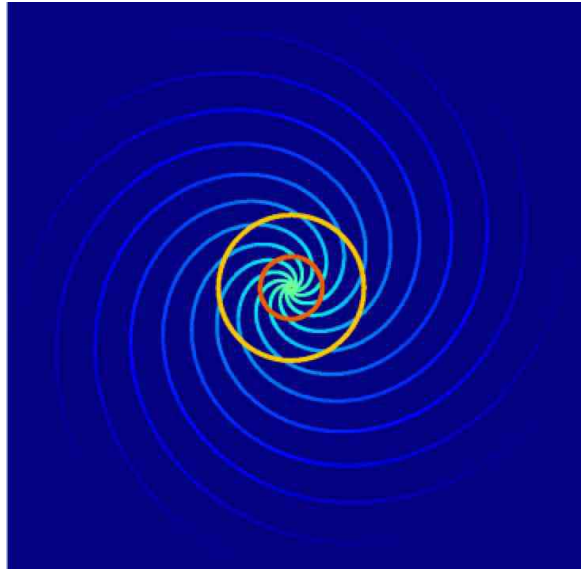}}}
\caption[Collapsing vortex geometry.]{A collapsing vortex geometry (draining bathtub): The multi-colored spirals denote streamlines of the fluid flow. Following a streamline towards the center of the vortex the velocity increases. (The brighter the blue, the faster the velocity on the streamline.) The outer golden circle represents the ergosurface while the inner orange circle represents the (outer) event horizon. }
\end{figure}
%

The continuity equation (\ref{cy_eq}) for this cylindrically symmetric problem is
\begin{equation} \label{eu_ti_eq}
\nabla \cdot (\rho \;\vec{v})=0,
\end{equation}
and the rearranged Euler equation (\ref{euler-rearranged}) for a pressure $p$ which depends only on the radial-coordinate $r$ is
\begin{equation} \label{vec_f}
\vec f = \rho \, (\vec{v}\cdot \nabla) \vec{v} + c^2 \; {\partial}_r\rho \; \hat{r}.
\end{equation}
At this stage we note that $\rho$ is in general not an independent variable.  Because Eq.~\refb{eu_ti_eq} corresponds to a divergence-free field, integration over any closed circle in the two-dimensional plane yields
\begin{equation}
 \oint \rho(r)\;\vec{v}(r) \cdot \hat r \; \d s
= 2\pi\;\rho(r)\; v_{\hat r}(r)\; r = 2 \pi\; k_1.
\end{equation}
Then \emph{provided} $v_{\hat r}\neq 0$,
\begin{equation} \label{alpha_eq2}
{\rho(r)} =\frac{k_1}{ r \;v_{\hat r}(r) }.
\end{equation}
Substituting into the rearranged Euler equation gives
\begin{equation}  
\vec f ={k_1\over r v_{\hat r}} \; (\vec{v}\cdot \nabla) \vec{v} + c^2 \; {\partial}_r \left( {k_1\over r v_{\hat r}}\right) \; \hat{r},
\end{equation}
where $c$ is a function of $\rho$, and thus a function of $v_{\hat r}$.  This now completely specifies the force profile in terms of the desired velocity profile, $ v_{\hat r}$, $ v_{\hat \phi}$, the equation of state, and a single integration constant $k_1$. 
Calculating the fluid acceleration leads to
\begin{equation}
\label{vec_a}
(\vec{v}\cdot \nabla) \vec{v} = \left\{v_{\hat r}(r) \partial_r v_{\hat r}(r)- \frac{v_{\hat \phi}(r)^2}{r} \right\}\hat{r} + \left\{ v_{\hat r}(r) \partial_r v_{\hat \phi}(r)  + \frac{ v_{\hat r}(r)\;v_{\hat \phi}(r)}{r}  \right\} \hat{\phi},
\end{equation}
which can be rearranged to yield
\begin{equation}
(\vec{v}\cdot \nabla) \vec{v} = \left\{ {1\over 2} \partial_r [v_{\hat r}(r)^2]
- \frac{v_{\hat \phi}(r)^2}{r} \right\}\hat{r} + \left\{ {v_{\hat r}(r)\over r}  \partial_r [r\; v_{\hat \phi}(r)]  
\right\} \hat{\phi}.
\end{equation}

Finally, decomposing the external force into radial and tangential (torque-producing directions) we have
\begin{equation} \label{f_r}
f_{\hat r} =  \vec f \cdot \hat r = k_1 \left\{ {1\over r v_{\hat r}} \;
\left[  {1\over 2} \partial_r [v_{\hat r}(r)^2]- \frac{v_{\hat \phi}(r)^2}{r} \right]+ c^2 \; {\partial}_r \left( {1\over r v_{\hat r}}\right)  \right\},
\end{equation}
and
\begin{equation} \label{f_theta}
f_{\hat\phi} =  \vec f \cdot \hat \phi  = k_1 
\left\{ {1\over r^2}  \partial_r [r\; v_{\hat \phi}(r)]  \right\}.
\end{equation}
The radial equation can undergo one further simplification to yield
\begin{equation} \label{f_r_2}
f_{\hat r} = k_1 \left\{ {1\over r}  \, \left[ 1 - {c_s^2\over v_{\hat r}^2} \right] \,\partial_r v_{\hat r}
-  {c_s^2 + v_{\hat\phi}^2 \over r^2 \; v_{\hat r}} \right\}.
\end{equation}
%

\subsubsection{Summary\label{Sec:Summary.General.Vortex.Flow}}
In the geometric acoustics regime the acoustic line-element for the most general (time-independent cylindrically symmetric collapsing\,/\,expanding) line vortex is
\begin{eqnarray}
\d s^2 &\propto&
-\left( c^2-v_{\hat r}^2-v_{\hat\phi}^2\right) \d t^2  
- 2v_{\hat r}\; \d r \; \d t
- 2 v_{\hat\phi} \; r \; \d\phi \; \d t
+\d r^2 + r^2 \d\phi^2 +\d z^2.
\end{eqnarray}
We again reiterate that --- given a barotropic equation of state $p(\rho)$ --- once the velocity profile $ v_{\hat r}$ and $ v_{\hat  \phi}$ is specified, then up to a single integration constant $k_1$, the density and speed of sound are no longer free but are fixed by the continuity equation and the equation of state respectively. Furthermore, the Euler equation then tells you exactly how much external force is required to set up the fluid flow. There are still \emph{two} freely specifiable functions which we can take to be the
two components of velocity.

In the physical acoustics regime the acoustic line-element is
\begin{eqnarray}
\d s^2&=&\left( \frac{\rho}{c} \right) 
        \Big[ -\left( c^2-v_{\hat r}^2-v_{\hat\phi}^2\right) \d t^2  
        - 2v_{\hat r}\; \d r \;\d t
        - 2 v_{\hat\phi} \; r \; \d\phi \;\d t
+\d r^2 + r^2 \d\phi^2 +\d z^2 \Big].
\end{eqnarray}
The major difference for physical acoustics is that for technical reasons the massless curved-space Klein--Gordon equation (d'Alembertian wave equation) can only be derived if the flow has zero vorticity. This requires $v_{\hat\phi}(r) =k_2/r$ and hence $f_{\hat\phi}=0$, so that the flow would be un-torqued.  More precisely, the d'Alembertian wave equation is a good approximation as long as the frequency of the wave is high compared to the vorticity. However, in the presence of significant torque and vorticity, a more complicated wave equation holds~\cite{Perez-Bergliaffa:2001lq}, but that wave equation requires additional geometrical structure beyond the effective metric, and so is not suitable for developing general relativity analogue models.  (Though this more complicated set of coupled PDEs is of direct physical interest in its own right.)~\footnote{We shall subsequently see that the ``equivalent Kerr vortex'' is not irrotational --- but the vorticity is proportional to the angular momentum, so there is a large parameter regime in which the effect of vorticity is negligible.}
%

\subsubsection{Special cases\label{Sec:Special.Cases.General.Vortex}}
There are several special cases of particular interest:
\begin{itemize}
\item[-] No vorticity, $\nabla\times\vec v=0$;
\item[-] No angular torque, $f_{\hat\phi}=0$;
\item[-] No radial flow $v_{\hat r}=0$;
\item[-] No radial force, $f_{\hat r}=0$;
\item[-] No external force, $\vec f = 0$;
\end{itemize}
which we will now explore in more detail.
%
\paragraph{Zero vorticity\,/\,zero torque}
If we assume zero vorticity, $\nabla\times\vec v=0$, the above calculation simplifies considerably, since then
\begin{equation}
v_{\hat{\phi}}= \frac{k_{2}}{r} \; ,
\end{equation} 
which implies $f_{\hat \phi}=0$. Conversely, if we assume zero torque, then (assuming $k_1\neq 0$) the vorticity is zero.

But note that the simple relationship \emph{zero torque} $\iff$ \emph{zero vorticity} requires the assumption of nonzero radial flow. With zero radial flow the torque is always zero for a time independent
flow, regardless of whether or not the flow is vorticity free.
%
\paragraph{Zero radial force:}
Assuming zero radial force, $f_{\hat r}=0$, and assuming $k_1\neq 0$, one finds
\begin{equation} \label{f_r_5}
{1\over r v_{\hat r}} \; \left[  {1\over 2} \partial_r [v_{\hat r}(r)^2] - \frac{v_{\hat \phi}(r)^2}{r} \right]
+ c^2(r) \; {\partial}_r \left( {1\over r v_{\hat r}}\right)  =0 .
\end{equation}
Thus the radial and angular parts of the background velocity are now 
dependent on each other. Once you have chosen, \emph{e.g.},
$v_{\hat{\theta}}(r)$ and $c(r)$, a differential equation constrains
$v_{\hat{r}}$:
\begin{equation} \label{conditon_zero_radial_force}
 v_{\hat{r}} \; \partial_r v_{\hat{r}} \left[1-
   \frac{c^2}{v_{r}^2}\right] 
=  \frac{c^2 + v_{\hat{\theta}}^2}{r} \; .
\end{equation} 

\paragraph{Zero external force:}
If we assume zero external force, $\vec f=0$, both the radial and
angular external forces are zero.  Then assuming $k_1\neq0$
\begin{equation}
v_{\hat{r}} \; \partial_r v_{\hat{r}} \left[1-  \frac{c^2}{v_{r}^2}\right] = 
\frac{c^2}{r}+\frac{k_{2}^2}{r^3}\; .
\end{equation} 
Since $c$ depends on $\rho$ via the barotropic equation, and $\rho$ depends on $v_r$ via the continuity equation, this is actually a rather complicated nonlinear ODE for $v_r(r)$. (For zero radial flow this reduces to the tautology $0=0$ and we must adopt the analysis of the previous subsection.)

In general the ergosurface is defined by the location where the flow goes supersonic
\begin{equation}
v_{\hat r}^2 + v_{\hat\phi}^2 = c^2
\end{equation}
while the horizon is defined by the location where
\begin{equation}
v_{\hat r}^2 =  c^2
\end{equation}
Note that horizons can form in three rather different ways:
\begin{itemize}
\item $v_{\hat r} = -c \neq 0$ --- a black hole horizon.
\item $v_{\hat r} = +c\neq 0$ --- a white hole horizon.
\item $v_{\hat r} = \pm c = 0$ --- a bifurcate horizon.
\end{itemize}
In analogue models it is most usual to keep $c\neq0$ and use fluid flow to generate the horizon, this is the case for instance in the Painl\'eve--Gullstrand version of the Schwarzschild geometry~\cite{Visser:1998gn,Matt.-Visser:2002ot}. The alternate possibility of letting both $v_{\hat r}\to0$ and $c\to0$ to obtain a horizon is not the most obvious construction from the point of view of acoustic geometries, but cannot \emph{a priori} be excluded on either mathematical or physical grounds. Indeed, it is this less obvious manner of implementing the acoustic geometry that most closely parallels the analysis of Schwarzschild black holes in curvature coordinates or isotropic coordinates, and we shall soon see that this route is preferred when investigating the Kerr spacetime. 
%
%
\section{Rotating black holes \label{Sec:Rotating.Black.Holes}}
%
To begin with we briefly summarize some aspects of rotating black holes, \eg, the line element for a rotating black hole in Boyer--Lindquist coordinates, the similarities and differences compared to non-rotating black holes (in Schwarzschild coordinates), and finally we point out some classical and quantum effects in the vicinity of rotating black holes that makes them interesting for the analogue model programme.

\subsection{Kerr black holes\label{Sec:Kerr.Black.Holes}}
The \emph{Kerr black hole}, named after the New Zealand mathematician\,/\,physicist Roy Kerr, is a solution for the vacuum Einstein equations for a rotating black hole. The Kerr black hole only depends on \emph{two} physical parameters, the total mass $m$, and the angular momentum $J$. From an astrophysical point of view the Kerr black hole is of immense importance, since any \emph{realistic} gravitational collapse is not spherically symmetric, and therefore leads to the formation of a black hole with non-zero angular momentum. An observer collapsing with the matter very rapidly approaches the ``center'' of the black hole, where a singularity is expected. From the perspective of a distant observer, the remnant of a collapsed star is a black hole only characterized by $m$ and $J$. Thus any information about the collapse, or the original star, is lost. It has been conjectured that the singularity resulting from either spherical or non-spherical collapse is concealed by the event horizon, and therefore the formation of naked singularities is conjectured to be prohibited, see the discussion of \emph{cosmic censorship} in Refs~\cite{Wald:1984rg,Misner:1973aa,DInverno:1992aa}.\\

The line element for a rotating black hole of mass $m$ and angular momentum $J$ in Boyer--Lindquist coordinates is given by
\begin{eqnarray}
ds^{2} =  &-& \left( 1 - \frac{2\, m\, r}{\rho^{2}} \right)\, c^{2} \, dt^{2} - \frac{4\, m\, a\, r\sin^{2}\theta}{\rho^{2}} \, d\phi \, dt \\ &+& \frac{\rho^{2}}{\Delta} \, dr^{2} + \rho^{2} \, d\theta^{2}  
\label{Eq:Kerr.Boyer.Lindquist}
+ \left( r^{2} + a^{2} + \frac{2\, m\, r\, a^{2} \, \sin^{2} \theta}{\rho^{2}} \right) \, \sin^{2}\theta \, d\phi^{2} \, .
\end{eqnarray} 
Here $a$ represents the specific angular momentum of the black hole, 
\begin{equation}
a = J/m \, ,
\end{equation}
the ratio between the angular momentum $J$, and the total mass $m$, and we also define the two quantities $\rho$ and $\Delta$ as functions of $a$, $m$, $r$ and $\theta$:
\begin{eqnarray}
\rho^{2} &=&  r^{2} + a^{2} \cos^{2}\theta \, ;  \\
\Delta &=& r^{2} - 2\, m\, r +a^{2} \, .
\end{eqnarray}

As already pointed out in the beginning of the section, from the perspective of an observer following the collapse all matter will fall towards the \emph{intrinsic} singularity.  Evaluating 
\begin{equation}
\label{Eq:Riemann.Invariant}
R^{abcd} R_{abcd} = \frac{48 \, m^{2} \, (r^{2}-a^{2} \, \cos^{2}\theta) \, [\rho^{2} - (4\,r\,a\,\cos\theta)^{2}]}{\rho^{12}} 
\end{equation}
explicitly, it can be seen that the Riemann invariant is infinite at $\rho^{2}=0$. Thus the singularity for a Kerr black hole is associated with
\begin{equation}
\rho^{2} \equiv  r^{2} + a^{2} \cos^{2}\theta = 0 \, .
\end{equation}
While for non-rotating black holes the singularity lies at a particular point $r=0$, we now encounter a \emph{ring singularity} at $r=0$, $\theta=\pi/2$. In terms of ``cartesian'' coordinates 
\begin{eqnarray}
x&=&\sqrt{r^{2}+a^{2}} \, \sin\theta\, \cos\phi \, , \\
y&=&\sqrt{r^{2}+a^{2}} \, \sin\theta \, \sin \phi \, , \\ 
z&=&r \,\cos\theta \, ,
\end{eqnarray}
the ring singularity is in the $z=0$ plane for $x^{2}+y^{2}=a^{2}$,
in the $z=0$ plane for $x^{2}+y^{2}=a^{2}$. Note, in cases where $a \to 0$ the ring singularity is shrinking to a point singularity at $r=0$, in agreement with the non-rotating case.

%
\begin{figure}[ht]
 \begin{center}
 \input{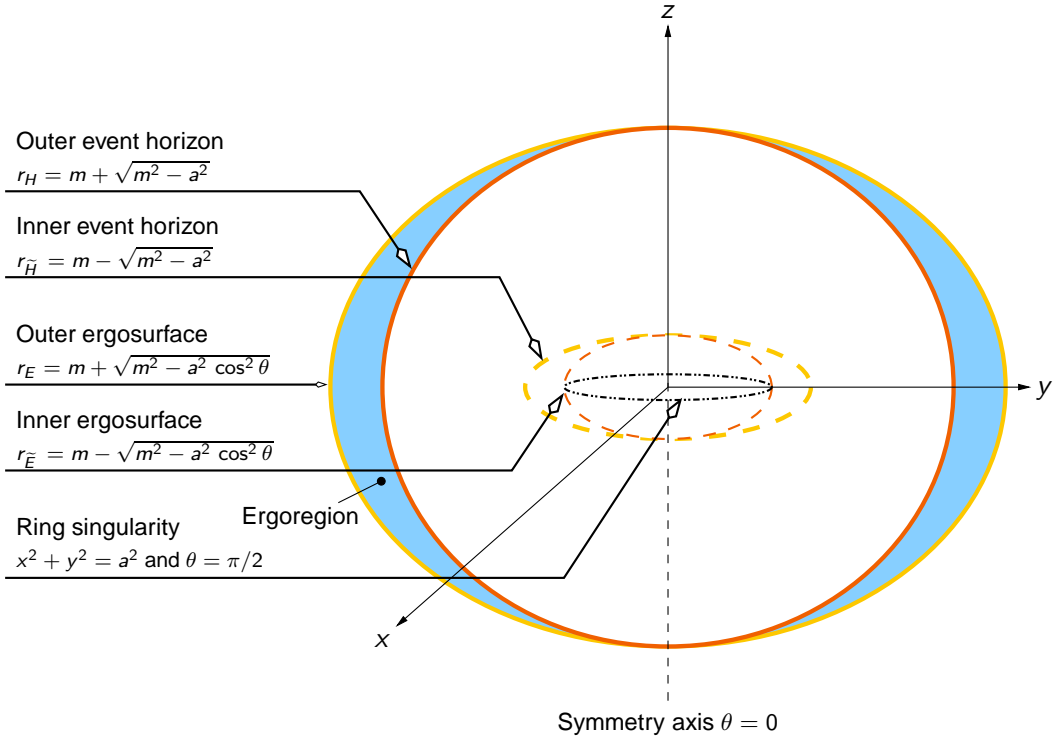}
 \caption[Rotating Kerr black hole.]  {\label{Fig.Kerr}
 The Kerr solution for a rotating black hole exhibits various surfaces of interest~\cite{DInverno:1992aa}.}
 \end{center}
\end{figure}
%

The coordinate singularity, $g_{11}=\infty$ is related to
\begin{equation}
\Delta \equiv r^{2} - 2\, m\, r +a^{2} = 0 \, ,
\end{equation}
such that we obtain from $\Delta = 0$ two surfaces of infinite red-shift, 
\begin{equation}
\label{Eq:Outer.Inner.Event.Horizon}
r_{\pm}=m \pm \sqrt{m^{2} -a^{2}} . 
\end{equation}
Here $r_{H}=r_{+}$ denotes the \emph{outer event horizon}, and $r_{\widetilde{H\,}}=r_{-}$ the \emph{inner event horizon}. Notice that for $a \to 0$ the outer event horizon simplifies to $r_{H} \to 2m$ and $r_{\widetilde{H \,}} \to 0$. This is also in agreement with a non-rotating black hole, where we expect a single event horizon at $2m$.

It can be seen that for $a \to 0$ 
\begin{equation}
ds^{2} = - \left( 1 - \frac{2m}{r} \right) \, dt^{2} + \left( 1 - \frac{2m}{r} \right)^{-1} \, dr^{2} + r^{2} \, \left(d\theta^{2} + \sin^{2} \theta \, d\phi^{2}   \right) \, ,
\end{equation}
the Kerr solution takes the form of a non-rotating black hole in Schwarzschild coordinates. Note also that for $m \to 0$ both --- the Schwarzschild and the Kerr --- line elements represent flat Minkowski spacetimes.
While this is obvious for the Schwarzschild line element, it is less obvious for the Kerr solution, unless we consider the limit $m \to 0$ of the Riemann invariant~\refb{Eq:Riemann.Invariant}, $R^{abcd} R_{abcd} \to 0$, or more precisely not that in an orthonormal frame $R_{\hat{a}\hat{b}\hat{c}\hat{d}}\to 0$. \\

As the Kerr solution is an extension of the non-rotating black hole, we expect similar classical and quantum effects to appear. However, the Kerr black hole also offers one completely \emph{new} feature that does not appear in non-rotating black holes, the \emph{ergosurface}.

The \emph{ergoregion} is the region bounded between the outer event horizon and the \emph{ergosurface}. The latter is determined by the coordinate condition $g_{00}=0$, therefore at 
\begin{equation}
r_{\pm}= m \pm \sqrt{m^{2} - a^{2} \, \cos^{2}\theta}. 
\end{equation}
Again, we obtain two surfaces, the \emph{outer ergosphere} at $r_{E}=r_{+}$, and the \emph{inner ergosphere} at $r_{\widetilde{E\,}}=r_{-}$.
The relationship between the event horizons and ergospheres is as follows $r_{\widetilde{E\,}} < r_{\widetilde{H\,}} < r_{H}  \leq r_{E}$, see Eq.~\refb{Eq:Outer.Inner.Event.Horizon}. We illustrated the different layers of a rotating black hole in Fig.~\ref{Fig.Kerr}. (Note, that for $a \to 0$ the outer event horizon and the ergosurface are on top of each other, consequently a non-rotating black hole \emph{cannot} exhibit an ergoregion.)

While from an astrophysical point of view the Kerr black hole is of immense interest as the remnant of a collapsed star, the acoustic analogue focuses on the classical and quantum effects in the vicinity of the \emph{ergoregion}.

\subsection{The ergosphere \label{Sec:Ergosphere}}
The \emph{ergosphere} represents a spacetime region where the angular velocity of the rotating black hole is high enough to ``drag the surrounding space along with the velocity of light''. Any observer or particle entering this region can no longer remain in a non-rotating orbit --- regardless of how much force is applied. It will be dragged along with the rotating spacetime. However, as the observer\,/\,object is still outside the event horizon $r \geq r_{+}$, it is in principle possible to escape to infinity. \\

In 1969 Roger Penrose~\cite{Penrose:1969pc} discovered that it is possible to extract energy from rotating black holes. This mechanism is referred to as the \emph{Penrose effect} (its field theory analogue is referred to as \emph{superradiant scattering}) and can be understood as follows~\cite{Wald:1984rg}.
%

%
\begin{figure}[ht]
 \begin{center}
 \input{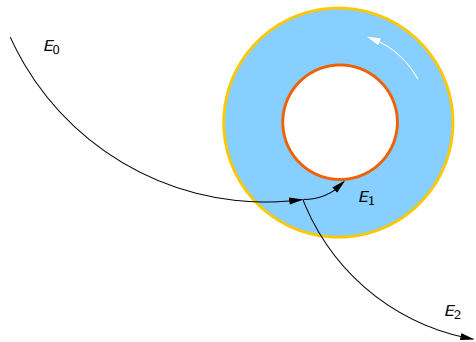}
 \caption[Penrose Process.]  {\label{Fig.Superradiance}
 Illustration of the superradiance scattering in the ergoregion of a rotating black hole. (The figure shows a projection of the black hole onto some $\theta=\mbox{constant}$ plane, for $\theta = \pi/2$.)}
 \end{center}
\end{figure}
%
Consider a particle --- starting far away from the rotating black hole --- that is freely falling into the ergoregion. 
The Kerr geometry is time-independent, and therefore posesses a time like Killing vector field $\xi^{a}$. 
It is possible to establish a relationship between time translation in the Kerr geometry, and a conserved and well-defined energy far away (due to asymptotic flatness) from the black hole, such that 
$E_{0}=-p_{a}\xi^{a}$; see Noether's theorem for example in~\cite{Peskin:1995aa} . 
Here $p^{a}$ is the $4$-momentum of the test particle. As pointed out in the introduction, there are no forces acting on a freely falling particle, so that the energy remains constant when the particle is approaching the black hole. The particle has been prepared such that it will be broken up into two fragments (\eg, employing explosives and a timing device), once it enters the ergoregion. Conservation of energy-momentum $p_{0}^{a} = p_{1}^{a} + p_{2}^{a}$, shows that $E_{0}=E_{1} + E_{2}$. The break-up is arranged so that, $J_{\mathrm{particle}}/m_{\mathrm{particle}} < J/m$, implying that one piece has negative energy, $E_{1}<0$, and the corresponding particle falls into the black hole (crosses the event horizon), see Fig.~\ref{Fig.Superradiance}. The remaining piece will leave the ergoregion with greater energy than it entered, $E_{0}> E_{2}$. The difference $\vert E_{1} \vert $ has been subtracted from the black hole. In~\cite{Wald:1984rg} it is also explained how much energy can be subtracted, and that this process will not convert the Kerr black hole into a naked singularity. \\

The Penrose effect describes the behavior of classical objects in the vicinity of ergosphere. \emph{Superradiance} describes the corresponding quantum process. In some sense the Penrose effect occurs already at the level of classical wave-packets, while the superradiance effect shows particle enhancement already for a single-mode analysis in terms of non-interacting vacuum fluctuations.\\

Altogether the rotating black hole exhibits two interesting quantum effects. The existence of the outer horizon at $r_{H}$ guaranties Hawking radiation. Similar to the Hawking radiation from non-rotating black holes, the Kerr black hole has a thermal radiation associated with the surface gravity at $r_{H}$.
Beyond that rotating black holes feature the Penrose process and superradiance~\cite{Williams:1995aa,Winstanley:2001aa,Glampedakis:2001aa,Williams:2004aa,Lepe:2005aa,Gariel:2007aa,Jiang:2007aa}. 

In many cases Hawking radiation and superradiance are expected to be small enough not to back-react significantly with the gravitational field, so that curved-spacetime quantum-field-theory is a sufficiently good approximation to cover the quantum effects without (explicitly) invoking the Einstein equation. 
Effective superradiant scattering has been studied in the context of analogue spacetime in~\cite{Basak:2003aa,Basak:2003ab,Basak:2005aa,Slatyer:2005aa}.

To conclude the introduction, we notice that \emph{in principle} both effects are expected to occur at the level of the acoustic analog for a rotating black hole. \\

Before we turn to study the possibility of finding an acoustic analogue for the Kerr black hole, we would like to stress that a special slicing, where we take $\theta = \mbox{constant}$, but $\theta \neq 0$, contains all the important features --- ergosurfaces and horizons --- of the rotating black hole. (Best is to choose $\theta = \pi/2$, where the radial distance between the ergosurface and outer horizon, $r_{E} - r_{H}$, is maximal.) 

\subsection{The Kerr equator\label{Sec:Kerr.Black.Holes.Equator}}
To compare the vortex acoustic geometry to the physical Kerr geometry
of a rotating black hole~\cite{Kerr:1963aa}, consider the equatorial slice
$\theta=\pi/2$ in Boyer--Lindquist coordinates~\cite{Boyer:1967aa}:
\begin{equation}
(\d s^2)_{(2+1)} = -\d t^2 + {2m\over r} (\d t - a \;\d\phi)^2 
+{\d r^2\over 1-2m/r+a^2/r^2} +  (r^2+a^2)\;\d\phi^2.
\end{equation}
We would like to put this into the form of an ``acoustic metric''
\begin{equation}
g_{\mu\nu} = \left({\rho\over c}\right)\;
\left[\begin{array}{c|c}{-\{c^2-h^{mn}\;v_n\;v_n}\} & -v_j\\
\hline
-v_i & h_{ij} 
\end{array}\right].
\end{equation}
If we look at the 2-d $r$-$\phi$ plane, the metric is
\begin{equation}
(\d s^2)_{(2)} = 
{\d r^2\over 1-2m/r+a^2/r^2} +  \left(r^2+a^2+ {2ma^2\over r}\right)\d\phi^2.
\end{equation}
Now it is well-known that any 2-d geometry is locally conformally
flat~\cite{Wald:1984rg}, though this fact is certainly not manifested in these
particular coordinates.  Introduce a new radial coordinate $\tilde r$
such that:
\begin{equation}
{\d r^2\over 1-2m/r+a^2/r^2} +  \left(r^2+a^2+ {2ma^2\over r}\right)\d\phi^2
=
\Omega(\tilde r)^2 \; [\d\tilde r^2 +\tilde r^2\;\d\phi^2].
\end{equation}
This implies
\begin{equation}
 \left(r^2+a^2+ {2ma^2\over r}\right) = \Omega(\tilde r)^2 \; \tilde r^2,
\end{equation}
and
\begin{equation}
{\d r^2\over 1-2m/r+a^2/r^2} = \Omega(\tilde r)^2\; \d\tilde r^2,
\end{equation}
leading to the differential equation
\begin{equation}
{1\over \tilde r(r)}\; {\d \tilde r(r)\over\d r} =
 {1\over \sqrt{1-2m/r+a^2/r^2} \sqrt{r^2+a^2+ {2ma^2/ r}}},
\end{equation}
which is formally solvable as
\begin{equation}
\tilde r(r) = \exp\left\{
\lint  {\d r\over \sqrt{1-2m/r+a^2/r^2} \sqrt{r^2+a^2+ {2ma^2/r}}}
\right\}.
\end{equation}
The normalization is most easily fixed by considering the $m=0=a$
case, in which case $\tilde r=r$, and then using this to write the
general case as
\begin{equation}
\tilde r(r) = r \exp\left[ -
\int_r^\infty  \left\{
{1\over \sqrt{1-2m/\bar r+a^2/\bar r^2} \sqrt{\bar r^2+a^2+ {2ma^2/\bar r}}} 
-{1\over \bar r} \right\} \d\bar r
\right],
\end{equation}
where $\bar r$ is simply a dummy variable of integration.  If $a=0$
this integral can be performed in terms of elementary functions
\begin{equation}
\int  {\d r\over r \sqrt{1-2m/r}} = \ln \left( \sqrt{r^2 - 2mr} + r - m \right)
\qquad\qquad [r>2m],
\end{equation}
so that
\begin{equation}
\tilde{r} (r)= \half \left( \sqrt{r^2 - 2mr} + r - m \right),
\end{equation}
though for $m\neq 0$ and general $a$ no simple analytic form holds.
Similarly, for $m=0$ and $a\neq 0$ it is easy to show that
\begin{equation}
\tilde r(r) = \sqrt{r^2+a^2}, 
\end{equation}
though for $a\neq 0$ and general $m$ no simple analytic form holds.  

Nevertheless, since we have an exact [if formal] expression for
$\tilde r(r)$ we can formally invert it to assert the existence of a
function $r(\tilde r)$. It is most useful to write $\tilde r=r
\;F(r)$, with $\lim_{r\to\infty}F(r)=1$, and to write the inverse
function as $r = \tilde r \; H(\tilde r)$ with the corresponding limit
$\lim_{\tilde r\to\infty}H(\tilde r)=1$. Even if we cannot write
explicit closed form expressions for $F(r)$ and $H(\tilde r)$ there is
no difficulty in calculating them numerically, or in developing series
expansions for these quantities, or even in developing graphical
representations.

We now evaluate the conformal factor as
\begin{equation}
\Omega(\tilde r)^2 = 
{r^2+a^2+ {2ma^2/ r} \over \tilde r^2} = 
H(\tilde r)^{2} \; \left(1+ {a^2\over r^2} + {2ma^2\over r^3}\right),
\end{equation}
with $r$ considered as a function of $\tilde r$, which now yields
\begin{equation}
(\d s^2)_{(2+1)} = -\d t^2 + {2m\over r} (\d t^2  - 2a \;\d\phi\;\d t) +
\Omega(\tilde r)^2 \; [\d\tilde r^2 +\tilde r^2\;\d\phi^2].
\end{equation}
Equivalently
\begin{eqnarray}
(\d s^2)_{(2+1)} &=& \Omega(\tilde r)^2 
\Bigg\{ -\Omega(\tilde r)^{-2} \left[1- {2m\over r}\right]\d t^2  
-\Omega(\tilde r)^{-2}  {4am\over r} \;\d\phi\;\d t 
+ [\d\tilde r^2 +\tilde r^2\;\d\phi^2] \Bigg\}.
\end{eqnarray}

%
\section{Acoustic Kerr equator \label{Sec:Acoustic.Kerr.Equator}}
%
This now lets us pick off the coefficients of the equivalent acoustic
metric.  For the overall conformal factor
\begin{equation}
{\rho\over c} = \Omega(\tilde r)^2 = 
H^2(\tilde r) \; \left(1+ {a^2\over r^2} + {2ma^2\over r^3}\right).
\end{equation}
For the azimuthal ``flow''
\begin{equation}
v_\phi =  \Omega(\tilde r)^{-2} \; {2am\over r} 
=  {2am\over r} \; H^{-2}(\tilde r) \; 
\left(1+ {a^2\over r^2} + {2ma^2\over r^3}\right)^{-1}.
\end{equation}
In terms of orthonormal components
\begin{equation}
v_{\hat\phi} = {v_\phi\over\tilde r}  =  
  {2am\over r^2} \; H^{-1}(\tilde r) \; 
\left(1+ {a^2\over r^2} + {2ma^2\over r^3}\right)^{-1}.
\end{equation}
This is, as expected, a vortex geometry. 
Finally for the ``coordinate speed of light'', corresponding to the
speed of sound in the analogue geometry
\begin{eqnarray}
c^2 &=& \Omega(\tilde r)^{-2} \;\left[1- {2m\over r}\right] + 
\Omega(\tilde r)^{-4} \; {4a^2m^2\over\tilde r^2 \; r^2}, 
\\
&=&  \Omega(\tilde r)^{-4} \left\{ 
\Omega(\tilde r)^{2}\;\left[
1- {2m\over r}\right] +  \; {4a^2m^2\over\tilde r^2 \; r^2} 
\right\}.
\end{eqnarray}
The speed of sound can be rearranged a little
\begin{equation}
c^2 
=  \Omega(\tilde r)^{-4} \;H^2(\tilde r) \; \left\{ 
\left[1+ {a^2\over r^2} + {2ma^2\over r^3}\right]
\;\left[1- {2m\over r}\right] +  \; {4a^2m^2\over r^4} 
\right\}.
\end{equation}
This can now be further simplified to obtain
\begin{equation}
c^2 
=  \Omega(\tilde r)^{-4} \;H^2(\tilde r) \; 
\left\{ 1- {2m\over r} + {a^2\over r^2}
\right\},
\end{equation}
and finally leads to
\begin{equation}
\rho  = H(\tilde r) \; \sqrt{ 1- {2m\over r} + {a^2\over r^2} },
\end{equation}
with $r$ implicitly a function of $\tilde r$. Note that the velocity field of the ``equivalent Kerr vortex'' is not irrotational. Application of Stokes' theorem quickly yields
\begin{equation}
\omega
= {1\over 2\pi \tilde r} {\partial\over\partial\tilde r}\left(2\pi \tilde r \; v_{\hat\phi} \right)
= {1\over \tilde r} {\partial\over\partial\tilde r}\left( v_{\phi} \right)
= {1\over \tilde r} {\partial\over\partial\tilde r}\left[ {2am \tilde r^2\over r^3} \; 
\left(1+ {a^2\over r^2} + {2ma^2\over r^3}\right)^{-1}\right].
\end{equation}
which we can somewhat simplify to
\begin{equation}
\omega
= {4am \over r^3} \; 
\left(1+ {a^2\over r^2} + {2ma^2\over r^3}\right)^{-1} 
+ {\Omega\;\tilde r}\; \sqrt{1-2m/r+a^2/r^2} \;
 {\partial\over\partial r}\left[ {2am \over r^3} \; 
\left(1+ {a^2\over r^2} + {2ma^2\over r^3}\right)^{-1}\right]. 
\end{equation}
Note that (as should be expected), the vorticity is proportional to an 
overall factor of $a$, and so is proportional to the angular momentum 
of the Kerr geometry we are ultimately interested in.

Note that $\tilde r$ is the radial coordinate in which the space part of
the acoustic geometry is conformally flat, so that $\tilde r$ is the
``physical'' radial coordinate that corresponds to distances measured
in the laboratory where the vortex has been set up. Unfortunately this particular
radial coordinate is also mathematically rather difficult to work
with.  For some purposes it is more useful to present the coefficients
of the acoustic metric as functions of $r$, using the relationship
$F(r) = 1/H(\tilde r)$. Then we have:
\begin{equation}
{\rho(r)\over c(r)} = \Omega^2(r) = 
F^{-2}(r) \; \left(1+ {a^2\over r^2} + {2ma^2\over r^3}\right).
\end{equation}
\begin{equation}
v_{\hat\phi}(r)
=   {2am\over r^2} \; F(r) \; 
\left(1+ {a^2\over r^2} + {2ma^2\over r^3}\right)^{-1}.
\end{equation}
\begin{equation}
c(r)
=  \sqrt{ 1- {2m\over r} + {a^2\over r^2} } \; F(r) \;   
\left(1+ {a^2\over r^2} + {2ma^2\over r^3}\right)^{-1}.
\end{equation}
Finally this yields
\begin{equation}
\rho(r) = F^{-1}(r) \; \sqrt{ 1- {2m\over r} + {a^2\over r^2} },
\end{equation}
and the explicit if slightly complicated result that
\begin{equation}
F(r) = \exp\left[ -
\int_r^\infty  \left\{
{1\over \sqrt{1-2m/\bar r+a^2/\bar r^2} \sqrt{\bar r^2+a^2+ {2ma^2/\bar r}}} 
-{1\over \bar r} \right\} \d\bar r
\right].
\end{equation}
One of the advantages of writing things this way, as functions of $r$,
is that it is now simple to find the locations of the horizon and
ergosphere. 

The ergo-surface is defined by $c^2-v^2=0$, equivalent to the
vanishing of the $g_{tt}$ component of the metric. This occurs at
\begin{equation}
r_E=2m.
\end{equation}
The horizon is determined by the vanishing of $c$, (recall that $v_{\hat
  r}\equiv0$), this requires solving a simple quadratic with the result
\begin{equation}
r_H =  m+ \sqrt{m^2-a^2} < r_E.
\end{equation}
These results agree, as of course they must, with standard known
results for the Kerr metric. It is easy to check that $F(r_H)$ is
finite (the function inside the exponential is integrable as long as
the Kerr geometry is non-extremal), and that $F(r_E)$ is finite. This
then provides some simple consistency checks on the geometry:
\begin{eqnarray}
\rho(r_H)&=&0 \, =  \, c(r_H);
\\
\Omega(r_H)&=& \hbox{finite};
\\
F(r_H) &=& G(r_H)^{-1} =  \hbox{finite};
\\
\Omega(r_H) \; v_{\hat\phi}(r_H) &=& {a\over r_H};
\\
\omega(r_H) &=& {a\over 2 m r_H};
\\
c(r_E)&=& |v_{\hat\phi}(r_E)|.
\end{eqnarray}
In particular if we compare the vorticity at the horizon, 
$\omega(r_H) = {a/( 2 m r_H)}$, with the surface gravity 
\begin{equation}
\kappa_H = {2m\over r_H^2} -{2a^2\over r_H^3},
\end{equation}
we find a large range of parameters for which the peak frequency 
of the Hawking spectrum is high compared to the frequency scale 
set by the vorticity --- this acts to suppress any complications 
specifically coming from the vorticity. (Of course in the extremal 
limit, where $r_H\to 0$, the vorticity at the horizon becomes infinite.
In the extremal limit one would need to make a more careful analysis 
of the effect of nonzero vorticity on the Hawking radiation.)

From the point of view of the acoustic analogue the region inside the
horizon is unphysical.  In the unphysical region the density of the 
fluid is zero and the concept of sound meaningless.  In the physical 
region outside the horizon the flow has zero radial velocity, and zero torque, but
is \emph{not} irrotational.  Since by fitting the equatorial slice of
Kerr to a generic acoustic geometry we have fixed $\rho(r)$ and
$c^2(r)$ as functions of $r$ [and so also as functions of $\tilde r$]
it follows that $p(r)$ is no longer free, but is instead determined by
the geometry. From there, we see that the EOS $p(\rho)$ is determined,
as is the external force $f_{\hat r}(r)$.  The net result is that we
can [in principle] simulate the Kerr equator exactly, but at the cost
of a very specific \emph{fine-tuning} of both the equation of state
$p(\rho)$ and the external force $f_{\hat r}(r)$.

\begin{figure}[htbp]
\begin{center}
\mbox{
\subfigure[$\,$ $a=1$ ; $m=2$. \label{Fig:}]{\includegraphics[width=0.45\textwidth]{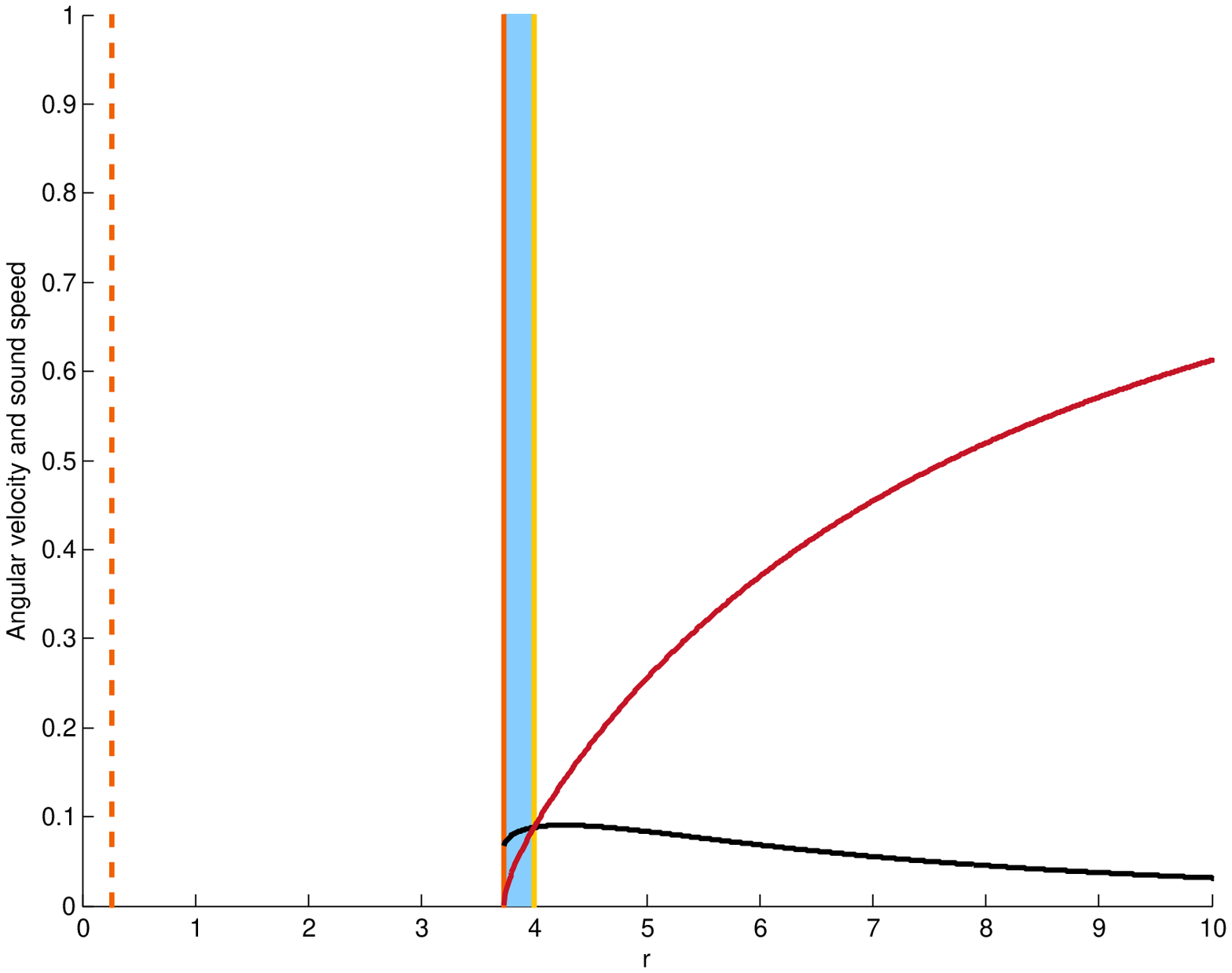}}
\hspace{0mm}
\subfigure[$\,$ $a=2$ ; $m=2$. \label{Fig:}]{\includegraphics[width=0.45\textwidth]{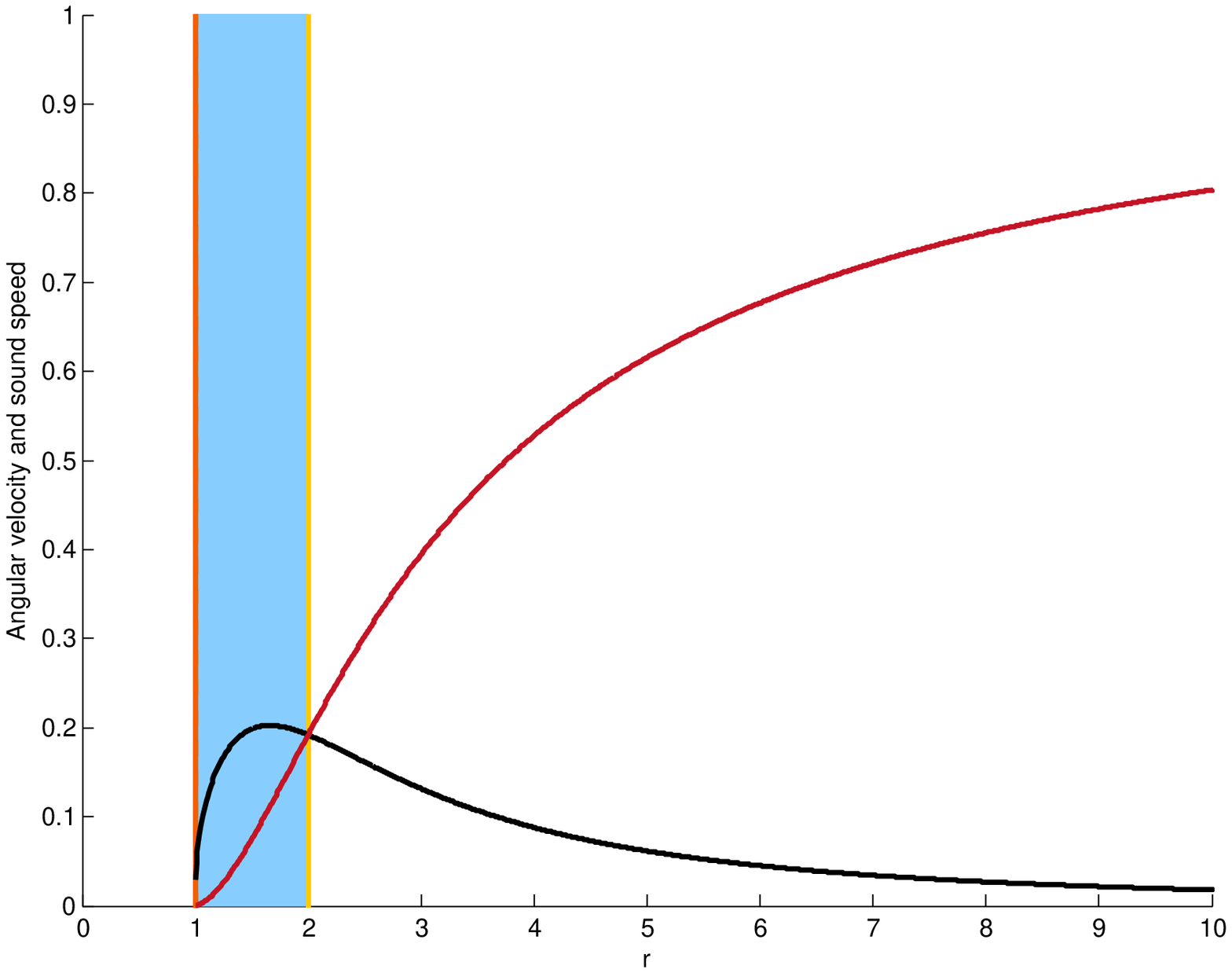}}
}
\end{center}
\caption[Sound speed and rotational background velocity in acoustic Kerr geometry.]{
{ Illustrative graph of the speed of sound (red) and rotational velocity (black)
  of the vortex as a function of $r$. }}
\label{F:cv}
\end{figure}

In Fig.~\ref{F:cv} we present an illustrative graph of the speed of
sound and velocity of the fluid.  Key points to notice are that the
speed of sound (red curve) and velocity of flow (black curve)
intersect on the ergosurface (golden line), that the speed of sound
goes to zero at the horizon (orange line), and that the velocity of
fluid flow is finite on the horizon. Finally the inner horizon is
represented by the dashed orange vertical line. While the region inside the
outer horizon is not physically meaningful, one can nevertheless
mathematically extend some (not all) of the features of the flow
inside the horizon.  This is another example of the fact that analytic
continuation across horizons, which is a standard tool in general
relativity, can sometimes fail for physical reasons when one considers
Lorentzian geometries based on physical models that differ from
standard general relativity. This point is more fully discussed in
\cite{Barcelo:2004aa,Jacobson:1998tp,Jacobson:2002ab}.

\begin{figure}[htbp]
\mbox{
\subfigure[$\,$ $a=1$ ; $m=2$. \label{Fig:}]{\includegraphics[width=0.45\textwidth]{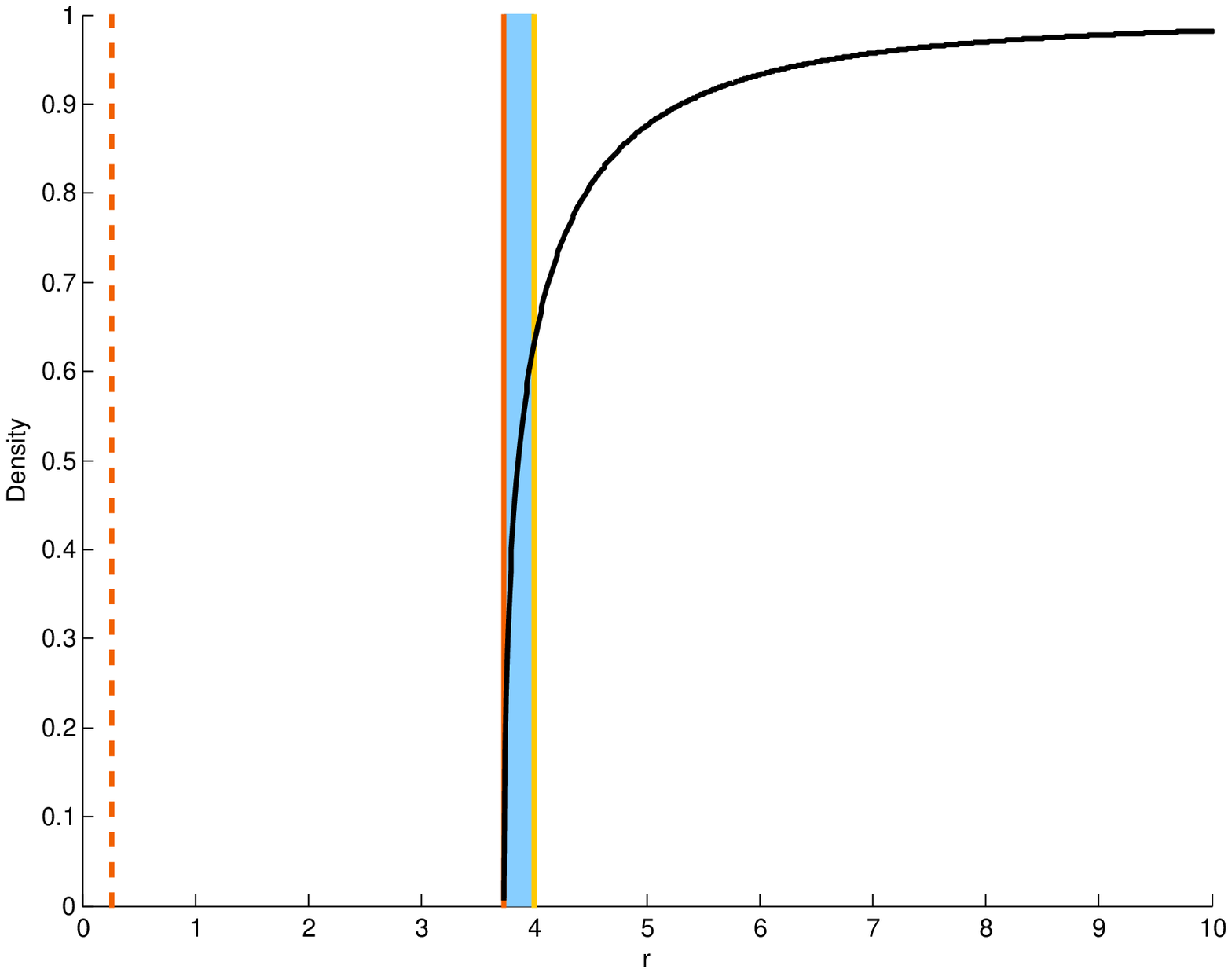}}
\hspace{0mm}
\subfigure[$\,$ $a=2$ ; $m=2$. \label{Fig:}]{\includegraphics[width=0.45\textwidth]{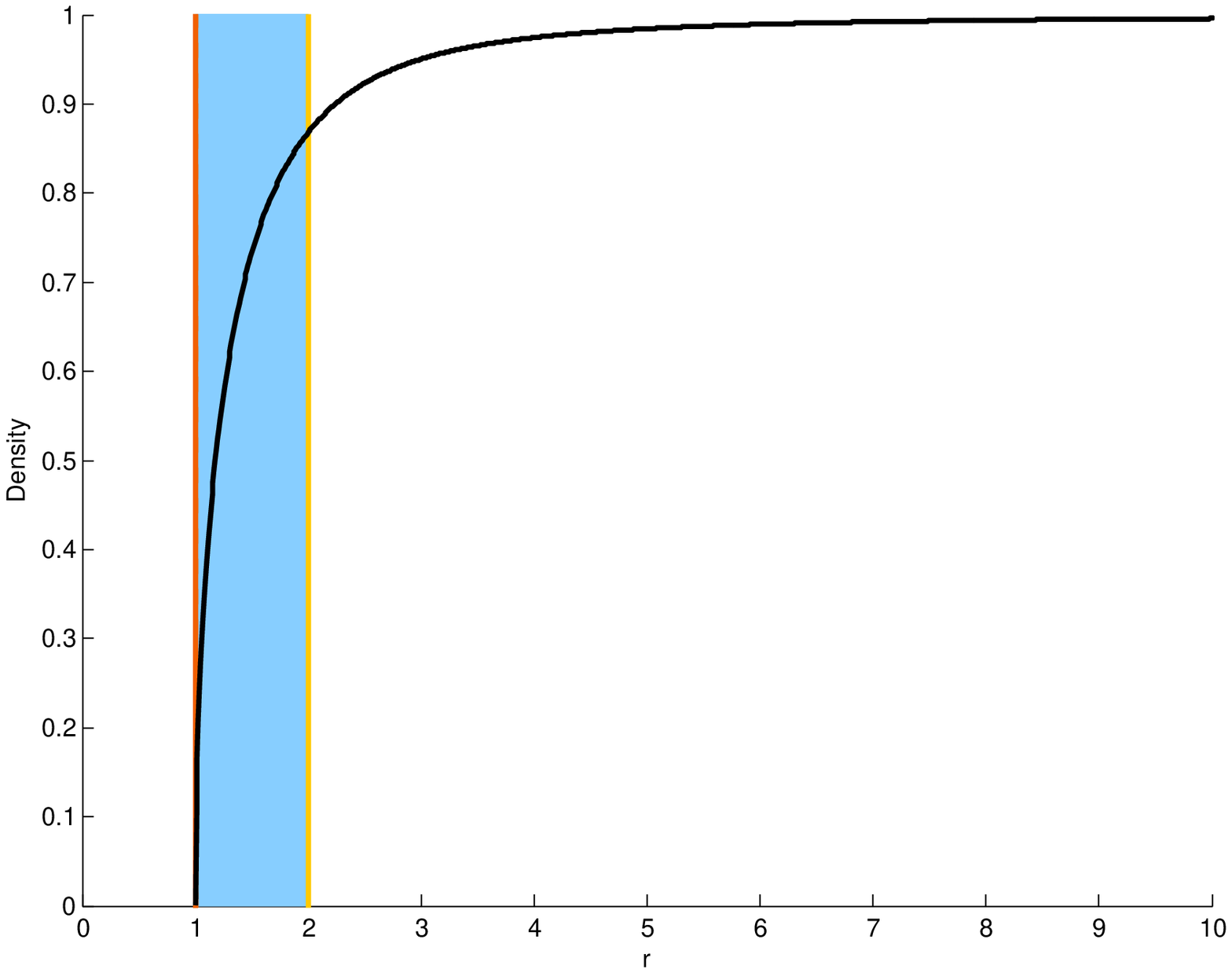}}
}
\caption[Fluid density in acoustic Kerr geometry.]{
 { Illustrative graph of the fluid density as a function of $r$. }}
\label{F:density}
\end{figure}

\begin{figure}[htbp]
\mbox{
\subfigure[$\,$ $a=1$ ; $m=2$. \label{Fig:}]{\includegraphics[width=0.45\textwidth]{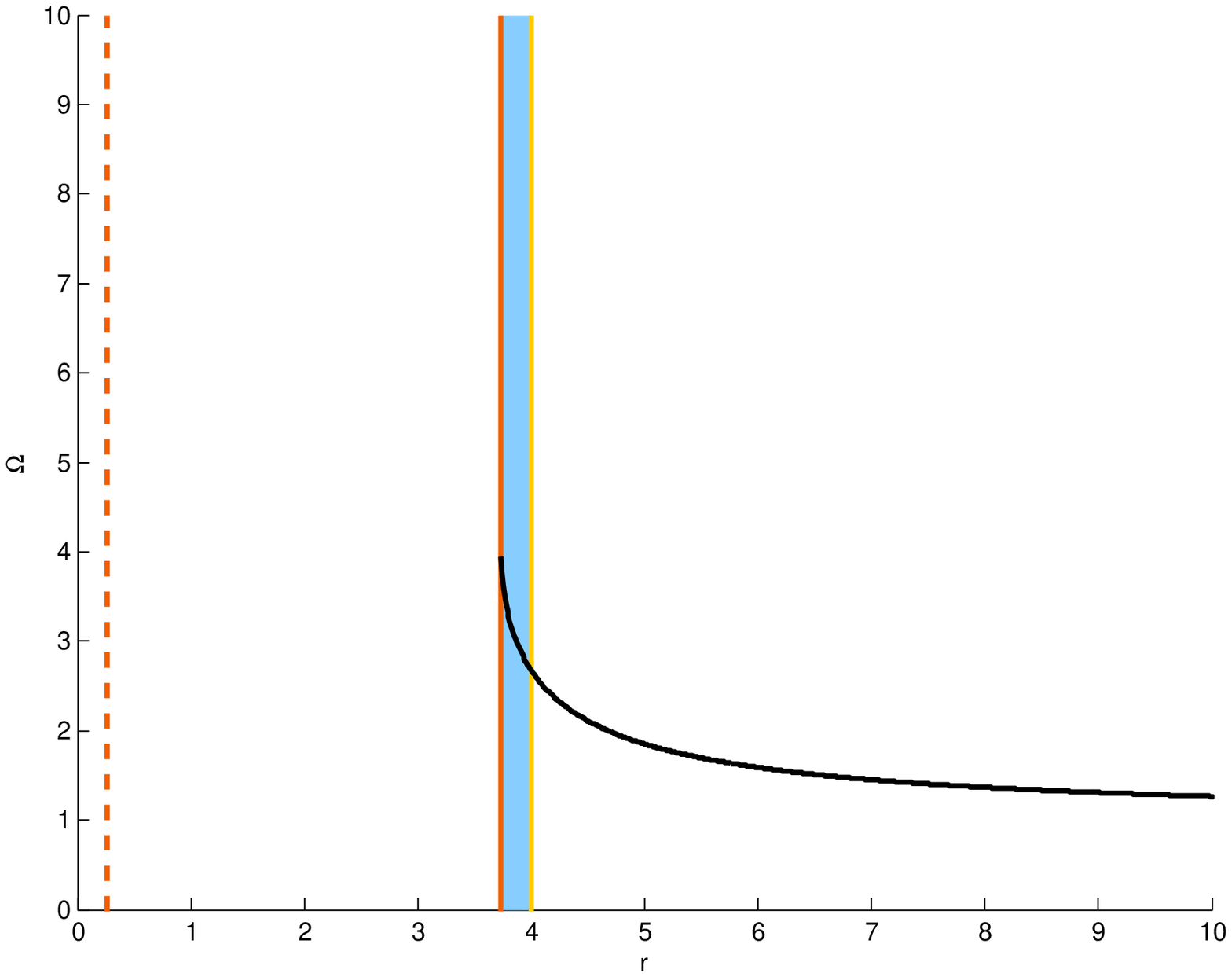}}
\hspace{0mm}
\subfigure[$\,$ $a=2$ ; $m=2$. \label{Fig:}]{\includegraphics[width=0.45\textwidth]{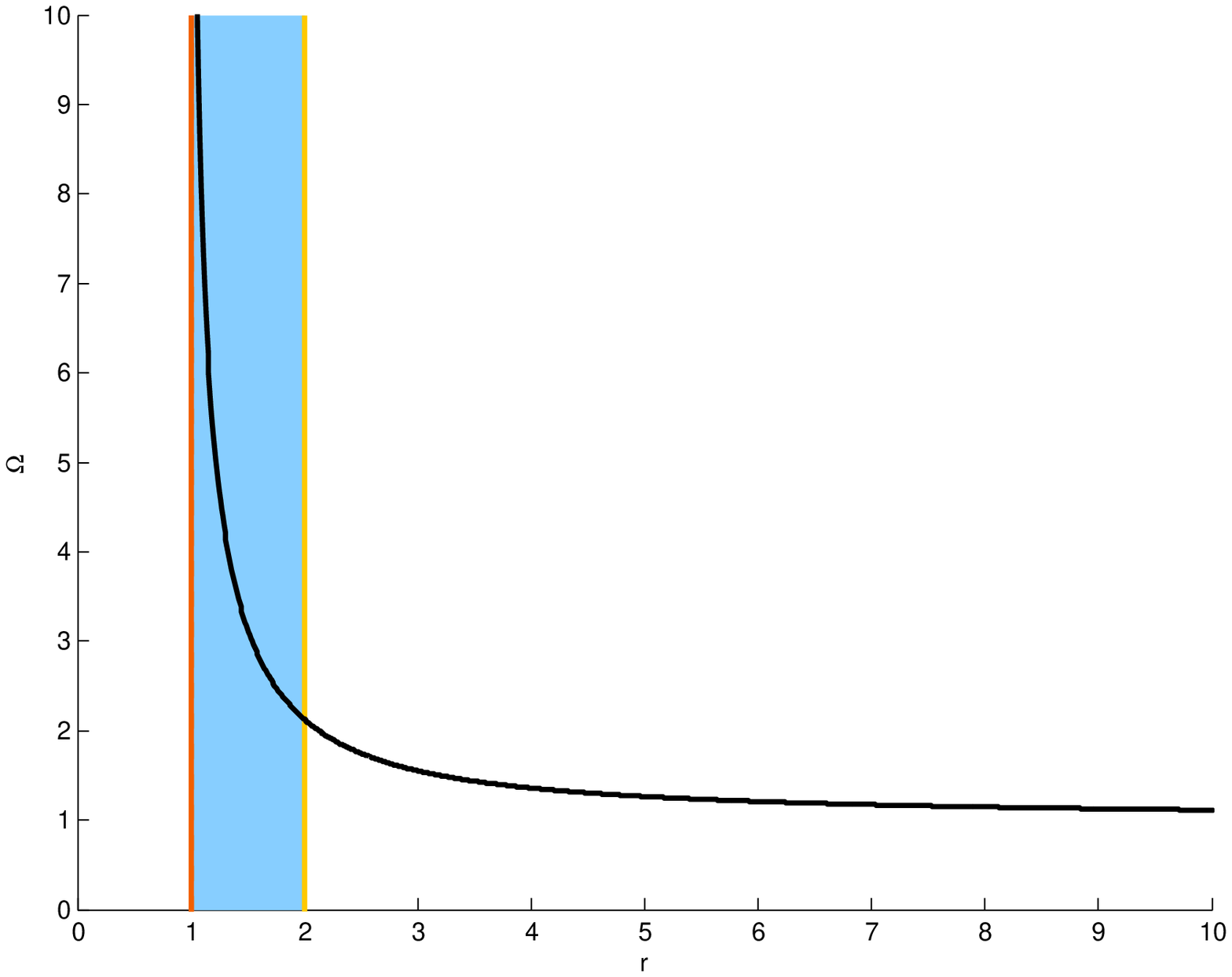}}
}
\caption[Conformal factor in acoustic Kerr geometry.]{
 { Illustrative graph of the conformal factor $\Omega$ as a function of $r$. }}
\label{F:Omega}
\end{figure}

In Fig.~\ref{F:density} we present an illustrative graph of the
fluid density. Note that the density asymptotes to a constant at large
radius, and approaches zero at the horizon. Finally, in figure
\ref{F:Omega} we present an illustrative graph of the conformal factor
$\Omega$, which remains finite at the horizon and ergosurface, and
asymptotes to unity at large distances from the vortex core. 
Figures~\ref{F:cv}--\ref{F:Omega} all correspond to $(a)$ a ``normal'' rotating black hole for $m=2$ and $a=1$ and $(b)$ an \emph{extremal} black hole, where $m=2=a$.

%
\section{Discussion\label{Sec:Discussion.Kerr.Vortex}}		
%
We have shown that the Kerr equator can [in principle] be exactly
simulated by an acoustic analogue based on a vortex flow with a very
specific equation of state and subjected to a very specific external
force. Furthermore we have as a result of the analysis also seen that
such an analogue would have to be very specifically and deliberately
engineered. Thus the results of this investigation are to some extent
mixed, and are more useful for theoretical investigations, and for the
gaining of insight into the nature of the Kerr geometry, than they are
for actual laboratory construction of a vortex simulating the Kerr
equator.

One of the technical surprises of the analysis was that the
Doran~\cite{Doran:2000aa,Hamilton:2004aa} coordinates (the natural generalization of the
Painl\'eve--Gullstrand coordinates that worked so well for the
Schwarzschild geometry) did not lead to a useful acoustic metric even
on the equator of the Kerr spacetime.  Ultimately, this can be traced
back to the fact that in Doran coordinates the space part of the Kerr
metric is non-diagonal, even on the equator, and that no simple
coordinate change can remove the off-diagonal elements. In
Boyer--Lindquist coordinates however the space part of the metric is
at least diagonal, and the $\tilde r$ coordinate introduced above
makes the spatial part of the equatorial geometry conformally
flat. This $\tilde r$ coordinate is thus closely related to the radial
coordinate in the isotropic version of the Schwarzschild solution and
$\tilde r$ in fact reduces to that isotropic radial coordinate as
$a\to0$.

However, in a more general physical setting the
Doran~\cite{Doran:2000aa,Hamilton:2004aa} coordinates might still be useful for
extending the acoustic analogue away from the equator, and into the
bulk of the domain of outer communication.  To do this we would have
to extend and modify the notion of acoustic geometry. The fact that
the spatial slices of the Kerr geometry are \emph{never} conformally
flat~\cite{Garat:2000aa,Kroon:2003aa,Kroon:2004aa} forces any attempt
at extending the acoustic analogy to consider a more general class of
acoustic spacetimes. A more general physical context that may prove
suitable in this regard is anisotropic fluid media, and we are currently 
investigating this possibility. The simplest anisotropic fluid media are 
the classical liquid crystals~\cite{Chandrasekhar:1992aa,Collings:1997aa}, which in the present 
context suffer from the defect that they possess considerable viscosity and exist chiefly at room temperature --- this makes them unsuitable for the construction of analogue horizons, and particularly unsuitable for the investigation of quantum aspects of analogue horizons.  Much more promising in this regard are the quantum liquid crystals. These are anisotropic superfluids such as, in particular, 3He--A~\cite{Volovik:2003jn,Volovik:2001qu}.  The anisotropic superfluids are generally characterized by the presence of a non-scalar order parameter; they behave as superfluid liquid crystals, with zero friction at $T=0$. The premier example of an anisotropic superfluid is fermionic  3He--A, where the effective gravity for fermions is described by vierbeins, though 
several types of Bose and Fermi condensates of cold atoms also exhibit similar behavior~\cite{Volovik:2003jn,Volovik:2001qu}.


In the system we have presented above we have found that two of the
main features of the Kerr geometry appear: the horizon (outer horizon)
and ergo-region.  By slight modification of standard arguments, these geometric features are expected to lead
to the quantum phenomena of super-radiance and Hawking
radiation~\cite{Unruh:1981bi,Unruh:1995aa,Visser:1993tk,Visser:1998gn,Matt.-Visser:2002ot}. The super-radiance
corresponds to the reflection and amplification of a wave packet at
the ergo surface~\cite{Slatyer:2005aa}. (See also~\cite{Basak:2003ab,Basak:2003aa}.) The
model, because it contains a horizon, also satisfies the basic
requirements for the existence of Hawking radiation, which would now
be a thermal bath of phonons emitted from the
horizon~\cite{Unruh:1981bi}. These
are two of the most fundamental physics reasons for being interested
in analogue models~\cite{Matt.-Visser:2002ot}~\footnote{Furthermore, there is a strong
  feeling that the cosmological constant problem in elementary
  particle physics might be related to a mis-identification of the
  fundamental degrees of freedom~\cite{Volovik:2003jn,Volovik:2001qu}.}.
A subtlety in the argument that leads to Hawking radiation arises from the way that the choice of coordinates \emph{seems} to influence the choice of quantum vacuum state. In the presence of a horizon, the choice of quantum vacuum is no longer unique, and standard \emph{choices} for the quantum vacuum are the Unruh, Hartle--Hawking, and Boulware states. It is the Unruh vacuum that corresponds (for either static or stationary black holes) to Hawking radiation into an otherwise empty spacetime, while the Hartle--Hawking vacuum (for a static black hole) corresponds to a black hole in thermal equilibrium with its environment (at the Hawking temperature). For a stationary [non-static] black hole the Hartle--Hawking vacuum state does not strictly speaking exist~\cite{Wald:2001aa}, but there are quasi-Hartle--Hawking quantum states that possess most of the relevant features~\cite{Iyer:1979aa}. In general relativity, because physics is coordinate independent, the choice of vacuum state is manifestly independent of the choice of coordinate system. In the analogue spacetimes considered in this article, because the preferred choice of coordinate system is intimately tied to the flowing medium, the situation is perhaps less clear.  The coordinate system we have adopted is invariant under combined time reversal and parity, which might tempt one to feel that a pseudo-Hartle--Hawking vacuum is the most natural one. On the other hand, if we think of building up the vortex from a fluid that is initially at rest, then there is clearly a preferred time direction, and the ``white hole'' component of the maximally extended horizon would exist only as a mathematical artifact.  In this physical situation the Unruh vacuum is the most natural one. We feel that there is an issue worth investigating here,  to which we hope to return in some future article.


Finally, since astrophysically all black holes are expected to exhibit some
degree of rotation, it is clear that an understanding of the influence
of rotation on analogue models is important if one wishes to connect
the analogue gravity programme back to astrophysical
observations.~\footnote{An interesting attempt in the opposite
  direction~\cite{Das:2005aa,Das:2004aa} is the recent work that analyzes
  accretion flow onto a black hole in terms of a dumb hole
  superimposed upon a general relativity black hole.}  In conclusion,
there are a number of basic physics reasons for being interested in
acoustic analogues of the Kerr geometry, and a number of interesting
directions to which the present analysis might be extended.

\chapter{Emergent particle masses in Lorentzian and Finslerian geometries \label{Chap:Geometries.Masses}}




In this chapter we will discuss an analogue
spacetime\index{analogue~spacetime} based on the propagation of
excitations in a 2-component
Bose--Einstein\index{Bose--Einstein~condensate}
condensate (BEC)~\cite{Visser:2004zn,Visser:2005ai,Liberati:2006kw,Weinfurtner:2006iv,Weinfurtner:2006iv,Liberati:2006sj}. This analogue
spacetime has a very rich and complex structure. 
The 2-BEC\index{2-BEC~system} system permits us to
provide a mass-generating \index{mass-generating~mechanism} mechanism
for the quasi-particle\index{quasi-particle}
excitations~\cite{Visser:2004zn,Visser:2005ai}. 
Additionally, in certain portions
of parameter\index{parameter~space} space the most natural
interpretation of the geometry is in terms of a specific class of
pseudo--Finsler \index{pseudo--Finsler~spacetime} spacetimes, and
indeed we will see how more generally it is possible to associate a
pseudo--Finsler \index{pseudo--Finsler~spacetime} spacetime with the
leading symbol of a wide class of hyperbolic \index{hyperbolic~system}
partial differential equations.  In other parts of
parameter\index{parameter~space} space, the most natural
interpretation of the geometry is in terms of a bi-metric
\index{bi-metric~spacetime} spacetime, where one has a manifold that
is simultaneously equipped with two distinct pseudo-Riemannian metric
tensors. Further specialization in parameter\index{parameter~space}
space leads to a region where a single pseudo-Riemannian metric tensor
is encountered --- this mono-metric \index{mono-metric~spacetime}
regime corresponds to Lorentzian spacetimes of the type encountered in
standard general relativity\index{general~relativity} and
cosmology~\cite{Schutzhold:2005mc,Fischer:2004iy,Weinfurtner:2005aa,Barcelo:2003ia}.  Thus
the analogue spacetime\index{analogue~spacetime} based on 2-component
BECs provides models not just for standard general relativistic
spacetimes, but also for the more general bi-metric
\index{bi-metric~spacetime}, and even more general pseudo--Finsler
\index{pseudo--Finsler~spacetime} spacetimes.


\section{Theory of the 2-component BEC}                          		               


The basis for our analogue model is an ultra-cold dilute atomic gas of
$N$ bosons, which exist in two single-particle states $\vert \A
\rangle$ and $\vert \B \rangle$.  For example, we consider two
different hyperfine\index{hyperfine~states} states, $\vert
F=1,m_{F}=-1 \rangle$ and $\vert F=2,m_{F}=1 \rangle$ of $^{87}\Rb$
\cite{Jenkins:2003aa,Trippenbach:2000aa}.  They have different total angular momenta
$F$ and therefore slightly different energies. That permits us, from a
theoretical point of view, to keep $m_{\A} \neq m_{\B}$, even if they
are very nearly equal (to about one part in $10^{16}$). At the assumed
ultra-cold temperatures and low densities the atoms interact only via
low-energy collisions, and the 2-body atomic potential can be replaced
by a contact potential. That leaves us with with three atom-atom
coupling constants, $U_{\A\A}$, $U_{\B\B}$, and $U_{\A\B}$, for the
interactions within and between the two
hyperfine\index{hyperfine~states} states. For our purposes it is
essential to include an additional laser field, that drives transition
between the two single-particle states.

\begin{figure}[!htb]
 \begin{center}
 \input{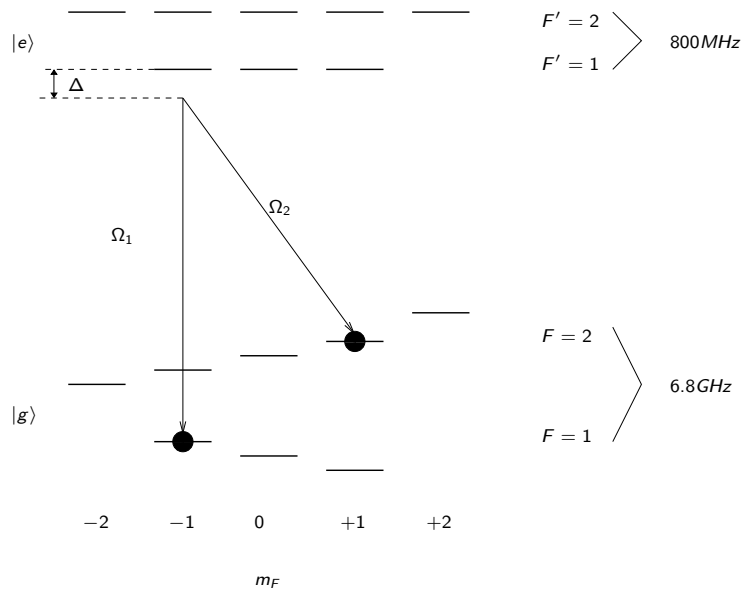}
 \caption[Energy levels for $\Rb^{87}$.]  {\label{energylevel}The
   horizontal lines indicate the hyperfine  states of $^{87}\Rb$. The
   arrows represent two laser fields --- with the two frequencies
   $\Omega_{1}$ and $\Omega_{2}$ --- necessary to drive transitions
   between the two trapped states $\vert F=1,m_{F}=-1 \rangle$ and
   $\vert F=2,m_{F}=1 \rangle$, where the frequency difference
   corresponds to the energy difference of the two hyperfine states.
   This is realized by a three-level atomic system, because the
   hyperfine states must be coupled over an intermediate level, that
   has to lie somewhat below the excited $\vert e \rangle$ states, as
   indicated by $\Delta$.}
 \end{center} 
 \index{hyperfine~states} 
 \end{figure}

 In Fig.~\ref{energylevel} the energy levels for different
 hyperfine\index{hyperfine~states} states of $^{87}\Rb$, and possible
 transitions involving three-level processes, are schematically
 explained. A more detailed description on how to set up an external
 field driving the required transitions can be found in \cite{Bloch:2000aa}.

\subsection{Gross--Pitaevskii equation}

The rotating-frame\footnote{A rotation in state-space.} Hamiltonian for our closed 2-component system is
given by:~\footnote{In general, it is possible that the collisions
  drive coupling to other hyperfine states. Strictly speaking the
  system is not closed, but it is legitimate to neglect this
  effect~\cite{Jenkins:2002aa}. }\index{hyperfine~states}
\begin{eqnarray}
\hat H = \int \d \r \;
\Bigg\{ & \sum_{i = \A,\B} & \left(-\hat \Psi_{i}^\dag \frac{\hbar^2 \nabla^2}{2 m_{i}} \hat \Psi_{i}  
+ \hat \Psi_i^\dag V_{ext,i} (\r) \hat \Psi_i \right) 
\nonumber \\
+ \frac{1}{2}& \sum_ {i,j = \A,\B}& \left(
 U_{i j} \hat \Psi_i^\dag  \hat \Psi_j^\dag  \hat \Psi_i  \hat \Psi_j  
 + \lambda  \hat \Psi_i^\dag (\mathbf{\sigma}_{x})_{i j}  \hat \Psi_j
\right)
\Bigg\} \, ,
\end{eqnarray}
with the transition rate $\lambda$ between the two
hyperfine\index{hyperfine~states} states. Here $\hat \Psi_i(\r)$ and
$\hat \Psi_i^{\dag}(\r)$ are the usual boson field annihilation and
creation operators for a single-particle state at position $\r$, and
$\mathbf{\sigma}_x$ is the usual Pauli matrix.
For temperatures at or below the critical BEC temperature, almost all
atoms occupy the spatial modes $\Psi_\A(\r)$ and $\Psi_\B(\r)$. The
mean-field description for these modes,
\begin{equation}  \label{2GPE} 
 i \, \hbar \, \partial_{t} \Psi_{i} = \left[
   -\frac{\hbar^2}{2\,m_{i}} \nabla^2 + V_{i}-\mu_{i} + U_{ii}
   \, \betrag{\Psi_{i}}^2 + U_{ij} \betrag{\Psi_{j}}^2
   \right] \Psi_{i} + \lambda \, \Psi_{j} \, , 
\end{equation}
are a pair of coupled
Gross--Pitaevskii\index{Gross--Pitaevskii~equation} equations (GPE):
$(i,j)\rightarrow (\A,\B)$ or $(i,j)\rightarrow (\B,\A)$.

\subsection{Dynamics}

In order to use the above 2-component BEC as an analogue model, we
have to investigate small perturbations (sound waves) in the
condensate cloud.\footnote{The perturbations amplitude has to be
small compared to the overall size of the condensate could, so that
the system remains in equilibrium.}  The excitation spectrum is
obtained by linearizing around some background densities $\rho_{i0}$
and phases $\theta_{i0}$, using:
\begin{equation}
\Psi_{i}= \sqrt{\rho_{i0}+ \varepsilon \, \rho_{i1} }\,
e^{i(\theta_{i0}+ \varepsilon \, \theta_{i1} )}
\quad\hbox{for}\quad i=\A,\,\B \, .
\end{equation}
To keep the analysis as general as possible, we allow the two initial
background phases to be independent from each other, and define
\begin{equation}
\delta_{\A\B} \equiv \theta_{\A0} - \theta_{\B0},
\end{equation}
as their difference. \\

A tedious calculation \cite{Visser:2004zn,Visser:2005ai,Liberati:2006kw} shows that it is
convenient to introduce the following $2 \times 2$ matrices: An
effective coupling matrix,
\begin{equation}
\hat{\Xi}=\Xi+\hat X, 
\end{equation}
where we introduced the energy-independent matrix
\begin{equation}
\Xi \equiv \frac{1}{\hbar}
\MatrixB{\tilde{U}_{\A\A}}{\tilde{U}_{\A\B}}{\tilde{U}_{\A\B}}{\tilde{U}_{\B\B}}.
\end{equation}
This matrix contains the quantities 
\begin{eqnarray}
\tilde{U}_{\A\A} &
&\equiv U_{\A\A}-{\lambda \cos \delta_{\A\B} \; \sqrt{\rho_{\A0}\rho_{\B0}}\over2} {1\over \rho_{\A0}^2} , 
\\
\tilde{U}_{\B\B} &
&\equiv U_{\B\B}-{\lambda \cos \delta_{\A\B} \; \sqrt{\rho_{\A0}\rho_{\B0}}\over2} {1\over \rho_{\B0}^2}, 
\\
\tilde{U}_{\A\B} &
&\equiv U_{\A\B}+{\lambda \cos \delta_{\A\B} \; \sqrt{\rho_{\A0}\rho_{\B0}}\over2} {1\over \rho_{\A0}\,\rho_{\B0}}.
\label{U_tilde_AB}
\end{eqnarray}
A second matrix, denoted $\hat X$, contains differential operators
$\hat Q_{X1}$ --- these are the second-order differential operators
obtained from linearizing the quantum
potential:\index{quantum~potential}
\begin{eqnarray}
 V_{\rm Q}(\rho_X) 
&\equiv&  
-  {\hbar^2\over2m_X} \left( {\nabla^2\sqrt{\rho_X}\over\sqrt{\rho_X}}\right) 
=
-  {\hbar^2\over2m_X} \left( {\nabla^2\sqrt{\rho_{X0}+\varepsilon\rho_{X1}}
\over\sqrt{\rho_{X0}+\varepsilon\rho_{X_1}} }\right) 
\\
&=& 
-  {\hbar^2\over2m_X} 
\left(  \hat Q_{X0}(\rho_{X0}) + \varepsilon \;\hat Q_{X1} (\rho_{X1}) \right).
\end{eqnarray}
The quantity $ \hat Q_{X0}(\rho_{X0})$ corresponds to the background
value of the quantum \index{quantum~pressure} pressure, and
contributes only to the background equations of motion --- it does not
affect the fluctuations. Now in a general background
\begin{equation}
\hat Q_{X1} (\rho_{X1})= {1\over2} \left\{ 
{ (\nabla\rho_{X0})^2-(\nabla^2\rho_{X0})\rho_{X0}\over\rho_{X0}^3} 
- {\nabla \rho_{X0}\over\rho_{X0}^2} \nabla 
+ {1\over\rho_{X0}} \nabla^2
\right\} \rho_{X1},
\end{equation}
and we define the matrix $\hat X$ to be
\begin{eqnarray}
\hat X &\equiv&
 -{\hbar\over2} \left[\begin{array}{cc} {\hat Q_{\A1}\over m_\A} & 0
    \\ 0 & {\hat Q_{\B1}\over m_\B}\end{array}\right].
\end{eqnarray}

Given the background homogeneity that will be appropriate for later
parts of the current discussion, this will ultimately simplify to
\begin{equation}
\hat Q_{X1} (\rho_{X1})=  {1\over2\rho_{X0}} \nabla^2 \rho_{X1},
\end{equation}
in which case
\begin{eqnarray}
\hat X &=&
-{\hbar\over4} \left[\begin{array}{cc} {1\over m_\A\;\rho_{\A0}} & 0
    \\ 0 & {1\over m_\B\;\rho_{\B0}} \end{array}\right]\nabla^2
=
 - X \; \nabla^2
\, .
\end{eqnarray}

Without transitions between the two hyperfine\index{hyperfine~states}
states, when $\lambda=0$, the matrix $\Xi$ only contains the coupling
constants $\Xi_{ij} \rightarrow U_{ij}/\hbar$. While $\Xi$ is
independent of the energy of the perturbations, the importance of
$\hat X$ increases with the energy of the perturbation. In the
so-called hydrodynamic \index{hydrodynamic~limit} approximation $\hat
X$ can be neglected, effectively $\hat X \rightarrow 0$ and $\hat \Xi
\rightarrow \Xi$.

Besides the interaction matrix, we also introduce a transition matrix,
\begin{equation}
\Lambda\equiv -\frac{2\lambda\, \cos \delta_{\A\B} \; \sqrt{\rho_{i0}\,\rho_{j0}} }{\hbar} \,
\MatrixB{+1}{-1}{-1}{+1}
\end{equation}
and a mass-density matrix,
\begin{equation} \label{m-d-matrix}
D \equiv \hbar \,  \MatrixB{\frac{ \rho_{\A0} } {m_{\A} }}{0}{0}{\frac{ \rho_{\B0} } {m_{\B} }}  
\equiv  \hbar \, \MatrixB{d_{\A}}{0}{0}{d_{\B}}.
\end{equation} 

The final step is to define two column vectors,  
\begin{equation}
{\bar{\theta}} \equiv [\theta_{\A1},\theta_{\B1}]^T,
\end{equation}
and  
\begin{equation}
{\bar{\rho}} \equiv [\rho_{\A1},\rho_{\B1}]^T.
\end{equation}
We then obtain two compact equations for the perturbation in the
phases and densities:
\begin{eqnarray} \label{thetavecdot}
\dot{\bar{\theta}}&=&  -\,\hat{\Xi} \; \bar{\rho} - \vec V  \cdot \nabla \bar{\theta} + \ipd \, \bar{\theta},
\\
\label{rhovecdot}
\dot{\bar{\rho}}&=& \, - \nabla \cdot \left( D \; \nabla \bar{\theta} +  \bar{\rho} \; \vec V \right) 
- \Lambda \;\bar{\theta} - \ipd^{\T} \bar{\rho} \, .
\end{eqnarray} 
Here the background velocity matrix simply contains the two
background velocities of each condensate,
\begin{equation}
\vec{V}= 
\left[
\begin{array}{cc}
{\vec{v}}_{\A0} & 0  \\
0 & {\vec{v}}_{\B0}  \\
\end{array}
\right] ,
\end{equation} 
with two possibly distinct background velocities,
\begin{equation}
\begin{split}
\vec{v}_{\A0} = & \frac{\hbar}{m_{\A}} \nabla \theta_{\A0}, \\
\vec{v}_{\B0} = & \frac{\hbar}{m_{\B}} \nabla \theta_{\B0} .\\
\end{split} 
\end{equation}
Additionally we also introduce the matrix $\ipd$, which depends on
the difference of the initial phases and is defined as
\begin{equation}
\ipd \equiv \frac{\lambda \sin \delta_{\A\B}}{\hbar} 
\left[
\begin{array}{cc}
+\sqrt{\frac{\rho_{\B0}}{ \rho_{\A0}}} & -\sqrt{\frac{\rho_{\B0}}{\rho_{\A0}}}  \\
+\sqrt{\frac{\rho_{\A0}}{\rho_{\B0}}} & -\sqrt{\frac{\rho_{\A0}}{\rho_{\B0}}} \\
\end{array}
\right].
\end{equation}

Now combine these two equations into one:
\begin{eqnarray} \label{Eq:GHWE} 
\partial_{t} (\hat{\Xi}^{-1} \; \dot{\bar{\theta}} ) =
& - &\partial_{t} \left(\hat{\Xi}^{-1} \; \vec V \cdot \nabla \bar{\theta} \right) 
 - \nabla (\vec V \; \hat{\Xi}^{-1} \; \dot{\bar{\theta}} )
 + \nabla \cdot \left[ \left(D - \vec V \; \hat{\Xi}^{-1} \; \vec V \right) \nabla \bar{\theta} \, \right] 
 + \Lambda  \; \bar{\theta} \\ 
 &+& \K  \, \bar{\theta} + 
{1\over2}\left\{\G^{a} \partial_{a} \bar{\theta} + \partial_a (\G^a \bar{\theta} ) \right\} , 
\nonumber
\end{eqnarray}
where the index $a$ runs from $0$--$3$ (that is, over both time and
space), and we now define
\begin{eqnarray}
\G^{t} &=& \hat{\Xi}^{-1} \ipd - \ipd^{\T} \hat{\Xi}^{-1},  \\
\G^{i} &=& \vec V \hat{\Xi}^{-1} \ipd - \ipd^{\T} \hat{\Xi}^{-1} \vec V, 
\end{eqnarray}
and 
\begin{eqnarray}
\K &=& \ipd^{\T} \hat{\Xi}^{-1} \ipd + 
{1\over2} \partial_{t} \left( \hat{\Xi}^{-1} \ipd + \ipd^\T \hat{\Xi}^{-1} \right) + 
{1\over2} \nabla \left( \vec V \hat{\Xi}^{-1} \ipd + \ipd^\T \hat{\Xi}^{-1} \vec V \right).
\nonumber\\
&&
\end{eqnarray}
Note that the $\G^a$ matrices are antisymmetric in field-space ($\A \leftrightarrow \B$),
while the matrix $\K$ is symmetric. Also, both $\G^a\to 0$ and
$\K\to0$ as $\delta_{\A\B}\to0$.

Our first goal is to show that Eqs.~\refb{Eq:GHWE}, which
fundamentally describes quasi-particle\index{quasi-particle}
excitations interacting with a condensed matter system in the
mean-field approximation, can be given a physical and mathematical
interpretation in terms of a classical background geometry for
massless and massive particles propagating through an analogue
spacetime~\cite{Barcelo:2001ay,Barcelo:2002dp,Visser:2002vp,Barcelo:2001gt}.\index{analogue~spacetime}
This analogy only holds (at least in its cleanest form) in the
so-called hydrodynamic \index{hydrodynamic~limit} limit
$\hat\Xi\to\Xi$, which limit is directly correlated with the
healing\index{healing~length} length which we shall now introduce.

\subsection{Healing length\label{sec:heal}}

The differential operator $\hat Q_{X1}$ that underlies the origin of
the $\hat X$ contribution above is obtained by linearizing the
quantum\index{quantum~potential} potential
\begin{equation}
 V_{\rm Q}(\rho_X) \equiv  -  {\hbar^2\over2 m_X}
 \left( {\nabla^2 \sqrt{\rho_X}\over\sqrt{\rho_X}}\right)
\end{equation}
which appears in the Hamilton--Jacobi\index{Hamilton--Jacobi~equation}
equation of the BEC flow.  This quantum
potential\index{quantum~potential} term is suppressed by the smallness
of $\hbar$, the comparative largeness of $m_X$, and for sufficiently
uniform density profiles.  But of course in any real system the
density of a BEC must go to zero at the boundaries of its electro-magnetic trap
(given that $\rho_X=|\psi_X(\vec x,t)|^2$).  In a 1-component BEC the
healing\index{healing~length} length characterizes the minimal
distance over which the order parameter goes from zero to its bulk
value. If the condensate density grows from zero to $\rho_0$ within a
distance $\xi$ the quantum\index{quantum~potential} potential term
(non local) and the interaction energy (local) are respectively
$E_{\rm kinetic}\sim \hbar^2/(2m\xi^2)$ and $E_{\rm interaction}\sim
4\pi\hbar^2 a \rho_0/m$. These two terms are comparable when
 \begin{equation}
\xi=(8\pi \rho_0 a)^{-1/2},
  \label{heal}
 \end{equation}
 where $a$ is the $s$-wave scattering\index{scattering~length} length
 defined as
\begin{equation}
a = {m \; U_0\over4\pi \hbar^2}.
\end{equation}
Note that what we call $U_0$ in the above expression is just the
coefficient of the non-linear self-coupling term in the
Gross--Pitaevskii\index{Gross--Pitaevskii~equation} equation, \ie,
just $U_{\A\A}$ or $U_{\B\B}$ if we completely decouple the 2 BECs
($U_{\A\B}=\lambda=0$).

Only for excitations with wavelengths much larger than the
healing\index{healing~length} length is the effect of the quantum
potential\index{quantum~potential} negligible. This is called the
hydrodynamic \index{hydrodynamic~limit} limit because the
single--BEC\index{Bose--Einstein~condensate} dynamics is then
described by the continuity and Hamilton--Jacobi equations of a
super-fluid, and its excitations behave like massless phononic modes.
In the case of excitations with wavelengths comparable with the
healing\index{healing~length} length this approximation is no longer
appropriate and deviations from phononic behavior will arise.

Such a simple discrimination between different regimes is lost once
one considers a system formed by two coupled
Bose--Einstein\index{Bose--Einstein~condensate} condensates. One is
forced to introduce a generalization of the healing $\xi$
length\index{healing~length} in the form of a ``healing matrix''. If
we apply the same reasoning used above for the definition of the
``healing\index{healing~length} length'' to the 2-component
BEC\index{2-BEC~system} system we again find a functional form like
that of Eq.~\refb{heal} however we now have the crucial
difference that both the density and the
scattering\index{scattering~length} length are replaced by matrices.
In particular, we generalize the scattering\index{scattering~length}
length $a$ to the matrix $\mathcal{A}$:
 \begin{equation}
 \mathcal{A} = {1\over4\pi\hbar^2} 
 \left[\begin{array}{cc}\sqrt{m_\A}&0\\0&\sqrt{m_\B}\end{array}\right]
 \;
 \left[\begin{array}{cc}
 \tilde U_{\A\A}&\tilde U_{\A\B}\\ \tilde U_{\A\B} &\tilde U_{\B\B} 
 \end{array}\right]
 \;
 \left[\begin{array}{cc}\sqrt{m_\A}&0\\0&\sqrt{m_\B}\end{array}\right].
 \end{equation}
 Furthermore, from~\refb{heal} a healing\index{healing~length} length
 matrix $Y$ can be defined by
 \begin{equation}
Y^{-2} = {2\over\hbar^2} \left[\begin{array}{cc}
\sqrt{\rho_{\A0} m_\A}&0\\0&\sqrt{\rho_{\B0} m_\B}\end{array}\right]
 \;
 \left[\begin{array}{cc}
 \tilde U_{\A\A}&\tilde U_{\A\B}\\ \tilde U_{\A\B} &\tilde U_{\B\B} 
 \end{array}\right]
 \;
 \left[\begin{array}{cc}
 \sqrt{\rho_{\A0} m_\A}&0\\0&\sqrt{\rho_{\B0} m_\B}\end{array}\right].
 \end{equation}
 That is, in terms of the matrices we have so far defined:
 \begin{equation}
 Y^{-2} = {1\over2} \;X^{-1/2} \; \Xi \; X^{-1/2};
 \qquad
 Y^2 = 2\; X^{1/2} \; \Xi^{-1} \; X^{1/2}.
 \end{equation}
 Define ``effective'' scattering\index{scattering~length} lengths and
 healing\index{healing~length} lengths for the 2-BEC
 \index{2-BEC~system} system as
\begin{equation}
a_\mathrm{eff} = {1\over2}\;\tr[{\mathcal{A}}] =
 {m_\A \tilde U_{\A\A} + m_\B \tilde U_{\B\B}\over 8\pi\hbar^2},
\end{equation}
and
\begin{equation}
\xi_\mathrm{eff}^2 = {1\over2}\;\tr[Y^2] = \tr[X \Xi^{-1}] 
= {\hbar^2 [\tilde U_{\B\B}/(m_\A \rho_{\A0}) + \tilde U_{\A\A}/(m_\B \rho_{\B0} )]\over
4 (\tilde U_{\A\A} \tilde U_{\B\B} - \tilde U_{\A\B}^2 ) }.
\end{equation}
That is
\begin{equation}
\xi_\mathrm{eff}^2 = 
{\hbar^2 [m_\A \rho_{\A0} \tilde U_{\A\A}+ m_\B \rho_{\B0} \tilde U_{\B\B}]
\over
4 m_\A m_\B \rho_{\A0} \rho_{\B0} \;(\tilde U_{\A\A} \tilde U_{\B\B} - \tilde U_{\A\B}^2 ) }.
\end{equation}
Note that if the two components are decoupled and tuned to be
equivalent to each other, then these effective
scattering\index{scattering~length} and healing\index{healing~length}
lengths reduce to the standard one-component results.


\section{Emergent spacetime at low energies}
             		                            
\begin{figure}[!htb]
 \begin{center}
 \input{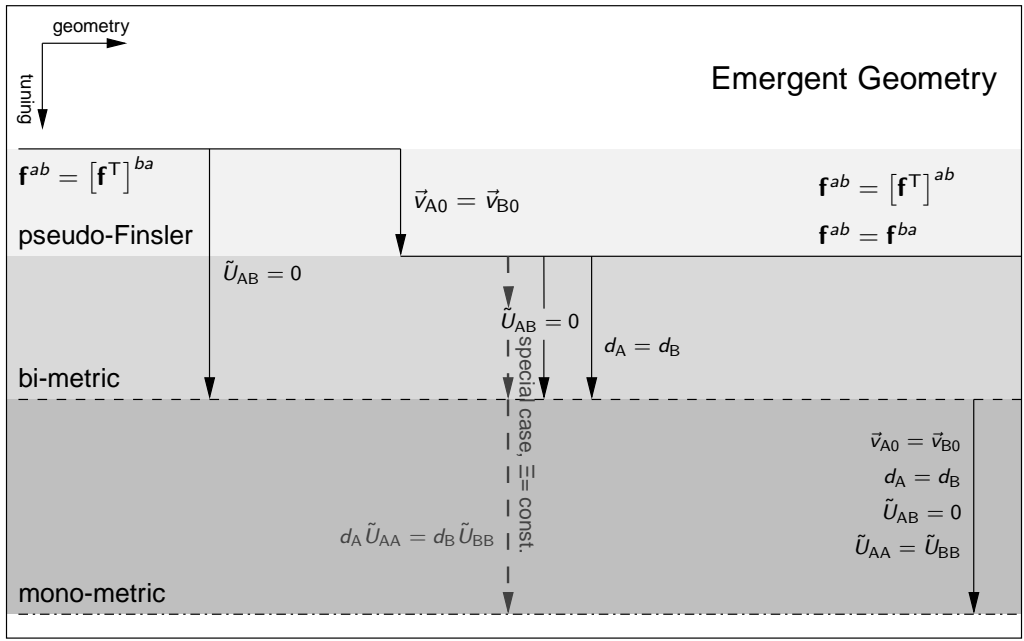}
\caption[Overview of fine tuning.]
{\label{Geometry}The dependence of the emergent geometry on the 2-component BEC parameters.}       
 \end{center}         
\end{figure}

The basic idea behind analogue models is to re-cast the equation for
excitations in a fluid into the equation describing a massless or
massive scalar field embedded in a pseudo--Riemannian geometry.
Starting from a two component superfluid we are going to show that it
is not only possible to obtain a massive scalar field from such an
analogue model, in addition we are also able to model much more
complex geometries.  In Fig.~\ref{Geometry} we illustrate how
excitations in a 2-component BEC are associated with various types of
emergent \index{emergent~geometry} geometry.

Most generally, we show that excitations in a 2-component BEC (in the
hydrodynamic \index{hydrodynamic~limit} limit) can be viewed as
propagating through a specific class of pseudo--Finsler
\index{pseudo--Finsler~spacetime} geometry.
As additional constraints are placed on the BEC
parameter\index{parameter~space} space, the geometry changes from
pseudo--Finsler \index{pseudo--Finsler~spacetime}, first to bi-metric,
and finally to mono-metric \index{mono-metric~spacetime}
(pseudo--Riemannian, Lorentzian) geometry.
This can be accomplished by tuning the various BEC parameters, such as
the transition rate $\lambda$, the background velocities $\vec
v_{\A0}$, $\vec v_{\B0}$, the background densities $\rho_{\A0}$,
$\rho_{\B0}$, and the coupling between the atoms $U_{\A \A}$, $U_{\B
  \B}$ and $U_{\A\B}$.

At first, it might seem to be quite an artificial thing to impose such
constraints onto the system. But if one considers that the two
macroscopic wave functions represent two interacting classical fields,
it is more or less obvious that this is the \emph{only} way in which
to enforce physical constraints onto the fields themselves, and on the
way they communicate with each other.
%

\subsection{Pseudo-Finsler geometry \label{GHWE}}

In the hydrodynamic \index{hydrodynamic~limit} limit ($\hat{\Xi} \rightarrow \Xi$), it is
possible to simplify Eq.~\ref{Eq:GHWE} --- without enforcing
any constraints on the BEC parameters --- if we adopt a
(3+1)-dimensional ``spacetime'' notation by writing $x^a=(t,x^i)$, with
$i\in\{1,2,3\}$ and $a\in\{0,1,2,3\}$. Then 
Eq.~\refb{Eq:GHWE} can be very compactly rewritten as~\cite{Barcelo:2001ay,Barcelo:2002dp}:
\begin{equation}
\label{E:fab_general}
\partial_a \left( \mathbf{f}^{ab} \; \partial_b  \bar\theta \right) 
+ \left(\Lambda +  \K \right)  \; \bar{\theta} 
+ {1\over2}\left\{ 
\G^{a} \; \partial_{a} {\bar{\theta}}  + \partial_a (\G^a \bar\theta) 
\right\}= 0.
\end{equation}
The object $\mathbf{f}^{ab}$ is a $4\times 4$ spacetime matrix
(actually a tensor density), each of whose components is a $2\times 2$
matrix in field-space --- equivalently this can be viewed as a
$2\times 2$ matrix in field-space each of whose components is a
$4\times 4$ spacetime tensor density. By inspection this is a
self-adjoint second-order linear system of PDEs. The spacetime
geometry is encoded in the leading-symbol of the PDEs, namely the
$\mathbf{f}^{a b}$, without considering the other subdominant
terms. That this is a sensible point of view is most easily seem by
considering the usual curved-spacetime d'Alembertian equation for a
charged particle interacting with a scalar potential in a standard
pseudo--Riemannian geometry
\begin{equation}
{1\over\sqrt{-g}} [\partial_a - i A_a] 
\left( \phantom{\Big{|}} \sqrt{-g} g^{ab} [\partial_b-iA_b] \theta \right) + V \theta = 0
\end{equation}
from which it is clear that we want to make the analogy
\begin{equation}
\mathbf{f}^{ab} \sim  \sqrt{-g} \; g^{ab}
\end{equation}
as the key quantity specifying the geometry. In addition
\begin{equation}
\G^a \sim i A^a \qquad \hbox{and} \qquad \Lambda + \K \sim V - g^{ab} A_a A_b
\end{equation}
so that $\G^a$ is analogous to a vector potential and $\Lambda$ (plus
corrections) is related to the scalar potential $V$ --- in a
translation invariant background this will ultimately provide a mass
term.

Specifically in the current 2-BEC \index{2-BEC~system} system we have
\begin{equation} \label{Finsler_1}
\mathbf{f}^{a b}=
\left(
\begin{array}{c|c}
-\Xi^{-1}            & - ( \vec{V} \Xi^{-1} )^{\T} \\
\hline
\vphantom{\Big|}
- \vec{V} \Xi^{-1}  & 
D - \vec{V}\Xi^{-1} \vec{V}^{\,T}
\end{array}
\right),
\end{equation}
where
\begin{equation}
\vec{V}^{\,\T} = \left[
\begin{array}{cc}
\vec{v}_{\A0}^{ \T} & 0 \\ 0 & \vec{v}_{\B0}^{\T}
\end{array} \right]
\end{equation}
is a $2\times2$ matrix in field space that is also a row vector in
physical 3-space.  Overall, this does look like a rather complicated
object. However, it is possible to re-write the $4\times4$ geometry
containing $2\times 2$ matrices as its elements, in form of a single
$(2\cdot4) \times (2\cdot 4)$ matrix.\footnote{This result can be
  generalized for $n$-component systems. Any $4 \times 4$ geometry
  obtained from a $n$-component system can be re-written as a single
  $(n\cdot4) \times (n\cdot 4)$ matrix.}  Explicitly we
have\footnote{Note $\Xi_{12}= \Xi_{21}$, so $\Xi_{12}^{-1}=
  \Xi_{21}^{-1}$.}
\begin{eqnarray} \nonumber
&&
\mathbf{f}^{a b}=
\left[ \begin{array}{c|c}
{\Xi}^{-1}_{11} 
\left(\begin{array}{c|c} -1  & - \vec{v}_{\A0}^{\T} \\ \hline
	- \vec{v}_{\A0} & \frac{ d_{\A}}{{\Xi}^{-1}_{11} }\delta_{ij} - \vec{v}_{\A0} \vec{v}_{\A0}^{\T}
	\end{array} \right)
& 
{\Xi}^{-1}_{12} 
\left(\begin{array}{c|c} 1  &  \vec{v}_{\B0}^{\T} \\ \hline
	\vec{v}_{\A0} &  \vec{v}_{\A0} \vec{v}_{\B0}^{\T}
	\end{array} \right)
\\ \hline
{\Xi}^{-1}_{21} 
\left( \begin{array}{c|c} 1  &  \vec{v}_{\A0}^{\T} \\ \hline \vec{v}_{\B0} &  \vec{v}_{\B0} \vec{v}_{\A0}^{\T}
	\end{array} \right)
& 
{\Xi}^{-1}_{22} 
\left( \begin{array}{c|c} -1  & - \vec{v}_{\B0}^{\T} \\ \hline - \vec{v}_{\B0} & \frac{d_{\B} }{{\Xi}^{-1}_{22}}\delta_{ij} - \vec{v}_{\B0} \vec{v}_{\B0}^{\T} \end{array}
\right) \end{array} \right],
\\
&&
\end{eqnarray}
which we can re-write as
\begin{equation} 
\mathbf{f}=
\left[
\begin{array}{c|c}
-{\Xi}^{-1}_{11} \, \mathcal{V}_{1}\mathcal{V}_{1}^{\T} + D_{11} h
& 
-{\Xi}^{-1}_{12} \, \mathcal{V}_{1}\mathcal{V}_{2}^{\T}+ D_{12} h
\\
\hline
-{\Xi}^{-1}_{21} \, \mathcal{V}_{2}\mathcal{V}_{1}^{\T}+ D_{21} h
& 
-{\Xi}^{-1}_{22} \, \mathcal{V}_{2}\mathcal{V}_{2}^{\T}+ D_{22} h
\end{array}
\right],
\end{equation}
where
\begin{eqnarray}
\mathcal{V}_{1}^a &:=& \left(1,\vec{v}_{\A0}^i \right), \\
\mathcal{V}_{2}^a &:=& \left(1,\vec{v}_{\B0}^i \right),
\end{eqnarray}
and
\begin{equation}
h^{ab}:= \diag(0,1,1,1).
\end{equation}

Even simpler is the form
\begin{equation} 
\mathbf{f}=
\left[
\begin{array}{c|c}
-{\Xi}^{-1}_{11} \, \mathcal{V}_{1}\mathcal{V}_{1}^{\T} 
& 
-{\Xi}^{-1}_{12} \, \mathcal{V}_{1}\mathcal{V}_{2}^{\T}
\\
\hline
-{\Xi}^{-1}_{21} \, \mathcal{V}_{2}\mathcal{V}_{1}^{\T} 
& 
-{\Xi}^{-1}_{22} \, \mathcal{V}_{2}\mathcal{V}_{2}^{\T} 
\end{array}
\right]
+
D \otimes h.
\end{equation}
The key point is that this allows us to write
\begin{equation} 
\label{pseudo--Finsler-metric}
\mathbf{f}^{a b}=
\MatrixB{f_{11}^{a b}}{f_{12}^{a b}}{f_{21}^{a b}}{f_{22}^{a b}},
\end{equation}
where
\begin{equation}
\begin{split}
f_{11}^{a b}=& 
-{\Xi}^{-1}_{11} \, \mathcal{V}_{1}^{a}\mathcal{V}_{1}^{b} + D_{11} h^{a b}, \\
f_{12}^{a b}=&
-{\Xi}^{-1}_{12} \, \mathcal{V}_{1}^{a}\mathcal{V}_{2}^{b}, \\
f_{21}^{a b}=&
-{\Xi}^{-1}_{12} \, \mathcal{V}_{2}^{a}\mathcal{V}_{1}^{b}, \\
f_{22}^{a b}=&
-{\Xi}^{-1}_{22} \, \mathcal{V}_{2}^{a}\mathcal{V}_{2}^{b} + D_{22} h^{a b}. \\
\end{split}
\end{equation}

It is also possible to separate the representation of
$\mathbf{f}^{ab}$ into field space and position space as follows
\begin{equation} 
\label{Finsler_product}
\begin{split}
\mathbf{f}^{a b} =&
\MatrixB{{\Xi}^{-1}_{11}}{0}{0}{0}  \mathcal{V}_{1}^{a} {\mathcal{V}_{1}^{b}}
+\MatrixB{0}{0}{0}{{\Xi}^{-1}_{22}} \mathcal{V}_{2}^{a} {\mathcal{V}_{2}^{b}} \\
&+\MatrixB{0}{{\Xi}^{-1}_{12}}{0}{0}  \mathcal{V}_{1}^{a} {\mathcal{V}_{2}^{b}} +
\MatrixB{0}{0}{{\Xi}^{-1}_{21}}{0} \mathcal{V}_{2}^{a} {\mathcal{V}_{1}^{b}} +
 D  h^{a b}.
\end{split}
\end{equation}

Why do we assert that the quantity $\mathbf{f}^{ab}$ defines a
pseudo--Finsler \index{pseudo--Finsler~spacetime} geometry? (Rather
than, say, simply a $2\times2$ matrix of ordinary Lorentzian
geometries?) To see the reason for this claim, recall the standard
result~\cite{Courant:1989aa} that the leading symbol of a system of PDEs
determines the ``signal speed'' (equivalently, the characteristics, or
the causal structure)~\cite{Barcelo:2002dp}. Indeed if we consider the
eikonal\index{eikonal} approximation (while still remaining in the
realm of validity of the hydrodynamic \index{hydrodynamic~limit}
approximation) then the causal structure is completely determined by
the leading term in the Fresnel\index{Fresnel~equation} equation
\begin{equation}
\det[ \mathbf{f}^{ab} k_a k_b] = 0,
\end{equation}
where the determinant is taken in field space. (The quantity
$\mathbf{f}^{ab} k_a k_b$ is exactly what is called the leading symbol
of the system of PDEs, and the vanishing of this determinant is the
statement that high-frequency modes can propagate with wave vector
$k_a$, thereby determining both characteristics and causal structure.)
In the 2-BEC \index{2-BEC~system} case we can explicitly expand the
determinant condition as
\begin{equation}
(f_{11}^{ab} k_a k_b) (f_{22}^{cd} k_c k_d) 
-
(f_{12}^{ab} k_a k_b) (f_{21}^{cd} k_c k_d) = 0.
\end{equation}
Define a completely symmetric rank four tensor
\begin{equation}
Q^{abcd} \equiv f_{11}^{(ab} f_{22}^{cd)}
-
f_{12}^{(ab} f_{21}^{cd)},
\end{equation}
then the determinant condition is equivalent to
\begin{equation}
Q^{abcd}  k_a k_b k_c k_d = 0,
\end{equation}
which now defines the characteristics in terms of the vanishing of the
pseudo-co--Finsler \index{co--Finsler~geometry} structure 
\begin{equation}
Q(k) = Q^{abcd}  k_a k_b k_c k_c,
\end{equation}
defined on the cotangent bundle. As explained in Appendix~\ref{Sec:Finsler}, this
pseudo-co--Finsler \index{co--Finsler~geometry} structure can be
Legendre transformed to provide a pseudo--Finsler
\index{pseudo--Finsler~spacetime} structure, a Finslerian notion of
distance
\begin{equation}
ds^4 = g_{abcd} \; dx^a dx^b dx^c dx^d.
\end{equation}
Here the completely symmetric rank 4 tensor $g_{abcd}$ determines the
``sound cones'' through the relation $ds=0$. It is interesting to note
that a distance function of the form
\begin{equation}
ds = \sqrt[4]{g_{abcd} \; dx^a dx^b dx^c dx^d}
\end{equation}
first made its appearance in Riemann's inaugural lecture of
1854~\cite{Riemann:1873aa,Clifford:1873aa}, though he did nothing further with it, leaving it
to Finsler \index{Finsler!geometry} to develop the branch of geometry
now bearing his name~\cite{Finsler:1918aa}. The present discussion is
sufficient to justify the use of the term ``pseudo--Finsler'' in the
generic 2-BEC \index{2-BEC~system} situation, but we invite the more
mathematically inclined reader to see Appendix~\ref{Sec:Finsler} for a sketch of how
much further these ideas can be taken.

The pseudo--Finsler \index{pseudo--Finsler~spacetime} geometry
implicit in~\refb{pseudo--Finsler-metric} is rather complicated
compared with the pseudo-Riemannian geometry we actually appear to be
living in, at least as long as one accepts standard general relativity
\index{general~relativity} as a good description of reality.
To mimic real gravity, we need to simplify our model.
It is now time to use the major advantage of our analogue model, the
ability to tune the BEC parameters, and with it the 2-field background
configuration.  The first order of business is to decouple
$\mathbf{f}^{ab}$ in field space.

\subsection{Bi-metric geometry \label{bi-metric}}

The reduction of Eq.~\ref{Finsler_product} to a diagonal
representation in field space (via an orthogonal rotation on the
fields),
\begin{equation} \label{f_diagonal}
\mathbf{f}^{ab} \to \diag \left[ f^{ab}_{11},f^ {ab}_{22}\right] 
= \diag \left[ \sqrt{-g_{11}}\, g^{ab}_{11},\sqrt{-g_{22}}\, g^{ab}_{22} \right],
\end{equation}
enforces a bi-metric \index{bi-metric~spacetime} structure onto the
condensate.  There are two ways to proceed.
\subsubsection{Distinct background velocities}
For
\begin{equation}
\mathcal{V}_{1} \neq \mathcal{V}_{2},
\end{equation}
we require all five $2\times2$ matrices appearing in
(\ref{Finsler_product}) to commute with each other. This has the
unique solution $\Xi^{-1}_{12}=0$, whence
\begin{equation}
\tilde{U}_{\A\B} = 0.
\end{equation}
We then get
\begin{equation} 
\label{bi-metric_product}
\mathbf{f}^{a b} =
\MatrixB{{\Xi}^{-1}_{11}}{0}{0}{0}  \mathcal{V}_{1}^{a} {\mathcal{V}_{1}^{b}}
+\MatrixB{0}{0}{0}{{\Xi}^{-1}_{22}} \mathcal{V}_{2}^{a} {\mathcal{V}_{2}^{b}} + D  h^{a b}.
\end{equation}
Since $D$ is a diagonal matrix this clearly represents a bi-metric
\index{bi-metric~spacetime} geometry.  The relevant parameters are
summarized in Table \ref{bi-metricity_results}.

\subsubsection{Equal background velocities}
For
\begin{equation}
\mathcal{V}_{1} =\mathcal{V}_{2} \equiv \mathcal{V},
\end{equation}
we are still dealing with a pseudo--Finsler
\index{pseudo--Finsler~spacetime} geometry, one which is now
independently symmetric in field space
($\mathbf{f}^{ab}=[\mathbf{f}^\T]^{ab}$), and position space
$\mathbf{f}^{ab}=\mathbf{f}^{ba}$.\footnote{The most general
  pseudo--Finsler geometry is symmetric under simultaneous exchange of
  field space and position space:
  $\mathbf{f}^{ab}=[\mathbf{f}^\T]^{ba}$.}\index{pseudo--Finsler~spacetime}
In terms of the BEC parameters that means we must set equal the two
background velocities, $\vec{v}_{\A0}=\vec{v}_{\B0}\equiv\vec{v}_{0}$,
and Eq.~(\ref{Finsler_product}) is simplified to:
\begin{equation} \label{Finsler_tuned}
\mathbf{f}^{ab} = - \Xi^{-1} \mathcal{V}^a \mathcal{V}^b + D h^{ab}.
\end{equation}
{From} the above, diagonalizability in field space now additionally
requires the commutator of the interaction and mass-density matrix to
vanish:
\begin{equation}
[\Xi,D]=0 \quad \implies \quad \tilde{U}_{\A\B}(d_{\A}-d_{\B})=0.
\end{equation}
Here, we have a choice between two tuning conditions that do the job:
\begin{equation}
\tilde{U}_{\A\B}=0 \qquad \hbox{or} \qquad d_{\A}=d_{\B}.
\end{equation}

Under the first option, where $\,\tilde{U}_{\A\B}=0$, the two
off-diagonal elements in Eq.~(\ref{Finsler_tuned}) are simply
zero, and we get the desired bi-metricity in the form\footnote{We
  would like to stress that this constraint can be easily fulfilled,
  at least in the special case $\delta_{\A\B}=0$, by tuning the
  transition rate $\lambda$, see Eq.~(\ref{U_tilde_AB}).}
\begin{equation} 
\label{bi-metric_product2}
\mathbf{f}^{a b} =
\MatrixB{{\Xi}^{-1}_{11}}{0}{0}{{\Xi}^{-1}_{22}}  \mathcal{V}^{a} {\mathcal{V}^{b}}
+ D  h^{a b}.
\end{equation}

Under the second option, for $d_{\A}=d_{\B}\equiv d$, we have $D =
d\,\mathbf{I}$. The situation is now a bit trickier, in the sense
that one has to diagonalize ${\Xi}^{-1}$:
\begin{eqnarray}
\tilde{\Xi}^{-1} &=& O^{\T} \, \Xi^{-1} O  \\ 
\nonumber
&=& \diag
\left[
\frac{\tilde{U}_{\A\A}+\tilde{U}_{\B\B}+\sqrt{(\tilde{U}_{\A\A}-\tilde {U}_{\B\B})^{2}+4\tilde{U}_{\A\B}^{2}}}
       {2\,(\tilde{U}_{\A\A}\tilde{U}_{\B\B}-\tilde{U}_{\A\B}^{2})} ,
\frac{\tilde{U}_{\A\A}+\tilde{U}_{\B\B}-\sqrt{(\tilde{U}_{\A\A}-\tilde {U}_{\B\B})^{2}+4\tilde{U}_{\A\B}^{2}}}
      {2\,(\tilde{U}_{\A\A}\tilde{U}_{\B\B}-\tilde{U}_{\A\B}^{2})}
\right]  \, .
\end{eqnarray}
%
Once this is done, the way to proceed is to use the elements of
$\tilde{\Xi}^{-1}$ instead of ${\Xi}^{-1}$ in Eq.~(\ref
{bi-metric_product2}).
The relevant parameters are summarized in Table \ref{bi-metricity_results}.

\begin{table}[htdp]
\begin{center}
\tiny
\begin{tabular}{|c||c|c|c|}
\hline
          \multicolumn{4}{|c|}{\rule[-3mm]{0mm}{8mm}\small \textsf{Bi-metric tuning  scenarios}}   \\
\hline
      &\rule[-2mm]{0mm}{6mm} \small $\vec{v}_{\A0}\neq \vec{v}_{\B0}$  &  \multicolumn{2}{|c|}{\small $ \vec{v}_{\A0}=\vec{v}_{\B0}$}  \\
\hline
     &\rule[-2mm]{0mm}{6mm}  \small $\tilde{U}_{\A\B}=0$ & \small $\tilde{U}_{\A\B}=0$ & \small $d_{\A} =d_{\B}$  \\
\hline
\hline
& & &  \\
\small $f_{11}^{ab}\propto$
&
$ \left( \begin{array}{c|c}
- 1 & - \vec{v}_{\A0}^{\T} \\ \hline
-\vec{v}_{\A 0} & \tilde{U}_{\A\A}  d_{\A} \, h^{ij} -  \vec{v}_{\A0} \vec{v}_{\A0}^{\T} 
\end{array} \right)$
&
$ \left( \begin{array}{c|c}
- 1 & - \vec{v}_{0}^{\T} \\ \hline
-\vec{v}_{0} & \tilde{U}_{\A\A}  d_{\A} \, h^{ij} -  \vec{v}_{0}\vec{v} _{0}^{\T}
\end{array} \right)$
&
$ \left( \begin{array}{c|c}
- 1 & - \vec{v}_{0}^{\T} \\ \hline
-\vec{v}_{0} & \tilde{\Xi}^{-1}_{11}  d \, h^{ij} -  \vec{v}_{0}\vec {v}_{0}^{\T}
\end{array} \right)$\\
& & &  \\
\small $f_{22}^{ab} \propto$ &
$ \left( \begin{array}{c|c}
- 1 & - \vec{v}_{\B0}^{\T} \\ \hline
-\vec{v}_{\B 0} & \tilde{U}_{\B\B}  d_{\B} \, h^{ij} -  \vec{v}_{\B0} \vec{v}_{\B0}^{\T} 
\end{array} \right)$
&
$ \left( \begin{array}{c|c}
- 1 & - \vec{v}_{0}^{\T} \\ \hline
-\vec{v}_{0} & \tilde{U}_{\B\B}  d_{\B} \, h^{ij} -  \vec{v}_{0}\vec{v} _{0}^{\T}
\end{array} \right)$
&
$ \left( \begin{array}{c|c}
- 1 & - \vec{v}_{0}^{\T} \\ \hline
-\vec{v}_{0} & \tilde{\Xi}^{-1}_{22}  d \, h^{ij} -  \vec{v}_{0}\vec {v}_{0}^{\T}
\end{array} \right)$\\
& & &  \\
\hline
& & \multicolumn{2}{|c|}{}  \\
\small ${g_{11}}_{ab}\propto$  &
\small $\left( \begin{array}{c|c}
-(c_{11}^{2} - v_{\A0}^{2}) & - \vec{v}_{\A0}^{\T} \\ \hline
-\vec{v}_{\A0} &  h^{ij} \end{array} \right) $&
\multicolumn{2}{|c|}{
\small $\left( \begin{array}{c|c}
-(c_{11}^{2} - v_{0}^{2}) & - \vec{v}_{0}^{\T} \\ \hline
-\vec{v}_{0} &  h^{ij} \end{array} \right) $}
\\
& & \multicolumn{2}{|c|}{}  \\
\small ${g_{22}}_{ab}\propto$ &
\small $\left( \begin{array}{c|c}
-(c_{22}^{2} - v_{\B0}^{2}) & - \vec{v}_{\B0}^{\T} \\ \hline
-\vec{v}_{\B0} &  h^{ij} \end{array} \right) $&
\multicolumn{2}{|c|}{
\small $\left( \begin{array}{c|c}
-(c_{22}^{2} - v_{0}^{2}) & - \vec{v}_{0}^{\T} \\ \hline
-\vec{v}_{0} &  h^{ij} \end{array} \right) $}
\\
& & \multicolumn{2}{|c|}{}  \\
\hline
& \multicolumn{2}{|c|}{ } &  \\
\small $c_{11}^{2}=$ &
\multicolumn{2}{|c|}{ \small
$\tilde{U}_{\A\A} d_{\A} = \frac{ U_{\A\A} \rho_{\A0} + U_{\A\B}\rho_ {\B0}}{m_{\A}}$}&
\small $\tilde{\Xi}^{-1}_{11} \, d$
\\
& \multicolumn{2}{|c|}{ } &  \\
\small $c_{22}^{2}=$ &
\multicolumn{2}{|c|}{ \small
$\tilde{U}_{\B\B} d_{\B} = \frac{ U_{\B\B} \rho_{\B0} + U_{\A\B}\rho_ {\A0}}{m_{\B}}$}&
\small $\tilde{\Xi}^{-1}_{22} \, d $
\\
& \multicolumn{2}{|c|}{ } &  \\
\hline
\end{tabular}
\caption[Tuning conditions.]{\label{bi-metricity_results} If the pseudo--Finsler geometry
  decouples into two independent Lorentzian geometries $f^{ab}_{11} =
  \sqrt{-g_{11}} g_{11}$ and $f^{ab}_{11} = \sqrt{-g_{11}} g_{11}$,
  with two distinct speed of sounds $c_{11}$ and $c_{22}$, we are
  effectively dealing with a bi-metric Lorentzian metric. The table
  shows the results from three different tuning scenarios, that are
  sufficient to drive the 2-component BEC from Finsler to
  bi-Lorentzian spacetime. The rightmost column $d_A=d_B$ is addressed in \cite{Fischer:2004iy}
  where the authors analyze cosmic inflation in such a bi-metric system.}
\end{center}
\index{bi-metric~spacetime}
\index{pseudo--Finsler~spacetime}
\end{table}

There is a subtlety implicit in setting the background velocities equal
that should be made explicit. If $\mathcal{V}_{1} =\mathcal{V}_{2} $
so that $\vec v_{\A0} = \vec v_{\B0}$, then since the masses appear in
the relationship between phase and velocity we deduce
\begin{equation}
m_\B \theta_{\A0}(t,\vec x) - m_\A \theta_{\B0}(t,\vec x) = f(t).
\end{equation}
If $m_\A\neq m_\B$, and if the background velocity is nonzero, we must
deduce that $\delta_{\A\B}(t,x)$ will be at the very least be position
dependent, and we will be unable to set it to zero. Alternatively, if
we demand $\delta_{\A\B}=0$, and have $ \nabla \theta_{\A0}(t,\vec x)=
\nabla \theta_{\B0}(t,\vec x) \neq 0$, then we cannot set $\vec
v_{\A0} = \vec v_{\B0} \neq 0$. Fortunately this will not seriously
affect further developments.

Last, but certainly not least, we present the conditions for a
mono-metric \index{mono-metric~spacetime} geometry in a 2-component BEC.

\subsection{Mono-metric geometry \label{mono-metric}}

Despite the fact that there are three different routes to
bi-metricity, once one demands mono-metricity, where
\begin{equation}
\mathbf{f}^{ab} =\diag \left[ f^{ab}_{11},f^{ab}_{11} \right]
= \diag \left[ \sqrt{-g_{11}}\, g^{ab}_{11},\sqrt{-g_ {11}}\, g^{ab}_{11} \right],
\end{equation}
then one ends up with one set of unique constraints to reduce from
pseudo--Finsler \index{pseudo--Finsler~spacetime} to a single-metric
Lorentzian geometry, namely:
\begin{equation}
\begin{split}
&\vec{v}_{\A0} = \vec{v}_{\B0} = \vec v_0; \\
&\tilde{U}_{\A\B}=0 ;\\
&\tilde{U}_{\A\A}=\tilde{U}_{\B\B} = \tilde U ; \\
&d_{\A} = d_{\B} = d.
\end{split}
\end{equation} 
This tuning completely specifies the spacetime geometry, in that
\begin{equation}
f_{11}^{ab}=f_{22}^{ab} \propto
\left(
\begin{array}{c|c}
- 1 & - \vec{v}_{0}^{\T} \\
\hline
-\vec{v}_{0} & \tilde{U}  d \, h^{ij} -  \vec{v}_{0}\vec{v}_{0}^{\T}  ,
\end{array}
\right)
\end{equation}
and after a small calculation we get
\begin{equation}
{g_{11}}^{ab}={g_{22}}^{ab}\propto
\left(
\begin{array}{c|c}
-(c^{2} - v_{0}^{2}) & - \vec{v}_{0}^{\T} \\
\hline
-\vec{v}_{0} &  h^{ij} 
\end{array},
\right)
\end{equation}
where we have defined
\begin{equation}
c^{2} = \tilde{U} \, d,
\end{equation}
as the speed of sound.\footnote{The speed of sound for quasi-particle
  excitations is of course our analogue for the speed of light in real
  gravity.}\index{quasi-particle}

Throughout the preceding few pages we have analyzed in detail the
first term in Eq.~(\ref{E:fab_general}), and identified different
condensate parameters with different emergent
\index{emergent~geometry} geometries. Since there is more than one
term in the wave \index{wave~equation} equation describing excitations
in a two-component system, this is not the end of the story.  The
remaining terms in Eq.~(\ref{E:fab_general}), which we might
generically view as ``mass'' and ``vector potential'' terms, do not
directly affect the spacetime geometry as such.  But when an
excitation propagates through a specific analogue
spacetime\index{analogue~spacetime} geometry, these terms will
contribute to the kinematics. It then becomes useful to consider the
``mass eigenmodes'' \index{eigenmodes} in field-space.

\subsection{Merging spacetime geometry with mass eigenmodes \label{MergingGeoEig}}

The eigenmodes \index{eigenmodes} we are interested in are eigenmodes
of the field-space matrices occurring in the sub-dominant terms of the
wave \index{wave~equation} equation. These eigenmodes (when they
exist) do not notice the presence of multiple fields --- in our
specific case a 2-field system --- and therefore propagate nicely
through the effective curved spacetime.  As promised in the abstract
and the motivation, we are striving for an analogue model representing
a massive scalar field in a mono-metric \index{mono-metric~spacetime}
Lorentzian structure.  By using the results from section
\ref{mono-metric} we are able to decouple the first term of equation
(\ref{E:fab_general}).

In the following we are focusing on two issues: First, we decouple the
remaining terms in Eq.~(\ref{E:fab_general}), and subsequently we
check that these eigenmodes \index{eigenmodes} do not recouple the
geometric term.  There is however one more (technical) problem, and
that is the fact that the terms we want to associate with the
effective mass of the scalar field still contain partial derivatives
in time and space, which ultimately implies a dependence on the energy
of the propagating modes.\footnote{This can be easily be seen by going
  to the eikonal\index{eikonal} approximation where
  $\nabla \to - i\vec{k}$ and $\partial_{t}\to i\omega$.}
Luckily, this problem can be easily circumvented, for equal background
phases,\footnote{Note that $\delta_{\A\B}=0$ plus mono-metricity
  implies either $m_\A=m_\B$ with arbitrary $\vec v_0\neq 0$, or
  $m_\A\neq m_\B$ with zero $\vec v_0= 0$. These are exactly the two
  situations we shall consider below.}
\begin{equation}
\theta_{\A0} = \theta_{\B0},
\end{equation}
in which case
\begin{equation}
\K=\G^{t}=\G^{i}=0.
\end{equation}
This has the effect of retaining only the matrix $\Lambda$ among the
sub-dominant terms, so that the wave equation \index{wave~equation} becomes
\begin{eqnarray} 
\partial_a ( \mathbf{f}^{ab} \partial_b \bar\theta) 
 + \Lambda  \; \bar{\theta} =0.
\end{eqnarray}
Due to the fact that the structure of the coupling matrix $\Lambda$
cannot be changed, its eigenmodes \index{eigenmodes} determine the
eigenmodes of the overall wave \index{wave~equation} equation. The
eigenvectors of $\Lambda$ are given by
\begin{equation}
\begin{split}
\mathrm{EV1} &:= [+1,+1] \\
\mathrm{EV2} &:= [-1,+1]
\end{split}
\end{equation}
The final step is to make sure that our spacetime geometry commutes
with the eigenvectors \index{eigenvectors} of $\Lambda$, that is
\begin{equation}
\left[ \mathbf{f}^{ab},\Lambda \right] = 0.
\end{equation}
This constraint is only fulfilled in the mono-metric
\index{mono-metric~spacetime} case, where we are dealing with two
identical classical fields, that effectively do not communicate with
each other.\footnote{While $\tilde{U}_{\A\B}=0$, $U_{\A\B}\neq 0$.}
That is, all field matrices are proportional to the identity matrix.

\subsection{Special case:  $\Xi=\mbox{constant}$.}

There is one specific class of geometries we are particularly
interested it, and that is when $\Xi$ is a position independent and
time independent constant. In the next section we will focus
exclusively on this case, and apply it to quantum
gravity\index{quantum~gravity!phenomenology} phenomenology.  This case
is however, also of interest as an example of an alternate interplay
between fine tuning and emergent geometry.\index{emergent~geometry}
Under the assumption that $\Xi$ is position and time independent, we
are able to directly manipulate the overall wave equation
\index{wave~equation} for the excitations and as a consequence obtain
slightly milder tuning conditions for mono-metricity.

Let us define
\begin{equation} \label{tildephase}
\tilde\theta = \Xi^{-1/2}\; \bar\theta,
\end{equation}
and multiply the whole wave equation~(\ref{E:fab_general}) with
$\Xi^{1/2}$ from the left. What we are doing is a transformation in
field space onto a new basis $\tilde{\theta}$, and in the new basis the
wave \index{wave~equation} equation is given by,
\begin{equation}
\label{E:fab_general_constXi}
\partial_a \left( \tilde{\mathbf{f}}^{ab} \; \partial_b  \tilde\theta  \right)
+ \left(\tilde{\Lambda} +  \tilde{\K} \right)  \; \tilde{\theta}
+ {1\over2}\left\{
\tilde{\G}^{a} \; \partial_{a} {\tilde{\theta}}  + \partial_a (\tilde {\G}^a \tilde\theta)
\right\}= 0,
\end{equation}
where the matrices in field space transform as:
$\tilde\Lambda=\Xi^{1/2} \Lambda \Xi^{1/2}$, $\tilde\K=\Xi^{1/2} \K
\Xi^{1/2}$, $\tilde {\G}^{a}= \Xi^{1/2} \G^{a} \Xi^{1/2}$, and the
tensor-density as
\begin{equation}
\tilde{\mathbf{f}}^{ab} = \Xi^{1/2} \, \mathbf{f}^ {ab} \,\Xi^{1/2}.
\end{equation}
In general, the transformation matrix $\Xi^{1/2}$ is non-diagonal,
though always symmetric:\footnote{See Appendix~\ref{Sec:MatrixIdentities}.}
\begin{equation}
\Xi^{1/2} =
\frac{\Xi +\sqrt{\det \Xi} \;\; \mathbf{I} }{\sqrt{\tr[\Xi]+ 2  \sqrt {\det \Xi}}}.
\end{equation}
A close look at Eq.~(\ref{Finsler_product}), now using the
tensor-density $\tilde{\mathbf{f}}^{ab}$, makes it obvious that for
\begin{equation}
\tilde{U}_{\A\B} = 0,
\end{equation}
the geometry reduces from pseudo-Finsler to bi-metric.\index{bi-metric~spacetime} 
For the sake of keeping the discussion short and easy to
follow, we set the background velocities equal, and now get
\begin{equation}
\label{bi-metric_product_constXi}
\tilde{\mathbf{f}}^{a b} =
\mathcal{V}^{a} {\mathcal{V}^{b}} + \tilde{D}  h^{a b}.
\end{equation}
In view of the tuning, $\tilde{U}_{\A\B}=0$, we see
\begin{equation}
\tilde{D} = \diag (\tilde{U}_{\A\A}\, d_{\A},\tilde{U}_{\B\B}\, d_{\B}).
\end{equation}
The new mass-density matrix, and therefore the overall geometry, is
diagonal in field space, hence we are now dealing with the required
bi-metric structure.

So far we are in complete agreement with what we have obtained in our
previous analysis, see Fig.~\ref{Geometry}.  However, if we now ask
for mono-metricity, \index{mono-metric~spacetime} we obtain a slightly
milder constraint:
\begin{equation}
\tilde{U}_{\A\A}\, d_{\A} = \tilde{U}_{\B\B}\, d_{\B}.
\end{equation}

Last but not least, we show in detail the results we obtain for this
tuning scenario when including the $\Lambda$ term (the mass term). To
avoid confusion, we re-define a few matrices,
\begin{equation} \label{Omega2}
C_{0}^2 = \Xi^{1/2}\; D \;\Xi^{1/2} \, ;
\qquad \hbox{and} \qquad
\Omega^2 =   \Xi^{1/2} \;\Lambda\; \Xi^{1/2}.
\end{equation}
Both $C_{0}^2$ and $\Omega^2$ are symmetric matrices. If $ [C_{0}^2,
\; \Omega^2] = 0$, which is equivalent to the matrix equation $D \;
\Xi \; \Lambda = \Lambda \; \Xi \; D$, and is certainly satisfied in
view of the above constraint, then they have common eigenvectors. \index{eigenvectors}
Decomposition onto the eigenstates of the system results in a pair of
independent Klein--Gordon \index{Klein--Gordon equation} equations
\begin{equation} \label{KGE}
\frac{1}{\sqrt{-g_{\mathrm{I/II}}}}\,
\partial_{a} \left\{   \sqrt{-g_{\mathrm{I/II}}} \; (g_{\mathrm{I/ II}})^{ab} \;
\partial_{b} \tilde{\theta}_{\mathrm{I/II}} \right\} +
\omega_{\mathrm{I/II}}^2 \;
\tilde{\theta}_{\mathrm{I/II}} = 0 \; ,
\end{equation}
where the ``acoustic metrics'' are given by
\begin{equation} \label{metric}
(g_{\mathrm{I/II}})_{ab} \propto
\left[
\begin{array}{ccc}
-\left( c^2-v_0^2 \right)       &|& -\vec{v_0}^{\,T} \\
\hline
-\vec{v_0}  &|& \mathbf{I}_{d\times d}
\end{array}
\right].
\end{equation}
The metric components depend only on the background velocity
$\vec{v}_{0}$ and the common speed of sound $c$.\; It is also possible
to calculate the eigenfrequencies \index{eigenfrequencies} of the two
phonon\index{phonon} modes,
\begin{equation}
\omega_{I}^2 = 0;
\qquad
\omega_{II}^2 = \tr[\Omega^2] \, .
\end{equation}
A zero\,/\,non-zero eigenfrequency \index{eigenfrequencies} corresponds
to a zero\,/\,non-zero mass for the phonon\index{phonon} mode. %

In the eikonal\index{eikonal} limit we see that the in-phase perturbation will
propagate with the speed of sound,
\begin{equation}
\vec{v}_{s} =\vec{v}_0 + \hat k \;c\, ,
\end{equation}
while the anti-phase perturbations propagates with a lower group velocity given by:
\begin{equation}
\vec{v}_{g}= \frac{\partial \omega}{\partial \vec{k}}
=\vec{v}_0 + \hat k \;\frac{c^2}{\sqrt{\omega_{II}^2\; + c^2\; k^2}}.
\end{equation}
Here $k$ is the usual wavenumber. The dispersion relation we obtain
for the mono-metric \index{mono-metric~spacetime} structure is Lorentz
invariant.\index{Lorentz~invariance}

The fact that we have an analogue model representing both massive and
massless particles is promising for quantum gravity
phenomenology\index{quantum~gravity!phenomenology} if we now extend
the analysis to high-energy phonon\index{phonon} modes where the
quantum \index{quantum~pressure} pressure term is significant, and
where we consequently expect a breakdown of Lorentz
\index{Lorentz~invariance} invariance.  For the following, we
concentrate on the generalization of flat Minkowski spacetime, which
implies a constant $\Xi$ and zero background velocities, $\vec{v}_0$.
In the language of condensed matter physics, we are thinking of a
uniform condensate at rest.


\section{Outlook, summary and discussion}                                     
               

So where can (and should) we go from here? If 2-component BECs provide
such a rich mathematical and physical structure, are 3-component BECs,
or general multi-component BECs even better? That depends on what you
are trying to do:
\begin{itemize}
\item If one wishes to actually \emph{build} such an
  analogue\index{analogue~spacetime} spacetime in the laboratory, and
  perform actual experiments, then iteration through 1-BEC
  \index{1-BEC~system} and 2-BEC \index{2-BEC~system} systems seems
  the most promising route in terms of our technological capabilities.
\item For $n$-component BECs we sketch the situation in figure
  \ref{fig:5_{comp_system}}. The key point is that due to overall
  translation invariance one again expects to find one massless
  quasi-particle, with now $n-1$ distinct massive modes. Unfortunately
  the matrix algebra is now considerably messier --- not intrinsically
  difficult (after all we are only dealing with $n\times n$ matrices
  in field space) --- but extremely tedious. Physical insight remains
  largely intact, but (except in some specific particularly simple
  cases), computations rapidly become lost in a morass of technical
  detail.
\item However, if one wishes to draw general theoretical lessons from
  the analogue spacetime\index{analogue~spacetime} programme, then
  multi-component systems are definitely the preferred route ---
  though in this case it is probably better to be even more abstract,
  and to go beyond the specific details of BEC-based systems to deal
  with general hyperbolic \index{hyperbolic~system} systems of PDEs.
\item In Appendix~\ref{Sec:Finsler} we have sketched some of the key features of the
  pseudo--Finsler \index{pseudo--Finsler~spacetime} spacetimes that
  naturally emerge from considering the leading symbol of a hyperbolic
  \index{hyperbolic~system} system of PDEs. While it is clear that
  much more could be done based on this, and on extending the field
  theory ``normal modes'' of~\cite{Barcelo:2001ay,Barcelo:2002dp}, such an analysis
  would very much move outside the scope of the current thesis.
  \end{itemize}
In short the 2-BEC \index{2-BEC~system} system is a good compromise
between a system complex enough to exhibit a mass-generating
\index{mass-generating~mechanism} mechanism, and still simple enough
to be technologically tractable, with good prospects for laboratory
realization of this system in the not too distant future.
\begin{figure}[!htb]
\centering
\input{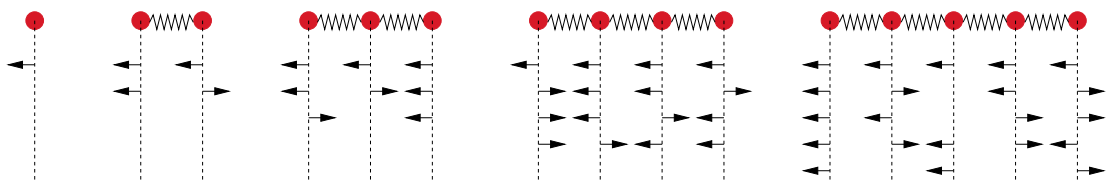}
\caption[Mass generation in $n$-component system.]{The figure captures the key features of possible
  eigenmodes\index{eigenmodes} for a small perturbation (circles) in a
  1 (left side), 2, 3, 4, and 5-component (right side) BEC. In a
  1-component system only one kind of perturbation is allowed, which
  corresponds to a massless particle propagating through an effective
  curved spacetime, while in the 2-component case two different kinds
  of mode appear, the one in-phase (massless particle) and one in
  anti-phase (massive particle).  For a three-component system we
  again expect to find one massless particle, when all perturbations
  are in phase, and now in addition to that two massive particles.}
\label{fig:5_{comp_system}}       
\end{figure}
%

The key features we have emphasised in this chapter have been:
\begin{itemize}
\item A general analysis of the 2-BEC \index{2-BEC~system} system to
  see how perturbations on a 2-BEC \index{2-BEC~system} background
  lead to a system of coupled wave \index{wave~equation} equations.
\item Extraction of the geometric notion of pseudo--Finsler
  \index{pseudo--Finsler~spacetime} spacetime from this
  wave\index{wave~equation} equation, coupled with an analysis of how
  to specialize pseudo--Finsler \index{pseudo--Finsler~spacetime}
  geometry first to a bi-metric \index{bi-metric~spacetime} Lorentzian
  geometry and finally to the usual mono-metric
  \index{mono-metric~spacetime} Lorentzian geometry of most direct
  interest in general relativity and cosmology.
\item The mass-generating \index{mass-generating~mechanism} mechanism
  we have identified in suitably coupled 2-component BECs is an
  essential step in making this analogue
  spacetime\index{analogue~spacetime} more realistic; whatever one's
  views on the ultimate theory of ``quantum
  gravity'',\index{quantum~gravity} any realistic low-energy
  phenomenology\index{quantum~gravity!phenomenology} must contain
  \emph{some} mass-generating \index{mass-generating~mechanism}
  mechanism.
\item Use of the ``quantum \index{quantum~pressure} pressure'' term in
  the 2-BEC \index{2-BEC~system} system to mimic the sort of Lorentz
  \index{Lorentz~invariance!violation} violating physics that (based
  on the relatively young field of ``quantum
  gravity\index{quantum~gravity!phenomenology} phenomenology'') is
  widely expected to occur at or near the Planck
  scale.\index{Planck~scale}
\item Intriguingly, we have seen that in our specific model the
  mass-generating \index{mass-generating~mechanism} mechanism
  interacts with the Lorentz \index{Lorentz~invariance!violation}
  violating mechanism, naturally leading to a situation where the
  Lorentz \index{Lorentz~invariance!violation} violations are
  suppressed by powers of the quasi-particle\index{quasi-particle}
  mass scale divided by the analogue of the Planck
  scale.\index{Planck~scale}
\end{itemize}

In summary, while we do not personally believe that the real universe
\emph{is} an analogue spacetime,\index{analogue~spacetime} we are
certainly intrigued by the fact that so much of what is normally
viewed as being specific to general relativity and/or particle physics
can be placed in this much wider context. We should also be forthright
about the key weakness of analogue models as they currently stand: As
we have seen, obtaining an analogue
spacetime\index{analogue~spacetime} geometry (including spacetime
curvature) is straightforward --- but what is not straightforward is
obtaining the Einstein equations.\index{Einstein~equations} The
analogue models are currently analogue models of quantum field theory
on curved spacetime, but not (yet?) true analogue models of Einstein
gravity.\index{Einstein~gravity} Despite this limitation, what can be
achieved through the analogue spacetime\index{analogue~spacetime}
programme is quite impressive, and we expect interest in this field,
both theoretical and hopefully experimental, to continue unabated.

\chapter{Analogue quantum gravity phenomenology \label{Chap:QGP}}

As we have seen in the previous chapter, the 2-BEC\index{2-BEC~system} system permits us to
provide a mass-generating \index{mass-generating~mechanism} mechanism
for the quasi-particle\index{quasi-particle}
excitations~\cite{Visser:2004zn,Visser:2005ai}.  The specific mass-generating
\index{mass-generating~mechanism} mechanism arising herein is rather
different from the Higgs mechanism of the standard model of particle
physics, and provides an interesting counterpoint to the more usual
ways that mass-generation\index{mass-generating~mechanism} is
achieved.  Furthermore, at short distances, where the ``quantum
\index{quantum~pressure} pressure'' term can no longer be neglected,
then even in the mono-metric \index{mono-metric~spacetime} regime one
begins to see deviations from ``Lorentz \index{Lorentz~invariance}
invariance'' --- and these deviations are qualitatively of the type
encountered in ``quantum gravity\index{quantum~gravity!phenomenology}
phenomenology'', with the interesting property that the Lorentz
\index{Lorentz~invariance!violation} violating physics is naturally
suppressed by powers of the quasi-particle\index{quasi-particle} mass
divided by the mass of the fundamental bosons that form the
condensate~\cite{Liberati:2006kw,Weinfurtner:2006nl,Weinfurtner:2006iv,Liberati:2006sj}. So in these analogue systems
the mass-generating \index{mass-generating~mechanism} mechanism is
related to the ``hierarchy problem'' and the suppression of
Lorentz-violating physics.  The 2-BEC model also allows us to probe
the ``universality'' (or lack thereof) in the Lorentz
\index{Lorentz~invariance!violation} violating
sector~\cite{Liberati:2006kw,Weinfurtner:2006nl,Weinfurtner:2006iv,Liberati:2006sj}. More generally, as one moves beyond
the hydrodynamic\index{hydrodynamic~limit} limit in generic
pseudo--Finsler\index{pseudo--Finsler~spacetime} parts of
parameter\index{parameter~space} space, one can begin to see hints of
geometrical structure even more general than the pseudo--Finsler
\index{pseudo--Finsler~spacetime} geometries.

While we do not wish to claim that the 2-BEC\index{2-BEC~system}
analogue spacetime of this chapter is necessarily a good model for the
real physical spacetime arising from the putative theory of ``quantum
gravity''\index{quantum~gravity} (be it string-model, loop-variable,
or lattice based), it is clear that the 2-BEC \index{2-BEC~system}
analogue spacetime\index{analogue~spacetime} is an extraordinarily
rich mathematical and physical structure that provides many
interesting hints regarding the sort of kinematics and dynamics that
one might encounter in a wide class of models for ``quantum
gravity\index{quantum~gravity!phenomenology} phenomenology''. This is
the fundamental reason for our interest in this model, and we hope we
can likewise interest the reader in this system and its relatives.


\section{Application to  quantum gravity phenomenology}                                   

\begin{figure}[htb]
 \begin{center}
 \input{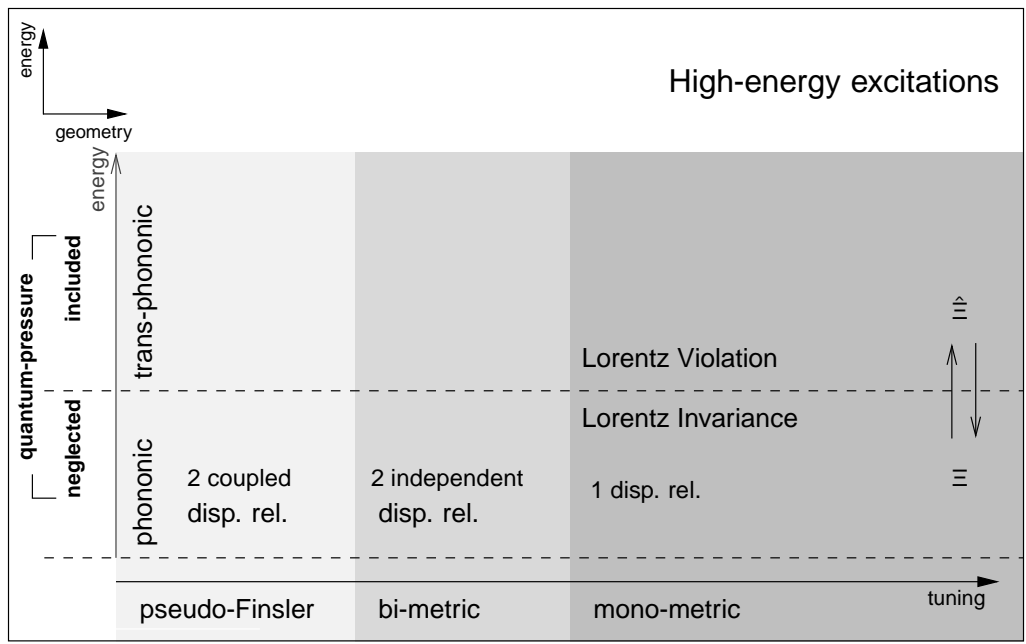}
 \caption[Emergent Lorentz symmetry.]  {\label{Concept}How to tune the
   system to exhibit breakdown of Lorentz symmetry.}
 \end{center} 
\index{Lorentz~invariance}      
\end{figure}

In using this 2-BEC \index{2-BEC~system} model to probe issues of
interest to the ``quantum gravity\index{quantum~gravity!phenomenology}
phenomenology'' community it behooves us to simplify as much as
possible the parts of the model not of direct interest for current
considerations. Specifically, we wish to use the ``quantum
\index{quantum~pressure} pressure'' term as a model for the type of
Lorentz \index{Lorentz~invariance!violation} violating physics that
might occur in the physical universe at or near the Planck
\index{Planck~scale} scale~\cite{Visser:2001ix}. Since we are then interested in high
energies, and consequently short distances, one might expect the
average spacetime curvature to be negligible --- that is, we will be
interested in looking for ``quantum \index{quantum~pressure}
pressure'' induced deviations from special
relativity,\index{special~relativity} and can dispense with the notion
of curved spacetimes for now. (``Flat'' pseudo--Finsler
\index{pseudo--Finsler~spacetime} spaces are already sufficiently
complicated to lead to interesting physics.)  In terms of the BEC
condensates this means that in this section of the chapter we will
concentrate on a spatially-homogeneous time-independent background, so
that in particular all the matrices $\mathbf{f}^{ab}$ will be taken to
be position-independent. (And similarly, $\Xi$, $\Lambda$, $D$,
\emph{etc.} are taken to be position independent and we set $\vec
v_0=0$, so the background is at rest.) We now consider Lorentz invariance breakdown in a flat mono-metric spacetime geometry as indicated in Fig. \ref{Concept}.  This greatly simplifies the
calculations (though they are still relatively messy), but without
sacrificing the essential pieces of the physics we are now interested
in.

%
%

Now the purpose of quantum
gravity\index{quantum~gravity!phenomenology} phenomenology is to
analyze the physical consequences arising from various models of
quantum gravity.  One hope for obtaining an experimental grasp on
quantum gravity is the generic prediction arising in many (but not
all) quantum gravity\index{quantum~gravity} models that ultraviolet
physics at or near the Planck scale,\index{Planck~scale}
$M_{\mathrm{Planck}} = 1.2 \times 10^{19} \; \mathrm{GeV/c^2}$, (or in
some models the string scale), typically induces violations of Lorentz
\index{Lorentz~invariance!violation} invariance at lower
scales~\cite{Mattingly:2005aa,Jacobson:2005aa}.  Interestingly most investigations, even
if they arise from quite different fundamental physics, seem to
converge on the prediction that the breakdown of Lorentz
\index{Lorentz~invariance} invariance can generically become manifest
in the form of modified dispersion relations
\begin{equation}
 \label{disp}
\omega^2 = \omega_0^2 + \left(1 + \eta_{2} \right) \, c^2 \; k^2 
+ \eta_{4} \, \left(\frac{\hbar}{M_{\mathrm{Lorentz~violation}}} \right)^2  \; k^4 + \dots \; ,
\end{equation}
where the coefficients $\eta_{n}$ are dimensionless (and possibly
dependent on the particle species considered), and we have restricted
our expansion to CPT\index{CPT~invariance} invariant terms (otherwise
one would also get odd powers in $k$). The particular inertial frame
for these dispersion relations is generally specified to be the frame
set by the cosmological microwave background, and $M_{\mathrm{Lorentz~violation}}$
is the scale of Lorentz \index{Lorentz~invariance} symmetry breaking
which furthermore is generally assumed to be of the order of
$M_{\mathrm{Planck}}$.\index{Planck~scale}

Although several alternative scenarios have been considered in the
literature in order to justify the modified kinematics discussed
above, to date the most commonly explored avenue is an effective field
theory (EFT)\index{effective~field~theory} approach.  In the present
chapter we focus on the class of non-renormalizable
EFTs\index{effective~field~theory} with Lorentz
\index{Lorentz~invariance!violation} violations associated to
dispersion relations like equation~(\ref{disp}).  Relaxing our
CPT\index{CPT~invariance} invariance condition this class would
include the model developed in~\cite{Myers:2003aa}, and subsequently studied by
several authors, where an extension of quantum electrodynamics including only mass
dimension five Lorentz-violating operators was considered. (That
ansatz leads to order $k^3$ Lorentz
\index{Lorentz~invariance!violation} and CPT\index{CPT~violation}
violating terms in the dispersion relation.)  Very accurate
constraints have been obtained for this model using a combination of
experiments and observations (mainly in high energy astrophysics). See
\emph{e.g.}\/~\cite{Jacobson:2005aa,Jacobson:2002aa,Jacobson:2003aa,Jacobson:2004aa,Jacobson:2003ab}.
In spite of the remarkable success of this framework as a ``test
theory'', it is interesting to note that there are still significant
open issues concerning its theoretical foundations.  Perhaps the most
pressing one is the so called {\em naturalness problem} which can be
expressed in the following way: Looking back at our ansatz
(\ref{disp}) we can see that the lowest-order correction, proportional
to $\eta_{2}$, is not explicitly Planck suppressed.\index{Planck
  suppressed} This implies that such a term would always be dominant
with respect to the higher-order ones and grossly incompatible with
observations (given that we have very good constraints on the
universality of the speed of light for different elementary
particles).  Following the observational leads it has been therefore
often assumed either that some symmetry (other than Lorentz
\index{Lorentz~invariance} invariance) enforces the $\eta_2$
coefficients to be exactly zero, or that the presence of some other
characteristic EFT\index{effective~field~theory} mass scale $\mu\ll
M_{\Pl}$ (\emph{e.g.}, some particle physics mass scale)
associated with the Lorentz \index{Lorentz~invariance} symmetry
breaking might enter in the lowest order dimensionless coefficient
$\eta_{2}$ --- which will be then generically suppressed by
appropriate ratios of this characteristic mass to the Planck
\index{Planck~scale} mass: $\eta_2\propto (\mu/M_{\Pl})^\sigma$
where $\sigma\geq 1$ is some positive power (often taken as one or
two). If this is the case then one has two distinct regimes: For low
momenta $p/(M_{\Pl}c) \ll (\mu/M_{\Pl})^\sigma$ the lower-order
(quadratic in the momentum) deviations in~(\ref{disp}) will dominate
over the higher-order ones, while at high energies $p/(M_{\Pl}c)
\gg (\mu/M_{\Pl})^\sigma$ the higher order terms will be dominant.

The naturalness problem arises because such a scenario is not well
justified within an EFT\index{effective~field~theory} framework; in
other words there is no natural suppression of the low-order
modifications in these models. In fact we implicitly assumed that
there are no extra Planck\index{Planck suppressed} suppressions hidden
in the dimensionless coefficients $\eta_n$ with $n>2$.
EFT\index{effective~field~theory} cannot justify why \emph{only} the
dimensionless coefficients of the $n\leq 2$ terms should be suppressed
by powers of the small ratio $\mu/M_{\Pl}$.  Even worse,
renormalization group arguments seem to imply that a similar mass
ratio, $\mu/M_{\Pl}$ would implicitly be present also in \emph{all}
the dimensionless $n>2$ coefficients --- hence suppressing them even
further, to the point of complete undetectability.  Furthermore it is
easy to show~\cite{Collins:2004aa} that, without some protecting symmetry, it
is generic that radiative corrections due to particle interactions in
an EFT\index{effective~field~theory} with only Lorentz
\index{Lorentz~invariance!violation} violations of order $n>2$ in
(\ref{disp}) for the free particles, will generate $n= 2$ Lorentz
\index{Lorentz~invariance!violation} violating terms in the dispersion
relation, which will then be dominant.  Observational
evidence~\cite{Mattingly:2005aa} suggests that for a variety of standard model
particles $|\eta_2|\lesssim 10^{-21}$. Naturalness in
EFT\index{effective~field~theory} would then imply that the higher
order terms are at least as suppressed as this, and hence beyond
observational reach.

A second issue is that of ``universality'', which is not so much a
``problem'', as an issue of debate as to the best strategy to adopt.
In dealing with situations with multiple particles one has to choose
between the case of universal (particle-independent) Lorentz
\index{Lorentz~invariance!violation} violating coefficients $\eta_n$,
or instead go for a more general ansatz and allow for
particle-dependent coefficients; hence allowing different magnitudes
of Lorentz \index{Lorentz~invariance!violation} symmetry violation for
different particles even when considering the same order terms (same
$n$) in the momentum expansion. The two choices are equally
represented in the extant literature (see \emph{e.g.}~\cite{Amelino-Camelia:2001aa}
and \cite{Jacobson:2003aa} for the two alternative ans\"atze), but it
would be interesting to understand how generic this universality might
be, and what sort of processes might induce non-universal Lorentz
\index{Lorentz~invariance!violation} violation for different
particles.

\subsection{Specializing the wave equation\label{sec:wave-equation}}%

For current purposes, where we wish to probe violations of Lorentz
\index{Lorentz~invariance!violation} invariance in a flat analogue
spacetime,\index{analogue~spacetime} we start with our basic wave
equation (\ref{Eq:GHWE}) and make the following specializations:
$\delta_{\A\B}\to0$ (so that $\G^a\to0$ and $K\to0$). We also set all
background fields to be homogeneous (space and time independent), and
use the formal operators $\hat \Xi^{1/2}$ and $\hat \Xi^{-1/2}$ to
define a new set of variables
\begin{equation}
\tilde\theta = \hat \Xi^{-1/2} \;\bar\theta,
\end{equation}
in terms of which the wave \index{wave~equation} equation becomes
\begin{equation}
\label{E:simplified-wave-equation}
\partial_t^2\tilde\theta = 
\left\{ \hat \Xi^{1/2}\; [ D \nabla^2 - \Lambda] \; \hat \Xi^{1/2} \right\} \tilde\theta,
\end{equation}
or more explicitly
\begin{equation}
\label{eq:last}
\partial_t^2\tilde\theta = 
\left\{ [\Xi- X \nabla^2]^{1/2}\;  [ D \nabla^2 - \Lambda] \; 
 [\Xi-X\nabla^2] ^{1/2} \right\} \tilde\theta.
\end{equation}
This is now a (relatively) simple PDE to analyze.  The objects $\hat
\Xi^{1/2}$ and $\hat \Xi^{-1/2}$ are $2\times2$ matrices whose
elements are pseudo-differential operators, but to simplify things it
is computationally efficient to go directly to the eikonal\index{eikonal} limit
where\footnote{Once we are in the eikonal\index{eikonal} approximation the
  pseudo-differential operator $\hat \Xi^{1/2} \to \sqrt{ \Xi + k^2 X}$
  can be given a simple and explicit meaning in terms of the
  Hamilton--Cayley theorems of Appendix~\ref{Sec:MatrixIdentities}.}
\begin{equation}
\hat \Xi \to \Xi + X\; k^2.
\end{equation}
This finally leads to a dispersion relation of the form
\begin{eqnarray}
\label{fresnel}
&&\det\big\{ \omega^2 \; \mathbf{I} -
 [\Xi+ X k^2]^{1/2}\;  [ D k^2  + \Lambda] \; 
 [\Xi+ X k^2] ^{1/2}  \big\} =0\,,
\end{eqnarray}
and ``all'' we need to do for the purposes of this chapter, is to
understand this quasiparticle excitation spectrum in detail.

\subsection{Hydrodynamic approximation\label{sec:hydro}}

The hydrodynamic \index{hydrodynamic~limit} limit consists of formally
setting $\hat X\to0$ so that $\hat \Xi \to \Xi$.  (That is, one is
formally setting the healing\index{healing~length} length matrix to
zero: $Y \to 0$. More precisely, all components of the healing length
matrix are assumed small compared to other length scales in the
problem.)  The wave \index{wave~equation} equation (\ref{eq:last}) now
takes the form:
\begin{eqnarray}
\partial_t^2\tilde\theta = 
\left\{ \Xi^{1/2}\; [ D \nabla^2 - \Lambda] \; \Xi^{1/2} \right\} \tilde\theta.
\end{eqnarray}
Since this is second-order in both space and time derivatives, we now
have at least the possibility of obtaining an exact ``Lorentz
\index{Lorentz~invariance} invariance''.  We can now define the
matrices
\begin{equation}
\Omega^2 = \Xi^{1/2}\;\Lambda\;\Xi^{1/2}; \qquad C_0^2 =   \Xi^{1/2}\; D\; \Xi^{1/2}; 
\end{equation}
so that after Fourier transformation
\begin{equation}
\omega^2 {\tilde{\theta}}  = \left\{ C_0^2 \;
k^2+\Omega^2\right\} \; \tilde{\theta}  \equiv H(k^2) \; 
\tilde{\theta},
\end{equation}
leading to the Fresnel\index{Fresnel~equation} equation
\begin{equation}
\det\{ \omega^2 \;\mathbf{I} - H(k^2) \} =0.
\end{equation}
That is
\begin{equation}
\omega^4 - \omega^2 \; \hbox{tr}[H(k^2)] + \det[H(k^2)] =
0,
\end{equation}
whence
\begin{equation}
\omega^2 = { \hbox{tr}[H(k^2)] \pm \sqrt{
    \hbox{tr}[H(k^2)]^2 - 4\; \det[H(k^2)] }\over 2}.
\label{eq:disp-rel-hydro}
\end{equation}
Note that the matrices $\Omega^2$, $C_0^2$, and $H(k^2)$ have now
carefully been arranged to be \emph{symmetric}. This greatly
simplifies the subsequent matrix algebra. Also note that the matrix
$H(k^2)$ is a function of $k^2$; this will forbid the appearance of
odd powers of $k$ in the dispersion relation --- as should be expected
due to the parity invariance of the system.

\subsubsection{Masses}
We read off the ``masses'' by looking at the special case of
space-independent oscillations for which
\begin{equation}
 \partial_{t}^2 {\bar{\theta}}  = -\Omega^2 \; \bar{\theta},
\end{equation}
allowing us to identify the ``mass'' (more precisely, the natural oscillation
frequency) as
\begin{equation}
\hbox{``masses''} 
\propto \hbox{eigenvalues of}\;(\Xi^{1/2}\;\Lambda\; \Xi^{1/2})
 =  \hbox{eigenvalues of}\;(\Xi\;\Lambda).
\end{equation}
Since $\Lambda$ is a singular $2\times2$ matrix this
automatically implies
\begin{equation}
\omega_I^2=0; \qquad \omega_{II}^2 = \hbox{tr}\,(\Xi\;\Lambda).
\label{eq:masses}
\end{equation}
So we see that one mode will be a massless phonon\index{phonon} while
the other will have a non zero mass. Explicitly, in terms of the
elements of the underlying matrices
\begin{equation}
\omega_I^2=0; \qquad  \omega_{II}^2 =
-\frac{2\sqrt{\rho_{\A0}\,\rho_{\B0}} \;\lambda }{\hbar^2}
\{ \tilde U_{\A\A} + \tilde U_{\B\B} - 2 \tilde U_{\A\B} \}
\label{eq:masses2}
\end{equation}
so that (before any fine-tuning or decoupling)
\begin{equation}
\omega_{II}^2 = 
-\frac{2\sqrt{\rho_{\A0}\,\rho_{\B0}} \;\lambda}{\hbar^2} 
 \times \left\{ U_{\A\A} +  U_{\B\B} - 2 U_{\A\B}
-{\lambda\over2\sqrt{\rho_{\A0}\,\rho_{\B0}}}  \left[
  \sqrt{\rho_{\A0}\over \rho_{\B0}} 
+\sqrt{\rho_{\B0}\over
    \rho_{\A0}}\right]^2 \right\}.
\label{eq:m-ndiag}
\end{equation}
It is easy to check that this quantity really does have the physical
dimensions of a frequency.

\subsubsection{Mono-metricity conditions\label{sec:mono-metr}}
In order for our system to be a perfect analogue of
special relativity:\index{special~relativity}
\begin{itemize}
\item we want each mode to have a quadratic dispersion
  relation;
\item we want each dispersion relation to have the same
  asymptotic slope.
\end{itemize}
Let us start by noticing that the dispersion relation
(\ref{eq:disp-rel-hydro}) is of the form
\begin{equation}
\omega^2 = [\hbox{quadratic}_1] \pm
\sqrt{[\hbox{quartic}]}.
\end{equation}
The first condition implies that the quartic must be a
perfect square
\begin{equation}
[\hbox{quartic}] = [\hbox{quadratic}_2]^2,
\end{equation}
but then the second condition implies that the slope of
this quadratic must be zero. That is
\begin{equation}
[\hbox{quadratic}_2](k^2) = [\hbox{quadratic}_2](0),
\end{equation}
and so
\begin{equation}
[\hbox{quartic}](k^2) =[ \hbox{quartic}](0) 
\end{equation}
must be constant independent of $k^2$, so that the
dispersion relation is of the form
\begin{equation}
\omega^2 = [\hbox{quadratic}_1](k^2) \pm[\hbox{quadratic}_2](0).
\end{equation}
Note that this has the required form (two hyperbolae with the same
asymptotes, and possibly different intercepts). Now let us implement
this directly in terms of the matrices $C_0^2$ and $M^2$.

\noindent\emph{Step 1:}
Using the results of the Appendix~\ref{E:A-2-matrices}, specifically equation (\ref{E:A-2-matrices}):
\begin{eqnarray}
\det[H^2(k)] &=& \det[\Omega^2 + C_0^2\; k^2 ] \\ &=&
\det[\Omega^2] - \tr\left\{ \Omega^2\; \bar C_0^2\right\} \; k^2 + \det[C_0^2]\;(k^2)^2.
\end{eqnarray}
(This holds for any linear combination of $2\times2$ matrices. Note
that we apply trace reversal to the squared matrix $C_0^2$, we do not
trace reverse and then square.)
Since in particular $\det[\Omega^2]=0$,  we have:
\begin{equation}
\det[H^2(k)] = - \tr\left\{ \Omega^2\;\bar C_0^2 \right\} \; k^2 + \det[C_0^2]\;(k^2)^2.
\end{equation}

\noindent\emph{Step 2:}
Now consider the discriminant (the quartic)
\begin{eqnarray}
 \qquad \hbox{quartic} &\equiv&
\hbox{tr}[H(k^2)]^2 - 4\; \det[H(k^2)] 
\\ 
 &=& (\tr[\Omega^2]+\tr[C_0^2]\; k^2)^2 - 4 \big[
    -\tr\left\{ \Omega^2\;\bar C_0^2 \right\} \; k^2 
    + \det[C_0^2]\;(k^2)^2 \big] \\
 &=& \tr[\Omega^2]^2 + \{2 \tr[\Omega^2]\tr[C_0^2]+4
   \tr\left\{ \Omega^2\;\bar C_0^2 \right\} \} k^2 
   + \left\{ \tr[C_0^2]^2 - 4\det[C_0^2]\right\} (k^2)^2 \; \; \\
 &=& \tr[\Omega^2]^2 +
  2 \{2 \tr\left\{ \Omega^2\; C_0^2 \right\} - \tr[\Omega^2]\tr[C_0^2] \} k^2 
  + \left\{ \tr[C_0^2]^2 - 4\det[C_0^2]\right\} (k^2)^2. \; \;
\end{eqnarray}
So in the end the two conditions above for mono-metricity
take the form
\begin{equation}
\hbox{mono-metricity}\iff\left\{ 
\begin{array}{l}
\tr[C_0^2]^2 - 4\;\det[C_0^2] =0;
\\
2 \tr\left\{ \Omega^2\; C_0^2 \right\} -  \tr[\Omega^2] \;\tr[C_0^2]= 0.
\end{array}\right .
\end{equation}
Once these two conditions are satisfied the dispersion
relation is
\begin{equation}
\omega^2 = { \hbox{tr}[H(k^2)] \pm \tr[\Omega^2] \over 2} =
      {\tr[\Omega^2]\pm\tr[\Omega^2] + \tr[C_0^2]\; k^2\over 2}
\end{equation}
whence
\begin{equation}
\omega_1^2 = {1\over2} \tr[C_0^2]\; k^2=c_0^2k^2
\qquad
\omega_2^2 = \tr[\Omega^2] + {1\over2} \tr[C_0^2]\; k^2=\omega_{II}^2+c_0^2k^2,
\end{equation}
as required. One mode is massless, one massive with
exactly the ``mass'' previously deduced. One can now define the quantity
\begin{equation}
m_{II} = \hbar \omega_{II}/c_0^2,
\end{equation}
which really does have the physical dimensions of a mass.

\subsubsection{Interpretation of the mono-metricity conditions\label{S:C1C2}}
But now we have to analyse the two simplification
conditions
\begin{eqnarray}
C1:&&\qquad \tr[C_0^2]^2 - 4\;\det[C_0^2]= 0;
\label{eq:monometr1}
\\
C2:&&\qquad 2\; \tr\left\{ \Omega^2\; C_0^2 \right\} - \tr[\Omega^2]\tr[C_0^2]= 0;
 \label{eq:monometr2}
\end{eqnarray}
to see what they tell us.  The first of these conditions is equivalent
to the statement that the $2\times2$ matrix $C_0^2$ has two identical
eigenvalues.\index{eigenvalues} But since $C_0^2$ is symmetric this
then implies $C_0^2 = c_0^2 \; \mathbf{I}$, in which case the second
condition is automatically satisfied. (In contrast, condition $C2$
does not automatically imply condition $C1$.)  Indeed if $C_0^2 =
c_0^2 \; \mathbf{I}$, then it is easy to see that (in order to make
$C_0^2$ diagonal)
\begin{equation}
\tilde U_{\A\B}=0,
\end{equation}
(which is sufficient, by itself, to imply bi-metricity) and
furthermore that
\begin{equation}
{\tilde U_{\A\A} \;\rho_{\A0}\over m_\A} = c_0^2 = {\tilde
  U_{\B\B} \;\rho_{\B0}\over m_\B}.
  \label{eq:c0ft}
\end{equation}
Note that we can now solve
for $\lambda$ to get
\begin{equation}
\lambda = -2 \sqrt{\rho_{\A0}\;\rho_{\B0}} \; U_{\A\B},
\label{eq:lbd}
\end{equation}
whence
\begin{equation}
c_0^2 = {U_{\A\A\;}\rho_{\A0}+U_{\A\B}\; \rho_{\B0}\over m_\A} =
{U_{\B\B\;}\rho_{\B0}+U_{\A\B}\; \rho_{\A0}\over m_\B},
\end{equation} 
and
\begin{equation}
\omega_{II}^2 = {4\rho_{\A0}\rho_{\B0}U_{\A\B}\over \hbar^2}  \left\{
U_{\A\A} + U_{\B\B} - 2U_{\A\B} + U_{\A\B}  \left[
  \sqrt{\rho_{\A0}\over \rho_{\B0}} +\sqrt{\rho_{\B0}\over
    \rho_{\A0}}\right]^2 \right\}.
\label{eq:m-diag}
\end{equation}
Note that (\ref{eq:m-diag}) is equivalent to
(\ref{eq:m-ndiag}) with (\ref{eq:lbd}) enforced. But this then implies
\begin{equation}
\label{E:m2}
\omega_{II}^2 = {4\rho_{\A0}\rho_{\B0}U_{\A\B}\over \hbar^2}  \left\{
U_{\A\A} + U_{\B\B} + U_{\A\B}  \left[
  {\rho_{\A0}\over \rho_{\B0}} +{\rho_{\B0}\over
    \rho_{\A0}}\right] \right\}.
\end{equation}
\emph{Interpretation:} Condition $C2$ forces the two low-momentum
``propagation speeds'' to be the same, that is, it forces the two
$O(k^2)$ coefficients to be equal. Condition $C1$ is the stronger
statement that there is no $O(k^4)$ (or higher order) distortion to
the relativistic dispersion relation.

\subsection{Beyond the hydrodynamic approximation \label{sec:qp}}

At this point we want to consider the deviations from the previous
analogue for special relativity.\index{special~relativity} Our
starting point is again Eq.~(\ref{eq:last}), now retaining the
quantum \index{quantum~pressure} pressure term, which we Fourier
transform to get:
\begin{equation}
\omega^2 {\tilde{\theta}}  = 
\left\{
\sqrt{\Xi+X\; k^2} \;\; [D\; k^2+\Lambda]\;\; \sqrt{\Xi+X\;k^2} 
\right\}\;  \tilde{\theta}  \equiv H(k^2) \; \tilde{\theta}.
   \label{eq:new-disp-rel}
\end{equation}
This leads to the Fresnel\index{Fresnel~equation} equation
\begin{equation}
\det\{ \omega^2 \;\mathbf{I} - H(k^2) \} =0.
\end{equation}
That is
\begin{equation}
\omega^4 - \omega^2 \; \hbox{tr}[H(k^2)] + \det[H(k^2)] =
0,
\end{equation}
whence
\begin{equation}
\omega^2 = { \hbox{tr}[H(k^2)] \pm \sqrt{
    \hbox{tr}[H(k^2)]^2 - 4\;  \det[H(k^2)] }\over 2},
\label{eq:tot-disp-rel}
\end{equation}
which is now of the form
\begin{equation}
\omega^2 = [\hbox{quartic}_1] \pm \sqrt{[ \hbox{octic} ]}.
\end{equation}

\subsubsection{Masses}
The ``masses'', defined as the zero momentum oscillation frequencies,
are again easy to identify. Just note that the $k$-independent term in
the Fresnel\index{Fresnel~equation} equation is exactly the same mass matrix
$\Omega^2=\Xi^{1/2}\;\Lambda\;\Xi^{1/2}$ that was present in the
hydrodynamical limit. (That is, the quantum potential term $X$
does not influence the masses.)

\subsubsection{Dispersion relations}
Differently from the previous case, when the hydrodynamic
\index{hydrodynamic~limit} approximation held, we now have that the
discriminant of (\ref{eq:tot-disp-rel}) generically can be an
eighth-order polynomial in $k$.  In this case we cannot hope to
recover an exact analogue of special
relativity,\index{special~relativity} but instead can at best hope to
obtain dispersion relations with vanishing or {\em suppressed}
deviations from special relativity at low $k$; possibly with large
deviations from special relativity at high momenta. From the form of
our equation it is clear that the Lorentz
\index{Lorentz~invariance!violation} violation suppression should be
somehow associated with the masses of the atoms $m_{A/B}$. Indeed we
will use the underlying atomic masses to define our ``Lorentz
\index{Lorentz~invariance!violation} breaking scale'', which we shall
then assume can be identified with the ``quantum gravity scale''.  The
exact form and relative strengths of the higher-order terms will be
controlled by tuning the 2--BEC system and will eventually decide the
manifestation (or not) of the naturalness problem and of the
universality issue.

Our approach will again consist of considering derivatives of
(\ref{eq:tot-disp-rel}) in growing even powers of $k^2$ (recall that
odd powers of $k$ are excluded by the parity invariance of the system)
and then setting $k\to0$. We shall compute only the coefficients up to
order $k^4$ as by simple dimensional arguments one can expect any
higher order term will be further suppressed with respect to the $k^4$
one.

We can greatly simplify our calculations if before performing our
analysis we rearrange our problem in the following way. First of all
note that by the cyclic properties of trace
\begin{eqnarray}
\tr[H(k^2)] &=& \tr[ (Dk^2+\Lambda)\;(\Xi+k^2X)] 
\\
&=& \tr[ \Lambda \Xi + k^2(D\Xi+\Lambda X) + (k^2)^2 D X]
\\
&=& \tr[  \Xi^{1/2} \Lambda \Xi^{1/2} + k^2(\Xi^{1/2} D \Xi^{1/2} 
+ X^{1/2}\Lambda X^{1/2})
+ (k^2)^2 X^{1/2} D X^{1/2}].
\end{eqnarray}
Putting this all together, we can now define symmetric matrices
\begin{equation}
\Omega^2 =   \Xi^{1/2} \Lambda \Xi^{1/2}; 
\end{equation}
\begin{equation}
C_0^2 = \Xi^{1/2} D \Xi^{1/2}; \qquad
\Delta C^2 =  X^{1/2}\Lambda X^{1/2};
\end{equation}
\begin{equation}
 C^2 = C_0^2 + \Delta C^2 = \Xi^{1/2} D \Xi^{1/2} + X^{1/2}\Lambda X^{1/2}; 
\end{equation}
\begin{equation}
Z^2 =  2 X^{1/2} D X^{1/2} = {\hbar^2\over 2} M^{-2}.
\end{equation}
With all these definitions we can then write
\begin{equation}
\tr[H(k^2)] = \tr\left[ \Omega^2 + k^2 (C_0^2+\Delta C^2) + {1\over2} (k^2)^2 Z^2 \right],
\end{equation}
where everything has been done inside the trace. If we now define
\begin{equation}
H_s(k^2) = \Omega^2 + k^2 (C_0^2+\Delta C^2) + {1\over2} (k^2)^2 Z^2,
\end{equation}
then $H_s(k^2)$ is by definition both polynomial and symmetric and satisfies
\begin{equation}
\tr[H(k^2)] = \tr[H_s(k^2)],
\end{equation}
while in contrast, 
\begin{equation}
\det[H(k^2)] \neq \det[H_s(k^2)].
\end{equation}
But then
\begin{equation}
\omega^2 = {1\over2}\left[ \tr[H_s(k^2)] \pm \sqrt{ \tr[H_s(k^2)]^2-4\det[H(k^2)]}\right].
\end{equation}
Whence
\def\d{{\mathrm{d}}}
\begin{equation}
{\d\omega^2\over\d k^2} = {1\over2}\left[\tr[H_s'(k^2)] \pm 
{
\tr[H_s(k^2)]\tr[H_s'(k^2)] - 2\det'[H(k^2)] \over
\sqrt{ \tr[H_s(k^2)]^2- 4\det[H(k^2)] }
}\right],
\end{equation}
and at $k=0$
\begin{equation}
\left.{\d\omega^2\over\d k^2}\right|_{k\to0} = {1\over2}\left[\tr[C^2] \pm 
{
\tr[\Omega^2]\tr[C^2] - 2\det'[H(k^2)]_{k\to0} \over
 \tr[\Omega^2] 
}\right].
\end{equation}
But now let us consider 
\begin{eqnarray}
\det[H(k^2)] &=& \det[  (Dk^2+\Lambda)\;(\Xi+k^2X) ] 
\\
&=& 
\det[Dk^2+\Lambda]\;\det[\Xi+k^2X] 
\\
&=&
\det[ \Xi^{1/2} (Dk^2+\Lambda) \Xi^{1/2}] \;
\det[ I + k^2 \Xi^{-1/2} X \Xi^{-1/2} ] 
\end{eqnarray}
where we have repeatedly used properties of the determinant.
Furthermore
\begin{eqnarray}
 \det[ I + k^2 \Xi^{-1/2} X \Xi^{-1/2} ] &=& \det[I+k^2 \Xi^{-1} X] 
 \\
&=& \det[I+k^2 X^{1/2} \Xi X^{1/2}] 
 \\
 &=& \det[I+k^2 Y^2/2],
\end{eqnarray}
so that we have
\begin{equation}
\det[H(k^2)] = \det[\Omega^2+C_0^2 k^2] \; \det[I + k^2 Y^2/2].
\end{equation}
Note the matrix $Y^2$ is the ``healing\index{healing~length}
length matrix'' we had previously defined, and that the net result of
this analysis is that the full determinant is the product of the
determinant previously found in the hydrodynamic
\index{hydrodynamic~limit} limit with a factor that depends on the
product of wavenumber and healing\index{healing~length} length.

But now, given our formula (\ref{E:A-2-matrices}) for the determinant,
we see
\begin{eqnarray}
{\det}'[H(k^2)] &=& (-\tr(\Omega^2\bar C_0^2)+ 2 k^2 \det[C_0^2] ) \; \det[I + k^2 Y^2/2] 
\nonumber
\\
&&
+  \det[\Omega^2+C_0^2 k^2] \; (-\tr[\bar Y^2] +  k^2 \det[Y^2])/2,
\end{eqnarray}
whence
\begin{equation}
{\det}'[H(k^2)]_{k\to0} = -\tr(\Omega^2\bar C_0^2 ),
\end{equation}
and so
\begin{equation}
\left.{\d\omega^2\over\d k^2}\right|_{k\to0} = {1\over2}\left[\tr[C^2] \pm 
{
\tr[\Omega^2]\tr[C^2] + 2 \tr(\Omega^2\bar C_0^2 ) \over
 \tr[\Omega^2] 
}\right].
\end{equation}
That is:
\begin{equation}
\label{eq: varpi2}
\left.{\d\omega^2\over\d k^2}\right|_{k\to0} = 
{1\over2}\left[\tr[C^2] 
\pm 
\left\{ \tr[C^2] + 2{ \tr(\Omega^2\bar C_0^2 ) \over \tr[\Omega^2] }
\right\}
\right].
\end{equation}
Note that all the relevant matrices have been carefully symmetrized.
Also note the important distinction between $C_0^2$ and $C^2$.  Now
define
\begin{equation}
c^2 = {1\over2}\tr[C^2],
\end{equation}
then
\begin{equation}
\left.{\d\omega^2\over\d k^2}\right|_{k\to0} = c^2 (1\pm \eta_2),
\end{equation}
with
\begin{equation}
\eta_2 
= \left\{ 
{ \tr[C^2]\tr[\Omega^2] + 2 \tr(\Omega^2\bar C_0^2 ) 
\over \tr[\Omega^2] \tr[C^2]}
\right\}
 = \left\{ 1 + {\tr(\Omega^2\bar C_0^2 ) \over \omega_{II}^2 \; c^2}
\right\}.
\end{equation}
Similarly, consider the second derivative: 
\begin{eqnarray}
{\d^2\omega^2\over\d(k^2)^2} &=&  {1\over2}\Bigg[\tr[H_s''(k^2)] 
\pm 
{
\tr[H_s(k^2)]\tr[H_s''(k^2)] + \tr[H_s'(k^2)]\tr[H_s'(k^2)]- 2\det''[H(k^2)] \over
\sqrt{ \tr[H_s(k^2)]^2- 4\det[H(k^2)] }
}
\nonumber
\\
&&
\mp 
{
( \tr[H_s(k^2)]\tr[H_s'(k^2)] - 2\det'[H(k^2)] )^2\over
(\tr[H_s(k^2)]^2- 4\det[H(k^2)] )^{3/2}
}
\Bigg],
\end{eqnarray}
whence
\begin{eqnarray}
\left.{\d^2\omega^2\over\d(k^2)^2}\right|_{k\to0} &=&  {1\over2}\Bigg[\tr[Z^2] 
\pm 
{
\tr[\Omega^2]\tr[Z^2] + \tr[C^2]^2- 2\det''[H(k^2)]_{k\to0} \over
\tr[\Omega^2]
}
\nonumber
\\
&&
\mp 
{
( \tr[\Omega^2]\tr[C^2] - 2\det'[H(k^2)]_{k\to0} )^2\over
\tr[\Omega^2]^3
}
\Bigg].
\end{eqnarray}
The last term above can be related to $\d\omega^2/\d k^2$, while the
determinant piece is evaluated using
\begin{eqnarray}
{\det}''[H(k^2)] &=& (2\det[C_0^2] ) \; \det[I + k^2 Y^2/2] 
\\
&&
+ (-\tr(\Omega^2\bar C_0^2)+ 2 k^2 \det[C_0^2] ) \;  (-\tr[\bar Y^2] +  k^2 \det[Y^2])/2
\nonumber
\\
&&
+  \det[\Omega^2+C_0^2 k^2] \; ( \det[Y^2]/2)
\nonumber
\\
&&
+   (-\tr(\Omega^2\bar C_0^2)+ 2 k^2 \det[C_0^2] )  \; (-\tr[\bar Y^2] +  k^2 \det[Y^2])/2.
\nonumber
\end{eqnarray}
Therefore
\begin{eqnarray}
{\det}''[H(k^2)]_{k\to0} &=& (2\det[C_0^2] ) 
+ (-\tr(\Omega^2\bar C_0^2) ) \;  (-\tr[\bar Y^2] )/2
+  \det[\Omega^2] \; (\det[Y^2])/2
\nonumber
\\
&&
+   (-\tr(\Omega^2\bar C_0^2))  \; (-\tr[\bar Y^2])/2.
\end{eqnarray}
That is, (recalling $\tr[\bar A] = - \tr[A]$),
\begin{equation}
{\det}''[H(k^2)]_{k\to0} = (2\det[C_0^2] ) 
-(\tr(\Omega^2\bar C_0^2) ) \;  (\tr[Y^2] ),
\end{equation}
or
\begin{equation}
{\det}''[H(k^2)]_{k\to0} = -\tr[C_0^2 \bar C_0^2]
- \tr[\Omega^2\bar C_0^2] \;  \tr[Y^2] .
\end{equation}
Now assembling all the pieces, a little algebra yields
\begin{eqnarray}
 \left.{\d^2\omega^2\over\d(k^2)^2}\right|_{k\to0} =  
{\textstyle 1\over \textstyle 2} \Bigg[
\tr[Z^2]  \pm \tr[Z^2] 
\pm 2 \frac{\tr[\Omega^2\bar{C}^2_0]}{\tr[\Omega^2]}\tr[Y^2]
\pm {\tr[C^2]^2- 4\det[C^2_0] \over \tr[\Omega^2] }
\mp
{\tr[C^2]^2\over\tr[\Omega^2]} \eta_2^2
\Bigg].\label{eq: varpi4}
\end{eqnarray}

With the above formula we have completed our derivation of the
lowest-order terms of the generic dispersion relation of a coupled
2-BEC \index{2-BEC~system} system --- including the terms introduced
by the quantum potential at high wavenumber --- up to terms of order
$k^4$.  From the above formula it is clear that we do not generically
have Lorentz \index{Lorentz~invariance} invariance in this system:
Lorentz \index{Lorentz~invariance!violation} violations arise both due
to mode-mixing interactions (an effect which can persist in the
hydrodynamic \index{hydrodynamic~limit} limit where $Z\to0$ and
$Y\to0$) and to the presence of the quantum potential (signaled by
$Z\neq0$ and $Y\neq0$).  While the mode-mixing effects are relevant at
all energies the latter effect characterizes the discrete structure of
the effective spacetime at high energies. It is in this sense that the
quantum potential determines the analogue of quantum gravity effects
in our 2-BEC \index{2-BEC~system} system.
%

\subsection{The relevance for quantum gravity phenomenology\label{sec:qgp}}

Following this physical insight we can now easily identify a regime
that is potentially relevant for simulating the typical ans\"atze of
quantum gravity\index{quantum~gravity!phenomenology} phenomenology.
We demand that any violation of Lorentz
\index{Lorentz~invariance!violation} invariance present should be due
to the microscopic structure of the effective spacetime. This implies
that one has to tune the system in order to cancel exactly all those
violations of Lorentz \index{Lorentz~invariance!violation} invariance
which are solely due to mode-mixing interactions in the hydrodynamic
\index{hydrodynamic~limit} limit.

We basically follow the guiding idea that a good analogue of
quantum-gravity-induced Lorentz \index{Lorentz~invariance!violation}
violations should be characterized only by the ultraviolet physics of
the effective spacetime. In the system at hand the ultraviolet physics
is indeed characterized by the quantum potential, whereas possible
violations of the Lorentz \index{Lorentz~invariance!violation}
invariance in the hydrodynamical limit are low energy effects, even
though they have their origin in the microscopic interactions. We
therefore start by investigating the scenario in which the system is
tuned in such a way that no violations of Lorentz
\index{Lorentz~invariance} invariance are present in the hydrodynamic
\index{hydrodynamic~limit} limit. This leads us to again enforce the
conditions $C1$ and $C2$ which corresponded to ``mono-metricity" in
the hydrodynamic \index{hydrodynamic~limit} limit.

In this case (\ref{eq: varpi2}) and  (\ref{eq: varpi4}) take respectively the form
\begin{eqnarray}
\left.{\d\omega^2\over\d k^2}\right|_{k\to0} &=& 
{1\over2}\left[\tr[C_0^2] +(1\pm1)\,\tr[\Delta C^2]\right]
= c_0^2 + {1\pm1\over2}\tr[\Delta C^2],\label{eq:varpi2b}
\end{eqnarray}
and
\begin{eqnarray}
\left.{\d^2\omega^2\over\d(k^2)^2}\right|_{k\to0} &=&
{\tr[Z^2]  \pm \tr[Z^2] \over 2}\mp\tr[C^2_0]\tr[Y^2]
\pm{1\over 2} {\tr[\Delta C^2]^2+2\tr[C^2_0]\tr[\Delta C^2] \over \tr[\Omega^2] }
\mp {1\over 2}
{
\tr[\Delta C^2]^2\over
\tr[\Omega^2]
}
\nonumber
\\
&=&
{\tr[Z^2]  \pm \tr[Z^2] \over 2}\pm\tr[C^2_0]\left(-\tr[Y^2]+
{\tr[\Delta C^2] \over \tr[\Omega^2] }\right).
\label{eq:varpi4b}
\end{eqnarray}

Recall (see section \ref{S:C1C2}) that the first of the physical
conditions $C1$ is equivalent to the statement that the $2\times2$
matrix $C_0^2$ has two identical eigenvalues. But since $C_0^2$ is
symmetric this then implies $C_0^2 = c_0^2 \; \mathbf{I}$, in which
case the second condition is automatically satisfied.  This also leads
to the useful facts
\begin{eqnarray}
&&\tilde{U}_{\A\B}=0 \quad \Longrightarrow \quad \lambda = -2 \sqrt{\rho_{\A0}\;\rho_{\B0}} \; U_{\A\B}
\label{eq:lambdaUAB} ;\\
&& c_0^2 ={\tilde U_{\A\A} \;\rho_{\A0}\over m_\A} =  {\tilde
  U_{\B\B} \;\rho_{\B0}\over m_\B}.
    \label{eq:c0ft2}
\end{eqnarray}
Now that we have the fine tuning condition for the laser coupling we
can compute the magnitude of the effective mass of the massive
phonon\index{phonon} and determine the values of the Lorentz
\index{Lorentz~invariance!violation} violation coefficients. In
particular we shall start checking that this regime allows for a real
positive effective mass as needed for a suitable analogue model of
quantum gravity\index{quantum~gravity!phenomenology} phenomenology.

\subsubsection{Effective mass}
Remember that  the definition of $m_{II}$ reads 
\begin{equation}
m_{II}^2 = \hbar^2 \omega_{II}^2/c_0^4.
\label{eq:m2-c1}
\end{equation}
Using equation~(\ref{eq:lambdaUAB}) and equation~(\ref{eq:c0ft2})
we can rewrite $c^2_0$ in the following form
\begin{equation}
c_0^2 = [m_\B \rho_{\A0} U_{\A\A} + m_\A \rho_{\B0} U_{\B\B} 
         + U_{\A\B} (\rho_{\A0} m_\A + \rho_{\B0} m_\B)] /
         (2 m_\A m_\B).
         \label{eq:c0av}
\end{equation}
Similarly equation~(\ref{eq:lambdaUAB}) and equation~(\ref{eq:c0ft2})
when inserted in Eq.~(\ref{E:m2}) give
\begin{equation}
\omega_{II}^2 = \frac{4 U_{\A\B} (\rho_{\A0} m_\B + \rho_{\B0} m_\A) c_0^2}{ \hbar^2}.
\label{eq:om2}
\end{equation}
We can now estimate $m_{II}$ by simply inserting the above expressions
in equation~(\ref{eq:m2-c1}) so that
\begin{equation}
m_{II}^2 = {
8 U_{\A\B} (\rho_{\A0} m_\A+\rho_{\B0} m_\B) m_\A m_\B 
\over
          [m_\B \rho_{\A0} U_{\A\A} + m_\A \rho_{\B0} U_{\B\B} 
         + U_{\A\B} (\rho_{\A0} m_\A + \rho_{\B0} m_\B)]
}.
\end{equation}
This formula is still a little clumsy but a great deal can be
understood by doing the physically reasonable approximation $m_\A
\approx m_\B=m$ and $\rho_\A \approx \rho_\B$. In fact in this case one
obtains
\begin{equation}
m_{II}^2 \approx m^2  \; {
8 U_{\A\B}
\over [U_{\A\A}+2U_{\A\B}+U_{\B\B}]
}.
\end{equation}
This formula now shows clearly that, as long as the mixing term
$U_{\A\B}$ is small compared to the ``direct" scattering
$U_{\A\A}+U_{\B\B}$, the mass of the heavy phonon\index{phonon} will
be ``small" compared to the mass of the atoms.  Though experimental
realizability of the system is not the primary focus of the current
paper, we point out that there is no obstruction in principle to
tuning a 2-BEC \index{2-BEC~system} system into a regime where
$|U_{\A\B}| \ll |U_{\A\A}+U_{\B\B}|$.  For the purposes of this chapter
it is sufficient that a small effective phonon\index{phonon} mass
(small compared to the atomic masses which set the analogue quantum
gravity scale) is obtainable for some arrangement of the microscopic
parameters.
We can now look separately at the coefficients of the quadratic and
quartic Lorentz \index{Lorentz~invariance!violation} violations and
then compare their relative strength in order to see if a situation
like that envisaged by discussions of the naturalness problem is
actually realized.

\subsubsection{Coefficient of the quadratic deviation}
One can easily see from (\ref{eq:varpi2b}) that the $\eta_2$
coefficients for this case take the form
\begin{eqnarray}
\eta_{2,I} &=& 0;\\
\eta_{2,II}\; c_0^2  &=&\tr[ \Delta C^2]=\tr[X^{1/2} \Lambda X^{1/2}]=\tr[X\Lambda]\nonumber
\\
&=& -\frac{1}{2}\frac{\lambda}{m_\A m_\B}\left( 
\frac{m_\A\rho_{\A0}+m_\B\rho_{\B0}}
{\sqrt{\rho_{\A0}\rho_{\B0}}} \right).
\end{eqnarray}

So if we insert the fine tuning condition for $\lambda$,
equation~(\ref{eq:lambdaUAB}), we get
\begin{eqnarray}
\eta_{2,II}  
=\frac{U_{\A\B}\left( 
m_\A\rho_{\A0}+m_\B\rho_{\B0}\right)
}{m_\A m_\B c_0^2}.
\label{E:eta_2II}
\end{eqnarray}
Remarkably we can now cast this coefficient in a much more suggestive
form by expressing the coupling $U_{\A\B}$ in terms of the mass of the
massive quasi-particle\index{quasi-particle} $m_{II}^2$.  In order to do this we start from
Eq.~(\ref{eq:om2}) and note that it enables us to express
$U_{\A\B}$ in (\ref{E:eta_2II}) in terms of $\omega_{II}^2$, thereby
obtaining
\begin{equation}
\eta_{2,II} =\frac{\hbar^2}{4 c^4_0} \; 
\frac{\rho_{\A0} m_\A + \rho_{\B0} m_\B}{\rho_{\A0} m_\B + \rho_{\B0} m_\A} \;
\frac{\omega_{II}^2 }{m_\A m_\B}.
\end{equation}
Now it is easy to see that
\begin{equation}
\frac{\rho_{\A0} m_\A + \rho_{\B0} m_\B}{\rho_{\A0} m_\B + \rho_{\B0} m_\A}
\approx \mathcal{O} (1),
\end{equation}
and that this factor is identically unity if either $m_\A = m_\B$ or $\rho_{\A0} = \rho_{\B0}$.
All together we are left with
\begin{equation} 
\eta_{2,II}  = \bar{\eta} \left(
\frac{m_{II} }{\sqrt{m_\A m_\B}} \right)^2, 
\label{eq:eta2qfin}
\end{equation}
where $\bar{\eta}$ is a dimensionless coefficient of order unity.

The product in the denominator of the above expression can be
interpreted as the geometric mean of the fundamental bosons masses
$m_\A$ and $m_\B$. These are mass scales associated with the
microphysics of the condensate --- in analogy with our experience with
a 1-BEC \index{1-BEC~system} system where the ``quantum gravity
scale'' is set by the mass of the BEC atoms. It is then natural to
define an analogue of the scale of the breakdown of Lorentz
\index{Lorentz~invariance!violation} invariance as $M_{\rm
  eff}=\sqrt{m_\A m_\B}$.  (Indeed this ``analogue Lorentz breaking
scale'' will typically do double duty as an ``analogue Planck
\index{Planck~scale} mass".)

Using this physical insight it should be clear that
equation~(\ref{eq:eta2qfin}) effectively says
\begin{equation}
\eta_{2,II}\approx\left(\frac{m_{II}}{M_{\rm eff}}\right)^2,
\label{eq:eta2fin}
\end{equation}
which, given that $m_I=0$, we are naturally lead to generalize to
\begin{equation}
\eta_{2,X}\approx\left(\frac{m_X}{M_{\rm eff}}\right)^2= 
\left( {\hbox{mass scale of quasiparticle}
\over
\hbox{effective Planck scale}}\right)^2; \qquad X=I,II.
\label{eta2_final}
\end{equation}
The above relation is exactly the sort of dimensionless ratio
$(\mu/M)^\sigma$ that has been very often \emph{conjectured} in the
literature on quantum gravity\index{quantum~gravity!phenomenology}
phenomenology in order to explain the strong observational constraints
on Lorentz \index{Lorentz~invariance!violation} violations at the
lowest orders. (See earlier discussion.)
Does this now imply that this particular regime of our 2-BEC
\index{2-BEC~system} system will also show an analogue version of the
naturalness problem? In order to answer this question we need to find
the dimensionless coefficient for the quartic deviations, $\eta_4$,
and check if it will or won't itself be suppressed by some power of
the small ratio $m_{II}/M_{\rm eff}$.

\subsubsection{Coefficients of the quartic deviation}
Let us now consider the coefficients of the quartic term presented in
equation~(\ref{eq:varpi4b}).  For the various terms appearing in
(\ref{eq:varpi4b}) we get
\begin{equation}
\tr[Z^2]=2\tr[DX]=
\frac{\hbar^2}{2}\left(\frac{m^2_\A+m^2_\B}{m^2_\A m^2_\B}\right);
\end{equation}
\begin{equation}
\tr[\Delta C^2]=\tr[X \Lambda]=
-\frac{\lambda}{2}\frac{m_\A\rho_{\A0}+m_\B\rho_{\B0}}
{m_\A m_\B\sqrt{\rho_{\A0}\rho_{\B0}}}
=
U_{\A\B}\frac{m_\A\rho_{\A0}+m_\B\rho_{\B0}}{m_\A m_\B};
\end{equation}
\begin{equation}
\tr[Y^2]=2\tr[X\Xi^{-1}]=\frac{\hbar^2}{2}
\frac{\rho_{\A0}m_{\A}\tilde{U}_{\A\A}+\rho_{\B0}m_{\B}\tilde{U}_{\B\B}}
{\rho_{\A0}m_{\A}\rho_{\B0} m_{\B} \tilde{U}_{\A\A}\tilde{U}_{\B\B}};
\end{equation}
where in the last expression we have used the fact that in the current
scenario $\tilde{U}_{\A\B}=0$.
Now by definition 
\begin{equation}
\eta_4 = 
{1\over2} ({M_\mathrm{eff}^2}/{\hbar^2}) \left[{\d^2\omega^2\over (\d k^2)^2}\right]_{k=0}
\end{equation}
 is the dimensionless coefficient in front of the $k^4$. So
\begin{eqnarray}
\eta_4 &=& 
\frac{M_\mathrm{eff}^2}{2\hbar^2} \left[ {\tr[Z^2]  \pm 
\tr[Z^2] \over 2}\pm\tr[C^2_0]\left(-{\tr[Y^2]\over2}+
{\tr[\Delta C^2] \over \tr[\Omega^2] }\right)\right]\\
&=& 
\frac{ M_\mathrm{eff}^2\; c_0^2}{\hbar^2} 
\left[ 
{\tr[Z^2]  \pm \tr[Z^2] \over 2\tr[C^2_0]}\pm\left(-{\tr[Y^2]\over2}+
{\tr[\Delta C^2] \over \tr[\Omega^2] }\right)\right].
\end{eqnarray}
Whence 
\begin{eqnarray}
\eta_{4,I} &=& \frac{ M_\mathrm{eff}^2\;c_0^2}{\hbar^2}
\left[ {\tr[Z^2] \over \tr[C^2_0]}+\left(-{\tr[Y^2]\over2}+
{\tr[\Delta C^2] \over \tr[\Omega^2] }\right)\right];
\\
\eta_{4,II} &=& \frac{ M_\mathrm{eff}^2\;c_0^2}{\hbar^2}
\left[ \left({\tr[Y^2]\over2}-
{\tr[\Delta C^2] \over \tr[\Omega^2] }\right)\right].
\end{eqnarray}
Let us compute the two relevant terms separately:
\begin{eqnarray}
{\tr[Z^2] \over \tr[C^2_0]} &=&
\frac{\hbar^2}{4c_0^2}\left(\frac{m^2_\A+m^2_\B}{m^2_\A m^2_\B}\right)=
\frac{\hbar^2}{4 c_0^2 M_\mathrm{eff}^2}
\left(\frac{m^2_\A+m^2_\B}{m_\A m_\B}\right);
\end{eqnarray}
\begin{eqnarray}
-\tr[Y^2]/2+
{\tr[\Delta C^2] \over \tr[\Omega^2] } &=& 
-\frac{\hbar^2}{4 M_\mathrm{eff}^2}
\left[
\frac{\rho_{\A0}m_{\A}\tilde{U}_{\A\A}^2+\rho_{\B0}m_{\B}\tilde{U}_{\B\B}^2}
{\rho_{\A0}\rho_{\B0}\tilde{U}_{\A\A}\tilde{U}_{\B\B}
\left(\tilde{U}_{\A\A}+\tilde{U}_{\B\B}\right)}
\right]
\nonumber\\
&=& 
-\frac{\hbar^2}{4 M_\mathrm{eff}^2\;c_0^2}
\left[
\frac{m^2_{\A}\tilde{U}_{\A\A}+m^2_{\B}\tilde{U}_{\B\B}}
{m_{\A}m_{\B}\left(\tilde{U}_{\A\A}+\tilde{U}_{\B\B}\right)}
\right];
\end{eqnarray}
where we have used $\rho_{X0}\tilde{U}_{XX}=m_{X}c^2_0$ for $X=A,B$ as
in equation~(\ref{eq:c0ft2}).  Note that the quantity in square
brackets in the last line is dimensionless.  So in the end:
\begin{eqnarray}
 \eta_{4,I} &=&  \frac{1}{4}\left[
\left(\frac{m^2_\A+m^2_\B}{m_\A m_\B}\right)
-\frac{m^2_{\A}\tilde{U}_{\A\A}+m^2_{\B}\tilde{U}_{\B\B}}
{m_{\A}m_{\B}\left(\tilde{U}_{\A\A}+\tilde{U}_{\B\B}\right)}
\right]\\
&=&
\frac{1}{4}
\left[
\frac{m^2_{\A}\tilde{U}_{\B\B}+m^2_{\B}\tilde{U}_{\A\A}}
{m_{\A}m_{\B}\left(\tilde{U}_{\A\A}+\tilde{U}_{\B\B}\right)}\right];
\label{eq:eta4I}\\
 \eta_{4,II} &=& \frac{1}{4}
\left[
\frac{m^2_{\A}\tilde{U}_{\A\A}+m^2_{\B}\tilde{U}_{\B\B}}
{m_{\A}m_{\B}\left(\tilde{U}_{\A\A}+\tilde{U}_{\B\B}\right)}\right].
\label{eq:eta4II}
\end{eqnarray}
\emph{Note:} In the special case $m_\A=m_\B$ we recover identical
quartic deviations $\eta_{4,I}=\eta_{4,II}=1/4$, indicating in this
special situation a ``universal'' deviation from Lorentz
\index{Lorentz~invariance} invariance.  Indeed we also obtain
$\eta_{4,I}=\eta_{4,II}$ if we demand $\tilde{U}_{\A\A} =
\tilde{U}_{\B\B}$, even without fixing $m_\A=m_\B$.

Thus in the analogue spacetime\index{analogue~spacetime} we have
developed, the issue of {\em universality} is fundamentally related to
the complexity of the underlying microscopic system. As long as we
keep the two atomic masses $m_\A$ and $m_\B$ distinct we generically
have distinct $\eta_4$ coefficients (and the $\eta_2$ coefficients are
unequal even in the case $m_\A=m_\B$). However we can easily recover
identical $\eta_4$ coefficients, for instance, as soon as we impose
identical microphysics for the two BEC systems we couple.

\subsubsection{Avoidance of the naturalness problem}
We can now ask ourselves if there is, or is not, a naturalness problem
present in our system. Are the dimensionless coefficients
$\eta_{4,I/II}$ suppressed below their naive values by some small
ratio involving $M_{\rm eff}=\sqrt{m_\A m_\B}$~? Or are these ratios
unsuppressed?  Indeed at first sight it might seem that further
suppression is the case, since the square of the ``effective Planck
\index{Planck~scale} scale" seems to appear in the denominator of both
the coefficients (\ref{eq:eta4I}) and (\ref{eq:eta4II}). However, the
squares of the atomic masses also appear in the numerator, rendering
both coefficients of order unity.

It is perhaps easier to see this once the dependence of
(\ref{eq:eta4I}) and (\ref{eq:eta4II}) on the effective coupling
$\tilde{U}$ is removed. We again use the substitution
$\tilde{U}_{XX}=m_{X}c^2_0/\rho_{X0}$ for $X=\A,\B$, so obtaining:
\begin{eqnarray}
\eta_{4,I} &=&  \frac{1}{4}
\left[
\frac{m_{\A}\rho_{\A0}+m_{\B}\rho_{\B0}}{m_{\A}\rho_{\B0}+m_{\B}\rho_{\A0}}
\right];
\label{eq:eta4Ifin}\\
&&\nonumber\\
\eta_{4,II} &=& \frac{1}{4}
\left[
\frac{m_{\A}^3 \rho_{\B0}+m_{\B}^3 \rho_{\A0}} 
{
m_\A m_\B  \;( m_{\A}\rho_{\B0}+m_{\B}\rho_{\A0})}
\right].
\label{eq:eta4IIfin}
\end{eqnarray}
{From} these expressions it is clear that the $\eta_{4,I/II}$ coefficients
are actually of order unity.

That is, if our system is set up so that $m_{II}\ll m_{\A/\B}$ --- which
we have seen in this scenario is equivalent to requiring $U_{\A\B}\ll
U_{\A\A/\B\B}$ --- no naturalness problem arises as for $p > m_{II}\;c_0$
the higher-order, energy-dependent Lorentz-violating terms ($n\geq 4$)
will indeed dominate over the quadratic Lorentz-violating term.

It is quite remarkable that the quadratic coefficients
(\ref{eta2_final}) are {\em exactly} of the form postulated in several
works on non-renormalizable EFT with Lorentz
\index{Lorentz~invariance!violation} invariance violations (see
e.g.~\cite{Amelino-Camelia:2001aa}).  They are indeed the squared ratio of the
particle mass to the scale of Lorentz
\index{Lorentz~invariance!violation} violation.  Moreover we can see
from (\ref{eq:eta4I}) and (\ref{eq:eta4II}) that there is no further
suppression --- after having pulled out a factor $(\hbar /
M_{\mathrm{Lorentz~violation}})^2$ --- for the quartic coefficients
$\eta_{4,\mathrm{I/|I}}$. These coefficients are of order one and
generically non-universal, (though if desired they can be forced to be
universal by additional and specific fine tuning).

The suppression of $\eta_2$, combined with the \emph{non-suppression}
of $\eta_4$, is precisely the statement that the ``naturalness
problem'' does not arise in the current model. We stress this is not a
``tree level'' result as the dispersion relation was computed directly
from the fundamental Hamiltonian and was not derived via any EFT
reasoning.  Moreover avoidance of the naturalness problem is not
directly related to the tuning of our system to reproduce special relativity in the
hydrodynamic \index{hydrodynamic~limit} limit. In fact our conditions
for recovering special relativity at low energies do not \emph{a priori} fix the the
$\eta_2$ coefficient, as its strength after the ``fine tuning" could
still be large (even of order one) if the typical mass scale of the
massive phonon\index{phonon} is not well below the atomic mass scale.
Instead the smallness of $\eta_2$ is directly related to the
mass-generating \index{mass-generating~mechanism} mechanism.

The key question is now: Why does our model escape the naive
predictions of dominant lowest-dimension Lorentz
\index{Lorentz~invariance!violation} violations?
(In fact in our model for any $p\gg m_{II}$ the $k^4$ Lorentz
\index{Lorentz~invariance!violation} violating term dominates over the
order $k^2$ one.)  We here propose a nice interpretation in terms of
``emergent symmetry'':\index{emergent~symmetry} Non-zero $\lambda$
\emph{simultaneously} produces a non-zero mass for one of the
phonons,\index{phonon} \emph{and} a corresponding non-zero Lorentz
\index{Lorentz~invariance!violation} violation at order $k^2$.
(Single BEC systems have only $k^4$ Lorentz
\index{Lorentz~invariance!violation} violations as described by the
Bogoliubov dispersion relation.)  Let us now drive $\lambda\to 0$, but
keep the conditions $C1$ and $C2$ valid at each stage. (This also
requires $U_{\A\B}\to 0$.) One gets an
EFT\index{effective~field~theory} which at low energies describes two
non-interacting phonons\index{phonon} propagating on a common
background. (In fact $\eta_2\to0$ and $c_I=c_{II}=c_0$.) This system
possesses a $SO(2)$ symmetry.  Non-zero laser coupling $\lambda$
softly breaks this $SO(2)$, the mass degeneracy, and low-energy
Lorentz \index{Lorentz~invariance} invariance. Such soft Lorentz
\index{Lorentz~invariance!violation} violation is then characterized
(as usual in EFT)\index{effective~field~theory} by the ratio of the
scale of the symmetry breaking $m_{II}$, and that of the scale
originating the Lorentz \index{Lorentz~invariance!violation} violation
in first place $M_{\rm Lorentz~violation}$. We stress that the $SO(2)$ symmetry is
an ``emergent\index{emergent~symmetry} symmetry'' as it is not
preserved beyond the hydrodynamic \index{hydrodynamic~limit} limit:
the $\eta_4$ coefficients are in general different if $m_\A\neq m_\B$,
so $SO(2)$ is generically broken at high energies.  Nevertheless this
is enough for the protection of the {\em lowest}-order Lorentz
\index{Lorentz~invariance!violation} violating operators.  The lesson
to be drawn is that emergent symmetries are sufficient to minimize the
amount of Lorentz \index{Lorentz~invariance!violation} violation in
the lowest-dimension operators of the
EFT.\index{effective~field~theory} In this regard, it is intriguing to
realise that an interpretation of SUSY as an accidental symmetry has
indeed been considered in recent times~\cite{Goh:2005aa}, and that this is
done at the cost of renouncing attempts to solve the hierarchy problem
in the standard way. It might be that in this sense the smallness of
the particle physics mass scales with respect to the Planck scale
could be directly related to smallness of Lorentz violations in
renormalizable operators of the low-energy effective field theory we
live in. We hope to further investigate these issues in future work.

\chapter{Early universe cosmology in emergent spacetime \label{Chap:Early.Quantum.Cosmology}}
We now present an example of emergent cosmological spacetime as the hydrodynamic
limit of a more fundamental microscopic theory. The low-energy,
long-wavelength limit in our model is dominated by collective
variables that generate an effective Lorentzian metric. This
system naturally exhibits a microscopic mechanism allowing us to
perform controlled signature change between Lorentzian and
Riemannian geometries. We calculate the number of particles
produced from a finite-duration Euclidean-signature event, where
we take the position that to a good approximation the dynamics is
dominated by the evolution of the linearized perturbations, as
suggested by Calzetta and Hu [Phys. Rev. A {\bf 68} (2003)
043625]. We adapt the ideas presented by Dray {\emph{et al.}}
[Gen. Rel. Grav. {\bf 23} (1991) 967], such that  the
field and its canonical momentum are continuous at the signature-change event.\\
We investigate the interplay between the underlying microscopic
structure and the emergent gravitational field, focussing on its
impact on particle production in the ultraviolet regime. In
general, this can be thought of as the combination of
trans-Planckian physics and signature-change physics. Further we
investigate the possibility of using the proposed signature change
event as an amplifier for analogue ``cosmological particle
production'' in condensed matter experiments.
%
%
\section{Introduction and Motivation \label{sec:Motivation}}
%
%
Emergent spacetimes \cite{Hu:2005wu, Barcelo:2005ln,
Matt.-Visser:2002ot} allow us to approach the subject of curved
spacetime quantum field theory (CST-QFT) through the back door.
This has been demonstrated in detail by using ultra-cold
(non-relativistic), highly dilute Bose gases
\cite{Garay:2000ez,Garay:2001in,Barcelo:2001gt}. Under appropriate
conditions the fundamental microscopic theory can be replaced by a
classical mean-field, the Bose--Einstein condensate (BEC)
\cite{Castin:2001aa}. Collective excitations,  both classical and
quantum,  experience an effective spacetime whose entries are
purely macroscopic mean-field variables. The kinematic equations
for linearized perturbations --- neglecting back-reaction and
finite temperature effects --- are equivalent to covariant
minimally coupled scalar fields, with a d'Alembertian
\begin{equation}
\Delta \ldots ={1\over\sqrt{-g}} \, \partial_{a}(\sqrt{-g} \, g^{ab} \, \partial_{b} \ldots)
\end{equation}
 defined by an effective metric $g_{ab}$ \cite{Garay:2000ez,Garay:2001in,Barcelo:2001gt}.\\

Quantum field theory in curved spacetime is a good approximation
for semiclassical gravity --- at the level where back-reactions of
the quantum fields on the gravitational field are negligible
\cite{Birrell:1984aa}. As a consequence quantum effects in curved
spacetimes (CSTs) --- \eg, Hawking radiation
\cite{Unruh:1981bi,Visser:1993tk,Visser:1998gn}  and cosmological
particle production  --- do not require emergent Einstein gravity
\emph{per se}; the existence of an emergent spacetime, an
effective gravitational field $g_{ab}$, is sufficient. Of course
both systems, semiclassical gravity and any analogue model, must
involve some dynamics --- and so they will eventually diverge from
each other. The only possible loophole would be if Einstein
gravity were itself to be the ``hydrodynamics'' of some more
fundamental theory (of microscopic objects, for example strings,
molecules, or atoms). For our purposes a perfect match is not
required, and we refer the interested reader to
\cite{Sakharov:1968dy, Jacobson:1995eg, Barcelo:2001lu,
Visser:2002hf, Volovik:2003jn} for further details.

It has been shown that the repulsive or attractive nature of
atomic interaction in a Bose gas is directly related to the
signature of the low-energy emergent metric: Lorentzian $(-,+++)$
for repulsive interactions, Euclidean $(+,+++)$ for attractive
interactions \cite{Barcelo:2001gt}. In 2001 a BEC experiment
\cite{Donley:2001aa,Roberts:2001aa} was carried out that can be
viewed as the first analogue model experiment. By tuning through a
Feshbach resonance \cite{Inouye:1998aa} the atomic interactions
were driven into a weakly attractive regime, and triggered a
controlled condensate collapse. Two years later, the theoretical
work of Calzetta and Hu  \cite{Calzetta1:2003xb,Calzetta:2005yk},
connected the so-called Bose-nova phenomenon with the
amplification, mixing of positive and negative modes, and
squeezing of vacuum fluctuations due to a signature change event.
Perhaps surprisingly, the calculations carried out in
\cite{Calzetta1:2003xb,Calzetta:2005yk} did not include background
condensate dynamics and yet their theoretical predictions reflect
the experimental data relatively well. For short time-intervals of
attractive atom-atom interaction (\ie, a brief excursion into
Euclidean signature), the Bose-nova event is dominated by the
evolution of the quantum perturbations, and to a good
approximation independent of the background condensate dynamics.

Specifically, given that we do not as yet have any precise
detailed  model for emergent Einstein gravity, as opposed to
emergent curved spacetime, it seems necessary to focus on quantum
effects that are merely of kinematic rather than dynamic nature.
However, there are ways to study the influence of possible quantum
gravity candidates with CST-QFT. This branch of physics is called
quantum gravity phenomenology (QGP) \cite{Mattingly:2005aa}.
Emergent spacetimes can be used to analyze some portions of QGP,
where Lorentz invariance violations (LIV) are present at
ultraviolet scales. The LIV scale is supposed to be connected with
the Planck length, where new physics is expected. This idea is
naturally implemented in any emergent spacetime model. For
example, the BEC-based analogue models only recover Lorentz
invariance (LI) for the low-energy, long-wavelength phonon modes.
For trans-phononic modes microscopic corrections (\eg, quantum
pressure effects) have to be absorbed into the macroscopic
picture. The borderline between the two modes can be viewed as the
analogue LIV scale. Given that trans-phononic modes start to see
first signs of the fundamental microscopic theory, it makes sense
to speak of LIV at the analogue Planck scale. The key reason why
it is interesting to study modifications in the dispersion
relation is that many different effective field theories (EFTs)
already predict deviations at the kinematical level. A detailed
treatment of the analogue trans-Planckian model can be found in
\cite{Visser:2001ix,Liberati:2006sj,Liberati:2006kw,Weinfurtner:2006iv,Weinfurtner:2006nl}.

In ongoing work  \cite{Jain:2006ki} a classical phase-space method has
been used to numerically simulate cosmological particle production in BECs.
There it is shown that for a consistent treatment microscopic corrections
play an important role in the emergent spacetime picture. This leads to
emergent ``rainbow metrics'', with a Planck-suppressed momentum dependence
for the modes, and consequently leads to a modification in the quasi-particle
spectrum.

The main aim of this chapter is to 
merge all of the preceding points and address
the trans-Planckian problem for a signature
change event in a Bose gas. We are particularly interested in the
ultraviolet physics of the phonon modes and hence have chosen a
specific BEC set-up where the external trapping potential does not
interfere with the dynamics (\eg, hard-walled box). Step-by-step,
in section  \refb{sec:BEC.CST-QFT} we show how spacetime emerges
from a Bose gas; calculate the quasi-particle production from
sudden sound speed variations in section
\refb{sec:Sudden.Lorentzian.Variations}; extend this calculation
to sudden variations for finite regions with different signature
(Lorentzian $\leftrightarrow$ Euclidean) in section
\refb{sec:Sudden.Euclidean.Variations}; introduce ultraviolet
physics and re-calculate the quasi-particle production in section
\refb{sec:UV-corrections}; suggest in section \refb{sec:amplifier} to employ a finite Euclidean region as a particle amplifier for cosmological particle production in a BEC; and last but certainly not least, we address the theoretical and experimental impact of our results in section \refb{sec:EndHappy}.
%
\section{Emergent geometry from a Bose gas \label{sec:BEC.CST-QFT}}
%
The intent in this section is to give readers unfamiliar with the topic some deeper
understanding of BEC-based analogue spacetimes. Following the example of
\cite{Barcelo:2005ln} we introduce an ultra-cold (\ie, non-relativistic),
highly dilute and weakly interacting gas of Bosons, using the formalism of
canonical quantization, and derive the equation of motion for small quantum
fluctuations around some classical background; better known as a BEC, see
section \refb{subsec:Bose-gas}. The Bose--Einstein condensate is a state of
matter where the Bosons macroscopically occupy the lowest quantum state. In
section \refb{subsec:A-spacetime} we focus on the hydrodynamic case and recover
a covariant minimally coupled massless scalar field for small quantum
fluctuations in the BEC.
Subsequently, in section \refb{sec:UV-corrections}, we will revisit this
derivation and include trans-phononic modes (\eg, ultraviolet physics) into the
emergent spacetime picture.
%
\subsection{Ultra-cold, weakly interacting Bose gas\label{subsec:Bose-gas}}
Suppose we have a system of $N$ Bosons. In quantum field theory the field variables are quantum operators that act on quantum states (Hilbert space of states; \eg, Fock space).
Field operators either create, $\op{\psi}^{\dag}\tx$, or destroy, $\op{\psi}\tx$, an individual Boson at a particular point in space and time, and satisfy the commutators:
\begin{eqnarray}
\comBig{\op{\psi}\tx }{\op{\psi}\txp}=\comBig{\op{\psi}^{\dag}\tx}{\op{\psi}^{\dag}\txp}=0\,;
\label{Eq:com1}\\
\comBig{\op{\psi}\tx}{\op{\psi}^{\dag}\txp}=\delta(\mathbf{x}-\mathbf{x}')\,.
\label{Eq:com2}
\end{eqnarray}
For a gas of trapped, ultra-cold, highly dilute and weakly interacting Bosons the Hamiltonian is given by
\begin{eqnarray} \label{Eq:Hamiltonian}
\op{H \,}= \int{ d^{d}\mathrm{x} \left(
- \op{\psi}^{\dag} \frac{\hbar^{2}}{2m}\nabla^{2} \op{\psi}
+ \op{\psi}^{\dag} V_{\mathrm{ext}} \op{\psi}
+ \frac{U}{2} \, \op{\psi}^{\dag}\op{\psi}^{\dag}\op{\psi}\op{\psi} \right) } \, .
\end{eqnarray}
This is a sum of  the kinetic energy of the Boson field, and the
two potential energy contributions; the external trap
$V_{\mathrm{ext}}$, and the particle interactions. The extreme
dilution of the gas (\eg, $10^{13}-10^{15}\,
\mathrm{atoms}/\mathrm{cm}^{3}$) suppresses more-than-two-particle
interactions, and in the weakly interacting regime the actual
inter-atom potential has been approximated by a pseudo-contact
potential,
\begin{equation} \label{Eq:U}
U = \frac{4 \pi \hbar^{2} a}{m}\,.
\end{equation}
Here $m$ is the single-Boson mass, and $a$ the $s$-wave scattering length. The sign of the scattering length determines the qualitative behavior of the interactions,
\begin{equation} \label{Eq:nature_of_a}
\begin{array}{rl}
a > 0 & \quad\mathrm{repulsive}\, ; \\
a < 0 & \quad \mathrm{attractive}\, .
\end{array}
\end{equation}
Negative and positive values of $a$ are experimentally accessible by tuning external magnetic fields, that interact with the inter-atomic potential; this process is called Fesh\-bach resonance \cite{Inouye:1998aa}. We would like to emphasize the importance of Eq.~\refb{Eq:nature_of_a} for the remaining sections.

We now have all the necessary information about our system, encoded in Eqs.~\refb{Eq:com1}--\refb{Eq:Hamiltonian},  to calculate its dynamics. We use the Heisenberg equation of motion to get the time-evolution for the field operator;
\begin{equation} \label{Heisenberg}
i\hbar \frac{\partial \op{\psi}}{\partial t} = \left[ -\frac{\hbar^{2}}{2m} \nabla^{2} + V_{\mathrm{ext}} + U\, \op{\psi}^{\dag}\op{\psi}  \right] \op{\psi} \,.
\end{equation}
To apply this discussion to the emergent spacetime programme, we use the macroscopic occupation of the lowest quantum state below a critical temperature (\eg, for alkali gases, below $10^{-5}$ K). If the cooling process prohibits the gas to solidify, a new state of matter will occur, the Bose--Einstein condensate. The condensate is a complex-valued macroscopic mean-field
\begin{equation} \label{Eq:psi}
\langle \op{\psi}\tx \rangle =
\psi\tx = \sqrt{n_{0}\tx} \exp\left(i\theta_{0}\tx\right)\,,
\end{equation}
where the individual microscopic particles give way to collective variables. We will show that the condensate density, $n_{0}$, and phase, $\theta_{0}$, define the analogue spacetime for small quantum fluctuations in the BEC. The essential step in deriving the kinematics for the perturbation is to separate the perturbation  from the condensate:
\begin{eqnarray}
\label{Eq:lin1}
\op{\psi} &\simeq& \psi\tx + \delta \op{\psi}\tx\,; \\
\label{Eq:lin2}
\op{\psi}^{\dag} &\simeq& \psi^{\dag}\tx + \delta \op{\psi}^{\dag}\tx\,.
\end{eqnarray}
This transformation is canonical if the creation $\delta\op{\psi}^{\dag}$ and destruction $\delta\op{\psi}$ of perturbations is consistent with the commutator Eqs.~\refb{Eq:com1}-\refb{Eq:com2}. Therefore, the commutators for the linearized quantum fluctuations are
\begin{eqnarray}
\comBig{\delta\op{\psi}\tx }{\delta\op{\psi}\txp}=\comBig{\delta\op{\psi}^{\dag}\tx}{\delta\op{\psi}^{\dag}\txp}=0\,;
\label{Eq:delta_com1}\\
\comBig{\delta\op{\psi}\tx}{\delta\op{\psi}^{\dag}\txp}=\delta(\mathbf{x}-\mathbf{x}')\,.
\label{Eq:delta_com2}
\end{eqnarray}

The quantum perturbations are small perturbations which, as per Eq.~\refb{Eq:psi}, are represented by two collective parameters; the density $n$, and the phase $\theta$. Clearly, any quantum perturbation should be related to variations in these two parameters; $n \simeq n_{0} + \op{n}$ and $\theta \simeq \theta_{0} + \op{\theta}$. A straightforward expansion of $\psi$ and $\psi^{\dag}$ around $n_{0}$ and $\theta_{0}$ leads to
\begin{eqnarray}
\label{Eq:exp.psi_1}
\delta\op{\psi} &\simeq& \psi \left( \frac{1}{2}\frac{\op{n}}{n_{0}} + i \op{\theta}  \right)\,, \\
\label{Eq:exp.psi_2}
\delta\op{\psi}^{\dag} &\simeq& \psi^{*} \left( \frac{1}{2}\frac{\op{n}}{n_{0}} - i \op{\theta} \right)\,.
\end{eqnarray}
In this way Eqs.~\refb{Eq:lin1} and \refb{Eq:lin2} are
compatible with Eq.~\refb{Eq:psi}. Thus the density,
$\op{n}$, and phase, $\op{\theta}$, fluctuations operators are
Hermitian operators:
\begin{eqnarray}
\op{n} \simeq n_{0} \left[ \frac{\delta\op{\psi}}{\psi} + \left(\frac{\delta\op{\psi}}{\psi}\right)^{\dag}  \right]\,;\\
\op{\theta} \simeq - \frac{i}{2} \left[ \frac{\delta\op{\psi}}{\psi} - \left(\frac{\delta\op{\psi}}{\psi}\right)^{\dag}\right]\,.
\end{eqnarray}
It is easy to see that the new operators are a set of canonical variables:
\begin{eqnarray}
\comBig{\op{n}\tx }{\op{n}\txp}=\comBig{\op{\theta}\tx}{\op{\theta}\txp}=0\,;
\label{Eq:nn_psipsi_com1}\\
\comBig{\op{n}\tx}{\op{\theta}\txp}=i\delta(\mathbf{x}-\mathbf{x}')\,.
\label{Eq:n_psi_com2}
\end{eqnarray}
The latter can be further modified, and we will subsequently revisit Eq.~\refb{Eq:n_psi_com2}.
While the split into background plus perturbation is up to this
point exact, we now linearize by assuming the perturbation to be
small, allowing us to neglect quadratic and higher-order products
of the perturbation field $\delta \psi$. (There are also more
sophisticated calculational techniques available based on the
Hartree--Fock--Bogoliubov--Popov approximation, but they are an
unnecessary complication in the present situation.)

Merging equations (\ref{Eq:lin1})--(\ref{Eq:lin2}) with equations (\ref{Eq:exp.psi_1})--(\ref{Eq:exp.psi_2}), applying them to Eq.~\refb{Heisenberg}, we obtain two equations,
\begin{eqnarray}
\label{hydro.n.1}
\frac{\partial \hat n}{\partial t} + \nabla \left[ \frac{n_{0} \hbar}{m} \nabla \hat \theta + \hat n \cdot \mathbf{v} \right] = 0\,; \\
\label{hydro.n.2}
\frac{\partial \hat \theta}{\partial t} + \mathbf{v} \cdot \nabla \hat \theta + \frac{\mathcal{U}}{\hbar} \, \hat n = 0\,.
\end{eqnarray}
Here we introduce the background velocity of the condensate,
\begin{equation}
\mathbf{v} = \frac{\hbar}{m} \nabla \theta_{0}.
\end{equation}
The quantity $\mathcal{U}$ can be thought of as an effective atomic interaction, as seen by the collective excitations. For long-wavelength, low-energetic modes this simplifies to Eq.~\refb{Eq:U}, the usual pseudo-contact potential;
\begin{equation} \label{Eq.hydro.approx}
\mathcal{U} \rightarrow U \quad : \quad \mathrm{phononic}\; \mathrm{modes}\,.
\end{equation}
We will come back to this point in section \refb{sec:UV-corrections} where we shall write down the expression for $\mathcal{U}$ in general, and in the eikonal limit, in a manner appropriate to describe trans-phononic (or, ultraviolet) modes. In the language of condensed matter physics, we include ``quantum pressure'' effects. Without quantum pressure we restrict our analysis to the phononic regime, and assume that all collective excitations propagate with the same speed, the sound speed $c_{0}$;
\begin{equation} \label{Eq:sound.speed.hydro}
c_{0}^{2} = \frac{n_{0}U}{m}\,.
\end{equation}

Before we continue with our program, we would like to revisit the
commutator derived in Eq.~\refb{Eq:n_psi_com2}. In the
phononic regime, we are able to write Eq.~\refb{hydro.n.2} as
follows,
\begin{equation} \label{Eq:n.explicit}
\hat n = - \frac{\hbar}{U} \left( \frac{\partial}{\partial_{t}} + \mathbf{v} \cdot \nabla \right) \hat \theta
= -\frac{\hbar}{U} \frac{D\hat \theta}{Dt} \, .
\end{equation}
Thus the commutator \refb{Eq:n_psi_com2} can be written in terms of $\hat \theta$ and its fluid-following derivative (or material derivative) $D\hat\theta / Dt$:
\begin{equation} \label{Eq:Pi_psi_com2}
\comBig{\hat\theta\tx}{\hat \Pi_{\hat \theta}\txp}=i \delta(\mathbf{x}-\mathbf{x}')\,.
\end{equation}
Here we have defined,
\begin{equation} \label{Eq:conj.monmentum.1}
\hat \Pi_{\hat \theta} = \frac{\hbar}{U}  \frac{D\hat\theta}{Dt}\,,
\end{equation}
which can be viewed as the conjugate momentum of $\hat \theta$; see Eq.~\refb{Eq:Sig.Change.Paper.emergent.Lagrange.density} below.

The relation \refb{Eq:n.explicit} enables us to read density
perturbations as fluid-following derivatives of the phase
perturbations, and hence supplies us with the necessary tool to
eliminate all occurrences of $\hat n$ from equation
\refb{hydro.n.1}:
\begin{equation}
\label{Eq:phase.evolution}
-\partial_{t} \left[ \frac{\hbar}{U} \, \partial_{t} \hat \theta \right]
-\partial_{t} \left[ \frac{\hbar}{U} \, \mathbf{v} \, \nabla \hat \theta \right]
-\nabla \left[ \frac{\hbar}{U} \, \mathbf{v} \, \partial_{t} \hat \theta  \right]
+\nabla \left[ \frac{n_{0} \hbar}{m} \, \nabla \hat \theta -\frac{\hbar}{m} \, \mathbf{v} \left(\nabla\hat \theta\right) \mathbf{v} \right] =0 \,.
\end{equation}

This equation, governing the kinematics for the phase perturbations $\hat \theta$, is the connection between condensed matter physics and emergent quantum field theory in curved spacetimes.

\subsection{Analogue spacetime \label{subsec:A-spacetime}}
A  compact and insightful way to express the evolution of phase
perturbations \refb{Eq:phase.evolution} is
\begin{equation} \label{Eq:KLG.1}
\frac{1}{\sqrt{\left\vert \det(g_{ab}) \right\vert}} \; \partial_{a} \left( \sqrt{ \left\vert \det(g_{ab}) \right\vert }\, g^{ab} \, \partial_{b} \hat \theta  \right)  =0\;, \\
\end{equation}
where we introduce
\begin{equation} \label{Eq:g.1}
g_{ab} = \left( \frac{c_{0}}{U/\hbar} \right)^{\frac{2}{d-1}}
\left[
\begin{array}{cccc}
-\left(c_{0}^{2}-\mathbf{v}^{2}\right) & -v_{x} & -v_{y} & -v_{z} \\
-v_{x} & 1 & 0 & 0 \\
-v_{y} & 0 & 1 & 0 \\
-v_{z} & 0 & 0 & 1
\end{array}
\right]\, ;
\end{equation}
a covariant metric rank two tensor, whose entries are purely collective variables. The conformal factor depends on the spatial dimensionality, $d$, of the condensate cloud.
To derive Eq.~\refb{Eq:KLG.1}, we first write Eq.~\refb{Eq:phase.evolution} as,
\begin{equation} \label{Eq:KLG.f}
\partial_{a} \left( f^{ab} \; \partial_{b} \hat \theta \right)=0\,,
\end{equation}
where $f^{ab}$ is easily found to be,
\begin{equation} \label{Eq:f.1}
f^{ab} :=
\left[
\begin{array}{cccc}
-\frac{\hbar}{U} & -\frac{\hbar}{U} v_{x} & -\frac{\hbar}{U} v_{y} & -\frac{\hbar}{U} v_{z} \\
-\frac{\hbar}{U} v_{x} &  \frac{n_{0}\hbar}{m}-\frac{\hbar}{U}v_{x}^{2} &  -\frac{\hbar}{U}v_{x}v_{y} & -\frac{\hbar}{U}v_{x}v_{z} \\
-\frac{\hbar}{U} v_{y} & -\frac{\hbar}{U}v_{x}v_{y} & \frac{n_{0}\hbar}{m}-\frac{\hbar}{U}v_{y}^{2} & -\frac{\hbar}{U}v_{y}v_{z} \\
-\frac{\hbar}{U} v_{z} & -\frac{\hbar}{U}v_{x}v_{z} & -\frac{\hbar}{U}v_{y}v_{z} & \frac{n_{0}\hbar}{m}-\frac{\hbar}{U}v_{z}^{2} \\
\end{array}
\right]\,.
\end{equation}
The two equations, \refb{Eq:KLG.f}, and \refb{Eq:KLG.1}, are equivalent if
\begin{equation}
f^{ab} =\sqrt{\left\vert \det(g_{ab}) \right\vert } \; g^{ab}.
\end{equation}
Here $g^{ab}$ is a contravariant tensor, and since $g^{ab}g_{cb}=\delta^{a}{}_{c}$, it is only a question of matrix inversion to find its covariant equivalent, \refb{Eq:g.1}. For considerably more details and a thorough derivation we suggest the following literature: \cite{Barcelo:2005ln,Visser:2005hs,Barcelo:2001gt,Garay:2001in}.

At this stage we would like to comment on the physical
implications of the results presented so far. The motivation to
write the differential Eq.~\refb{Eq:phase.evolution},
governing the  excitation spectrum in the form presented in
\refb{Eq:KLG.1}, is to find an analogy for the curved spacetime
Klein--Gordon equation which describes minimally coupled spin-zero
Bosons in curved spacetime.
It is appropriate to define an emergent Lagrange density,
\begin{equation} \label{Eq:Sig.Change.Paper.emergent.Lagrange.density}
\mathcal{L}=-\frac{1}{2} f^{ab} \; \partial_{a}\hat\theta \; \partial_{b}\hat\theta\, ,
\end{equation}
such that Eq.~\refb{Eq:KLG.1} can be obtained as the
Euler-Lagrange equations justified by the principle of least
action. The momentum conjugate to $\hat \theta$ is specified by
\begin{equation}
\Pi_{\hat\theta} := {\partial \mathcal{L}\over\partial (\partial_{t} \hat\theta)}
=-f^{tb} \; \partial_{b}\hat\theta,
\end{equation}
and hence is in agreement with equation
\refb{Eq:conj.monmentum.1}. We see that quantum phase and density
perturbations in a Bose--Einstein condensate are a canonical set
of field and conjugate field operators on the emergent spacetime.

We now temporarily set aside this analogy and apply our model to particle production in non-smooth emergent geometries.
%
\section{Sudden changes in spacetime geometry \label{sec:Sudden.Lorentzian.Variations}}
%
In this section we calculate the mixing of positive and negative frequencies due to ``sudden'' step-wise variations in the sound speed, Eq.~\refb{Eq:sound.speed.hydro}. The initial and final emergent geometries are now flat Minkowski spacetimes, which are discontinuously connected at the step. Physically the step is generated by a very rapid change in the magnetic field, which very rapidly drives one through a Feshbach resonance, which in turn very rapidly changes the scattering length $a$, and so finally induces a  rapid change in the speed of sound.

\subsection{Quantum fields as harmonic oscillators \label{Sec:Quantum.Fields.Harmonic.Oscillators}}
An idealized Bose--Einstein condensate trapped in a finite quantization box of volume $L^{d}$
is comparable to flat Minkowski spacetime. Here the macroscopic parameters are zero background velocity, $\mathbf{v}=\mathbf{0}$, and constant sound speed \refb{Eq:sound.speed.hydro}. That is, one considers a uniform number density, $n_{0}\tx = \mathrm{constant}$, and a fixed scattering length, $a(t)=\mathrm{constant}$. Therefore the emergent metric given in \refb{Eq:g.1} is a diagonal tensor whose entries are time- and space independent;
\begin{equation} \label{Eq:eta.1}
f^{ab} =
\left[
\begin{array}{cccc}
-\frac{\hbar}{U} & 0 & 0 & 0 \\
0 & \frac{n_{0} \hbar}{m} & 0 & 0 \\
0 & 0 & \frac{n_{0} \hbar}{m} & 0 \\
0 & 0 & 0 & \frac{n_{0} \hbar}{m}
\end{array}
\right]\, .
\end{equation}

We employ the canonical variables on our effective relativistic spacetime to write the Klein--Gordon Eq.~\refb{Eq:KLG.1} as
\begin{equation} \label{Eq:KLG.2}
\partial_{t} \left( \frac{\hbar}{U} \; \partial_{t} \hat \theta   \right) - \frac{n_{0} \hbar}{m} \nabla^{2} \hat\theta = 0\,.
\end{equation}
It is possible to decouple the Klein--Gordon field \refb{Eq:KLG.2} into independent Harmonic oscillators. To show this, we make use of
\begin{equation}
\hat \theta\tx = \frac{1}{L^{d/2}}\sum_{k}{ \frac{1}{\sqrt{2}}
\left[ \hat a_{k} v_{k}^{*}(t) e^{i\mathbf{kx}}+\hat a_{k}^{\dag} v_{k}(t) e^{-i\mathbf{kx}} \right]}\, .
\end{equation}
Note that for a hard-walled box the modes fulfill non-periodic boundary conditions.
In Minkowski spacetime there exists a natural set of mode functions,
\begin{equation} \label{Eqn:Mode.Functions.Lorentzian}
v_{k}(t) =\sqrt{\frac{U}{\hbar}} \, \frac{1}{\sqrt{2\omega_{k}}}
\, e^{i\omega_{k} t}\, ,
\end{equation}
associated with the Poincar\'e group, a symmetry group of the Minkowski line-element. Here $\partial_{t}$ is a time-translation Killing vector which can also be thought of as a differential operator with eigenvalues $(-i\omega_{k})$, where $\omega_{k}>0$ are said to be positive frequency modes. Hence the vacuum is invariant under the action of the Poincar\'e group and all observers agree on existence, or non-existence, of particles in flat spacetime. The physics is observer-independent, as expected. For a more detailed treatment see reference \cite{Birrell:1984aa}, and the appendix below.

The decoupled equations for the mode-operators are
\begin{equation} \label{Eq:KLG.3}
\dot{\hat a}_{k}(t) = - i \omega_{k} \, \hat a_{k}(t)\, ; \quad  \mathrm{and} \quad \dot{\hat a}_{k}^{\dag}(t) = + i \omega_{k} \, \hat a_{k}^{\dag}(t)\, ,
\end{equation}
where ${\hat a}_{k}(t) = \hat a_{k} v_{k}^{*}(t)$ and
${\hat a}_{k}^{\dag}(t) = \hat a_{k}^{\dag} v_{k}(t)$. For
now, we are working in the low-energy, long-wavelength regime and
obtain a ``relativistic'' dispersion relation for the modes:
\begin{equation} \label{Eq.disp.rel.hydro}
\omega_{k} = c \, k\, .
\end{equation}
Later on, in section \refb{sec:UV-corrections}, we will include trans-phononic modes into our picture, and see how the microscopic structure induces LIV breaking terms for high-energy, short-wavelength perturbations.

The canonical creation and destruction operators obey the usual commutator
\begin{equation} \label{Eq:Pi_psi_com2.Minkowski.new}
\comBig{\hat a_{k}(t)}{\hat a_{-k'}^{\dag}(t)}= \delta_{kk'}\,,
\end{equation}
and acting on the particle basis for the Hilbert space of states, the Fock states $\vert \;\rangle$, they are a powerful tool to calculate the number of particles $n_{k}$ in the mode labeled by $k$.
The normalized basis vectors can be obtained from the vacuum, or zero-particle state, $\vert 0 \rangle$. This is the state that is destroyed by operators $\hat a_{k} \vert 0 \rangle = 0$, for all modes labeled by $k$.
In general
\begin{equation}
\hat a_{k}^{\dag} \vert n_{k} \rangle = (n+1)^{1/2} \vert (n+1)_{k}\rangle \, ,
\end{equation}
 and
\begin{equation}
\hat a_{k} \vert n_{k} \rangle = n^{1/2} \vert (n-1)_{k} \rangle \, ,
\end{equation}
so that we can define the number operator $\hat N_{k}$,
\begin{equation}
\hat N_{k} \vert n_{k} \rangle = \hat a_{k}^{\dag} a_{k} \vert n_{k} \rangle = n_{k} \vert n_{k} \rangle \, .
\end{equation}

We now use the tools we have presented above to calculate the
quasi-particle production in a BEC-based emergent spacetime due to
sudden changes in the sound speed; such that we patch two flat
spacetimes in a step-wise fashion, by suitably changing the
microscopic parameters.

\subsection{Particle production \label{Sec:Sig.Change.Paper.Particle.Production}}
From the emergent spacetime point of view step-wise changes in the microscopic parameters induce sudden variations in the collective variables, and enforce discontinuous interchanges between different Minkowski spacetimes. In terms of our emergent spacetime from a Bose gas, this can be achieved through an external magnetic field, that adjusts the atomic interactions, and consequently the scattering length $a$ and the sound speeds $\sound{i}$.

Following the ideas of Dray \emph{et al.} \cite{Dray:2004aa,Dray:1991zz}, we assume that the fields are continuous,
\begin{equation} \label{Eq.C1}
\left. \hat\theta^{\reg{i}} - \hat\theta^{\reg{j}}  \right\vert_{_{\sum}}= 0 \, ,
\end{equation}
on the space like hypersurface {\small $\sum$} at  fixed time
$t=t^{\reg{i}\reg{j}}$. Here $t=t^{\reg{i}\reg{j}}$ is the time at
which a transition occurs from $c^{\reg{i}}$ to $c^{\reg{j}}$. The
Klein--Gordon equation in each region can be written in terms of
exterior derivatives,
\begin{equation}
(-1)^{n}\partial_{a} \left( f^{ab}\,\partial_{b} \hat\theta^{\reg{i}} \right)
= *d(*d\hat\theta^{\reg{i}})
=: *dF^{\reg{i}}
=0\, ;
\end{equation}
hence in each region $dF^{\reg{i}}=0$, where $F^{\reg{i}}$ is an exact $(n-1)$-form. These forms are connected discontinuously;
\begin{equation}
F = \Theta \, F^{\reg{i}} + (1-\Theta) F^{\reg{j}}\, ,
\end{equation}
where $\Theta$ is the usual Heavyside function. Thus, we get
\begin{eqnarray}
dF &=& \Theta \; dF^{\reg{i}} + (1-\Theta) \; dF^{\reg{j}} + \delta(t)\, dt\wedge [F]  \\
&=&  \delta(t)\, dt\wedge [F] \, ,
\end{eqnarray}
where $[F]=F{^{\reg{i}}}-F{^{\reg{j}}}$. This supplies us with a connection condition for the canonical momentum,
\begin{equation}
\delta(t)\; dt\wedge [F] =0\, ;
\end{equation}
\ie,
\begin{equation} \label{Eq.C2}
\left. \hat\Pi_{\hat \theta}{}^{\reg{i}} - \hat\Pi_{\hat \theta}{}^{\reg{j}}  \right\vert_{_{\sum}}= 0 \, .
\end{equation}
Thus the field and its canonical momentum must be continuously connected across the spatial hypersurface of sudden variation. In the following we apply the connection conditions we have just found, and calculate the number of quasi-particles produced by sudden sound speed variations in our BEC.

In each region $\Reg{i}$ the mode expansion,
\begin{equation}
\hat \theta^{\reg{i}} \tx = \frac{1}{\sqrt{2 L^{d}}} \sum_{\mathbf{k}} e^{i\mathbf{kx}} \left[ \mfc{i}(t) \aop{i} + \mf{i}(t) \aopdagneg{i} \right]
\end{equation}
involves a distinct set of creation $\aopdag{i}$ and destruction $\aop{i}$ operators. The mode functions $\mf{i}$ and $\mfc{i}$ obey the curved-spacetime Klein--Gordon equation,
\begin{equation} \label{Eq:KLG.4}
\partial_{t} \left[ \frac{\hbar}{U^{\reg{i}}} \; \partial_{t}  \mf{i}(t)   \right] - \frac{n_{0} \hbar}{m}\, k^{2} \; \mf{i}(t) = 0\,;
\end{equation}
they form a complete basis for the two dimensional solution space. At the transition time $t^{\reg{i}\reg{j}}$ we connect the mode functions and their first derivatives,
\begin{eqnarray}
\label{Eq.C1.M}
\Mf{i}(t^{\reg{i}\reg{j}}) &=& M^{\reg{i}\reg{j}} \; \Mf{j}(t^{\reg{i}\reg{j}})\, , \\
\label{Eq.C2.M}
\frac{\hbar}{U^{\reg{i}}}\; \partial_{t}\Mf{i}(t^{\reg{i}\reg{j}}) &=&
M^{\reg{i}\reg{j}} \; \frac{\hbar}{U^{\reg{j}}} \; \partial_{t}\Mf{j}(t^{\reg{i}\reg{j}}) \,.
\end{eqnarray}
Here we have combined both mode functions into one mode vector $\Mf{i}$,
\begin{equation}
\Mf{i}(t) =\left( \begin{array}{l} \mf{i}(t) \\ \mfc{i}(t) \end{array} \right)\, .
\end{equation}
In the cases we are interested in the transition matrix $M^{\reg{i}\reg{j}} = M(t^{\reg{i}\reg{j}})$ is time-independent, and for a pair of complex conjugate mode functions in each region $\Reg{i}$ and $\Reg{j}$, the matrix is of the form
\begin{equation} \label{Eq.M}
M^{\reg{i}\reg{j}} = \left( \begin{array}{cc}
\Ba{i}{j} & \Bb{i}{j} \\ \Bbc{i}{j} & \Bac{i}{j}
\end{array} \right)\, .
\end{equation}
The Wronskian $W[\mf{i},\mfc{i}]=(\partial_{t}\mf{i})\mfc{i}-\mfc{i}(\partial_{t}\mfc{i})$ of the mode functions is time-independent, see Eq.~\refb{Eq:KLG.4}, which implies a constraint on the Bogliubov coefficients $\Ba{i}{j}$ and $\Bb{i}{j}$;
\begin{equation} \label{Eq.KG.condition}
\vert \Ba{i}{j} \vert^{2} - \vert \Bb{i}{j} \vert^{2} = 1\, .
\end{equation}
To start with, consider a no-particle state in the initial Minkowski spacetime (region $\Reg{i}$), such that $\aop{i}\vstate{i}=0$. The correlation between the mode operators can easily be constructed from $\hat \theta^{\reg{i}}(t^{\reg{i}\reg{j}})=\hat \theta^{\reg{j}}(t^{\reg{i}\reg{j}})$, considering Eq.~\refb{Eq.C1.M} and Eq.~\refb{Eq.KG.condition} we get
\begin{eqnarray}
\aop{i} &=& \Ba{i}{j} \; \aop{j} - \Bb{i}{j} \; \aopdagneg{j}\, , \\
\aopdag{i} &=& \Bac{i}{j} \; \aopdag{j} - \Bbc{i}{j} \; \aopneg{j} \, .
\end{eqnarray}
Consequently, the mean number of $\Reg{i}$-particles in the $\Reg{j}$-vacuum,
\begin{equation}
\vstatebra{j} \hat N_{k}^{i} \vstate{j} = \vert \Bb{i}{j} \vert^{2} \; \delta^{d}(0)\, ,
\end{equation}
depends only on $|\Bb{i}{j}|$.

Explicitly, the elements for the transition matrix in a sudden change from region $\Reg{i}$ to region $\Reg{j}$ at a time $t^{\reg{i}\reg{j}}$ are given by:
\begin{eqnarray}
\label{Eq.alpha.one.step}
\Ba{i}{j} = \frac{1}{2} \left( \sqrt{X^{\reg{i}\reg{j}}} + \frac{1}{\sqrt{X^{\reg{i}\reg{j}}}} \right) \, e^{i (+\omega_{k}^{\reg{i}} - \omega_{k}^{\reg{j}})t^{\reg{i}\reg{j}}}\, , \\
\label{Eq.beta.one.step} \Bb{i}{j} = -\frac{1}{2} \left(
\sqrt{X^{\reg{i}\reg{j}}} - \frac{1}{\sqrt{X^{\reg{i}\reg{j}}}}
\right) \, e^{i (+\omega_{k}^{\reg{i}} +
\omega_{k}^{\reg{j}})t^{\reg{i}\reg{j}}} \, ,
\end{eqnarray}
where to simplify formulae it is convenient to introduce the ratio of change in the dispersion relations $X^{\reg{i}\reg{j}}$,
\begin{equation} \label{Eq.X.hydro}
X^{\reg{i}\reg{j}} = \frac{\omega_{k}^{\reg{j}}}{\omega_{k}^{\reg{i}}} = \frac{\sound{j}}{\sound{i}} \, .
\end{equation}
The last expression in \refb{Eq.X.hydro} is only valid in the hydrodynamic approximation, where the number of particles produced from a single step is $k$-independent. The mean number of particles in each mode $k$ is given by
\begin{equation}
\vert \Bb{1}{2} \vert^{2} ={1\over4}
\left\vert \sqrt{\sound{2}\over\sound{1}} - \sqrt{\sound{1}\over \sound{2}}  \right\vert^{2} \, .
\end{equation}

The advantage of this representation lies in the simple way that it can be extended for $m$ sudden variations in a row
\begin{equation} \label{Eq.M.1m}
\Mf{1} = \underbrace{M^{\reg{1}\reg{2}} \cdot M^{\reg{2}\reg{3}} \cdots M^{(\reg{m}-\reg{1})\reg{m}}}_{M^{\reg{1}\reg{m}}} \, \Mf{m} \, ,
\end{equation}
and each single transition matrix $M^{\reg{i}(\reg{i}+\reg{1})}$ is of the form \refb{Eq.M} evaluated at $t^{\reg{i} (\reg{i}+\reg{1})}$. The resulting matrix $M^{\reg{1}\reg{m}}$ carries the final Bogoliubov coefficients $\Ba{1}{m}$ and $\Bb{1}{m}$ for the whole chain of events. We would like to point out that the choice for the mode functions in the intermediate regimes does not have any influence on the final outcome, as can be seen in Eq.~\refb{Eq.M.1m}. However for the validity of our calculation it is necessary to choose a pair of complex conjugate mode functions.
%
\begin{figure}[htb]
 \begin{center}
 \input{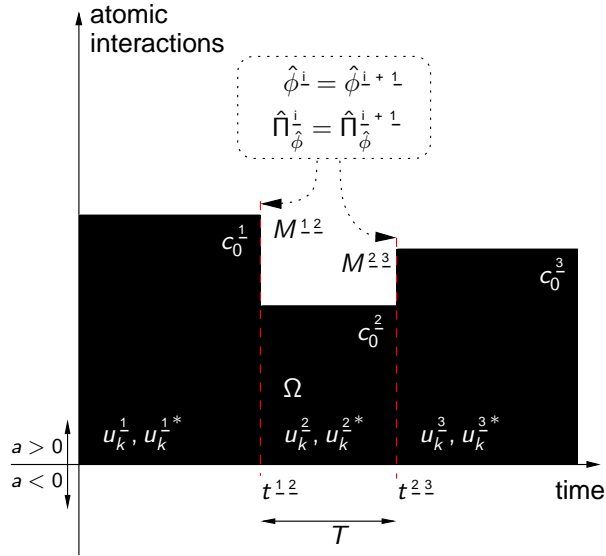}
 \caption[Schematic outline for sudden Lorentzian changes.]  {\label{Fig.Sudden.Lorentzian}
 Two sudden variations in the strength of the atomic interactions is analogous to two space-like hypersurfaces connecting three different Minkowski spacetimes emerging from a Bose gas.}
 \end{center}
\end{figure}
%
The rest of the chapter further investigates a particular scenario, that is two sudden variations in a row, see Figure~\ref{Fig.Sudden.Lorentzian}. In this situation
\begin{eqnarray}
\Ba{1}{3} &=& \Ba{1}{2}\;\Ba{2}{3} + \Bb{1}{2}\;\Bbc{2}{3} \, , \\
\Bb{1}{3} &=& \Ba{1}{2}\;\Bb{2}{3} + \Bb{1}{2}\;\Bac{2}{3} \, .
\end{eqnarray}
To obtain the Bogoliubov coefficients for a two-step process, we
use the single-step results \refb{Eq.alpha.one.step} and
\refb{Eq.beta.one.step} to obtain:
\begin{equation}
\label{Eq.Bog.alpha.hydro}
\Ba{1}{3} = \exp\left\{i\omega_{k}^{\underline{1}}t^{\reg{1}\reg{2}}-i\omega_{k}^{\underline{3}}(t^{\reg{1}\reg{2}}+T)\right\}  \times 
 \left(\frac{ X^{\reg{1}\reg{3}}+1}{2\sqrt {X^{\reg{1}\reg{3}}}} \cos(\Omega_{k} T) - \frac {X^{\reg{1}\reg{2}}+X^{\reg{2}\reg{3}}}{i\,2\sqrt{X^{\reg{1}\reg{3}}}}
\sin ( \Omega_{k} T )\right)   \, , \;
\end{equation}
and 
\begin{equation}
\label{Eq.Bog.beta.hydro} 
\Bb{1}{3} = -\exp\left\{+i\omega_{k}^{\underline{1}}t_{1}+i\omega_{k}^{\underline{3}}(t_{1}+dt)\right\}   \times 
\left( \frac{ X^{\reg{1}\reg{3}}-1}{2\sqrt {X^{\reg{1}\reg{3}}}} \cos(\Omega_{k} T ) +  \frac {X^{\reg{1}\reg{2}}-X^{\reg{2}\reg{3}}}{i\,2\sqrt{X^{\reg{1}\reg{3}}}} \sin ( \Omega_{k} T ) \right) \, ; \;
\end{equation}
where for the finite duration intermediate region $\Reg{2}$ we
have defined the time interval
$T=t^{\reg{2}\reg{3}}-t^{\reg{1}\reg{2}}$, and the dispersion
relation $\Omega_{k}  = \omega_{k}^{\reg{2}}$. The mean number of
particles produced during this process is then given by
\begin{equation}  \label{Eq.mean.number.LLL}
\vert \Bb{1}{3} \vert^{2} =
 \frac{(X^{\reg{1}\reg{3}}-1)^{2}}{4X^{\reg{1}\reg{3}}} 
-\frac{1}{4}\left[X^{\reg{1}\reg{2}}- \frac{1}{X^{\reg{1}\reg{2}}}\right]\left[X^{\reg{2}\reg{3}}- \frac{1}{X^{\reg{2}\reg{3}}}\right] \sin(\Omega_{k} T)^{2}.\quad
\end{equation}

It is easy to see that the mean number of particles produced in such a process oscillates between
two single step solutions of the form
\begin{equation}
N_{k} = \frac{(X-1)^{2}}{4X} \, .
\end{equation}
The upper bound replaces two up-down\,/\,down-up steps with a single step (here given by $X=X^{\reg{1}\reg{2}}/X^{\reg{2}\reg{3}}$), and the lower bound replaces two up-down\,/\,down-up steps with a single step  (here given by $X=X^{\reg{1}\reg{2}} X^{\reg{2}\reg{3}}$). We have illustrated the quasi-particle production in the hydrodynamic limit obtained by two-sudden steps in a row in Figure~\ref{Fig.Hydro.LLL}.

In the following we show how to extend the calculations for finite intervals during which the microscopic atoms experience negative (attractive) interactions. For small negative values of the scattering length $a$ the condensate description continues to exist for short periods of time, and hence motivates the study of Euclidean emergent geometries from a Bose gas. A sign change in the atomic interactions corresponds to a signature change in general relativity. As a first step we investigate quasi-particle production from such an event in the hydrodynamic limit.
%
%
\section{Signature change events \label{sec:Sudden.Euclidean.Variations}}
%
%
In this section we investigate the behavior of emergent spacetimes arising from a Bose gas with variations in the principal nature of the interactions; for a finite amount of time we switch to attractive atomic interactions. For short time-scales and small absolute values of the attractive $s$-wave potential, it is possible to hold on to the concept of an emergent spacetime. With the nature of the microscopic interactions switching from repulsive to attractive, the geometric hydrodynamics also changes from Lorentzian to Riemannian. This supplies us with a toy model for signature change events, which we are going to investigate next.

We first present the standard general relativity point of view, before adapting our previous calculations for the particle production due to sudden variations on finite duration Euclidean regions.
%
\subsection{Classical aspects \label{subsec:ClassicalAspects}}
Overall, we investigate manifolds that allow both Riemannian and Lorentzian regions. The latter is a generalization of Minkowski spacetimes (special relativity), while Riemannian geometries are Euclidean signature spaces including curvature (\eg, the surface of an orange). Distances in Riemannian spacetimes are positive definite, while Lorentzian distances can be imaginary (time-like), zero (light-like), or positive (space-like).

In Minkowski spacetimes the signature can easily be read off as the sign of positive and negative eigenvalues of $\eta_{ab}$. From Eq.~\refb{Eq:eta.1} we get Lorentzian signature $(-,+++)$. Arbitrary curved spacetimes are locally flat, and the signature can be read off from the pattern of eigenvalues of the metric tensor $g_{ab}$ at each point on the manifold. In Lorentzian signature the ``time'' coordinate can be chosen to have a different sign from the ``spatial'' coordinates. This is in contrast to the Riemannian, or Euclidean signature $(+,+++)$, where a distinction between space and time as such does not exist \cite{Wald:1984rg}.\\

We are mainly interested in the interface between these two spacetime geometries, and we investigate the physics around a space-like hypersurface ${\tiny \sum}$ that separates the two spaces. There are two ways to be driven through a signature change, continuously or discontinuously.

We are in favour of a non-smooth signature change, to avoid degeneracies of the effective gravitational field at the surfaces of separation. For continuous signature changes the metric volume element, and hence the existence of an orthonormal frame vanish at ${\tiny \sum}$; while for discontinuous signature changes the metric volume element (hence, the orthonormal frame) is well behaved \cite{Dray:2004aa,Dray:1996dc,Ellis:1992aa}.

Thus we are allowed to transfer the connection conditions, as derived in our previous calculations for sudden changes in purely Lorentzian geometries (see Eq.~\refb{Eq.C1}, Eq.~\refb{Eq.C2}, and the discussion in the appendix), to now connect regions of Minkowskian and Euclidean spacetimes \cite{Dray:1991zz}.
\subsection{Particle production from a finite-duration Euclidian region \label{subsec:EEE}}
%
\begin{figure}[!htb]
 \begin{center}
 \input{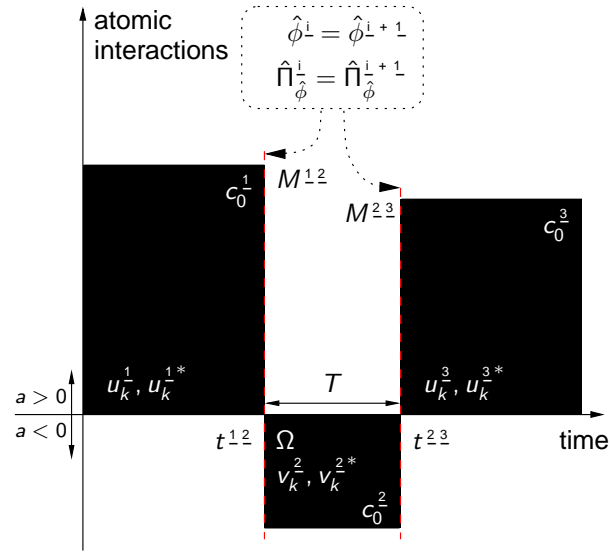}
 \caption[Schematic outline for sudden Euclidean changes.]  {\label{Fig.Sudden.LEL}Two sudden variations in the strength of the atomic interactions are analogues to two space-like hypersurfaces connecting three different spacetimes emerging from a Bose gas. While the initial and final geometries are Lorentzian, we allow the atomic interactions to be attractive in the intermediate region, corresponding to an Euclidean geometry.}
 \end{center}
\end{figure}
%
In the following we repeat our calculations from section \refb{sec:Sudden.Lorentzian.Variations}, but this time we choose an Euclidean geometry for region $\Reg{2}$; see figure \refb{Fig.Sudden.LEL}. As already pointed out above, the connection conditions remain the same. We further choose the mode functions in the Lorentzian regimes $\Reg{1}$, and $\Reg{3}$ to be as defined in Eq.~\refb{Eqn:Mode.Functions.Lorentzian}, where the sound speeds are well behaved; $(\sound{L})^{2}\sim U > 0$. For the intermediate Euclidean region we pick a special set of mode functions for $(\sound{E})^{2}\sim U < 0$, such that,
\begin{eqnarray}
\mfE{2}=\frac{1}{\sqrt{2 \vert \Omega_{k} \vert}} \left( \cosh(\vert \Omega_{k} \vert t) + i  \sinh(\vert \Omega_{k} \vert t) \right) \, ; \\
\mfEc{2}=\frac{1}{\sqrt{2 \vert \Omega_{k} \vert}} \left( \cosh(\vert \Omega_{k} \vert t) - i  \sinh(\vert \Omega_{k} \vert t) \right) \, .
\end{eqnarray}
The two mode functions remain a complex conjugate pair.
Here we make use of the purely imaginary dispersion relation $\Omega_{k}= i \abs{\Omega_{k}}$. In general, with the transformation matrix $S$,
\begin{equation}
S = \frac{\sqrt{i}}{2} \left(
\begin{array}{cc}
(1-i) & (1+i) \\ (1+i) & (1-i)
\end{array} \right)
\end{equation}
we can map between the two different sets of mode functions,
\begin{equation}
\MfE{2} = S \; \Mf{2} \, .
\end{equation}
Notice that $S^{-1}=S^{\dag}$ and $\det(S)=1$.

It is easy to see that $M^{\reg{1}\reg{2}} \, \Mf{2}$ transforms to $M^{\reg{1}\reg{2}} \, S \, \MfE{2}$, while $M^{\reg{2}\reg{3}} \, \Mf{3}$ transforms to $S^{-1} \, M^{\reg{2}\reg{3}} \, \Mf{3}$.
Altogether
\begin{equation}
M^{\reg{1}\reg{3}}=
M^{\reg{1}\reg{2}} S \, S^{-1} M^{\reg{2}\reg{3}}=
M^{\reg{1}\reg{2}} \, M^{\reg{2}\reg{3}}
\end{equation}
 is independent of the choice for the mode functions in the intermediate regime.

Thus we are allowed to use our previous results, replacing $\Omega{k}\rightarrow i \vert \Omega{k} \vert$, and $X^{\reg{1}\reg{2}}\rightarrow i \abs{X^{\reg{1}\reg{2}}}$ and $X^{\reg{2}\reg{3}}\rightarrow -i \abs{X^{\reg{2}\reg{3}}}$, and keeping everything else. For the mean number of particles we now obtain
\begin{equation} 
\label{Eq.mean.number.LEL}
\vert \Bb{1}{3} \vert^{2} =
 \frac{(X^{\reg{1}\reg{3}}-1)^{2}}{4X^{\reg{1}\reg{3}}} 
 +\frac{1}{4}\left[\abs{X^{\reg{1}\reg{2}}}+ \frac{1}{\abs{X^{\reg{1}\reg{2}}}}\right]\left[\abs{X^{\reg{2}\reg{3}}}+ \frac{1}{\abs{X^{\reg{2}\reg{3}}}}\right] \sinh(\abs{\Omega_{k}} T)^{2}.
\end{equation}
Our results are compatible with Dray \emph{et al.} \cite{Dray:2004aa, Dray:1991zz}, to the extent that the calculations and physical models overlap, but we extend the calculation for arbitrary values of the sound speeds $\sound{i}$. Furthermore we show that one can keep the calculations for purely sudden variation, as long as one picks a pair of complex conjugate mode functions in the intermediate regime. The standard general relativity calculation corresponds to $X^{\reg{1}\reg{2}}=i$, $X^{\reg{2}\reg{3}}=-i$, $X^{\reg{1}\reg{3}}=1$, so that
\begin{eqnarray} \
\vert \Bb{1}{3} \vert^{2} \to \sinh(\abs{\Omega_{k}} T)^{2}.
\end{eqnarray}
The basic reason for this tremendous simplification is that in pure general relativity (with, by definition, a single unique spacetime metric) one always has the freedom to choose coordinates such that $c=1$ in the Lorentzian region, and $c=i$ in the Euclidean region. This is a freedom we do not have in our BEC-based analogue spacetime ---
the way this shows up in our calculations is that a rapid change in the scattering length $a$ has \emph{two} effects in the condensed Bose gas: First the speed of sound is changed, modifying the ``signal cones''; and secondly the dimension-dependent conformal factor shifts by a finite amount. This second effect is absent in the traditional general relativity calculation of Dray \emph{et al.} --- for those authors it is sufficient to \emph{posit} a specific and simple change in the metric tensor $g_{ab}$ and calculate the resulting particle production. In our present situation, we first \emph{derive} a specific (dimension-independent) change in the tensor density $f^{ab}$  induced by changing the scattering length, and then \emph{derive} the corresponding (dimension-dependent) change in the metric tensor $g_{ab}$. The two situations are very closely related, but they are not quite identical.

We have plotted the mean number of particles produced in the Euclidean region in figure \ref{Fig.Hydro.LEL}. The graph shows the quasi-particle spectrum as a function of $k$. The number of particles produced depends on $\Omega_{k} T$. The longer the duration of Euclidian period, the more particles will be produced during this process.

However, there is a fundamental problem with the quasi-particle production in our effective spacetime.
Given that it is possible to connect actual condensate excitations with the calculated quasi-particle spectrum, we expect the total number of particles produced to be finite. But in both cases, for sudden variations with and without signature changes, our results imply an infinite number for the total quasi-particle production;
\begin{equation} \label{Eq.total.number.general}
N=2^{d-1}\pi \, \int_{k} dk \,k^{d-1} \; N_{k}  \, .
\end{equation}
In the next section we show how this problem resolves itself once microscopic corrections to the emergent spacetime picture are taken into account.
%
%
\section{Ultraviolet corrections \label{sec:UV-corrections}}
%
%
Up to now, we have restricted our calculations to the hydrodynamic limit, which is appropriate to describe the infrared behavior of the system. Low-energy excitations in the BEC are longitudinal phonon modes approximately propagating with the same speed; see the dispersion relation given in Eq.~\refb{Eq.disp.rel.hydro}. It is well known that so-called trans-phononic modes show a non-linear relation between excitation energy and wavelength.
In references \cite{Visser:2001ix,Weinfurtner:2006nl, Liberati:2006sj, Liberati:2006kw, Weinfurtner:2006iv} it has been pointed out that this kind of behavior might be viewed as ultraviolet corrections at the ``analogue Planck scale'', the borderline between phononic and trans-phononic modes.

Our next task is to use the eikonal approximation to include ultraviolet modes into the emergent spacetime picture.

\subsection{Rainbow geometries \label{subsec:Rainbow}}
Calculations up to this point were based on the assumption that spatial variations in the overall condensate are small. More specifically, variations in the kinetic energy of the condensate are considered to be negligible, compared to the internal potential energy of the Bosons,
\begin{equation} \label{Eq.hydrodynamic.limit}
\frac{\hbar^{2}}{2m} \, \frac{\nabla^{2} \sqrt{n_{0} + \hat n}}{\sqrt{n_{0} + \hat n}} \ll U \, .
\end{equation}
The left hand term in relation \refb{Eq.hydrodynamic.limit} is the quantum pressure term and is approximated to zero in the hydrodynamic limit.

We now keep the quantum pressure term. A straightforward computation shows that it is possible to absorb all first-order corrections into:
\begin{equation} \label{Eq.U.qp}
\mathcal{U}=U-\frac{\hbar^{2}}{4m n_{0}}\left\{
\frac{(\nabla n_{0})^{2}-(\nabla^{2}n_{0})n_{0}}{n_{0}^{2}} -\frac{\nabla n_{0}}{n_{0}^{2}}\nabla
+ \nabla^{2}
\right\} \, .
\end{equation}
That is, the effect of quantum pressure can be absorbed into an effective atom-atom interaction, see Eq.~\refb{Eq.hydro.approx}. For a uniform condensate this further simplifies to
\begin{equation}
\mathcal{U}= U-\frac{\hbar^{2}}{4m n_{0}} \nabla^{2}  \, .
\end{equation}

In order to obtain an emergent gravitational field, we apply the eikonal approximation,
\begin{equation}
\left.\mathcal{U}\right\vert_{\nabla \rightarrow -ik} \rightarrow U_{k} = U+ \frac{\hbar^{2}}{4m n_{0}} k^{2} \,.
\end{equation}
Thus also the speed of sound for ultraviolet modes has to be modified
\begin{equation} \label{Eq:sound.qp}
c_{k}^{2} = \frac{U_{k}n_{0}}{m}= c_{0}^{2}+ \epsilon_{\mathrm{qp}}^{2} k^{2} \, .
\end{equation}
Here we introduce,
\begin{equation}
\epsilon_{\mathrm{qp}} = \frac{\hbar}{2m} \, .
\end{equation}
We define the borderline between the phononic and trans-phononic modes,
\begin{equation}
\xi^{2} = \frac{1}{2} \frac{\epsilon_{\mathrm{qp}}^{2}}{c_{0}^{2}} \, ,
\end{equation}
to agree with the healing length $\xi$ of the condensate. The
healing length is the distance over which localized perturbations
in the condensate tend to smooth out \cite{Pethick:2001aa}.
Phononic excitations have wavelengths that are much larger than the healing length, $k \gg \xi$. These modes are relativistic modes in the sense of an emergent relativistic dispersion relation (see Eq.~\refb{Eq.disp.rel.hydro}). They propagate through an emergent gravitational field, given in \refb{Eq:g.1}.
Higher energy excitations around the healing scale, $k \sim \xi$, start to see deviations from the mean-field description. Such modes do not only experience collective condensate variables, they start to see the bigger picture behind the mean-field.
Consequently, they exhibit a non-relativistic dispersion relation,
\begin{equation} \label{Eq.disp.rel.qp}
\tilde{\omega}_{k} =c_{k} k=\sqrt{ c_{0}^{2} k^{2} + \epsilon_{\mathrm{qp}}^{2} k^{4}} \, .
\end{equation}
These microscopic corrections also influence the emergent spacetime, including an effective energy-dependent metric,
\begin{equation} \label{Eq:g.Rainbow}
g_{ab}^{k} = \left( \frac{c_{k}}{U_{k}/\hbar} \right)^{\frac{2}{d-1}}
\left[
\begin{array}{cccc}
-\left(c_{k}^{2}-\mathbf{v}^{2}\right) & -v_{x} & -v_{y} & -v_{z} \\
-v_{x} & 1 & 0 & 0 \\
-v_{y} & 0 & 1 & 0 \\
-v_{z} & 0 & 0 & 1
\end{array}
\right]\, ,
\end{equation}
which results in so-called rainbow geometries.

Before we continue with our program, that is to re-calculate the
mean number of particles produced including quantum pressure
effects, we would like to emphasize that all the corrections are
naturally small. For a detailed description of rainbow geometries,
and the suppression mechanism of LIV terms in Bose gases see
\cite{Liberati:2006sj,Liberati:2006kw,Weinfurtner:2006iv,Weinfurtner:2006iv,Weinfurtner:2006nl}.
%
\subsection{Lorentz symmetry violation and Particle Production \label{subsec:LIV_PP}}
As mentioned above, the spacetime description only holds in the eikonal approximation. Nevertheless, a careful analysis for the particle production can be made, where the atomic interaction is a function of differential operators, see Eq.~\refb{Eq.U.qp}. For an isotropic condensate, the two different approaches will lead to the same set of equations,
\begin{equation}
\partial_{t} \left( \frac{\hbar}{U_{k}^{\reg{i}}} \; \partial_{t}  \mf{i}(t)   \right) - \frac{n_{0} \hbar}{m}\, k^{2} \; \mf{i}(t) = 0\, .
\end{equation}
The mode functions $\mfqp{i}$ are still formally represented by
Eq.~\refb{Eqn:Mode.Functions.Lorentzian}, except that we
replace $\omega^{\reg{i}}$ with $\tilde{\omega}^{\reg{i}}$, which
now includes the $\epsilon_{\mathrm{qp}}$ quantum pressure term.

In the next two sections we connect those modes over three discontinuously patched spacetime geometries, and calculate the amplification and mixing of positive and negative modes living on such a manifold.
\subsubsection{Lorentz symmetry violation and L-L-L sudden \label{subsubsec:LLL}}
We start with purely Lorentzian geometries, but allow two sudden
variations in the speed of sound. The calculations are completely
in analogy with those presented in section \refb{subsec:EEE}. Here,
the sound speeds are given by Eq.~\refb{Eq:sound.qp}, and
from the definition for the ratio $X^{\reg{i}\reg{j}}$ in the
hydrodynamic approximation, we now define:
\begin{equation} \label{Eq.X.qp}
\tilde{X}^{\reg{i}\reg{j}} = \frac{\tilde{\omega}_{k}^{\reg{j}}}{\tilde{\omega}_{k}^{\reg{i}}} = \frac{c_{k}^{\reg{j}}}{c_{k}^{\reg{i}}} \, ,
\end{equation}
as the sound-speed ratio for trans-phononic modes. Formally, the results for the Bogoliubov coefficients --- $\Ba{1}{3}$ and $\Bb{1}{3}$, given in Eqs.~\refb{Eq.Bog.alpha.hydro} and \refb{Eq.Bog.beta.hydro} --- and consequently the mean number of particles produced during this two-step process as given in Eq.~\refb{Eq.mean.number.LLL}, all remain the same. The only effort required is to replace $X^{\reg{i}\reg{j}}$ with $\tilde{X}^{\reg{i}\reg{j}}$, and $\omega_{k}^{\reg{i}}$ with $\tilde{\omega}_{k}^{\reg{i}}$.\\

We have plotted the quasi-particle spectra in figure \ref{Fig.QP.LLL}. It can be seen that the quasi-particle spectrum for the two-step process is still oscillating between the two single-step processes represented by the black and green curve. Compared to the figure on the left, where we plotted the same process in the hydrodynamic limit, see figure \ref{Fig.Hydro.LLL}, we notice that the ultraviolet particle production rapidly approaches zero.

\subsubsection{Lorentz symmetry violation and L-E-L sudden \label{subsubsec:LEL}}
Finally, we are left to analyze sudden variations from Lorentzian, to Euclidean, and back to Lorentzian spacetimes for rainbow geometries. It is interesting to notice, that in the presence of quantum pressure corrections \emph{the meaning of signature is also energy-dependent}. We distinguish between the following cases:
\begin{equation} \label{Eq:signature_qp}\mathrm{Sig.\, \Reg{2}} = \left\{
\begin{array}
{r@{\quad : \quad}l}
 \abs{k} <   {\sound{2}}/{\epsilon_{\mathrm{qp}}} & \mathrm{Euclidean} \, ; \\
 \abs{k} >  {\sound{2}}/{\epsilon_{\mathrm{qp}}} & \mathrm{Lorentzian}  \, .
\end{array} \right.
\end{equation}
We would like to emphasize that $\sound{2}$ is the sound speed in region $\Reg{2}$ for $k\rightarrow 0$.

Again, we take the previous results found in the hydrodynamic limit, see equations (\ref{Eq.Bog.alpha.hydro}, \ref{Eq.Bog.beta.hydro}, \ref{Eq.mean.number.LLL}), where we need only replace $X^{\reg{i}\reg{j}}$ with $\tilde{X}^{\reg{i}\reg{j}}$, and $\omega^{\reg{i}}$ with $\tilde{\omega}^{\reg{i}}$. \\

Altogether we expect the mean number of particles produced for $
\abs{k} > {\sound{2}}/{\epsilon_{\mathrm{qp}}} $ to be equivalent
to those from the purely Lorentzian variations, while phononic
modes, with $\abs{k} \ll {\sound{2}}/{\epsilon_{\mathrm{qp}}}$,
should experience exponential growth similar to the result from
the hydrodynamic limit. We have plotted the mean number of
particles produced for the whole event, see figure
\ref{Fig.QP.LLL}. The plots are in agreement with our predictions.

One motivation for including microscopic corrections into our hydrodynamic calculations was to solve the problem of an infinite number of particles being produced in sudden variations; see Eq.~\refb{Eq.total.number.general}. To estimate the total number of particles, we need only to focus on the region $ \abs{k} > {\sound{2}}/{\epsilon_{\mathrm{qp}}} $. It can be seen that for large $k$ the mean number of particles scales as $N_{k} \sim k^{-4}$. Therefore, for an isotropic emergent spacetime (with two or three spatial dimensions $d$), we get
\begin{equation}
N_{\leqslant k}\sim 2^{d-1}\pi \, \int dk \,k^{d-5} \sim k^{d-4} .
\end{equation}
Therefore, the total number of particles produced is finite, as expected. However, it is easy to see that the total energy emitted,
\begin{equation} \label{Eq.total.engery.general}
E=2^{d-1}\pi \, \int_{k} dk \,k^{d-1} \; N_{k} \, \omega_{k}  \sim  2^{d-1}\pi \, \int_{k} dk \, k^{d-3} \, ,
\end{equation}
is still infinite. Note, that in the ultraviolet regime one has to
use the non-linear dispersion relation $\omega_{k}\sim k^{2}$; see
Eq.~\refb{Eq.disp.rel.qp}.  We will discuss this
further in the conclusion, after we investigate a possible application of signature change events for laboratory cosmology.
%
%
\section{Particle Amplifier for Cosmological Particle Production \label{sec:amplifier}}
%
In a Euclidean-signature emergent spacetime some modes (depending on the strength of the attractive interactions) grow and decay exponentially. As pointed out in \cite{Calzetta1:2003xb} this behavior is in analogy to cosmological particle production, where super--Hubble horizon modes (\ie, modes with frequencies that are smaller than the Hubble frequency) show similar kinematics. These modes are not free to oscillate, as they get dragged along with the spacetime fabric. 
This motivated us to investigate how pre-existing condensate perturbations are influenced by the existence of a short-duration Euclidean phase. To be more specific, we would like to determine whether significant amplification of a pre-existing particle spectrum occurs from exposing it to a short finite time of attractive interactions.
This might be of interest for laboratory cosmology in Bose gases. One of the outstanding problems involved with such experiments, as suggested in \cite{Fedichev:2006mc, Fedichev:2004on,
Barcelo:2003yk, Barcelo:2003ia, Fedichev:2004fi, Uhlmann:2005rx}, is the detectability of the quasi-particle spectrum in an emergent FRW-type universe. Due to the smallness of mode-population, common detection mechanisms would fail to measure the spectra. Recently there has been some attempt to solve this problem involving a more sophisticated detection mechanism \cite{Schutzhold:2006pv}. Another way to tackle this problem is to amplify the quasi-particle spectrum so that common detection mechanisms can be applied.
%
\begin{figure}[htb]
 \begin{center}
 \input{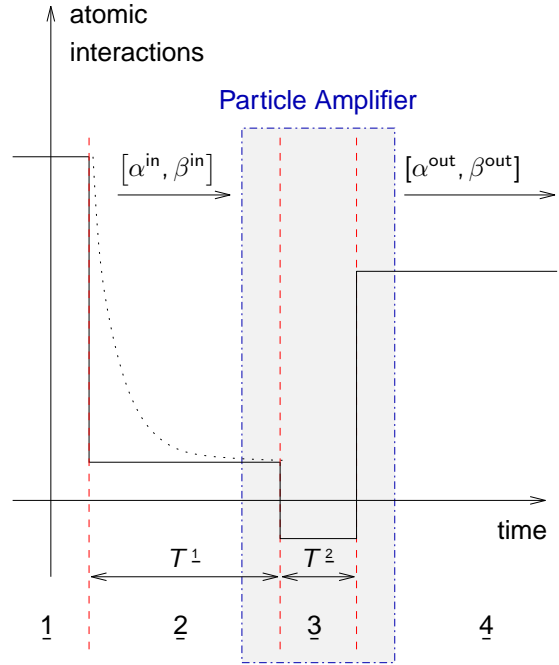}
 \caption[Particle Amplifier for quantum fluctuations.]  {\label{Fig.Particle.Amplifier}
The figure shows a particle production process starting from the vacuum in region $\Reg{1}$.  The bold line corresponds to particles produced in a single Lorentzian step from region $\Reg{1} \to \Reg{2}$, while the dotted line shows particles produced during region $\Reg{2}$.  Region $\Reg{3}$ depicts a subsequent amplification process of a finite time Euclidean interval.
Altogether the $\mathrm{OUT}$ signal in region $\Reg{4}$ can be viewed as an amplified $\mathrm{IN}$ signal.}
\end{center}
\end{figure}
%
\subsection{Main concept \label{Sec:Main.Concept.PA}}
In Figure~\refb{Fig.Particle.Amplifier} we illustrate the principle of our particle amplifier idea. The particle amplifier involves a $3$-step process. Region $\Reg{2}$ will create a quasi-particle spectrum $\Bb{1}{2}=\beta^{\mathrm{IN}}$, which we wish to amplify. The Euclidean region $\Reg{3}$ is the core of our particle amplifier process with $\Bb{2}{4}=\beta^{\mathrm{PA}}$, and region $\Reg{4}$ contains the amplified quasi-particle spectrum, $\Bb{1}{4}=\beta^{\mathrm{OUT}}$. For simplicity we focus on a Lorentzian sudden step process as the source of our initial quasi-particle spectrum,
however we stress that this method is more general and applies to any quasi-particle spectrum in region $\Reg{2}$ that is of a FRW-type universe.

We would like to determine an expression for our final amplified spectrum, in terms of the input spectrum, and the spectrum we would expect from the amplifying step. We know the connection
matrix after the amplifying step can be decomposed as (see Eq.~\refb{Eq.M.1m}):
\begin{equation} 
M^{\reg{1}\reg{4}} =M^{\reg{1}\reg{2}} \cdot  M^{\reg{2}\reg{4}} 
=
\begin{bmatrix}\Ba{2}{4}\Ba{1}{2}+\Bbc{2}{4}\Bb{1}{2}
& \Ba{1}{2}\Bb{2}{4}+\Bb{1}{2}\Bac{2}{4}\\
\Bbc{1}{2}\Ba{2}{4}+\Bac{1}{2}\Bbc{2}{4} &
\Bbc{1}{2}\Bb{2}{4}+\Bac{1}{2}\Bac{2}{4}
\end{bmatrix} \, .
\end{equation}
We now look at an expression for the final output spectrum in terms of the input spectrum and particle amplifier contribution. As we already have the Bogoliubov coefficients, this is straightforward to calculate. As $\det(M^{\reg{i}\reg{j}})=1$ it follows that:
\begin{equation}
\label{Eq:beta.out.14.general}
\vert\Bb{1}{4}\vert^{2}=\vert\Bb{2}{4}\vert^{2}+[2\vert\Bb{2}{4}\vert^{2}+1]\vert\Bb{1}{2}\vert^{2} 
+ 2 \, \mathrm{Re}\left(\Bac{2}{4}\Bb{1}{2}\Bbc{2}{4}\Bac{1}{2}\right) \, .
\end{equation}
This can be rewritten completely in terms of $\mathrm{IN}$, $\mathrm{OUT}$ and $\mathrm{PA}$ formalism,
\begin{equation}
\label{Eq:beta.out.IN.OUT.general}
\vert\beta^{\mathrm{OUT}}\vert^{2}=\vert\beta^{\mathrm{PA}}\vert^{2}+[2\vert\beta^{\mathrm{PA}}\vert^{2}+1]\vert\beta^{\mathrm{IN}}\vert^{2} 
+ 2 \, \sin(\gamma) \; \vert\alpha^{\mathrm{PA}}\vert  \vert\beta^{\mathrm{PA}}\vert \vert\alpha^{\mathrm{IN}}\vert  \vert\beta^{\mathrm{IN}}\vert \, ,
\end{equation}
where 
\begin{equation}
\sin(\gamma) = \mathrm{Re}\left(ph(\Bac{2}{4})ph(\Bb{1}{2})ph(\Bbc{2}{4})ph(\Bac{1}{2})\right) \, ,
\end{equation}
represents the relationship between the phases, $ph(...)$, of our Bogoliubov coefficients. Therefore the amplification process depends on $\gamma$, and it is necessary to find an explicit expression for it.

Although it would be possible to make further statements while keeping the analysis completely general, we now choose an explicit particle amplifier configuration and provide an example that demonstrates the parameters $\gamma$ depends on.
\subsection{Simple example \label{Sec:Simp.Ex.PA}}
For the simplest case, we look at a Euclidean amplifying step, with $X^{\reg{2}\reg{3}}=i, \, X^{\reg{3}\reg{4}}=-i,\, X^{\reg{2}\reg{4}}=1$ and write $\omega_{k}^{\underline{3}}=i|\omega_{k}^{\reg{3}}|$. This describes the scenario where the speeds of sound before and after the Euclidean step are equivalent. With these replacements the Bogoliubov coefficients for the Euclidean amplifying step can be split up as follows:
\begin{eqnarray}
\Ba{2}{4}&=&e^{i\omega_{k}^{\reg{2}}(t^{\reg{1}\reg{2}}+T^{\reg{1}})-i\omega_{k}^{\reg{4}}(t^{\reg{1}\reg{2}}+T^{\reg{1}}+T^{\reg{2}})} \,
\vert\Ba{2}{4}\vert \, , \\
\Bb{2}{4}&=&e^{+i\omega_{k}^{\reg{2}}(t^{\reg{1}\reg{2}}+T^{\reg{2}})+i\omega_{k}^{\reg{4}}(t^{\reg{1}\reg{2}}+T^{\reg{1}}+T^{\reg{2}})} \,
\vert \Bb{2}{4} \vert \, .
\end{eqnarray}
Using the notation for the Bogoliubov coefficients describing the input spectrum step to be;
\begin{eqnarray}
&\Ba{1}{2}=e^{i(\omega_{k}^{\reg{1}}-\omega_{k}^{\reg{2}})t^{\reg{1}\reg{2}}} \,
\vert\Ba{1}{2}\vert \, , \\ 
&\Bb{1}{2}=e^{i(\omega_{k}^{\reg{1}}+\omega_{k}^{\reg{2}})t^{\reg{1}\reg{2}}} \,
\vert\Bb{1}{2}\vert \, ,
\end{eqnarray} 
we then find the output spectrum after amplification becomes:
\begin{eqnarray}
\nonumber
\vert \Bb{1}{4}\vert^{2}&=&
\vert \Bb{2}{4}\vert^{2}+\left(2\vert \Bb{2}{4}\vert^{2}+1\right) \vert \Bb{1}{2}\vert \\
\label{Eq:PA.simple.example}
&&- 2 \, \sin\left(2\omega^{\reg{2}}_{k}\, T^{\reg{1}}\right) \times \sqrt{(\vert \Bb{1}{2}\vert^2+1)(\vert\Bb{2}{4}\vert^2+1)} \, \vert\Bb{2}{4}\vert \,  \vert \Bb{1}{2}\vert \, , \,
\end{eqnarray}
where we have employed the normalization; $\det(M^{\reg{i} \reg{j}})=1$.
In this particular case we found $\gamma = -2\omega^{\reg{2}}_{k}\, T^{\reg{1}}$, which therefore only depends on known parameters --- the time-interval $T^{\reg{1}}$ and the dispersion relation $\omega^{\reg{2}}_{k}$ in region $\Reg{2}$ ---  that are tunable (for a certain range of $k$-values) throughout a BEC experiment.

Depending on $\gamma$ the number of $\mathrm{OUT}$ particles oscillates around
\begin{eqnarray}
\vert \beta^{\mathrm{OUT}}\vert^{2}&=&
\left( 1 + 2\, \vert \beta^{\mathrm{IN}}\vert^{2}  \, + \frac{ \vert \beta^{\mathrm{IN}}\vert}{ \vert \beta^{\mathrm{PA}}\vert} \right) \;  \vert \beta^{\mathrm{PA}}\vert^{2} \, ,
\label{Eq:PA.simple.example.around}
\end{eqnarray}
and thus the difference between a process with and without initial particles is given by,
\begin{eqnarray}
\vert \beta^{\mathrm{OUT}}\vert^{2}&\approx&
 2\, \vert \beta^{\mathrm{IN}}\vert^{2}   \;  \vert \beta^{\mathrm{PA}}\vert^{2} \, .
\label{Eq:PA.simple.example.around.diff}
\end{eqnarray}
In the last step we assumed that $\vert \beta^{\mathrm{IN}}\vert \ll \vert \beta^{\mathrm{PA}}\vert$, which is plausible since we expect a much larger particle production in the Euclidean region. 

Nevertheless, we have shown that for the simplest example possible the particle amplifier does indeed amplify the initial spectrum, see Eq.~\refb{Eq:PA.simple.example.around.diff}. 
However, considerably more effort is required to analyse the particle amplifier in depth.  It will be important to explore the efficiency of our proposed model and how a more realistic initial particle spectrum resulting from cosmological particle production is amplified.  The difference between the maximum possible particle production from the amplifying step ($\sin(\gamma)=-1$) and the minimum particles produced ($\sin(\gamma)=1$) might be a possible way to increase the efficiency of the particle amplifier.
In the next section we will review our results, connect with the
existing literature, and discuss problems and open questions.
%
%
\section{Conclusions \label{sec:EndHappy}}
%
%
%
\begin{figure*}[!htb]
\begin{center}
\mbox{
\subfigure[$\,$ Hydrodynamic limit; L-L-L. \label{Fig.Hydro.LLL}]{\includegraphics[width=0.47\textwidth]{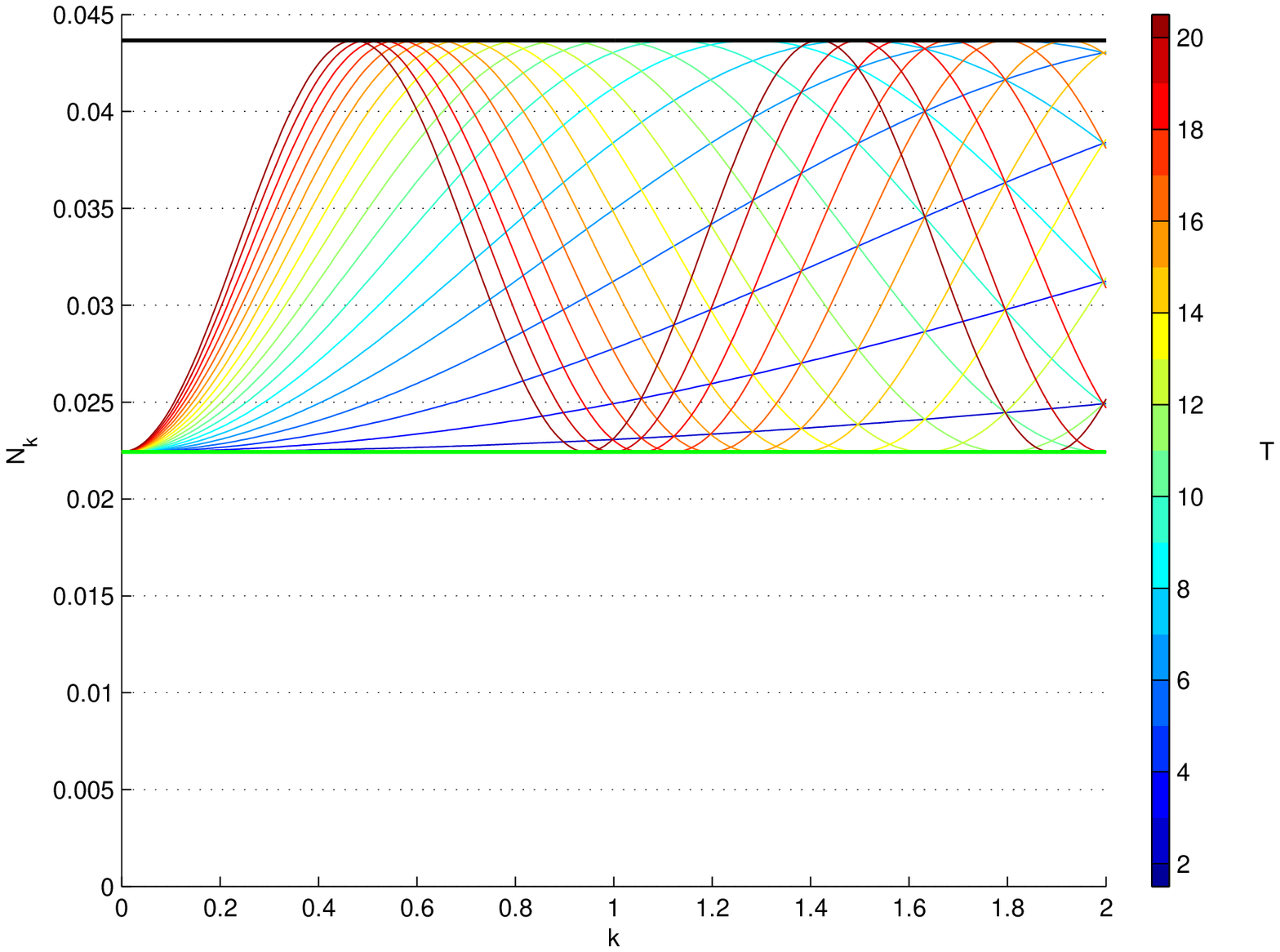}}
\hspace{0mm}
\subfigure[$\,$ Microscopic corrections; L-L-L. \label{Fig.QP.LLL}]{\includegraphics[width=0.47\textwidth]{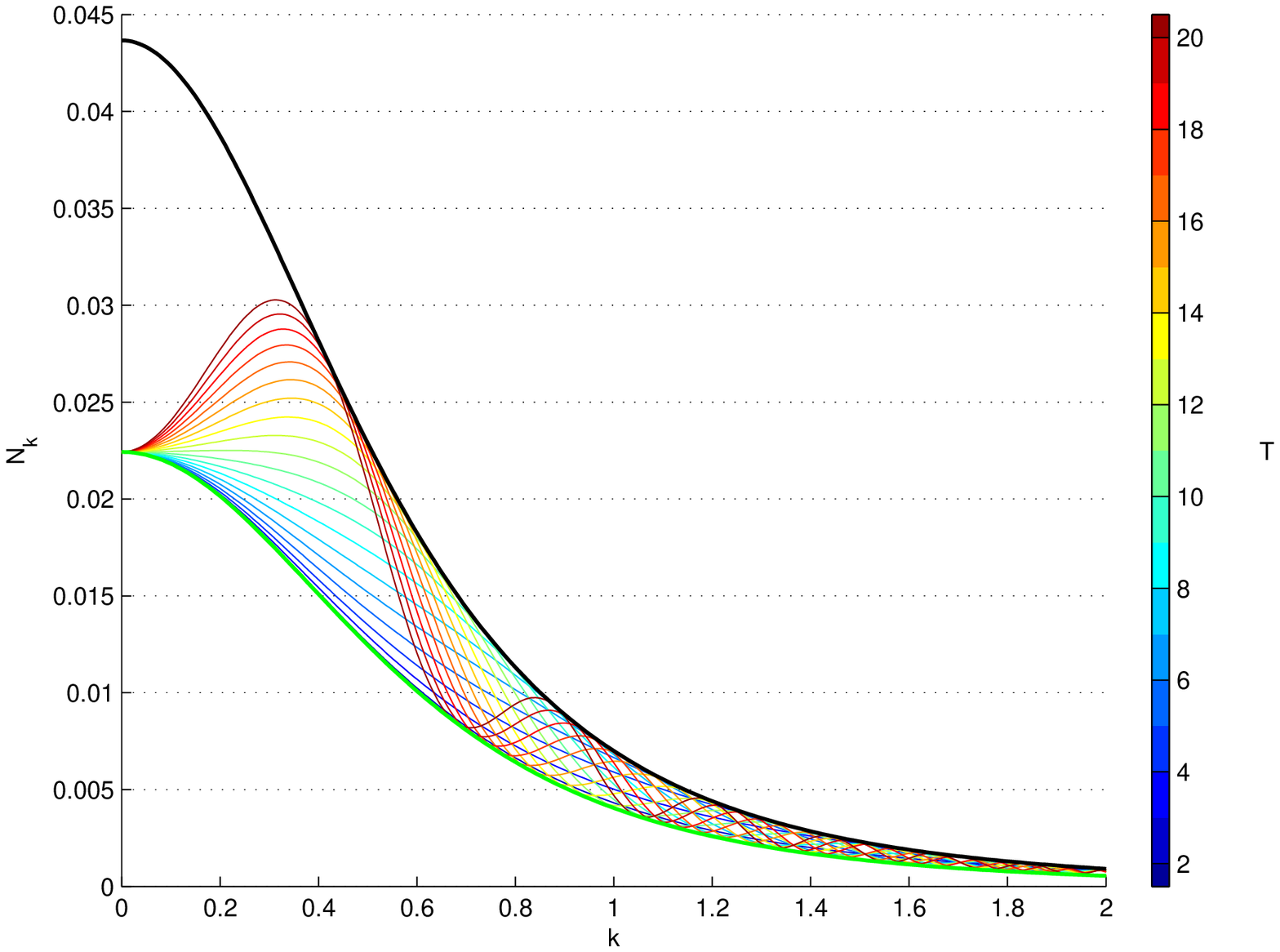}}
}
\mbox{
\subfigure[$\,$ Hydrodynamic limit; L-E-L. \label{Fig.Hydro.LEL}]{\includegraphics[width=0.47\textwidth]{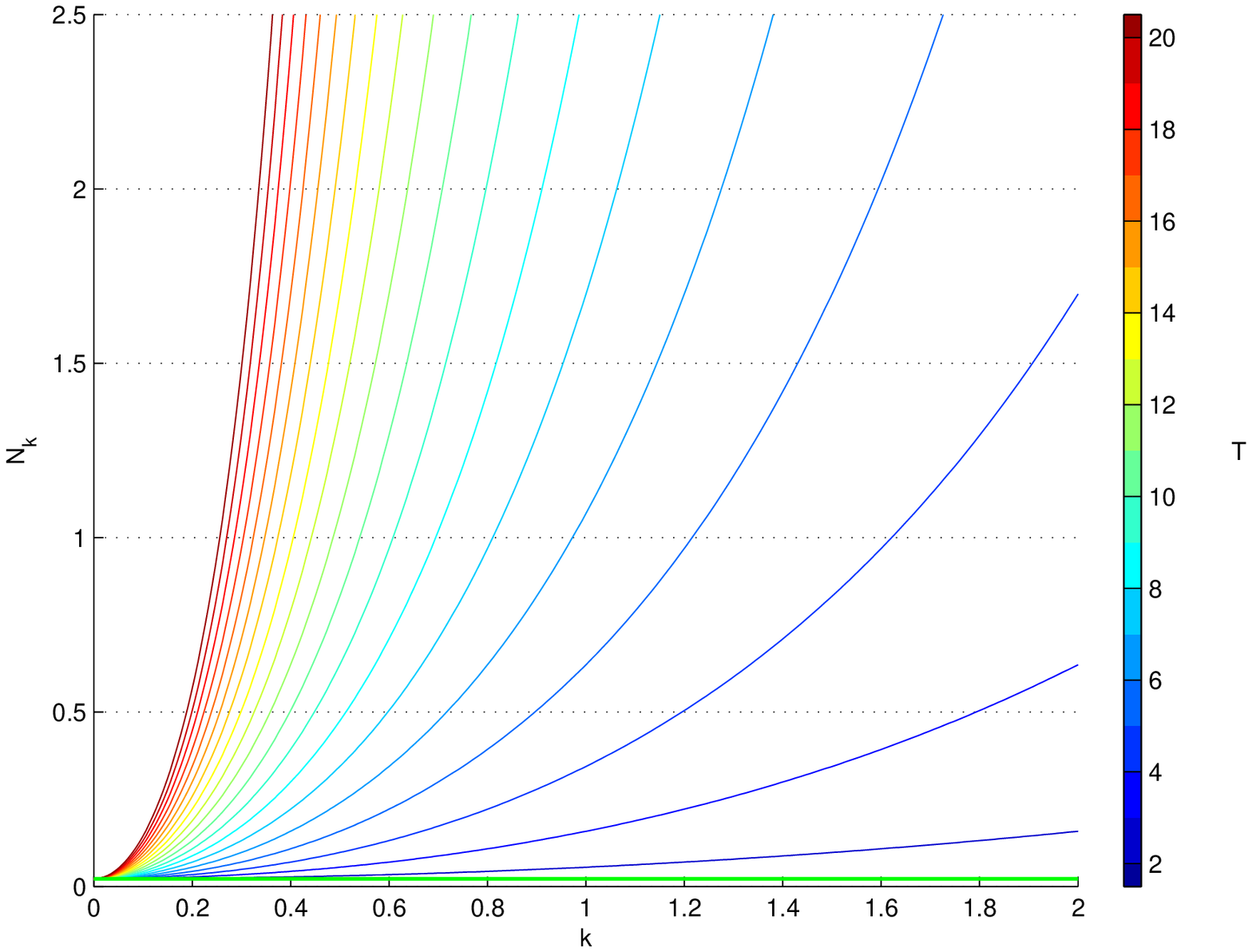}}
\hspace{0mm}
\subfigure[$\,$ Microscopic corrections; L-E-L. \label{Fig.QP.LEL}]{\includegraphics[width=0.47\textwidth]{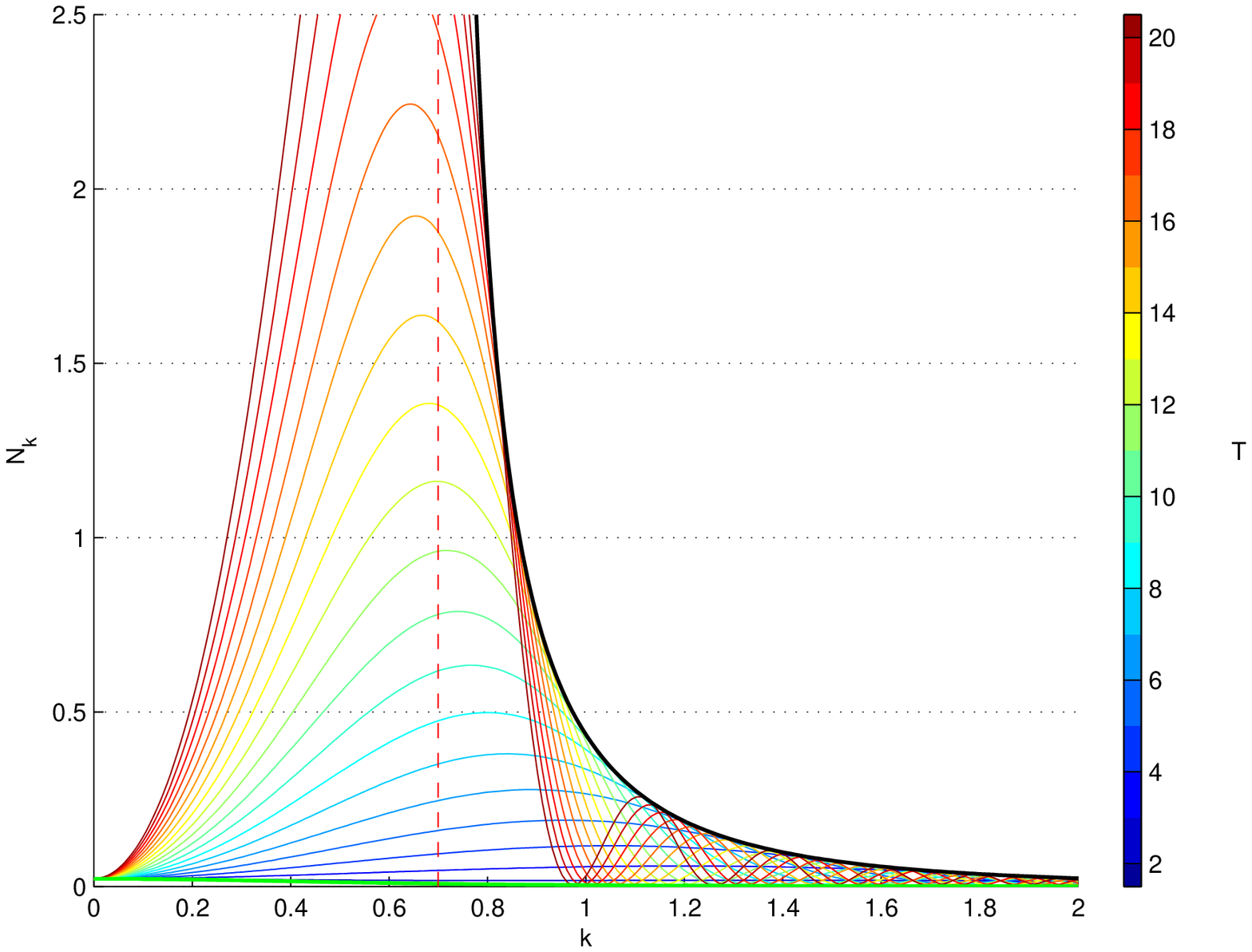}}     }
\caption[Particle spectra for sudden Lorentzian and Euclidean changes in emergent geometry.]{The figure displays the mean number of particles $N_{k}$ produced from two sudden variations in the strength of the atomic interactions. The different curves (color online) correspond to the time period $T$ for which the system is kept in the intermediate regime. The left column focusses on purely collective effects, while in the right column microscopic corrections are taken into account. The top two diagrams show variations between spacetimes emerging from a Bose gas with repulsive interactions ($a>0$). In the bottom row of the diagrams the intermediate regime exhibits an Euclidean geometry, where the underlying Bose gas shows attractive atomic interactions ($a<0$). (Please note the figures are \emph{not} based on ``real'' condensate parameters.)}\label{fig:desitterresults}.
\end{center}
\end{figure*}
%
In the first few pages of this chapter we introduced the reader to
the concept of emergent spacetimes. Under certain conditions, the
microscopic theory of an ultra-cold weakly interacting Bose gas
gives way to a mean-field description, the Bose--Einstein
condensate. A new state of matter has then been formed, described
by a complex classical field. Its behavior is dominated by
collective variables, rather than individual atoms. A careful
study of the quantum excitations around the mean-field reveals the
correspondence between the condensed matter physics and quantum
field theory in curved spacetimes.  The quantum perturbations are
massless spin-zero particles, that are controlled by an emergent
geometrical/gravitational field.

The actual excitations $\delta\hat\psi$ and $\delta\hat\psi^{\dag}$ are related to perturbations in the collective Hermitian variables, the phase $\hat\theta$ and density $\hat n$ of the complex mean-field. The condensed matter description for linearized perturbations supplies commutation relations for the phase and (up to a function) its material derivatives. We were able to show that these commutators are precisely the ones for the field $\hat\theta$ and its canonical momentum $\hat\Pi_{\hat\theta}$ on the emergent curved spacetime.

The effective gravitational field defined on  the Bose gas
naturally implements a mechanism to experimentally perform
signature-change events. The signature of the emergent
gravitational field is directly correlated to the sign of the
$s$-wave scattering amplitude of the microscopic collisions. From
a gas with repulsive\,/\,attractive interaction we expect an emergent
Lorentzian\,/\,Riemannian geometry. The existence of Riemannian
geometries in a Bose gas is limited to short duration time
intervals and weak interactions, since back-reaction and
instability issues would otherwise destroy the condensate; see
\cite{Hu:2005wu,Calzetta1:2003xb,Calzetta:2005yk,Hu:2003}.

Inspired by recent condensed matter experiments
\cite{Donley:2001aa,Roberts:2001aa}, and theoretical work based on
them \cite{Hu:2005wu,Calzetta1:2003xb,Calzetta:2005yk,Hu:2003}, we
investigated the quasi-particle production caused by an effective
gravitational field going through a sudden but finite duration
transition to Euclidean signature. We compared the signature
changing case with sudden variation between different Lorentzian
regimes, and showed that perturbations in the Euclidean regime
experience exponential growth, while for an intermediate
Lorentzian regime the modes oscillate between single steps. These
are given by the maxima and minima configuration of the two-step
process with $T=0$. Our results are in two ways a generalization
of the existing literature, (\eg, the standard general relativity
calculations of Dray \emph{et al.} \cite{Dray:2004aa,
Dray:1991zz}):

1) For the right choice of mode functions (a complex conjugate
pair), the calculation is formally equivalent to purely Lorentzian
sudden variation between arbitrary levels. This formalism can be
applied to any $n$-step process.

2) In addition, we went beyond the hydrodynamic description, which
is only sufficient to account for the infrared behavior of the
system, and included ultraviolet modes in our analysis. This leads
to modifications in both the emergent gravitational field and the
dispersion relation. For the gravitational field we obtained (in the
eikonal limit) a momentum-dependent rainbow geometry; and a
non-relativistic dispersion relation. The modification in the
dispersion relation can be classified as being within the boost
sub-group, and is of the form as suggested by some effective field
theories \cite{Mattingly:2005aa, Liberati:2006sj, Liberati:2006kw,
Weinfurtner:2006iv, Weinfurtner:2006nl}.

In figure \refb{fig:desitterresults} we illustrated our four different results, namely sudden variations without (first row) or with (second row) signature change events, within the hydrodynamic limit (left column), and beyond it (right column). As a result, the total number of particles produced is finite. One way to explain the behavior of the trans-Planckian modes is to understand the physical significance of the healing length of the Bose--Einstein condensate. The healing scale defines a length scale over which quantum fluctuations in the condensate tend to smooth out. In some sense, it defines the smallest possible size for a condensate, $V > \xi^{d}$, see \cite{Pethick:2001aa}. Perturbations with wavelengths that are much larger than the healing length experience a nice smooth mean-field, here our emergent gravitational field. High energy modes, with wavelengths at the magnitude of the healing length start to see fingerprints of the microscopic structure. Note that, perturbations of sufficiently high energy are not driven by collective variables. This can also be seen in figure \ref{Fig.QP.LEL}: While infrared modes behave like their hydrodynamic\,/\,relativistic neighbours to the left \ref{Fig.Hydro.LEL}, the ultraviolet modes do not notice the change in the signature of the gravitational field. They are behaving just like their upper neighbours \ref{Fig.QP.LLL} in a purely Lorentzian geometry.

Still, not all of our concerns are yet resolved. For example,
while the total number of particles produced can be made finite,
the total energy remains formally infinite even beyond the
hydrodynamic limit. This problem might be related to our choice of
ultra-high-energy description for the fundamental Bosons.
Perturbations of sufficiently high energy can excite single atoms
out of the condensate, and these are (in our description)
non-relativistic particles. Therefore, we suggest that to improve
the ultra-high-energy behavior it might be useful to start with a
fully relativistic description for the fundamental Bosons. (For a
description of relativistic Bose--Einstein condensates see, \eg,
\cite{Bernstein:1991aa}.)

Finally, we would like to comment on the possible connection between our theoretical results and the data from the Bose-nova experiment carried out by Donley \emph{et al.} \cite{Donley:2001aa, Roberts:2001aa}. There are fundamental differences between our approach, and the Bose-nova experiment. First, we have chosen a hard-walled box as an external potential, while in the experiment a harmonic trap has been used. In \cite{Hu:2005wu,Calzetta1:2003xb,Calzetta:2005yk,Hu:2003}, Hu and Calzetta have shown that the trapping energy delays the condensate collapse for a certain amount of time. Their calculations are in good accordance with the experimental values. In our case the trapping energy is zero, and hence we do not expect such a delay.

Another fundamental difference is that we kept our gravitational
field non-degenerate (apart from the actual instant of signature
change). 
In the experiment the initial scattering length was taken
to be zero and held there for a finite time, and hence Donley
\emph{et al.} started with a partly \emph{degenerate} gravitational
field. (Partly degenerate gravitational field, because for a harmonic trap the quantum pressure term exhibits a $k$-independent contribution resulting in a position-dependent correction to the effective atomic interaction; see first term inside the curly brackets in Eq.~\refb{Eq.U.qp}.)
In the experiment two different sequences were chosen
for the sudden variations, with two different outcomes. One, where
the scattering length has been driven from zero to attractive and
finally to large repulsive atomic interactions. The other set-up
was from zero to attractive, back to zero and then to large
repulsive atomic interactions. In both cases bursts
of atoms leaving the condensate were detected. But the bursts without the
intermediate regime of zero interactions were much stronger. These
are ``jets'' of atoms, while in the other case the bursts were
less strong. Hu and Calzetta approached this problem without
taking back-reaction effects --- of the quantum perturbations onto
the condensate --- into account. Naturally, their description
becomes less accurate with the length of the Euclidean
time-interval. We would like to emphasize the difference in the
behavior of the acoustic metric, the ``gravitational field'',
since in the first scenario it does not exhibit any finite
interval of degeneracy, while in the latter it does.

Furthermore, we would like to propose a rather different use for
our L-E-L-process, as a particle amplifier for cosmological
particle production in the laboratory. In our calculations so far
we always started from the vacuum with zero collective excitations
in the system. These calculations could easily be extended, such
as to start with a non-vacuum state, \eg, after cosmological
particle production in an emergent Friedmann--Robertson--Walker
type geometry \cite{Fedichev:2006mc, Fedichev:2004on,
Barcelo:2003yk, Barcelo:2003ia, Fedichev:2004fi, Uhlmann:2005rx}.
Recently this behavior has been studied numerically in a realistic
Bose--Einstein condensate, where the present authors and collaborators used a
classical-field-method approach, see \cite{Jain:2006ki}. To obtain
an effective expanding universe, the scattering length has to
decrease as a specific function of time. After a finite expansion
time, the quasi-particle spectrum obtained --- a very small number of particles
--- somehow has to be detected. Due to its smallness this
remains a significant experimental challenge. Recently, there has
been some theoretical effort regarding this problem, see
\cite{Schutzhold:2006pv}. That author suggests a rather
complicated detection mechanism, where only perturbations of one
wavelength at a time can be detected.
In section \refb{sec:amplifier} we suggested an alternative concept, that is to \emph{amplify} the quasi-particle spectrum, \eg, in a three-step process involving a brief finite-duration Euclidean region to amplify the signal, so that common detection mechanisms might be able to detect the amplified quasi-particle spectrum.  
For a simple example, we calculated the particle spectrum after the amplification process in terms of the initial spectrum. However, for any practical application in a real BEC experiment there are considerable technical issues that need to be explored. (For example, the possibility of using a L-L-L amplifier.)

To conclude this chapter we would like to briefly discuss the possibility
of real general relativistic signature change events. In 1983,
James Hartle and Steven Hawking proposed a signature change at
extremely early times, in the very early stages of the big bang,
when quantum gravity effects are expected to be dominant (see
reference \cite{Hartle:1983aa}). They suggested that the existence
of physical time, and hence the existence of our universe, is
associated with a signature change event from Euclidean to
Lorentzian geometry (the ``no boundary'' proposal).

In some sense it is possible to consider the reverse Bose-Nova experiment (attractive to repulsive) as the creation process for our emergent spacetime. At early times the Boson interactions are strongly attractive, such that the atoms do not show any collective\,/\,mean-field behavior. If now the interactions experience a smooth, or non-smooth change for the atomic interactions, from attractive to repulsive, the individual atoms have to give way to collective\,/\,mean-field variables. Therefore the existence of our emergent gravitational field might in some sense be associated with a pre-dating signature change event caused by a change in the underlying microscopic variables.
%

\chapter{Cosmological inflation in emergent rainbow spacetime \label{Chap:Late.Quantum.Cosmology}}
We now investigate cosmological particle production in spacetimes where Lorentz invariance 
emerges in the infrared limit, but is explicitly broken in the ultraviolet regime. Thus these 
models are similar to many (but not all) models of quantum gravity, where a breakdown 
of Lorentz invariance is expected for ultraviolet physics around the Planck\,/\,string scale. 
Our specific model focuses on the boost subgroup that supports CPT invariance and 
results in a momentum-dependent dispersion relation. Motivated by previous studies on 
spacetimes emerging from a microscopic substrate, we show how these modifications 
naturally lead to momentum-dependent rainbow metrics.

Firstly, we investigate the possibility of reproducing cosmological particle production in 
spacetimes emerging from real Bose gases. Several papers have been written on the 
analogy between the kinematics of linearized perturbations in Bose--Einstein condensates 
and effective curved-spacetime quantum field theory. Recently, we have studied 
the influence of non-perturbative ultraviolet corrections in time-dependent analogue 
spacetimes, leading to momentum-dependent emergent rainbow spacetimes. 
We show that models involving a time-dependent microscopic interaction are 
suitable for mimicking quantum effects in FRW spacetimes. Within certain limits the 
analogy is sufficiently good to simulate relativistic quantum field theory in time-dependent 
classical backgrounds, and the quantum effects are approximately robust against the
model-dependent modifications.
 
Secondly, we analyze how significantly the particle production process deviates from the common 
picture. While very low-energy modes do not see the difference at all, some modes ``re-enter the Hubble horizon'' during the inflationary epoch, and extreme ultraviolet modes 
are completely insensitive to the expansion. The analysis outlined here, because it is 
nonperturbative in the rainbow metric, exhibits features that cannot be extracted simply 
from the standard perturbative modification of particle dispersion relations. However, we 
also show how the final result, after many e-foldings, will approach a time-independent 
exponentially decaying particle spectrum.
%
%
\section{Introduction}
%
The use of Bose--Einstein condensates as analogue models for quantum field theory in spatially flat $k=0$ Friedmann--Robertson--Walker (FRW) geometries
\begin{equation}
\label{Eq:FRW.GR}
ds^{2} = g_{ab} \; dx^{a} \, dx^{b}= -d\tau^{2} + a(\tau)^{2} \sum_{i=1}^{d} (dx^{i})^{2} \, 
\end{equation}
(in $d$ spatial dimensions) has recently been extensively explored in \cite{Jain:2006ki}. The proposal~\cite{Barcelo:2003ia,Barcelo:2003yk,Fedichev:2004on,Fedichev:2004fi,Fischer:2004iy,Uhlmann:2005rx,Unruh:1981bi} is based on the analogy between the equation of motion for collective excitations around a macroscopically occupied ground state of an ultra-cold weakly interacting gas of Bosons (\ie, the novel state of matter referred to as the Bose--Einstein condensate), and
\begin{equation}
\label{Eq:KGE}
\frac{1}{\sqrt{\vert g \vert}} \; \partial_{a} \, \left(\sqrt{\vert g \vert} \, g^{ab} \, \partial_{b} \hat \theta \right)=0\, ,  
\end{equation}
the covariant free-field equation (Klein--Gordon equation) for spin-$0$ massless particles. (The indices $a$, $b$ run from $0$ to $d$ whereas $i$, $j$ run from $1$ to $d$ for the spatial coordinates only). Here $g^{ab}$ is a symmetric covariant rank two tensor whose entries include purely collective (mean-field) variables $c=c(t,\mathbf x)$, the speed of sound, $\mathbf{v}=\mathbf v(t,\mathbf x)$, and the background velocity. Using $g^{ab}\,g_{bc}=\delta^{a}{}_{c}$ we are able to define an effective line-element for the mean-field,
\begin{equation}
ds^{2} = \left(\frac{c}{U}\right)^{\frac{2}{d-1}} \left[ -(c^{2}-v^{2}) \, dt^{2} - dt \, \mathbf v^{\mathrm{T}}\, d\mathbf x + d\mathbf{x}^{\mathrm{T}} \, d\mathbf x   \right] \, .
\end{equation}
Here the quantity $U$ arises from the microscopic description of Bose--Einstein condensate, and represents the inter-atomic potential.

We have previously shown (see Chapter~\ref{Chap:Early.Quantum.Cosmology}) that Eq.~\refb{Eq:KGE} is an equation for perturbations in the collective variables, the phase $\hat \theta$ and its conjugate momentum $\hat \Pi_{\hat \theta}$ (see also \cite{Barcelo:2003ia,Barcelo:2003yk,Fedichev:2004on,Fedichev:2004fi,Fischer:2004iy,Uhlmann:2005rx,Unruh:1981bi}).
This kind of behavior is not restricted to Bose--Einstein condensates. Indeed the first modern paper on analogue models for gravity focussed on ordinary fluid mechanics; see \cite{Unruh:1981bi}. Since then numerous media have been suggested as substrates to develop analogue models for gravity~\cite{Barcelo:2005ln,Schutzhold:2007aa,Visser:1993tk,Visser:1998gn}. 
Nevertheless it is sometimes difficult to isolate the fundamental principles behind the analogy.

Inspired by the Bose gas, where both the microscopic and macroscopic theory is understood to an adequate extent, we suggest that it might be useful to transfer the experience gained from the various known examples for emergent spacetimes into a more general framework:

\begin{verse}
\emph{Emergent spacetimes} involve
$\mathrm{(i)}$ a microscopic system of fundamental objects (\eg, strings, atoms, or molecules); 
$\mathrm{(ii)}$ a dominant mean-field regime, where the microscopic degrees of freedom give way to collective variables;
$\mathrm{(iii)}$ a ``geometrical object'' (\eg, a symmetric tensor) dominating the evolution for infrared classical and quantum excitations around the mean field.
\end{verse}
Within certain limits we are free to choose our geometrical object to mimic Einstein's theory of gravity, (\eg, a symmetric rank two tensor $g_{ab}$ that is conformal to an exact solution obtained by solving the Einstein equation for a physically reasonable stress-energy tensor). 
\begin{verse}
\emph{Analogue models for gravity} are emergent spacetimes that are specifically adjusted to mimic as closely as possible Einstein's geometrical theory of gravity.
\end{verse}
The fundamental difference between emergent spacetimes and general relativity becomes obvious if we consider the following statement: Emergent spacetimes, as they appear in the analogue model programme, are ``a short and simple way'' to summarize the kinematics of linearized perturbations in a geometrical sense, without requiring the notion of a stress-energy tensor or the Einstein equations.

Thus, the analogy is (currently) restricted to the kinematic behavior of the system, and any back-reaction between the excitations and the mean-field might not be in analogy with Einstein's theory of gravity. Therefore it is advisable to restrict the use of analogue models for gravity as toy models for semi-classical quantum gravity at the level of curved spacetime quantum field theory, where the gravitational field is a purely classical field.

Indeed, analogue models for gravity are very idealized constructions, and require careful treatment with respect to the model-dependence to obtain the desired curved-spacetime quantum-field-theory effects. This has first been pointed out in \cite{Jacobson:1991sd,Unruh:2005aa}, where the authors analyzed the robustness of Hawking radiation against modifications in the excitation spectrum (most relevant to fluid dynamics: super- and sub- luminal dispersion).\\

In related previous work \cite{Jain:2006ki} we made use of classical phase space methods to numerically study ``cosmological particle production'' in a realistic BEC. In principle, there are two different ways to mimic an expanding universe in a Bose gas. In \cite{Fedichev:2004on,Fedichev:2004fi} the authors studied a freely expanding condensate cloud, such that the density is a function of time. This idea has been further investigated in \cite{Uhlmann:2005rx}. In both --- our previous paper \cite{Jain:2006ki} and in this chapter --- we pursue an alternative (more recent) idea introduced in \cite{Barcelo:2003ia,Barcelo:2003yk}, where the authors achieved FRW geometry through a time-dependence in the effective atom-atom potential. A change in the inter-atomic potential is related to a change in the speed of sound, and consequently relates to a time-dependent acoustic metric. As expected our simulation verified the analogy (in the hydrodynamic limit) to a good extent. However, the ultraviolet behavior showed that it is crucial to include quantum pressure effects to understand the ultraviolet\,/\,trans-phononic part of the quasi-particle spectrum. Thus we conclude that our specific model --- an ultra-cold gas of Bosons with time-varying atomic interactions --- shows a specific model-dependence that interferes with the low-momentum particle production process. One lesson that can be drawn from this new insight is that a careful choice of condensate parameters is necessary in order to mimic the desired curved-spacetime quantum field theory effect.

The main focus in this chapter is to understand the differences between ordinary quantum field theory in a ``real'' classical FRW-background and quantum field theory in an emergent FRW-background in a Bose gas. This question has been motivated by many effective field theories, where Lorentz invariance is broken at ultraviolet energies in a similar manner as in the emergent spacetimes we are investigating. 
\begin{verse}
Emergent spacetimes exhibit an \emph{emergent\,/\,effective Lorentz symmetry} for low-energy\,/\,infrared excitations around the macroscopic field. This symmetry will be broken in the high-energy\,/\,ultraviolet regime, that is at scales dominated by the underlying microscopic theory. These corrections are of a non-perturbative nature. 
\end{verse}

For now, we leave general questions aside, and focus on the specific emergent spacetime, investigating the influence on cosmological particle production due to ultraviolet corrections. We include quantum pressure effects --- which lead to a nonlinear excitation spectrum --- as nonperturbative ultraviolet corrections to the emergent geometry. We will show that this leads to momentum-dependent ``rainbow spacetimes''. Partly motivated by extant literature \cite{Magueijo:2004aa,Weinfurtner:2006eq} and our numerical results from the simulations of quantum effects in realistic Bose--Einstein condensates, we present a coherent interpretation for the quasi-particle spectrum produced in FRW rainbow spacetimes \cite{Weinfurtner:2006nl,Weinfurtner:2006eq,Liberati:2006kw,Liberati:2006sj,Weinfurtner:2006eq,Weinfurtner:2006iv,Visser:2001ix}.

The ``FRW rainbow metrics'' recover the ``ordinary'' FRW type geometries in the infrared (phononic) regime, but exhibit momentum-dependent modifications in the ultraviolet (trans-phononic) regime. We show that in such a geometry the scale factor for the universe $a(t)$ is effectively momentum-dependent $a(t) \to a_{k}(t)$, consequently leading to a momentum-dependent Hubble parameter: $H(t) \to H_{k}(t)$. In addition, we are dealing with an emergent spacetime that exhibits a time-dependent effective ``Planck-length''. Ultimately, the particle production process in our specific analogue model will show deviations from the one expected in ``standard'' curved-spacetime quantum field theory.

For the ``analogue gravity programme'' it is important to check the robustness of cosmological particle production against those modifications. Is there a window where the analogy is good enough, for example, to use the Bose gas as a toy model for inflation? \\

In the following we will show that in the present model the particle process is \emph{in general not} robust against the model-specific modifications. However, for short time duration expansion scenarios inflation can be simulated in a ``realistic'' Bose--Einstein condensate. \\

We perform a quantitative analysis to describe short time \emph{and} long-lasting expansions. We use the ratio between the mode frequency and the Hubble frequency (the inverse of the rate of change in the size of the emergent universe), to show that our specific model indeed shows significant deviations in the ultraviolet regime, \eg, both crossing \emph{and} re-entering of the ``Hubble horizon'' during the inflationary epoch; $a(t)$~$\sim$~$\exp(t)$.\\

By means of our numerical analysis we are able to ``read-off'' the final spectrum after a sufficiently long-lasting expansion. We explain why the final particle spectrum --- after an infinitely long-lasting inflationary epoch --- will be time-independent and finite.

%
%
\section[Emergent spacetimes]{Emergent spacetimes:\protect\\
Excitations in Bose--Einstein condensates}
\label{Sec:Emergent.Spacetimes} 
%
In the following we extend the calculations presented in our previous work \cite{Jain:2006ki}. The intent is to derive an emergent geometry that is able to simultaneously represent both the phononic and trans-phononic excitations in a Bose--Einstein condensate. It is a well established result --- both in theory and experiment --- that the energy-momentum relation for excitations in a  Bose--Einstein condensates is given by the nonlinear Bogoliubov dispersion relation. In \cite{Liberati:2006sj,Weinfurtner:2006nl,Weinfurtner:2006iv,Liberati:2006kw,Weinfurtner:2006eq,Visser:2001ix} it has been shown, (within the analogue model for gravity point of view), that phononic modes are relativistic modes, since they exhibit Lorentz symmetry. The corresponding emergent geometry is a flat spacetime with Lorentzian signature (Minkowski metric). Higher energy (trans-phononic) modes break Lorentz symmetry, the excitations being ``supersonic''. The full dispersion relation for excitations in a realistic Bose--Einstein condensate is similar to those expected to occur in certain effective field theories. For a more detailed description see \cite{Liberati:2006sj,Weinfurtner:2006nl,Weinfurtner:2006iv,Liberati:2006kw,Weinfurtner:2006eq,Visser:2001ix}. It is also known that this kind of modification is non-perturbative, since it originates in density fluctuations visible only at small scales (in the order of the healing\,/\,coherence length). These modes are trans-phononic modes, where quantum pressure effects are no longer negligible. We show that in a Bose--Einstein condensate with time-dependent condensate parameters, the borderline between phononic and trans-phononic modes is also time-dependent, and therefore requires a more accurate analysis at the level of the emergent geometry, leading to the concept of a rainbow metric. In other words we extend the treatment for ultraviolet modes from flat to curved spacetimes.

\subsection[Ultraviolet non-perturbative corrections]{Ultraviolet non-perturbative corrections:\protect\\
Quantum pressure effects
\label{Label}}
In \cite{Jain:2006ki} we derived a coupled pair of equations, 
\begin{eqnarray} 
\label{Eq:on_dynamics}
&& \partial_t \hat n+\nabla \cdot \left[\left(\frac{n_{0} \hbar}{m}\nabla\hat\theta\right)+(n_{0} \, \mathbf v)
\right] =0\, ,\\
\label{Eq:otheta_dynamics}
&&\partial_t \hat \theta+\mathbf v \cdot \nabla\hat\theta+\frac{\widetilde U}{\hbar}\;\hat n =0 \, ;
\end{eqnarray}
for quantum fluctuations in the condensate density $\hat n$ and phase $\hat \theta$. The background velocity $\mathbf v$ is given by
\begin{equation} 
\label{Eq:Background.Velocity}
\mathbf v = \frac{\hbar}{m} \, \nabla \theta_{0} \, ,
\end{equation}
as the gradient of the condensate phase $\theta_{0}$. Here $m$ is the mass of the fundamental Bosons, $n_{0}$ the number density, and $\hbar$ the Planck constant. The differential operator $\widetilde U$ has been defined as \cite{Weinfurtner:2006nl,Jain:2006ki}
\begin{equation}
\label{Eq.Effective.U}
\widetilde U = U - \frac{\hbar^2}{2 m}\;\widetilde D_{2}   \, ,
\end{equation}
where the differential operator
\begin{equation}
\label{QuantumPressure}
\widetilde D_{2} =\frac{1}{2} \left\{ \frac{(\nabla n_{0})^{2} -(\nabla^{2}n_{0})n_{0}}{n_{0}^{3}} -\frac{\nabla n_{0}}{n_{0}^{2}}\nabla +\frac{1}{n_{0}}\nabla^{2} \right\} \, ,
\end{equation}
accounts for the first-order correction obtained from linearizing the quantum potential term. Note the tilde notation is used to emphasize the fact that we are dealing with a differential operator. 

To develop the analogy between condensed matter physics and curved-spacetime quantum field theory, see Eq.~\refb{Eq:KGE}, it is necessary to combine Eq.~\refb{Eq:on_dynamics} and Eq.~\refb{Eq:otheta_dynamics} into a single equation for the phase fluctuations. For that it is important to note that we can rearrange \refb{Eq:otheta_dynamics} to make $\hat n$ the subject. That is
\begin{equation}
\label{Eq:n_of_theta} 
\hat n = \hbar \, \widetilde U^{-1} \, \left[ \partial_t \hat \theta+\mathbf v\cdot\nabla\hat \theta \right] = \hbar \, \widetilde U^{-1} \,
\frac{D \hat \theta}{Dt} \, ,
\end{equation}  
where $D\hat \theta/Dt$ is the rate of change of $\hat \theta$ following a
small volume of the fluid, the \emph{material derivative} of $\hat \theta$.
If the fluid is at rest, the material derivative reduces to
$D\hat \theta/Dt\rightarrow \partial_{t} \hat \theta$, but for a moving fluid it
represents a ``fluid-following'' derivative.
The integral differential operator $\widetilde U^{-1}$ can \emph{formally} be expanded as
\begin{eqnarray}
\label{Eq:Integral.Differential.Operator}
\widetilde U^{-1} = U^{-1} &+& \left[\frac{\hbar}{2m}\right] \, U^{-1} \widetilde D_{2} \, U^{-1}  \\ \nonumber
 &+& \left[\frac{\hbar}{2m}\right]^{2} \, U^{-1} \widetilde D_{2} \,  U^{-1} \widetilde D_{2} \, U^{-1} \\ \nonumber
&+& \left[\frac{\hbar}{2m}\right]^{3} \, U^{-1} \widetilde D_{2} \, U^{-1} \widetilde D_{2} \,  U^{-1} \widetilde D_{2} \, U^{-1} + ...
 \end{eqnarray}
where the \emph{formal} series converges only on the subspace of functions spanned by the eigenfunctions whose eigenvalues satisfy
\begin{equation}
\label{Eq:Convergence.Radius}
\lambda \left( \frac{\hbar^{2}}{2m} U^{-1} \widetilde D_{2} \right) < 1 \, .
\end{equation}
Since $\widetilde D_{2}$ and $\widetilde U$ are second-order linear differential operators, the inverse $\widetilde U^{-1}$ always exists as an integral operator (that is, in the sense of being a Green function). Expanding this Green function as in Eq.~\refb{Eq:Integral.Differential.Operator} above is a convenience that allows us to interpret $\widetilde U^{-1}$ as a sum of differential operators, but this is not a fundamental limitation on the formalism. 
%
\subsubsection{Acoustic and rainbow metrics \label{Sec:Rainbow.Metrics}}
We use equation \refb{Eq:n_of_theta} to eliminate $\hat n$ completely in our equations of motions, by substituting it into equation \refb{Eq:on_dynamics}. That yields a single equation for the perturbed phase
\begin{equation}
\label{Eq:WaveEquation}
\partial_{a} \left(f^{ab}\; \partial_{b} \hat \theta \right) = 0 \, ,
\end{equation}
where we have introduced the matrix $f^{ab}$ with inverse-differential-operator-valued entries:
\begin{equation}
\label{Eq:f}
f^{ab} = \hbar
\left[ \begin{array}{c|c} \vphantom{\Big|} -\widetilde U^{-1} & -\widetilde U^{-1} v^{j} \\ \hline \vphantom{\Big|} -v^{i} \widetilde U^{-1} & \frac{n_{0}}{m}\delta^{ij}- v^i \widetilde U^{-1} v^j \end{array}\right] \, .
\end{equation}
Note that in general $\widetilde U^{-1}$ is an integral operator so Eq.~\refb{Eq:WaveEquation} is an integro-differential equation.
If we additionally require that there
exists an (inverse) metric tensor $g_{ab}$ such that
\begin{equation}
\label{Eq:EmergentGeometry}
f^{ab} \equiv \sqrt{-g} \, g^{ab} \, ,
\end{equation}
where $g$ is the determinant of the metric tensor $g_{ab}$, then the
connection is formally made to the field equation for a minimally
coupled massless scalar field in a curved spacetime; see Eq.~\refb{Eq:KGE}.
Now in many situations of physical interest, the differential operator $\widetilde U$ can be
usefully approximated by a function --- for instance the hydrodynamic and eikonal limits.\\
%

\paragraph{Hydrodynamic approximation:}
In the hydrodynamic limit where the \emph{quantum pressure} is neglected one
has 
\begin{equation}
\label{Eq:Hydrodynamic.Limit}
\vert U \, n_{0} \vert  \gg \vert (\hbar^{2}/2m) \, \widetilde D_{2} \, n_{0} \vert 
\end{equation}
so that 
\begin{equation}
\widetilde U \to U  = {4\pi\hbar^2 a_{\mathrm{scatt}}(t)\over m}. 
\end{equation}
The scattering length $a_{\mathrm{scatt}}(t)$ represents the $s$-wave scattering term. (See, for instance, \cite{Barcelo:2001gt,Barcelo:2005ln}, and the discussion above.)  \\

Within this approximation we obtain
\begin{equation}
\label{f1}
f^{ab} =\frac{\hbar}{U} 
\left[ \begin{array}{c|c} \vphantom{\Big|} -1 & - v^{j} \\ \hline \vphantom{\Big|} -v^{i}  & \frac{n_{0} U}{m}\delta^{ij}- v^i  v^j \end{array}\right]
\, .
\end{equation} 
To mimic curved spacetime classical and quantum field theory effects the acoustic metric has to be formally in agreement with Einstein's theory of gravity.
For this low-momentum approximation we are able to define a common (for all wavelengths $k$) speed of sound in the condensate:
\begin{equation} 
\label{Eq:SoundSpeed}
c_k^{2} \rightarrow c^{2}= \frac{n_{0} U(t)}{m} \, .
\end{equation}
Applying Eq.~\refb{Eq:EmergentGeometry} we obtain,
\begin{equation}
\label{Eq:Acoustic.Metric} 
g_{a b} \equiv \left(
\frac{n_{0} \, \hbar}{c \, m } \right)^{\frac{2}{d-1}} \,
\left[ \begin{array}{c|c} -(c^{2}-v^{2}) & -v^j \\ \hline -v^i & \delta^{ij} \end{array}\right] \, ,
\end{equation}
as the \emph{acoustic metric} or \emph{analogue metric}.
In this case all collective excitations behave as
sound waves with the usual linear dispersion form 
\begin{equation}
\omega_{0} = c \, k
\end{equation}
and the quanta of excitations are thus phonons. \\

\paragraph{Eikonal approximation:}
%
Let us now invoke a different approximation that holds for trans-phononic modes, so that the
elements of the matrix $f^{ab}$ can be treated as (possibly momentum
dependent) functions, rather than differential operators:
Consider the \emph{eikonal limit} where $\widetilde U$ can usefully be approximated by a function
\begin{equation} \label{Eq:U.eikonal}
\widetilde U \to U_{k}(t,\mathbf x) = U(t,\mathbf x) + \frac{\hbar^2 k^2}{ 4 m n_0} \, , 
\end{equation}
which we shall conveniently abbreviate by writing $U_k$.
Beyond the hydrodynamic limit we obtain
\begin{equation}
\label{f2}
f^{ab} = \frac{\hbar}{U_k} 
\left[ \begin{array}{c|c} \vphantom{\Big|} -1 & - v^{j} \\ \hline \vphantom{\Big|} -v^{i}  & \frac{n_{0} U_k}{m}\delta^{ij}- v^i  v^j \end{array}\right]
\, .
\end{equation} 
Note that in the eikonal approximation the $k$ dependence hiding in $U_k$ will make this a momentum-dependent metric, a so-called \emph{rainbow metric}. The metric tensor is explicitly given by
\begin{equation}
\label{Eq:Rainbow.Metric}
g_{a b} \equiv
\left( \frac{n_{0}\, \hbar}{c_k \, m} \right)^{\frac{2}{d-1}} \,
\left[ \begin{array}{c|c} -(c_k^{2}-v^{2}) & -v^j \\ \hline -v^i & \delta^{ij} \end{array}\right] \, ,
\end{equation}
where we have introduced the quantity
\begin{equation}
\label{Eq:Eikonal.Sound.Speed}
c_{k}(t)^{2}=c(t)^{2}+\epsilon_{\mathrm{qp}}^{2} k^{2} \, ,
\end{equation}
for a uniform condensate density $n_{0}(t,\mathbf x)=n_{0}$.
It is convenient to define 
\begin{equation}
\label{qp}
\epsilon_{\mathrm{qp}}=\frac{\hbar}{2m} \, ,
\end{equation}
where $\epsilon_{\mathrm{qp}} \ll \vert c(t)/k \vert$ is a useful indication for the hydrodynamic limit; compare with Eq.~\refb{Eq:Hydrodynamic.Limit}. The dispersion relation in the eikonal limit is
\begin{equation}
\label{Eq:Omega.NonLinear}
\omega_{k}(t)=c_{k}(t)\, k=\sqrt{c(t)^{2}k^{2} + \epsilon_{\mathrm{qp}}^{2}k^{4}} \, ,
\end{equation}
and hence violates ``\emph{acoustic Lorentz invariance}''.  This is not surprising at all, since we know the quasi-particles become ``atom-like'', and so \emph{non-relativistic}, at high momentum~\cite{Visser:2001ix}. \\

\paragraph{Acoustic or rainbow spacetimes?}
%
It is only in the acoustic/hydrodynamic limit that the $k$ dependence of $U_k$ and $c_{k}$ vanish, so that the rainbow metric is reduced to an ordinary Lorentzian metric.

An interesting consequence of the Bogoliubov theory in Bose condensates is that in general the excitation spectrum displays nonlinear dispersion (see Eq.~\refb{Eq:Omega.NonLinear}),
being linear (\ie, phononic) for low $|\mathbf{k}|$ and becoming quadratic (\ie, free-particle like) at
large $|\mathbf{k}|$. When the nonlinear dispersion \refb{Eq:Omega.NonLinear} is incorporated into analogue models of gravity it is equivalent to breaking Lorentz invariance
\cite{Barcelo:2001gt,Barcelo:2005ln,Weinfurtner:2006iv,Liberati:2006kw,Liberati:2006sj,Weinfurtner:2006nl}.  

The hydrodynamic approximation is a statement about the \emph{smallness} of the quartic term in the dispersion relation for $\omega_{k}^{2}$~\refb{Eq:Omega.NonLinear}, with respect to the quadratic term.
The usual line of argument is that the smallness of $\epsilon_{\mathrm{qp}}$ makes it possible to neglect the second order in Eq.~\refb{Eq:Omega.NonLinear} for low-energy excitations, where $\epsilon_{\mathrm{qp}} \vert k \vert \ll c(t)$. Here Lorentz invariance is an emergent symmetry.
However, it is important to realise that the propagation speed can be a function of time, $c=c(t)$, and hence if $c(t)\to 0$ then one is dealing with a system that eventually violates Lorentz invariance at \emph{all} energy scales. There are no theoretical or experimental restrictions to prevent $U(t)\propto c(t)$ becoming arbitrarily small.  %

In the specific cases we are interested in, we are confronted with exactly this situation, and therefore a more subtle analysis is required as to whether the acoustic metric \refb{Eq:Acoustic.Metric} is a sufficient approximation, or whether we have to use the more sophisticated concept of a rainbow spacetime \refb{Eq:Rainbow.Metric}. We will elaborate on the point in Section~\ref{Sec:Inlation.FRW.Rainbow.Spacetimes}.\\
%

\paragraph{Limitations of the rainbow analogy?}
%
It is worth reiterating the assumptions used in making the above analogy.  The theory resulting from linearized quantum fluctuations leads to a \emph{free field theory} --- that is, the modes of the field $\hat \theta$ are non-interacting.  Since we are using a linearized theory, the interactions between quantum fluctuations themselves, and between quantum fluctuations  and the condensate mode are neglected. This is equivalent to assuming the metric tensor is an externally-specified classical quantity. 

Our numerical simulations carried out in \cite{Jain:2006ki} do not require these assumptions; there all modes of the system are included and these modes are able to interact via the nonlinear interaction term.  Thus
we were able to explore the validity of the assumptions of the free-field theory~\cite{Schutzhold:2005wt}.

We further note that while the present form of the analogy only holds for massless scalar (spin zero) quasi-particles, in general it is possible to modify the formalism to include massive minimally coupled scalar fields at the expense of dealing with more complex BEC configurations, \eg, a two-component BEC 
\cite{Weinfurtner:2006nl,Weinfurtner:2006eq,Liberati:2006kw,Liberati:2006sj,Weinfurtner:2006eq,Weinfurtner:2006iv,Visser:2005ai,Visser:2004zn}. In BEC language one would explain this situation in terms of a dispersion relation with a gap.
%
\subsubsection{Commutation relations \label{Sec:Commutation.Relations}}
To derive the analogy presented above, we approximated and transformed our field operators several times, see Chapter~\ref{Chap:Early.Quantum.Cosmology} and~\cite{Jain:2006ki}:
\begin{enumerate}
\item{$\hat\psi(t,\mathbf x)$ and $\hat\psi^{\dagger}(t,\mathbf x)$: The single Boson annihilation and creation operators; where
\begin{eqnarray} 
&&[\hat{\psi} (t,\mathbf{x}) , \hat{\psi} (t,\mathbf x')] =0 \, , \\
&&[\hat\psi^{\dag}(t,\mathbf{x}), \hat{\psi^{\dag}}(t,\mathbf x')] = 0 \, , \\
&&[\hat\psi^{\dag}(t,\mathbf{x}), \hat{\psi^{\dag}}(t,\mathbf x')] = \delta(\mathbf x  - \mathbf x') \, ;
\end{eqnarray}

}
\item{$\delta\hat\psi(t,\mathbf x)$ and $\delta\hat\psi^{\dagger}(t,\mathbf x)$: Decomposition into a single coherent mode $\psi(t,\mathbf x) = \langle \hat{\psi}(t,\mathbf x)\rangle$, and the quantum excitations $\delta\hat\psi(t,\mathbf x)$ around it. Altogether,
$\hat\psi(t,\mathbf x) = \psi(t,\mathbf x) + \delta\hat\psi(t,\mathbf x)$ and 
$\hat\psi^{\dag}(t,\mathbf x) = \psi^{*}(t,\mathbf x) + \delta\hat\psi^{\dag}(t,\mathbf x)$, where
\begin{eqnarray} 
&&[\delta\hat{\psi} (t,\mathbf{x}) , \delta\hat{\psi} (t,\mathbf x')] =0 \, , \\
&&[\delta\hat\psi^{\dag}(t,\mathbf{x}), \delta\hat{\psi^{\dag}}(t,\mathbf x')] = 0 \, \\
&&[\delta\hat\psi^{\dag}(t,\mathbf{x}), \delta\hat{\psi^{\dag}}(t,\mathbf x')] = \delta(\mathbf x  - \mathbf x') \, ;
\end{eqnarray}
} 
\item{$\hat n$ and $\hat\theta$: Mapping onto Hermitian phase and density fluctuation operators, as studied (for example) in \cite{Barcelo:2003ia,Barcelo:2003yk}.  Here we made use of the fact that the macroscopic
field $\psi(\mathbf x)$ is complex and so for topologically trivial regions ---
without zeros or singularities --- one can always express it as
$\psi(t,\mathbf x) = \sqrt{n(t,\mathbf x) } \; \exp(i\theta(t,\mathbf x) )$. Linearizing around the two parameters of the complex-valued field, $\theta \to \theta_{0} + \hat\theta$ and $n \to n_{0} + \hat n$, we can write
$\hat\psi \simeq \psi + \sqrt{n_{0}} \left( \frac{\hat n}{2 n_{0}} + i \, \hat\theta  \right)$, and its Hermitian conjugate, such that
\begin{eqnarray} 
&&\left[\hat n(t,\mathbf x),\hat n(t,\mathbf x')\right]=0 \, , \\
&&\left[ \hat\theta(t,\mathbf x),\hat\theta(t,\mathbf x')\right]=0 \, , \\
&&\left[\hat n(t,\mathbf x),\hat\theta(t,\mathbf x^{\prime})\right]=i\,\delta(\mathbf x-\mathbf x^{\prime}) \, .
\end{eqnarray}
}
\item{Finally, we are able to use the equation of motion \refb{Eq:n_of_theta} to formally express $\hat n$ in terms of $\hat \theta$:
\begin{eqnarray} 
&&\left[\frac{1}{\widetilde U} \, \frac{D\hat\theta(t,\mathbf x)}{Dt} ,\frac{1}{\widetilde U} \, \frac{D\hat\theta(t,\mathbf x')}{Dt} \right]=0 \, , \\
&&\left[ \hat\theta(t,\mathbf x),\hat\theta(t,\mathbf x')\right]=0 \, , \\
&&\left[ \frac{1}{\widetilde U} \, \frac{D\hat\theta(t,\mathbf x)}{Dt} ,\hat\theta(t,\mathbf x^{\prime})\right]=-\frac{i}{\hbar}\,\delta(\mathbf x-\mathbf x^{\prime}) \, ;
\end{eqnarray}
}
\end{enumerate}
There are two different ways to view the resulting commutator relations in the hydrodynamic and eikonal limit: A condensed matter point of view in terms of time-dependent commutators, or a commutator relationship for the phase perturbation $\hat\theta$ and its conjugate momentum $\hat\Pi_{\hat\theta}$ on the emergent spacetime. \\

\paragraph{Condensed matter point of view: \label{Sec:Condensed.Matter.Commutators}}
First we face the problem of how to deal with the differential operator $\widetilde D_{2}$, which involves studying two interesting limits where the commutation relation takes a simpler form. \\

In the \emph{hydrodynamic approximation} we get 
\begin{equation}
\left[ \frac{D\hat\theta(t,\mathbf x)}{Dt},\hat\theta(t,\mathbf x^{\prime})\right]=\frac{U(t)}{i\hbar}\,\delta(\mathbf x-\mathbf x^{\prime}),
\end{equation}
which is now a time-dependent commutation relation. \\

For $U(t)=0$ the
hydrodynamic commutator vanishes completely and we are left with
purely classical statements for $\hat\theta$.  This situation changes
significantly if one instead considers the eikonal approximation. \\

In the \emph{eikonal approximation} we get (in momentum space) 
\begin{equation}
\left[ \frac{D\hat\theta(t,\mathbf{k})}{Dt},\hat\theta(t,\mathbf{k}^{\prime})\right]
=\frac{U(t)+{\hbar^{2} k^2 \over{4m n_{0}}}}{i\hbar}\,\delta_{kk^{\prime}} \, ,
\end{equation}
and the commutator does not vanish for $U(t)\rightarrow 0$, though it can
vanish if $U(t)$ is driven negative.  \\

This suggests that the
presence of $\widetilde D_{2}$ cannot in general be neglected for a time-dependent atomic interaction $U(t)$. \\
%

\paragraph{Emergent spacetime point of view: \label{Sec:Emergent.Spacetime.Commutators}}
An alternative insight can be gained if we define an emergent Lagrange density,
\begin{equation} \label{Eq:emergent.Lagrange.density}
\mathcal{L}=-\frac{1}{2} f^{ab} \; \partial_{a}\hat\theta \, \partial_{b}\hat\theta\, ,
\end{equation}
in correspondence with Eq.~\refb{Eq:KGE}. 
The momentum conjugate to $\hat \theta$ is given by
\begin{equation}
\hat\Pi_{\hat\theta} :=\frac{ \partial \mathcal{L}}{\partial (\partial_{t} \hat\theta)}=-f^{tb} \, \partial_{b}\hat\theta,
\end{equation}
and hence we evaluate the conjugate momentum to $\hat\theta$ as,
\begin{equation}
\hat\Pi_{\hat\theta}=\frac{\hbar}{\widetilde U} \, \frac{D\hat\theta}{Dt} \, .
\end{equation}
With this new insight we are able to add another set of commutation relations, one that makes only sense after having introduced the emergent spacetime:
\begin{itemize}
\item[5.]{The phase and density operators are a canonical set of quantum field operators and conjugate field operators,
\begin{eqnarray}
\label{Eq:C1_5}&& \left[ \hat \theta(t,\mathbf x),\hat\theta(t,\mathbf x')\right] = 0 \, , \\ 
\label{Eq:C2_5}&& \left[ \hat\Pi_{\hat\theta}(t,\mathbf x),\hat\Pi_{\hat\theta}(t,\mathbf x')\right] =0 \, ,\\ 
\label{Eq:C3_5}&& \left[ \hat \theta(t,\mathbf x),\hat\Pi_{\hat\theta}(t,\mathbf x')\right] = i \delta(\mathbf x - \mathbf x') \, ; 
\end{eqnarray}
in an effective curved spacetime represented by Eq.~\refb{Eq:Acoustic.Metric}, for a massless spin-zero scalar field.}
\end{itemize}

In the \emph{hydrodynamic limit} $\widetilde U \to U$ we recover
\begin{equation}
\label{Eq:Conjugate.Momentum.Acoustic.Limit}
\hat\Pi_{\hat\theta}=\frac{\hbar}{U} \, \frac{D\hat\theta}{Dt} \, ,
\end{equation}
the standard result for the conjugate momentum in curved spacetime; for more details see \cite{Birrell:1984aa,Parker:1969aa}. \\

The conjugate momentum \emph{for ultraviolet modes} $U \to U_{k}$ is given by
\begin{equation}
\label{Eq:Conjugate.Momentum.Eikonal.Limit}
\hat\Pi_{\hat\theta}=\frac{\hbar}{U_{k}} \, \frac{D\hat\theta}{Dt} \, ,
\end{equation}
as the conjugate momentum in our rainbow geometry.\\

This is a significant result, since it shows that knowledge of the emergent spacetime picture provides a deeper insight into the full dynamics for the density and phase perturbations, and explains the explicit time-dependence in their commutation relations.

\subsection[FRW-rainbow geometries]{FRW-rainbow geometries:\protect\\ 
Specific time-dependence for atom-atom scattering
\label{Sec:FRW.type.geometries}}
Clearly time dependence can enter in any of the parameters $n_{0}$, $c$, and $\mathbf v$. We now focus on the case where $\mathbf{v}= 0$, and $n_{0}$ is constant throughout space and time, so that the system is homogeneous. This choice of parameters leads to the specific class of $k=0$ spatially flat FRW spacetimes \cite{Barcelo:2003yk,Duine:2002aa}.  Such geometries are always conformally flat and at any particular time the spatial geometry is simply that of flat Euclidean space. \emph{All} the time dependence is contained entirely in the speed of sound given by Eq.~\refb{Eq:SoundSpeed}. In a homogeneous condensate the sound speed and the scale factor are position independent, and within the acoustic\,/\,eikonal limit a separation of the field operators into time and position dependent parts is possible. \\

We introduce the dimensionless scale function $b(t)$ so that the interaction strength (or equivalently the scattering length) becomes time-dependent, 
\begin{equation}
U(t) = U_{0} \, b(t) \, ,
\end{equation}
and under consideration of Eq.~\refb{Eq.Effective.U} we get
\begin{equation}
\label{Eq.Effective.U.time}
\widetilde U = U_{0} \, \widetilde b(t)   \, ,
\end{equation}
and a time-dependent differential scaling operator $\widetilde b(t)$;
\begin{equation}
\label{Eq.Effective.b.time}
\widetilde b(t) = b(t) - \frac{\hbar^2}{2 m \, U_{0}}\;\widetilde D_{2} \, .
\end{equation}
It is important to realize that a change in the interaction strength, $U \to U(t)$, inevitably involves a shift in the border between the phononic and trans-phononic regime, and hence the nature of  collective excitations in the condensate. To see this relationship more clearly we write down the eikonal approximation of Eq.~\refb{Eq.Effective.b.time},
\begin{equation}
\label{Eq.Effective.b.time.eikonal}
U_{k}(t) =U_{0} \; b_{k}(t) = U_{0} \; \left(b(t)+ (k/K)^{2} \right) \, ,
\end{equation} 
where for further convenience it is useful to introduce $K$, 
\begin{equation}
\label{Eq:K}
K = \frac{c_{0}}{\epsilon_{\mathrm{qp}}} \, .
\end{equation}
We would like to stress the importance of the last two equations for the understanding of everything that follows below. Motivated by the ultraviolet deviations in our results obtained by the numerical simulation of time-dependent spacetimes in a realistic Bose--Einstein condensate \cite{Jain:2006ki}, we set out to find a spacetime description that includes trans-phononic modes into the spacetime picture. This forced us to generalize acoustic metrics to momentum-dependent (rainbow) metrics, that cover a larger $k$-range. These rainbow metrics, see Eq.~\refb{f2}, show a time- and  momentum-dependent scaling in the effective scale function for the interaction strength. To show this we take $b(t_0)=1$ at some arbitrary initial time $t_0$. 
From the convergence constraint \refb{Eq:Convergence.Radius} for a uniform condensate at rest at $t=0$, and working in the eikonal approximation,
\begin{equation}
\label{Eq:Convergence.Constraint.Uniform.Condensate}
\left\vert k \right\vert< \left\vert K \right\vert \, .
\end{equation}
This implies that for modes $k>K$ the spacetime picture begins to break down. This strongly suggests that one should consider $1/K$ as the analogue Planck length,
$\ell_{\mathrm{Planck}} = 1/K$. For modes with wavelength $k \ll K$ we recover the ``standard'' geometry (\ie, momentum-independent spacetimes). These modes are phononic modes. Making use of our rainbow geometries we extended the emergent spacetime picture for higher energetic (ultraviolet) modes with wavelengths $0< k < K$. For rainbow metrics the equation of motion for modes with wavelengths $k>K$ are most easily derived in momentum space. This can immediately be seen when one Fourier transforms the initial quantum fluctuations $\delta \hat \psi$ and $\delta \hat \psi^{\dag}$, since then it is not necessary to introduce $g_{ab}$ at all.
The situation is more complicated if we involve time-dependent atomic interactions. At any later time $t>0$ we get a different convergence constraint,
\begin{equation}
\left\vert k \right\vert< \left\vert  \sqrt{b(t)} \, K \right\vert \, ,
\end{equation}
and with it a time-dependent limit on the breakdown of the validity of the emergent geometry, 
\begin{equation}
\label{Eq.Time.Dependent.Effective.Planck.Length}
\ell_{\mathrm{Planck}}(t) = 1/(K \sqrt{b(t)}) \, . 
\end{equation}
In other words, the analogue Planck volume $\ell_{\mathrm{Planck}}(t)^{d}$ is --- depending on the form of $b(t)$ --- shrinking or expanding. We will return to this point when we explain particle production in a de Sitter-like universe, with a growing Planck volume. Note that for a time dependent condensate density the problem is different, and has been studied in \cite{Fedichev:2004on,Fedichev:2004fi,Fischer:2004iy,Uhlmann:2005rx}.
\\

It should be noted that in the acoustic approximation the scale factor,
\begin{equation} 
\label{Eq:U.acoustic.time}
U(t) = U_0 \; b(t) \,
\end{equation}
is indeed momentum-independent as expected. In practice a variation in the interaction strength is possible by using a Feshbach resonance \cite{Vogels:1997aa,Inouye:1998aa,Barcelo:2003ia,Barcelo:2003yk}. \\

From Eq.~\refb{Eq:Eikonal.Sound.Speed} we obtain the time-dependent speed of sound in the eikonal approximation,
\begin{equation}
\label{Eq:Eiknoal.Sound.Speed.time}
c_{k}(t) =  c_0  \; \sqrt{b(t)+ k^{2}/K^{2}} \, ,
\end{equation}
and thus
\begin{equation}
\label{Eq:Omega.FRW}
\omega_{k}(t) = \omega_{0}  \; \sqrt{b(t)+ k^{2}/K^{2}} = \omega_{0}  \; \sqrt{b_{k}(t)} \, .
\end{equation} 
Their acoustic counterparts are obtained in the limit $b_{k}(t) \to b(t)$. \\

The equation of motion \refb{Eq:n_of_theta}, which depends on $f^{ab}$, therefore shows no explicit dependence on the spatial dimensions $d$ of the condensate. 
In contrast, the emergent spacetime is given by $g_{ab}$, see Eq.~\refb{Eq:Acoustic.Metric} and Eq.~\refb{Eq:Rainbow.Metric}, where both depend explicitly on $d$.
In order to use a Bose--Einstein condensate (in the infrared limit) as an analogue model for Einstein's theory of gravity, we have to compare the Friedmann--Roberton--Walker universe line element given in Eq.~\refb{Eq:FRW.GR}, with the line element $g_{ab}\,dx^{a}\,dx^{b}$;
\begin{equation} 
\label{Eq:FRW.Acoustic}
d s^2 =  \left( \frac{n_0}{c_0}\right)^{\frac{2}{d-1}}
\left[ -c_0^2 \, b_{k}(t)^{\alpha} \, dt^2 + b_{k}(t)^{\alpha- 1} \, d\mathbf x^2 \right] \, .
\end{equation}
Here the exponent $\alpha$ is dimension-dependent and given as
\begin{equation}
\label{Eq:Acoustic.Lineelement.Exponent}
\alpha = \frac{d-2}{d-1} \, .
\end{equation}
This explains why the choice of $b(t)$ --- to mimic a specific FRW-type universe scale factor $a(t)$ --- depends on the spatial dimensions $d$. 

%
%
\section[Quantum field theory in rainbow geometries]{Quantum field theory in rainbow geometries:\protect\\
Parametric excitations}
\label{Sec:Cosmo.Particle.Production}			
%
Before we address problems of interest to the programme of cosmological particle production in a specific FRW-type universe, we prepare the necessary mathematical and physical formalism. Quantum field theory in rainbow geometries requires a careful treatment of the equation of motion and the commutation relations for the quantum field operators. 
There are three major differences between ordinary massless spin-$0$ particles in real curved spacetimes and BEC quasi-particles in the particle production process. These are: (1) The numerical finiteness of the particle production in all circumstances due to the non-perturbative ultraviolet corrections, as we will show below. (2) In addition, we are taking the point of view that any Bose--Einstein condensate experiment will have a finite time duration, and therefore well-defined initial and final vacuum states. (3) Finally, there exists a preferred frame, the laboratory frame in which the experiment (or at this stage gedanken-experiment) is implemented. The laboratory time therefore has to play an important role, and can be viewed as the most relevant coordinate choice.  \\

Conditions (1) and (2) allow us to employ the method of instantaneous Hamiltonian diagonalization to adequately calculate the particle production process.
The condition (3), that our particle detector is bound to the laboratory-frame simplifies the decision as to which time coordinate to choose. However, this limitation is also very disappointing. A major awareness resulting from ``conventional'' curved-spacetime quantum field theory has been the observer dependence of the particle spectrum~\cite{Unruh:1976aa}: Different coordinatizations motivate different vacuum choices. One might stretch the good nature of any condensed matter experimentalist, asking for a co-moving particle detector --- adjusted to a suitable vacuum in the chosen coordinate system --- in an infinitely long-lasting expanding emergent spacetime. Any unambiguous measurement requires one to stop the expansion, and to project onto a positive and negative plane wave basis.

%
\subsection[Particle production]{Particle production\protect\\
Linearized quantum excitations in condensate  
 \label{Sec:ParticleProduction}}
The equation of motion for quantum fluctuations in such a $d$-dimensional condensate are given by
\begin{equation}
\partial_{t} \left( \frac{\hbar}{\widetilde U(t)} \, \partial_{t} \hat \theta(t,\mathbf{x}) \right) - \frac{n_{0}\hbar}{m} \nabla^{2} \hat\theta(t,\mathbf{x}) =0 \, ;
\end{equation}
compare with Eq.~\refb{f2}.
In the following we are going to rewrite the equation of motion in a more suitable form, applying an auxiliary field, and simultaneously transform to momentum space.

\subsubsection{Auxiliary field operators in Fourier space \label{Sec:Auxiliary.Field.Fourier.space}}
We use the differential operator $\widetilde R = \widetilde U(t) / \hbar$ to write the field operators in terms of auxiliary field operators $\hat \chi$, where $\hat\theta = (\widetilde R)^{1/2} \,\hat\chi$. As long as $\widetilde U=\widetilde U(t)$, and with it $\hat \theta=\hat \theta(t)$ and $\hat \chi=\hat \chi(t)$, is position independent, we can always write
\begin{eqnarray}
\nonumber
\hat\theta(t,\mathbf x)&=& \int \frac{d^{d}k}{(2\pi)^{d/2}}  \sqrt{\frac{\widetilde U(t)}{\hbar}} \,
\hat\chi_{k}(t) \, \e^{i \, \mathbf{k} \mathbf{x}} \, , \\
\nonumber
&=& \frac{1}{\sqrt{\hbar}} \int \frac{d^{d}k}{(2\pi)^{d/2}} \hat\chi_{k}(t) \, \sum_{s=0}^{\infty} \frac{(-1)^{s} (2s)!}{(1-2s)(s!)^{2} 4^{s}} \, (\widetilde D_{2})^{s}  \e^{i \, \mathbf{k} \mathbf{x}} \, , \\
&=& \sqrt{\frac{U_{k}(t)}{\hbar}} \, \int \frac{d^{d}k}{(2\pi)^{d/2}}  
\hat\chi_{k}(t) \, \e^{i \, \mathbf{k} \mathbf{x}} \, .
\end{eqnarray}
The last statement is naively based on a Taylor series which only converges for modes with wavenumbers $\vert k \vert < \vert K \vert$. Within this radius of convergence the transformation is exact, but can be extended to arbitrary $k$-values in the eikonal approximation, where $\widetilde U(t) \to U_{k}(t)$.
The equation of motion for the mode operators $\hat \chi_{k}$,
\begin{equation}
\label{Eq:Equation.Motion.Chi}
\sqrt{\frac{\hbar}{U_{k}(t)}} \,
\left(
\ddot{\hat{\chi}}_{k}(t) +\Omega_{k}(t)^{2} \, \hat\chi_{k}(t) \right) = 0
\, , 
\end{equation}
and the equal time commutation relations,
\begin{eqnarray}
\label{Eq:C1_6}&& \left[ \hat \chi_{k}(t),\hat\chi_{k'}(t)\right] = 0 \, , \\ 
\label{Eq:C2_6}&& \left[ \partial_{t}\hat \chi_{k}(t),\partial_{t}\hat\chi_{k'}(t)\right] =0 \, ,\\ 
\label{Eq:C3_6}&& \left[ \hat \chi_{k}(t),\partial_{t}\hat\chi_{k'}(t)\right] = i \delta_{k,k'} \, ; 
\end{eqnarray}
are now slightly more convenient than~\refb{Eq:C1_5}-\refb{Eq:C3_5}.
The function $\Omega_{k}(t)$ is defined as
\begin{equation}
\label{Eq:Harmonic.Oscillator.Frequency}
\Omega_{k}(t)^{2} = c_{0}^{2}k^{2}\,b_{k}(t) -\frac{3}{4} \left(\frac{\dot b_{k}(t)}{b(t)}\right)^{2} +\frac{1}{2}\frac{\ddot b_{k}(t)}{b_{k}(t)} \, , 
\end{equation}
and therefore we have the connection between quantum field theory in FRW-type spacetimes and a parametrically excited harmonic oscillator. The notation $\Omega_{k}(t)$ will become quite obvious for time-independent cases, where it reduces to the usual dispersion relation $\Omega_{k} \to \omega_{k}$;  see Eq.~\refb{Eq:Omega.FRW}.
Notice the overall factor, $\sqrt{\hbar/U_{k}(t)}$, in Eq.~\refb{Eq:Equation.Motion.Chi}. This is of importance for the particle production process in cases of discontinuous, and continuous (but not differentiable) changes in $U(t)$. We will revisit this issue shortly in Section~\ref{Sec:Bogoliubov.Transformation.Time.Dependent.Emergent.Spacetime}.
\subsubsection{Mode expansion\label{Sec:Mode.Expansion}}
The equation of motion \refb{Eq:Equation.Motion.Chi} is a homogeneous differential equation, which can be written as
\begin{equation}
\label{Eq:Equation.Motion.Chi.L}
\widetilde L \, \hat \chi_{k} = 0 \, ,
\end{equation}
where
\begin{equation}
\widetilde L = \partial_{t}^{2} + \Omega_{k}(t)^{2} 
\end{equation}
is a linear second order differential operator with a $2$-dimensional solution space. 

A common tool in quantum mechanics is to describe quantum states --- with unknown or variable number of quasi-particles --- with respect to an orthonormal occupancy number basis (\ie, the Fock space basis in the infinite-dimensional function space, the Hilbert space of state).
A Fock state is a quantum state consisting of an ensemble of excited non-interacting particles,
\begin{equation}
\vert n_{\mathbf k1}, n_{\mathbf k2}, ... \rangle = \frac{1}{\sqrt{n_{\mathbf k1}! \, n_{\mathbf k2}! \, ...}} \left[ (\hat a_{\mathbf k1}^{\dag})^{n_{\mathbf k1}}  (\hat a_{\mathbf k2}^{\dag})^{n_{\mathbf k2}} ... \right]  \vert 0 \rangle \, ,
\end{equation}
with definite occupation numbers $(n_{\mathbf k1},n_{\mathbf k2},...)$ in the modes $(\hat\chi_{\mathbf k1},\hat\chi_{\mathbf k2},...)$.
The creation, $\hat a_{k}^{\dag}$, and annihilation, $\hat a_{k}$ operators create or destroy a single-particle in the mode $\chi_{\mathbf k}$;
\begin{eqnarray}
\hat a_{\mathbf ki}^{\dag} \vert ..., n_{\mathbf ki}-1, ... \rangle &=& \sqrt{n_{\mathbf ki}} \, \vert ..., n_{\mathbf ki}, ... \rangle \, , \\
\hat a_{\mathbf ki} \vert ..., n_{\mathbf ki}, ... \rangle &=& \sqrt{n_{\mathbf ki}} \, \vert ..., n_{\mathbf ki}-1, ... \rangle \, .
\end{eqnarray}
The state $\vert 0 \rangle$ --- short for $\vert 0_{\mathbf k1},0_{\mathbf k2},... \rangle$ --- is a special state, the vacuum state. It is defined as the eigenstate of all annihilation operators $\hat a_{k}$ with eigenvalue $0$, such that $\hat a_{k} \vert 0 \rangle = 0$. Thus any arbitrary quantum state $\vert \varphi \rangle $ in the Fock spaces is a linear combination of all excited states,
\begin{equation}
\vert \varphi \rangle = \sum_{n_{\mathbf k1}, n_{\mathbf k2}, ...} P_{{n_{\mathbf k1}, n_{\mathbf k2}, ...}} \, \vert n_{\mathbf k1}, n_{\mathbf k2}, ... \rangle \, ,
\end{equation}
that can be created out of the vacuum. Here $P_{{n_{\mathbf k1}, n_{\mathbf k2}, ...}} $ is the probability to measure the single Fock state with the mode occupation $(n_{\mathbf k1}, n_{\mathbf k2}, ...)$.

Within this framework we expand the mode operators in terms of destruction and creation operators,
\begin{equation}
\label{Eq:Mode.Expansion.1}
\hat \chi_{k}(t) = \frac{1}{\sqrt{2}} \,
\left[ v_{k}^{*}(t) \, \hat a_{k} + v_{k}(t) \, \hat a_{-k}^{\dag}\right] \, .
\end{equation}
The coefficients $v_{k}$ and $v_{k}^{*}$ are a set of linearly-independent mode functions and any linear combination is a complete solution of
\begin{equation}
\widetilde L v_{k}(t) = 0 \, .
\end{equation}
(Due to the isotropy of the $s$-wave scattering amplitude of the atomic interactions, the mode functions are also isotropic, \ie, $v_{\mathbf k} = v_{k}$ and $v_{\mathbf k}^{*}=v_{k}^{*}$. It therefore seems unlikely that we would need to extend the specific analogue set-up to mimic anisotropic expansion scenarios. The case is different for analogue models involving a changing condensate density, $n_{0}=n_{0}(t)$; where a non-uniform density expansion can easily be achieved, for example see \cite{Fedichev:2004on,Fedichev:2004fi,Fischer:2004iy,Uhlmann:2005rx}.)

In order to obtain a ``nice'' canonical set of operators, which obey the common equal time commutation relations (see Eq.~\refb{Eq:C1_5}-\refb{Eq:C2_5}),
\begin{eqnarray}
\label{Eq:C1_7}&& \Big[ \hat a_{k}^{},\hat a_{k'}^{}\Big] = 0 \, , \\ 
\label{Eq:C2_7}&& \left[ \hat a_{k}^{\dag},\hat a_{k'}^{\dag}\right] =0 \, ,\\ 
\label{Eq:C3_7}&& \left[ \hat a_{k},\hat a_{k'}^{\dag} \right] = \delta_{k,-k'} \, ,
\end{eqnarray}
the Wronskian $W$ of the mode functions has to be normalized as follows,
\begin{equation}
\label{Eq:Wronskian}
W[v_{k},v_{k}^{*}] = \dot v_{k} \, v_{k}^{*} - v_{k} \, \dot v_{k}^{*} = 2 i \,.
\end{equation}
It is easy to see that --- under the application of the equation of motion for the mode functions --- the Wronskian is always time-independent. 
Note, a spin-zero scalar particle is its own anti-particle, where $(\hat a_{k})^{\dag}=\hat a_{-k}^{\dag}$. We would also like to draw attention to the fact that with a particular choice of Fock space, we select a particular subspace of the Hilbert space of states (\ie, the subspace that can be created out of the vacuum state, determined by our choice of $\hat a_{k}$). This will be of further interest when we discuss the validity and limitations of the Bogoliubov transformation.
%
\subsubsection{Bogoliubov transformation\label{Sec:Bog.Trafo}}
In Eq.~\refb{Eq:Mode.Expansion.1} we can in principle pick some particular mode expansion,
$(\vec v_{k})^{\dag} \cdot \vec{A}_{k}$, with a specific set of mode functions $\vec v_{k}^{T}=(v_{k}(t),v_{k}^{*}(t))$, and set of mode operators $\vec{A}_{k}^{T}=(\hat a_{k},\hat a_{-k}^{\dag})$. This choice is not unique and any other mode expansion $(\vec u_{k})^{\dag} \cdot \vec{B}_{k}$, with mode functions $\vec u_{k}^{T}=(u_{k}(t),u_{k}^{*}(t))$, and mode operators $\vec{B}_{k}^{T}=(\hat b_{k},\hat b_{-k}^{\dag})$ would have been possible. The relation between these two different representations is referred to as Bogoliubov transformation.

As mentioned above, the mode expansion $(\vec v_{k})^{\dag} \cdot \vec{A}$ corresponds to an orthonormal basis in the infinite-dimensional function space, provided the $\vec v_{k}$ are normalized; that is $W[v_{k}(t),v_{k}^{*}(t)]=2i$ for all times $t$. If we restrict ourselves to exclusively mapping between orthonormal frames --- then the mode functions fulfill the normalization constraint $W[u_{k}(t),u_{k}^{*}(t)]=2i$, and the mode operators obey the commutation relations given in Eqs.~(\ref{Eq:C1_7}-\ref{Eq:C3_7}) ---
the Bogoliubov transformation is given by $\vec{v}_{k} = M \cdot \vec{u}_{k}$, where $M$ must be a $2\times 2$ matrix,
\begin{equation}
\label{Eq:M.Transformation.Matrix}
M =
\left(
\begin{array}{cc}
\alpha_{k}^{*} & \beta_{k}^{*} \\ \beta_{k} & \alpha_{k}
\end{array}
\right) \, .
\end{equation}
The complex-valued coefficients $\alpha_{k}$ and $\beta_{k}$ are called Bogoliubov coefficients, and the transformation preserves the normalization condition if $\det(M)=1$, that is for
\begin{equation}
\label{Eq:Bogoliubov.Normalization.Alpha.Beta}
\vert \alpha_{k} \vert^{2} - \vert \beta_{k} \vert^{2} = 1 \, .
\end{equation}
At this stage we notice a minor technical subtlety when we look at summing over discreet $k$ modes versus integrating over continuous $k$ modes. Equation \refb{Eq:Bogoliubov.Normalization.Alpha.Beta} is appropriate for summing over discreet modes whereas for continuous modes we should strictly speaking use $\vert \alpha_{k} \vert^{2} - \vert \beta_{k}\vert^{2} =\delta^{(d)}(\vec{0})=V/(2\pi)^{d}$. Here $\delta^{(d)}(\vec{0})$ is the $d$-dimensional momentum space Dirac function. We will work with continuous integrals in momentum space but will suppress unnecessary occurrence of the volume V, where including it would lead to unnecessary clutter that might not aid understanding. \\

The relationship between the ``old'' and ``new'' mode operators, where $(\vec v_{k})^{\dag} \cdot \vec{A}=(\vec u_{k})^{\dag} \cdot \vec{B}$ and $(\vec v_{k})^{\dag} = (\vec u_{k})^{\dag} \cdot M^{\dag}$, is given as $\vec{A}_{k}=(M^{\dag})^{-1} \cdot \vec{B}_{k}$ and $\vec{B}_{k}=(M^{\dag}) \cdot \vec{A}_{k}$.
Here
\begin{equation}
M^{\dag} =
\left(
\begin{array}{cc}
\alpha_{k} & \beta_{k}^{*} \\ \beta_{k} & \alpha_{k}^{*}
\end{array}
\right)  
~~\mbox{and}~~ 
(M^{\dag})^{-1} =
\left(
\begin{array}{cc}
\alpha_{k}^{*} & -\beta_{k}^{*} \\ -\beta_{k} & \alpha_{k}
\end{array}
\right)  \, ,
\end{equation}
and therefore
\begin{eqnarray}
\label{Eq:operator.ak}
\hat a_{k} &=& \alpha_{k}^{*} \hat b_{k} - \beta_{k}^{*} \hat b_{-k}^{\dag},~~\mbox{and}~~
\hat a_{-k}^{\dag} = \alpha_{k}\hat b_{-k}^{\dag} - \beta_{k} \hat b_{k}^{\dag}  \,  ;  \; \;\\
\label{Eq:operator.bk}
\hat b_{k} &=& \alpha_{k} \hat a_{k}       + \beta_{k}^{*} \hat a_{-k}^{\dag},~~\mbox{and}~~
\hat b_{-k}^{\dag} = \alpha_{k}^{*}\hat a_{-k}^{\dag} - \beta_{k} \hat a_{k}^{\dag}  \, . \; \;
\end{eqnarray}
Obviously, the two vacuum states, $\hat a_{k} \, \vert {}_{(a)} 0 \rangle = 0$ and $\hat b_{k} \, \vert {}_{(b)} 0 \rangle = 0$, are different, as $\hat b_{k} \, \vert {}_{(a)} 0 \rangle \neq 0$ for $\beta_{k}\neq 0$. 

\subsubsection{Particles and lowest energy eigenstate\label{Sec:Meaning.Particles}}
\emph{Quasi-particles} are excited states, and therefore they depend on the choice of (mode functions and the associated) vacuum state. 
Under the application of Eq.~\refb{Eq:operator.bk} we can \emph{formally} calculate the mean density of $b$-particles, $\hat n^{(b)}$, the occupation number for the mode $\hat \chi_{k}$, with respect to the $a$-vacuum $\vert {}_{(a)} 0 \rangle$ as follows:
\begin{equation}
\label{Eq:Occupation.Number.Desity}
n_{k}^{(b)} = \frac{\langle {}_{(a)} 0 \vert \, \hat b_{k}^{\dag} \hat b_{k} \, \vert {}_{(a)}0 \rangle}{\mbox{Volume}} = \vert \beta_{k} \vert^{2}  \,.
\end{equation}
The challenge is to find particular specified mode functions that represent physically meaningful vacuum states, that minimize the expectation value of the Hamiltonian. Only then does the chosen vacuum correspond to the ``actual'' physical vacuum, and its excitations describe ``real'' quasi-particles. (That is, we are looking for an unambiguous measure of particles relative to some vacuum state.)
The problem of finding the ``right'' vacuum state (``best'' orthonormal frame) has been controversial from the outset. However, as we shall see, it is straightforward to find the zero-particle vacuum state for Minkowski spacetimes, and the whole controversy arises only for time-dependent cases, where the Hamiltonian $\hat H(t)=\int_{\mathbf k} \hat{\mathcal{H}}_{k}(t)$ is explicitly time-dependent;
\begin{eqnarray}
\nonumber
\hat{\mathcal{H}}_{k}(t) &=&  \frac{   
\{ [\partial_{t} ( \sqrt{U_{k}(t)} \hat \chi_{k}(t) ) ] 
[\partial_{t} ( \sqrt{U_{k}(t)} \hat \chi_{-k}(t) ) ]\}
}{2 \, U_{k}(t)} \\
\label{Eq:Hamiltonian.Density}
&\;& +\frac{1}{2}\, \omega_{k}(t)^{2} \, \chi_{k}(t) \, \chi_{-k}(t) \, .
\end{eqnarray}
In combination with the mode expansion \refb{Eq:Mode.Expansion.1} the Hamiltonian in momentum space is given by
\begin{eqnarray}
\nonumber
\hat H(t) &=& \frac{1}{4} \, \int d^{d}\mathbf k \, F_{k}(t) \, \hat a_{k}^{\dag} \hat a_{-k} + F_{k}^{*}(t) \, \hat a_{k} \hat a_{-k}\\ 
\label{Eq:Hamiltonian.Mode.Expansion}
&\;& \quad \quad \quad \quad  + E_{k}(t) \, (2 \hat a_{k}^{\dag} \hat a_{k} + \delta^{(d)}(0) ) \, ,
\end{eqnarray}
where the factor $E_{k}(t)$ for the diagonal terms is,
\begin{eqnarray}
\nonumber
E_{k}(t) &=& \vert \dot{v}_{k}(t) \vert^{2} + \{ \omega_{k}^{2}(t) + 1/4 \, [ \dot U_{k}(t)/U_{k}(t)]^{2} \} \, \vert v_{k} \vert^{2} \\  
\label{Eq:Ekt}
&+& 1/2 \,  \dot U_{k}(t)/U_{k}(t)  \, \partial_{t} \vert v_{k}(t) \vert^{2} \, ,
\end{eqnarray}
and the off-diagonal term $F_{k}(t)$ is given by
\begin{eqnarray}
\nonumber
F_{k}(t) &=& ( \dot{v}_{k}(t) )^{2} + \{ \omega_{k}^{2}(t) + 1/4 \, [ \dot U_{k}(t)/U_{k}(t)]^{2} \} \, (v_{k})^{2} \\  
\label{Eq:Fkt}
&+& 1/2 \,  \dot U_{k}(t)/U_{k}(t)  \, \partial_{t} (v_{k}(t))^{2}  \, .
\end{eqnarray}
Therefore there are no time-independent eigenstates that represent the physical vacuum state, one that minimizes the expectation value for the Hamiltonian,
\begin{equation}
\langle {}_{(t)} 0 \vert \, \hat H_{k}(t) \, \vert {}_{(t)} 0 \rangle = \frac{1}{4} \, \int d^{d}\mathbf{k} \, E_{k}(t)  \, ,
\end{equation}
for all times.

Yet somehow, in order to obtain a meaningful statement about the particle production in our effective curved spacetime, we need to deal with or circumvent this problem. \\

\paragraph{Instantaneous Hamiltonian diagonalization:
\label{Sec:Instantaneous.Hamiltonian.Diagonalization}}
One possibility to find an approximate vacuum at an instant of time --- say $t=t_{0}$ ---  is to define the vacuum state $\vert {}_{(t_{0})} 0 \rangle$ of the \emph{instantaneous} Hamiltonian $\hat{H(t_{0})\,}$, where we project our mode functions instantaneously onto a plane-wave basis with the dispersion relation $\omega_{k}(t_{0})$. Within this approximation we find for the coefficients in Eq.~\refb{Eq:Hamiltonian.Mode.Expansion}:
\begin{eqnarray}
\label{Eq:Ekt.Minkowski}
E_{k} &=&  \vert \dot{v}_{k}(t) \vert^{2} +  \omega_{k}^{2}(t_{0})  \, \vert v_{k} \vert^{2}  \, ; \\
F_{k} &=&  ( \dot{v}_{k}(t) )^{2} + \omega_{k}^{2}(t_{0})  \, (v_{k})^{2} \, .
\end{eqnarray}
It can be shown that the expectation value for the instantaneous Hamiltonian --- up to an arbitrary phase --- at $t=t_{0}$ is minimal for mode functions $v_{k} = 1/\sqrt{\omega_{k}(t_{0})}$ and $\dot v_{k}=i \, \sqrt{\omega}=i\, \omega_{k} \, v_{k}(t_{0})$. 

Note that at $t_{0}$ the two coefficients $F_{k}(t_{0})=0$ and $E_{k}(t_{0})=2\omega_{k}(t_{0})$, and that under this condition the instantaneous Hamiltonian (see Eq.~\refb{Eq:Hamiltonian.Mode.Expansion}) is diagonal. Therefore it is referred to as the \emph{vacuum state of instantaneous Hamiltonian diagonalization}.

In a time-dependent problem the vacuum state changes, and therefore one has to give up on this particular definition of vacuum for times $t>t_{0}$. The association between the $\vert \beta_{k} \vert^{2}$ Bogoliubov coefficients (as outlined above) and physical quasi-particles requires that we approach --- either asymptotically or abruptly  --- flat initial and final state spacetimes. Only a time-independent spacetime perpetuates its vacuum state, where $\omega_{k}(t) = \omega_{k}$, and thus the normalized mode functions
\begin{eqnarray}
\label{Eq:Mode.Functions.Minkowski.v}
v_{k}(t) &=& \frac{1}{\sqrt{\omega_{k}(t)}} \, \exp(i\,\omega_{k} \, t) \, , \\
\label{Eq:Mode.Functions.Minkowski.vc}
v_{k}^{*}(t) &=& \frac{1}{\sqrt{\omega_{k}(t)}} \, \exp(-i\,\omega_{k} \, t) \, ,
\end{eqnarray}
in this case represent a physically meaningful vacuum state for all times $t$.  
%
\subsubsection{Bogoliubov transformation in emergent spacetimes \label{Sec:Bogoliubov.Tranformation.Emergent.Spacetimes}}
The Bogoliubov transformation outlined in Section~\ref{Sec:Bog.Trafo} can be applied to time-dependent problems, as long as we do not fool ourself and make any statement about the presence of ``real'' qusi-particles unless the expansion has started from an initial time-independent spacetime, and eventually approaches a time-independent spacetime. We also require $\omega_{k}^{2}>0$, such that at the beginning and the end the mode functions are simple (normalized) in- and out-going plane waves. For a constant, but imaginary frequency, $\omega_{k}^{2}<0$, the instantaneous Hamiltonian is still diagonal, but its expectation value no longer has a minimum, see Chapter~\ref{Chap:Early.Quantum.Cosmology}.

To apply the Bogoliubov transformation (see, Section~\ref{Sec:Bog.Trafo}, and see Eqs.~(\ref{Eq:operator.ak}-\ref{Eq:operator.bk})) to our specific problem it is necessary to consider that the original problem was defined in terms of two linearly independent field variables, the field operator $\hat \theta$, and its conjugate momentum $\hat \Pi_{\hat \theta}$.
The connection conditions arise from the necessity that the field operator,
\begin{equation}
\label{Eq:C1}
\left[ \hat \theta \right] = \lim_{\epsilon \to 0} \left\{ \hat \theta(t-\epsilon)  -  \hat \theta(t+\epsilon) \right\}  = 0 \, ,
\end{equation}
and its conjugate momentum (on the emergent spacetime),
\begin{equation}
\label{Eq:C2}
\left[ \hat \Pi_{\hat\theta}  \right] =  \lim_{\epsilon \to 0} \left\{   \hat \Pi_{\hat\theta}(t-\epsilon) -  \hat \Pi_{\hat\theta}(t+\epsilon) \right\} = 0 \, ,
\end{equation}
have to be continuous at all times. (A more detailed treatment of this problem in the context of emergent spacetimes from Bose gases can be found in \cite{Weinfurtner:2007aa}.)
Note, that these two conditions have to be fulfilled for any arbitrary (albeit physically reasonable) change in the contact potential $U(t)$.
In Section~\ref{Sec:Auxiliary.Field.Fourier.space} we have already pointed out that our emergent spacetime is special in the sense that all time-dependence (in our model) has to be implemented via $U(t)$, which also shows up in the overall conformal factor in the equation of motion, see 
Eq.~\refb{Eq:Equation.Motion.Chi}. \\

\paragraph{Bogoliubov coefficients for a time-dependent emergent spacetime: 
\label{Sec:Bogoliubov.Transformation.Time.Dependent.Emergent.Spacetime}}
Let us consider a contact potential defined as follows,
\begin{equation}
U_{k}(t) =  g_{k}(t) \, \Theta_{\mathrm{HS}}(t-t_{0}) +  h_{k}(t) \, \Theta_{\mathrm{HS}}(t_{0}-t)\, .
\end{equation}
We are dealing with different sets of mode functions in each region; $v_{k}^{T}=(v_{k},v_{k}^{*})$ for $t < t_{0}$, and $u_{k}^{T}=(u_{k},u_{k}^{*})$ for $t > t_{0}$. (Here $\Theta_{\mathrm{HS}}$ is the symbol for the Heaviside step function.) For the time being we do not make any further assumption except that the mode functions have to be a solution of the equation of motion~\refb{Eq:Equation.Motion.Chi} in their respective region.
In particular, we do \emph{not yet} assume the modes are normalized. The connection conditions, given in Eqs.~\refb{Eq:C1} and \refb{Eq:C2}, in combination with the Bogoliubov transformation --- see transformation matrix \refb{Eq:M.Transformation.Matrix}, mapping between two pairs of complex-conjugate mode functions (it is always possible to choose such solutions) --- provide us with two matrix equations (that is, four component equations), 
\begin{eqnarray}
\sqrt{\frac{g_{k}}{\hbar}} \, \vec{v}_{k} &=& M \, \sqrt{\frac{h_{k}}{\hbar}} \, \vec{u}_{k} \, , \\
\frac{\hbar}{g_{k}} \,  \partial_{t}\left(\sqrt{\frac{g_{k}}{\hbar}} \, \vec{v}_{k}\right) &=& \frac{\hbar}{h_{k}} \,  M \, \partial_{t}\left( \sqrt{\frac{h_{k}}{\hbar}} \, \vec{u}_{k} \right) \, ,
\end{eqnarray}
for the four unknown transmission coefficients $\alpha_{k}$, $\alpha_{k}^{*}$, $\beta_{k}$, and $\beta_{k}^{*}$. Fortunately, due to our specific choice of complex mode functions, we only need to calculate $\alpha_{k}$ and $\beta_{k}$, and get the others by calculating their complex conjugate. We obtain
\begin{equation}
\label{Eq.Alpha.General}
\alpha_{k}(t) = \frac{2 \, \left( g \, \dot u_{k}  v_{k}^{*} - h \, u_{k} \dot v_{k}^{*} \right) + v_{k}^{*} u_{k} \left( g \, \frac{\dot h}{h} - \frac{\dot g}{g} \, h \right)}{2 \sqrt{g\,h} \; W[u_{k},u_{k}^{*}]} \, ,
\end{equation}
and
\begin{equation}
\label{Eq.Beta.General}
\beta_{k}(t) = \frac{2 \, \left( h \, \dot v_{k}^{*} u_{k}^{*} - g \,  v_{k}^{*} \dot u_{k}^{*} \right) - v_{k}^{*} u_{k}^{*} \left( g \, \frac{\dot h}{h} - \frac{\dot g}{g} \, h \right)}{2 \sqrt{g\,h} \; W[u_{k},u_{k}^{*}]} \, .
\end{equation}
A somewhat time-consuming, but trivial calculation shows that further conditions are necessary to obtain a Bogoliubov transformation in the desired normalized form~\refb{Eq:Bogoliubov.Normalization.Alpha.Beta}, since in general~\refb{Eq.Alpha.General} and~\refb{Eq.Beta.General} lead to
\begin{equation}
\vert \alpha_{k}(t)\vert^{2} - \vert \beta_{k}(t)\vert^{2} = \frac{W[v_{k},v_{k}^{*}]}{W[u_{k},u_{k}^{*}]} \neq 1 \, .
\end{equation}
Hence it is necessary to choose a consistent normalization condition, $W[v_{k},v_{k}^{*}]=W[u_{k},u_{k}^{*}]=\mathrm{constant}$, so that $\vert \alpha_{k}(t)\vert^{2} - \vert \beta_{k}(t)\vert^{2} =1$. Also note that for multiple-step events, meaning multiple Bogoliubov transformations described by the compound matrix $M_{n} \cdot M_{n-1} \cdots M_{2} \cdot M_{1} $, the physics is independent of the particular choice for the normalization of the intermediate mode functions, since
\begin{equation}
\det(M_{n}  \cdots M_{1} ) =\det(M_{n})  \cdots \det(M_{1}) 
= \frac{W[v_{k}^{1},(v_{k}^{1})^{*}]}{W[(v_{k}^{n},v_{k}^{n})^{*}]} \, .
\end{equation}

Our main focus in this chapter is to calculate the particle production in a FRW-rainbow metric, and therefore we wish to restrict our problems to cases where we start and end in a physically meaningful vacuum state. This way we are able to connect $\vert \beta_{k} \vert^{2}$ with the occupation number for real condensate excitations in an effectively expanding spacetime. In our companion paper, see \cite{Jain:2006ki}, we investigated various expansion scenarios. We will now concentrate on the most relevant expansion scenario in terms of cosmology, \ie, de Sitter like inflation.

There are three principal different cases, in terms of $U(t)$, that are of interest when addressing the problem of inflation in emergent spacetimes:

\begin{description}
\item[Section~\ref{Sec:Finiteness.Particle.Production.Emergent.Spacetimes}:]{\emph{U(t) discontinuously, connects two flat spacetime regions:} This case is indirectly interesting for inflation, since the de Sitter expansion approaches the sudden case for infinitely fast expansion~\cite{Visser:1999aa}. But, we will show shortly that in emergent spacetimes --- due to the nonperturbative ultraviolet corrections --- the particle production for the extreme limit is finite, thus the Bogoliubov transformation for the de Sitter case is always well-defined.} 
\item[Section~\ref{Sec:Toy.Model.Conventional.Inflation}:]{\emph{U(t) continuously, but not continuously differentiably connects two flat spacetime regions with a finite de Sitter like phase in between:} This calculation can be carried out in the hydrodynamic limit, but fails to be a good approximation for infinitely long-lasting inflation in emergent spacetimes.
Due to an unsolvable second order differential equation in the eikonal limit, we present a qualitative analysis, that should be compared with the numerics presented in our companion paper \cite{Jain:2006ki}.}
\item[Section~\ref{Sec:Long.Lasting.Rainbow.Inflation}:]{\emph{U(t) is a smooth function everywhere, such that we are left with one de Sitter like region:} The qualitative analysis shows that in our particular emergent spacetime the de Sitter expansion has, in the infinite past and in the infinite future, two distinct physical vacua, and therefore we are able to predict the existence of time-independent real (unambiguous) quasi-particles created during an infinitely long-lasting expansion. This might at first seem of less interest for the condensed matter community, but as our simulations show, this can be realized for a sufficiently long expansion time. We will see that for the right parameter choice the particle spectrum approaches a characteristic final shape, for which we can numerically determine the form of the final particle spectrum.}
\end{description}

\subsection[Finiteness of particle production in emergent spacetimes]{Finiteness of particle production in emergent spacetimes \label{Sec:Finiteness.Particle.Production.Emergent.Spacetimes}}
There is a relatively simple way to calculate an upper bound on the particle production in a universe that is subjected to a (finite-size) expansion, in terms of a sudden variation in the size of the universe~\cite{Visser:1999aa}.

The case of a particle production for a sudden transition has been previously explored by Jacobson for a parametric oscillator \cite{Jacobson:2003ac}. (Though the underlying physics is rather different, there is also a model for \emph{sonoluminescence} that is based on a rapid change in refractive index --- that model shares many of the mathematical features encountered in the present calculation~\cite{Liberati:1998aa,Liberati:2000aa,Liberati:2000ab,Liberati:2000ac,Belgiorno:1999ha,Liberati:1999aa,Liberati:2006sj}.)
A more closely related work on sudden changes --- between Lorentzian\,/\,Euclidean signatures --- in emergent spacetime  has recently been carried out in \cite{Weinfurtner:2007aa}.

To calculate the particle production in the limiting case of a sudden expansion, we consider the situation where the atom-atom interaction is ``instantaneously'' switched from $U$ to $U/X$ at some time $t_0$. 
The scale function (see Eq.~\refb{Eq:U.acoustic.time}) is given by
\begin{eqnarray}
\label{bsudden}
b(t) = 1 - \left(1 - \frac{1}{X}\right) \Theta_{\mathrm{HS}}(t-t_{0}) \, .
\end{eqnarray}
This corresponds to a change between two regions with distinct dispersion relations, connected at $t_{0}$. We choose as mode functions for $t<t_{0}$,
\begin{eqnarray}
\label{Eq:Mode.Functions.Minkowski.in}
v_{k}(t)=\frac{\e^{+i\omega_{k}^{\mathrm{in}}t}}{\sqrt{\omega_{k}^{\mathrm{in}}}} 
~~\mbox{and}~~
v_{k}^{*}(t)=\frac{\e^{-i\omega_{k}^{\mathrm{in}}t}}{\sqrt{\omega_{k}^{\mathrm{in}}}} \, , 
\end{eqnarray}
and for  $t>t_{0}$,
\begin{eqnarray}
\label{Eq:Mode.Functions.Minkowski.out}
u_{k}(t)=\frac{\e^{+i\omega_{k}^{\mathrm{out}}t}}{\sqrt{\omega_{k}^{\mathrm{out}}}} 
~~\mbox{and}~~
u_{k}^{*}(t)=\frac{\e^{-i\omega_{k}^{\mathrm{out}}t}}{\sqrt{\omega_{k}^{\mathrm{out}}}} \, ; 
\end{eqnarray}
these are normalized (see Eq.~\refb{Eq:Wronskian}) positive and negative frequency modes. As explained in Section~\ref{Sec:Instantaneous.Hamiltonian.Diagonalization}, these mode functions represent physically meaningful vacua in flat spacetime regions. 

Straightforwardly, we can apply Eqs.~\refb{Eq.Alpha.General} and \refb{Eq.Beta.General} to calculate $\alpha$ and $\beta$. We get for the Bogoliubov coefficients: 
\begin{eqnarray}
\label{Eq:Alpha.Sudden}
\alpha_{k}(t_{0}) &=& \frac{1}{2} \left[ \sqrt{\frac{\omega_{k}^{\mathrm{out}}}{\omega_{k}^{\mathrm{in}}}} +\sqrt{\frac{\omega_{k}^{\mathrm{in}}}{\omega_{k}^{\mathrm{out}}}} \right]  \e^{i\,(\omega_{k}^{\mathrm{out}} - \omega_{k}^{\mathrm{in}}) \, t_{0}}  ; \quad \; \; \\
\label{Eq:Beta.Sudden}
\beta_{k}(t_{0}) &=& \frac{1}{2} \, \left[ \sqrt{\frac{\omega_{k}^{\mathrm{out}}}{\omega_{k}^{\mathrm{in}}}} -\sqrt{\frac{\omega_{k}^{\mathrm{in}}}{\omega_{k}^{\mathrm{out}}}} \right] \e^{-i\,(\omega_{k}^{\mathrm{out}} + \omega_{k}^{\mathrm{in}}) \, t_{0}}  . \quad \; \;
\end{eqnarray}
Using the transformation laws between the $\mathrm{in}$ and $\mathrm{out}$ Fock state operators, see Eq.~\refb{Eq:operator.bk}, and Eq.~\refb{Eq:Occupation.Number.Desity}, we obtain the occupation number density of $\mathrm{out}$-particles in the $\mathrm{in}$-vacuum, 
\begin{equation}
\label{Eq:nk.sudden}
n_{k}^{\mathrm{out}} =\frac{1}{4} \, \left\vert \frac{(\omega_{k}^{\mathrm{out}}-\omega_{k}^{\mathrm{in}})^{2}}{\omega_{k}^{\mathrm{out}}\omega_{k}^{\mathrm{in}}} \right\vert \, ,
\end{equation}
where we assumed that the eigenfrequencies are real. Note that this relation is valid beyond the eikonal limit, such that we are dealing with a nonlinear dispersion, see Eq.~\refb{Eq:Eiknoal.Sound.Speed.time}, in both regions;
\begin{eqnarray}
\omega_{k}^{\mathrm{in}} &=& \omega_{0}^{\mathrm{in}} \sqrt{1 + k^{2}/K^{2}} \, , \\
\omega_{k}^{\mathrm{out}} &=& \omega_{0}^{\mathrm{in}} \sqrt{1/X + k^{2}/K^{2}} \, . 
\end{eqnarray}
Thus we get
\begin{equation}
n_{k}^{\mathrm{out}} = \frac{ 
\left(\sqrt{X} \sqrt{K^{2} + k^{2}} -  \sqrt{K^{2} + X \, k^{2}} \right)^{2}}{4\, \sqrt{X} \sqrt{K^{2} + k^{2}} \, \sqrt{K^{2} + X \, k^{2}}} \, .
\end{equation}
It is perhaps surprising how simple the final result is, despite the additional technical machinery required to derive it.  Of course, particle production is momentum-independent only within the hydrodynamic
limit, when 
\begin{equation}
\omega_{k}^{\mathrm{in}/\mathrm{out}}\to \omega_{0}^{\mathrm{in} /\mathrm{out}} \, . 
\end{equation}
The number density in the infrared limit, when $k < K$, can be simplified to
\begin{equation}
n_{\mathrm{hydro}} = \frac{1}{4} \, \left\vert X^{1/4} - X^{-1/4} \right\vert^{2} \, ,
\end{equation}
and --- up to an overall (thus undetectable) minus sign --- is in agreement with the results obtained in ``ordinary'' momentum independent spacetimes with sudden jumps in the dispersion relation, instead of the scale factor; for example see \cite{Jacobson:2003ac}.
Bogoliubov transformations with infinite total particle production per unit volume are ill-defined, as
the $\mathrm{out}$-vacuum state $\vert {}_{\mathrm{out}} 0 \rangle$, can only be written in terms of a \emph{normalizable} linear-combination of $\mathrm{in}$-particle states, if $\vert \beta_{k} \vert^{2} \to 0$ faster $k^{-d}$ for large $k$. To see an explicit derivation consult, for example, Ref.~\cite{Mukhanov:2007aa}.
We shall now show how due to non-perturbative corrections in our FRW-rainbow spacetimes, this problem does not appear. Here the total number of quasi-particles produced in the ultraviolet limit remains finite under any circumstances. 
At large momenta
\begin{equation}
\lim_{k\gg K} n_{k} =   \frac{(X-1)^{2}}{X^{2}} \, \left( \frac{K}{2 \, k} \right)^{4} + O(1/k^6)  \, ,
\end{equation}
and separately we see that for an infinitely large expansion
\begin{equation}
\label{Eq:sudden.X.Infinity}
\lim_{X\to \infty} n_{k} = \frac{1}{4} \, \frac{(\sqrt{K^{2}+k^{2}}-k)^{2}}{k \, \sqrt{K^{2}+k^{2}}} \, . 
\end{equation}
We can also combine these two limits, and obtain $n_{k} \to [K/(2 \, k)]^{4}$.
The quantum pressure term supresses particle production at high momentum --- effectively because at high momenta the quasi particle are free and do not ``see'' the evolving spacetime. (The high-momentum limit of the rainbow metric is ``static''.)

As a consequence the number particle production per spacetime volume,
\begin{equation}
N_{\leqslant k} \sim 2^{d-1} \pi \int  d\mathbf k \; k^{d-5} \sim k^{d-4} \, ,
\end{equation}
for the here relevant cases of $2$ or $3$ spatial dimensions, is finite.

It is also possible, see Eq.~\refb{Eq:sudden.X.Infinity}, to exactly calculate the \emph{total} number of
quasi-particles produced during, $c_{0} \to c_{0}/\sqrt{X}$, for $X \to \infty$, an \emph{infinite} expansion:
\begin{equation}
N_{\infty} = \frac{2^{d-3} \, \pi}{d} \, K^{d} \, .
\end{equation}

Particle production in our emergent spacetime is a good example as to how corrections --- at the level of effective field theories --- from the underlying microscopic structure can circumvent the problem of an infinite number density caused by a finite process. On the other hand, it seems naive to assume the validity of common quantum field theory on all scales. If spacetime is indeed the infrared limit of a more fundamental theory, the presence of ultraviolet deviations seems to be unavoidable.
%
%
\section[FRW-rainbow spacetimes in $(2+1)$ dimensions]{FRW-rainbow spacetimes in $(2+1)$ dimensions:\protect\\
Controlled parametric excitations in $2$-dimensional condensate}
\label{Sec:Inlation.FRW.Rainbow.Spacetimes}			
%
In the following we investigate the robustness of the cosmological particle production process against model-dependent ultraviolet corrections in the dispersion relation. The modifications arise from the microscopic substructure, therefore are of non-perturbative nature and not the result of some perturbative loop-calculations. Related work on Hawking radiation from acoustic black holes has been carried out, where the various authors mainly studied the problem in media with subsonic and supersonic dispersion relations \cite{Jacobson:1991sd,Unruh:2005aa}. In addition, similar efforts have been made in the field of ``conventional'' cosmology, by implementing \emph{ad hoc} trans-Planckian modifications to the system, for example see~\cite{Martin:2003rp,Mersini-Houghton:2001aa,Bastero-Gil:2003aa}. This motivated us to extend our main intention --- testing the analogy --- to a detailed study of the deviations (from standard cosmology) in our specific model. 
We demonstrate this with a particular physically relevant model, a rapidly exponentially growing effective universe, how ``conventional'' and ``modified'' models naturally merge in the infrared limit at early times. Whereas, they show significant deviations at late times, yet for a sufficiently short-time expansion, the resulting particle spectra --- in systems with linear and nonlinear dispersion relations --- are very similar to each other.
%
\subsection[$(2+1)$ dimensional FRW-rainbow spacetimes]{$(2+1)$ dimensional FRW-rainbow spacetimes:\protect\\
$2$ dimensional condensates with time-dependent atomic interactions
 \label{Sec:d=2}}
In Eq.~\refb{Eq:FRW.Acoustic} we established the concept of a FRW-type rainbow universe to describe the $(2+1)$ or $(3+1)$ dimensional spacetime as seen by small quantum fluctuations in a $2$ or $3$ dimensional condensate. As pointed out in the text passage shortly after the line-element, besides the dimensionality, also the scale factor $b(t)$ is different for $2$ and $3$ dimensions. For example, if we wish to mimic the de Sitter geometry, where the scale factor in the FRW line element, see Eq.~\refb{Eq:FRW.GR}, takes the form 
\begin{equation}
\label{Eq:a.FRW.deSitter.hydrodynamic}
a(\tau) = \exp(H\, \tau) \, ,
\end{equation}
we have to choose the scale factor for the atomic interactions,
\begin{equation}
\label{SwitchedDeSitter}
b(t) = 
\left\{ 
\begin{array}{ll}
\exp(-t/t_{s})  & ~~\mbox{for}~~ d=2,\\
t_{s}/(t+t_{s}) & ~~\mbox{for}~~ d=3, \\ 
\end{array} \right.
\end{equation}
corresponding to the spatial dimension. A more detailed treatment of this problem --- that originates in the discrepancy of laboratory $t$ and proper time $\tau$ for the $3$-dimensional case ---  can be found in \cite{Barcelo:2003yk}. In $2$ spatial dimensions --- where  $\alpha(2)=0$ --- the effective line-element is given by
\begin{equation}
\label{met2Dconstrho}
\mathrm{d} s_{d=2}^2 
= \left( \frac{n_0}{c_0}\right)^{2} \left[ -c_0^2 \,  d t^2+ b_{k}(t)^{-1} d\mathbf x^2 \right] \, ,
\end{equation}
here laboratory time $t$ and proper time $\tau$ are of the same form, and we obtain the very simple relationship,
\begin{equation}
\label{Eq:FRWsaclefatcor_2d}
a_{k}(t) = b_{k}(t)^{-1/2}=\frac{1}{\sqrt{b(t) + (k/K)^{2}}} \, ,
\end{equation}
between the two scale factors. To connect $t_{s}$, the scale unit in the laboratory, to the Hubble parameter $H$, we need to choose
\begin{equation}
H = \frac{1}{2 \, t_{s}} \, .
\end{equation}
We will soon revisit the Hubble parameter, when we discuss its physical motivation, and how the Hubble parameter applies to our de Sitter-rainbow spacetime.
For practical reasons we shall be mainly interested in $(2+1)$ spacetime dimensions, that is, $d=2$ space dimensions and $1$ time dimension, simply because the numerical simulations we wish to compare the theory with are more easily carried out in 2 space dimensions.

We will always use $a(t)$ to refer to the \emph{FRW scale factor}; if we ever really need the scattering length  we will refer to it as $a_{\mathrm{scatt}}(t)$.

In what physical respect do we really have an ``expanding universe'', given
that the condensate is physically contained in a fixed volume $V$? A
decrease in the scattering length corresponds to a decrease in the
speed of sound propagating in the condensate; therefore any acoustic
excitations will propagate with decreasing speed in the condensate as
time passes. To an ``internal'' observer at rest in the effective
spacetime, and communicating by means of acoustic signals, a decrease
of the speed of sound is indistinguishable from an isotropic expansion
of the spatial dimensions.
%
\subsection[Inflation in emerging universe]{Inflation in emerging universe:\protect\\ 
Exponentially decreasing atomic interactions
\label{Sec:Exponential.Universe}}
Maybe the most interesting cosmological case to study in our emergent spacetime is the \emph{de Sitter} universe, where the scale factor is given by an exponentially expanding (or contracting) universe.
The concept of inflation was introduced simultaneously around 1981 and 1982 by Guth~\cite{Guth:1981aa}, Linde~\cite{Linde:1990aa}, and Albrecht and Steinhardt~\cite{Albrecht:1982aa} to explain the homogeneity of the temperature observed in our universe, beyond casually disconnected areas. Not long after (\eg, see Guth, Hawking~\cite{Hawking:1982aa}, Bardeen~\cite{PhysRevD.22.1882}, Turner~\cite{Turner:1993aa} and Brandenburger~\cite{Brandenberger:1983aa}) it has been realized that inflation also accounts for the existence of the perturbations in our universe today. These perturbations in the form of a slight deviation from a uniform temperature in our cosmological microwave background (CMB) have been measured in 2001 by Netterfield \emph{et al}.
To date, inflation --- a phase of a rapid de Sitter-like expansion --- seems to be the most plausible explanation for the CMB map. According to all observational evidence the inflationary epoch ends in a radiation-dominant area, where $a(t)\sim t^{1/2}$. This is due to the high temperatures after the so-called reheating process. Eventually (much later) the universe cooled down and turned into the current matter-dominated universe, with $a(t)\sim t^{2/3}$.  For more details see \cite{Rothman:1993aa,Dodelson:2003aa}.
While we expect deviations from the standard picture in our emergent rainbow spacetime, it seems desirable to hold on to the concept of a thermal spectrum resulting from an exponentially expanding universe. We will come back to this point at the end of this section and show that this hope might not be in vain.

Before we focus on the particle spectrum in our de Sitter-rainbow metric, we would like to introduce some physically important parameters. They will be of great value throughout the remaining part of this chapter, and are necessary to understand the particle production process for a $2d$ analogue FRW-type universe. Let us start with the obvious question:
\subsubsection{Why rainbow spacetimes?\label{Sec:Why.Rainbow.Spacetime}}
The modified rainbow spacetimes owe their names to their momentum-dependence. This kind of modification can be absorbed into the time and (now also) momentum-dependent scale factor. For a de Sitter like universe (see Eqn~\refb{Eq:a.FRW.deSitter.hydrodynamic} and \refb{SwitchedDeSitter}) in the hydrodynamic limit, we get
\begin{equation}
\label{Eq:Rainbow.scale.factor.deSitter.eikonal}
a_{k}(t) =\frac{1}{ \sqrt{\exp(-2H \, t)+ (k/K)^{2}} }\, ,
\end{equation}
for the modified scale factor Eq.~\refb{Eq:a.FRW.deSitter.hydrodynamic}; in $2$ spatial dimensions that is.
Thus the hydrodynamic,
\begin{equation}
\exp(-2H \, t) \gg \vert k/K \vert^{2} \, ,
\end{equation}
crossover,
\begin{equation}
\exp(-2H \, t) \sim  \vert k/K \vert^{2} \, ,
\end{equation}
and free particle,
\begin{equation}
\exp(-2H \, t) \ll  \vert k/K \vert^{2} \, ,
\end{equation}
limits are a matter of dividing the spectrum into appropriate energy regimes \emph{at a particular time $t$}. 
It is interesting that for early times --- when the interactions between the atoms are strong --- we naturally approach the hydrodynamic case,
\begin{equation}
\lim_{t \to - \infty} a_{k}(t) \to a(t) \, ,
\end{equation} 
in the sense that more and more modes are phononic, and therefore larger and larger $k$-values are covered by ``conventional'' FRW-type quantum-field-theory. 

Quite the contrary occurs after an infinitely long-lasting expansion, where all modes behave as free particles,
\begin{equation}
\lim_{t \to + \infty} a_{k}(t) \to \vert K/k \vert \, ,
\end{equation} 
and the universe, as seen by a mode with the wavelength $k$, will approach a final finite fixed size.

\begin{figure}[!htb]
\centering
\includegraphics[width=0.60\textwidth]{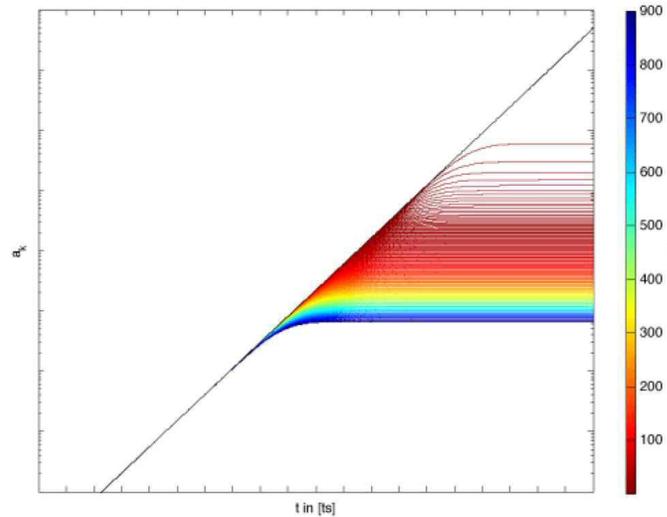}
\caption[Rainbow scale factor for effective inflation in Bose gas.]{In this figure we plot the logarithm of the scale function $a_{k}(t)$ for each $k$-value --- for $k \in [1,900]$ --- in a different color (colors online only). The different colors embody the energy of the modes: Gradually changing from low-energy\,/\,infrared (dark red) to high-energy\,/\,ultraviolet (dark blue). The straight black line shows the behavior of the scale function in the hydrodynamic limit. While the rainbow-scale function approaches the hydrodynamic limit for low-energy modes, the ultraviolet modes show strong deviations. Note, that in the infinite past all modes are phononic, and therefore $a_{k}(t)\to a(t)$.}
\label{Fig:deSitter.Rainbow.ScaleFactor}
\end{figure}

Due to this fundamental difference between our analogue model and the ``theory'' we wish to mimic, we know already that there will only be a finite time-period --- its length depends on $K$, and therefore on the tunable initial interaction strength --- beyond which the analogy breaks down. Note that the particle production process must naturally come to an end, when the expansion rate slows down to zero. That the effective expansion rate of the FRW-rainbow universe approaches zero will be shown next, but before we would like to answer the question posed in the headline of this subsection, with a simple illustration for the de Sitter-rainbow scale factor in Figure~\ref{Fig:deSitter.Rainbow.ScaleFactor}.

%
%

\subsubsection{What about the Hubble parameter?\label{Sec:What.About.Hubble.Paramter}}
The de Sitter universe is special in the sense that its ``rate of expansion''
\begin{equation}
H := \frac{\dot a(t)}{a(t)} \, ,
\end{equation}
is constant. The de Sitter universe is a solution of the Einstein equations with a positive cosmological constant, $\Lambda$; \ie, $H \propto \sqrt{\Lambda}$. If the acceleration of our universe can be put down to this cosmological constant, the universe will expand forever --- and further dilute the matter and radiation distribution in our universe --- until it approaches the de Sitter spacetime~\cite{Cattoen:2007aa}.

What is the situation in our emergent de Sitter universe? The rate of size change in the emergent de Sitter universe is given by,
\begin{equation}
\label{Eq:Hubble.general.2d.deSitter}
H_{k} = H \; \frac{\exp(-2H \, t)}{\exp(-2H \, t)+ (k/K)^{2}}  \, , 
\end{equation}
a momentum-dependent \emph{rainbow Hubble parameter}. At early times, or for phononic modes, when $a_{k}(t) \to a(t)$, the rainbow Hubble parameter,
\begin{equation}
\lim_{t \to - \infty} H_{k}(t) \to H \, ,
\end{equation} 
reduces to the conventional Hubble parameter, while for late times,
\begin{equation}
\lim_{t \to + \infty} H_{k}(t) \to H \, K^{2} \, (\exp(-H\,t)/ k)^{2} \to 0 \, .
\end{equation} 
This universe --- mode by mode --- stops expanding as the modes leave the phononic regime.

Therefore, in our particular set-up the long-time-kinematics of the particle production process is determined by the non-perturbative substructure, and not by the emergent spacetime picture. Any speculations with respect to an everlasting expanding universe are (within our model) in vain.
\subsubsection{Characteristic value for quantum process?\label{Sec:Characteristic.Value.Quantum.Process}}
To get a grasp on the cosmological particle production process, these two parameters (the effective scale function $a_{k}(t)$, and the effective Hubble parameter $H_{k}(t)$) are not quite enough. They are a good measure to describe the kinematics of the emergent gravitational field, but one also needs to know how the microscopic corrections effect the energy of the modes. Only the ratio between the mode frequency,
\begin{equation}
\label{Eq:Omega.FRW.2d.deSitter.eikonal}
\omega_{k}(t) = \omega_{0} \;  \sqrt{\exp(-2H \, t)+  (k/K)^{2}}   \, ,
\end{equation}
and the Hubble frequency can tell us whether the mode will be disturbed by the classical background or not. This result has been established for quantum field theory in conventional de Sitter spacetimes, and we review this point in the following section.

For now, we simply transfer the qualitative description known from conventional cosmological particle production, to our rainbow spacetimes. In this spirit we define \emph{the ratio} between the modified dispersion relation and the effective Hubble parameter as follows:
\begin{equation}
\label{Eq:Ratio.deSitter.eikonal}
\mathcal R_{k}(t) =\frac{\omega_{k}(t)}{H_{k}(t)}= \frac{\omega_{0}}{H}  \; \frac{(\exp(-2H \, t) +  (k/K)^{2} )^{3/2} }{\exp(-2H \, t)} \, .
\end{equation}
Within the hydrodynamic limit this ratio simplifies to,
\begin{equation}
\label{Eq:Ratio.deSitter.hydro}
R_{k}(t) = \frac{\omega_{0}}{H}  \; \exp(-H \, t)  \, ,
\end{equation}
a monotone decreasing function, as time goes on.
Again, we calculate limits for very early times, 
\begin{equation}
\lim_{t \to - \infty} \mathcal R_{k}(t) \to \frac{\omega_{0}}{H}  \; \exp( - H \, t) = R_{k}(t) \, ,
\end{equation} 
and very late times, 
\begin{equation}
\lim_{t \to + \infty} \mathcal R_{k}(t) \to \frac{\omega_{0}}{H} \;  (k/K)^{3}  \; \exp( + 2H \, t)  \, ,
\end{equation} 
to see once again, that for late times\,/\,trans-phononic modes the deviations in our emergent spacetime play 
an important rule. \\

We will show that this ratio is sufficient to understand, and therefore predict qualitatively, the particle production process in our analogue spacetimes.  

\subsection[Quantum field theory and rainbow inflation]{Quantum field theory and rainbow inflation:\protect\\ 
Excitations from exponentially changing interaction strength
\label{Sec:Quantum.Field.Theory.Rainbow.Inflation}}
The equation of motion in the presence of time-dependent atomic interactions is given in Eq.~\refb{Eq:Equation.Motion.Chi}. In the de Sitter rainbow universe, with the scale factor \refb{Eq:Rainbow.scale.factor.deSitter.eikonal}, and the dispersion relation \refb{Eq:Omega.FRW.2d.deSitter.eikonal}, the harmonic oscillator frequency \refb{Eq:Harmonic.Oscillator.Frequency} simplifies to
\begin{equation}
\label{Eq:Harmonic.Oscillator.Frequency.Eikanol.General}
\Omega_{k}(t)^{2} = \omega_{0}^{2} \, b_{k}(t) - H^{2} + \Delta_{k}(t)^{2} \, ,
\end{equation}
where the effective (or rainbow) scale factor is given by
\begin{equation}
\label{Eq:Rainbow.Scale.Factor.deSitter}
b_{k}(t) = \exp[-2Ht] + (k/K)^{2} \, ,
\end{equation}
and the last term is given as
\begin{equation}
\Delta_{k}(t)^{2} = H^{2} \, \, \frac{[4 \, \exp(-2Ht) +(k/K)^{2}] \, (k/K)^{2} }{[\exp(-2Ht) +(k/K)^{2}]^{2}} ,
\end{equation}
which can be neglected in hydrodynamic approximation. Therefore the equation of motion for our rainbow de Sitter spacetime is given by a rather complicated differential equation, where common techniques for solving second-order differential equations fail. 

However, this is not the end of the story, since at early times, or for phononic modes we get,
\begin{equation}
\label{Eq:Harmonic.Oscillator.deSitter.Hydrodynamic}
\lim_{t \to  -\infty} \;
\left\{ \begin{array}{rcl}
b_{k}(t) & \to & b(t) \\ &\,& \\
\Delta_{k}(t)& \to & 0
\end{array} \right\} \quad
\Omega_{k}(t)^{2} \to \omega_{0}^{2} \, b(t) - H^{2} \, ,
\end{equation}
where 
\begin{equation}
\label{Eq:Atomic.Scale.Factor.deSitter.H}
b(t)=\exp(-2Ht) \,  ,
\end{equation}
and within this limit the equation of motion simplifies to a differential equation solvable by Bessel functions. Within this regime, our emergent spacetimes can be regarded as a good toy model for quantum field theory in conventional de Sitter spacetimes. However, we need to apply special boundary conditions, that is to set limits on the time-duration and the $k$-range. Only then will the cosmological particle production process be robust against the model-specific non-perturbative corrections that are present in a spacetime emerging from a gas of Bosons. 

As the expansion continues, the more modes will cross-over from phononic to trans-phononic, and finally for an infinitely long-lasting expansion, we get
\begin{equation}
\label{Eq:Harmonic.Oscillator.Frequency.Early.Times}
\lim_{t \to  +\infty} \;
\left\{ \begin{array}{rcl}
b_{k}(t) & \to & (k/K)^{2} \\ &\,& \\
\Delta_{k}(t)& \to & H
\end{array} \right\} \quad
\Omega_{k}(t)^{2} \to \omega_{0}^{2} \, (k/K)^{2} \, ,
\end{equation}
and therefore are left with no phononic regime whatsoever. That is, in the infinite future all excitations of the fluctuations of the system are free-particle like. In this regime the quantum fluctuations are decoupled from the collective behavior, the emergent spacetime picture, and thus are static.

Therefore, without any further considerations, we deduce that in the infinite future we will always end up with well-defined $\mathrm{out}$-states, that represent the vacuum state of instantaneous Hamiltonian diagonalization, see Sec.~\ref{Sec:Instantaneous.Hamiltonian.Diagonalization}. The normalized mode functions in the infinite future are given by Eq.~\refb{Eq:Mode.Functions.Minkowski.out}, where $\omega_{k}^{\mathrm{out}}=\omega_{0} \vert k \vert / K$. \\

We will subsequently revisit this point in Sec.~\ref{Sec:Qualitative.Behavior.Quantum.Fluctuations}, but for now our concern will be to focus on a regime will be to focus on a regime --- the hydrodynamic limit --- where the problem of particle production is tractable, and there is a straightforward description of the physics.
%
%
\subsubsection{Toy-model for conventional inflation\label{Sec:Toy.Model.Conventional.Inflation}}
%
\begin{figure*}[!htb]
\begin{center}
\mbox{
\subfigure[$\,$ $t=0.25 \times t_{f}$ ; $t_{s}=5 \times 10^{-5}$. \label{Fig:timeslice_dsa_t_0.25tf_ts5e-5}]{\includegraphics[width=0.45\textwidth]{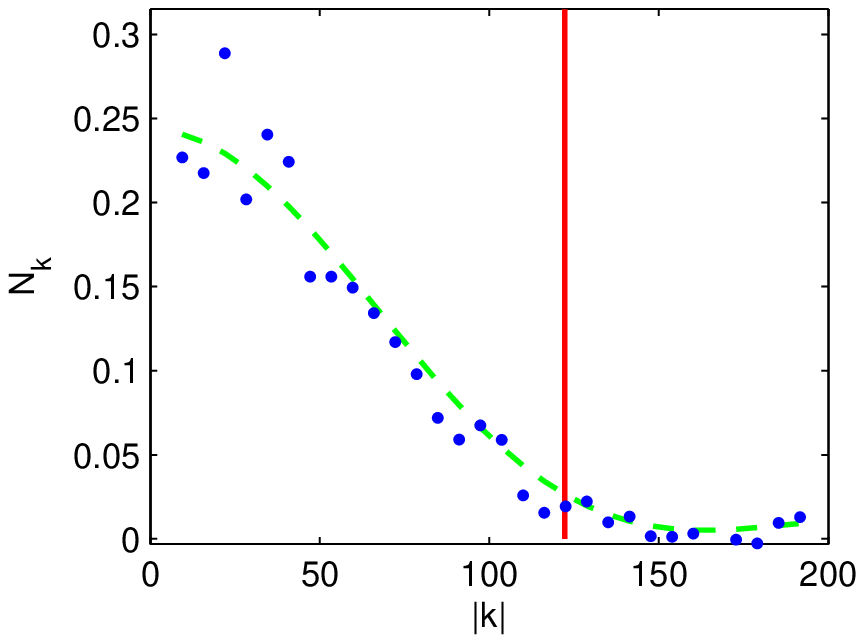}}
\hspace{0mm}
\subfigure[$\,$ $t=0.50 \times t_{f}$ ; $t_{s}=5 \times 10^{-5}$. \label{Fig:timeslice_dsa_t_0.50tf_ts5e-5}]{\includegraphics[width=0.45\textwidth]{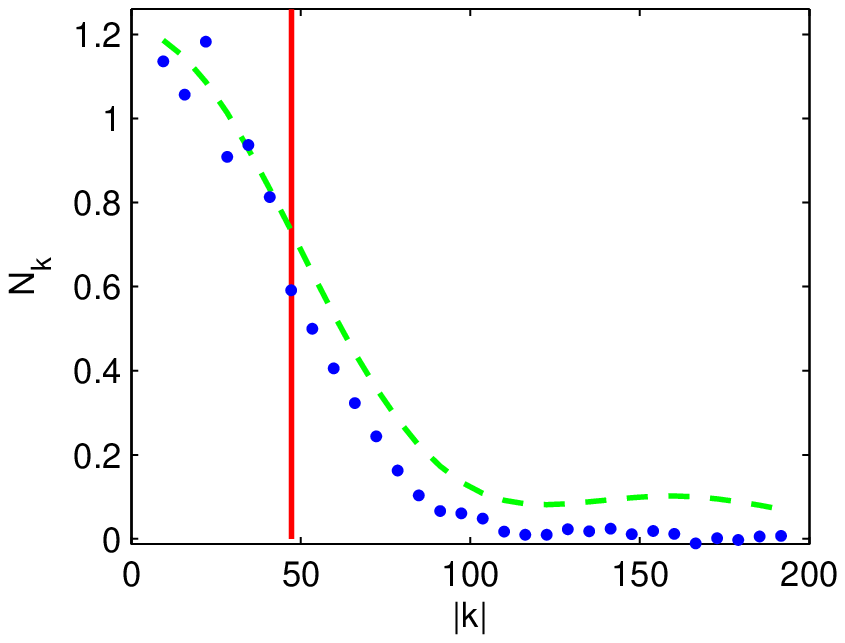}}
}
\mbox{
\subfigure[$\,$ $t=0.75 \times t_{f}$ ; $t_{s}=5 \times 10^{-5}$. \label{Fig:timeslice_dsa_t_0.75tf_ts5e-5}]{\includegraphics[width=0.45\textwidth]{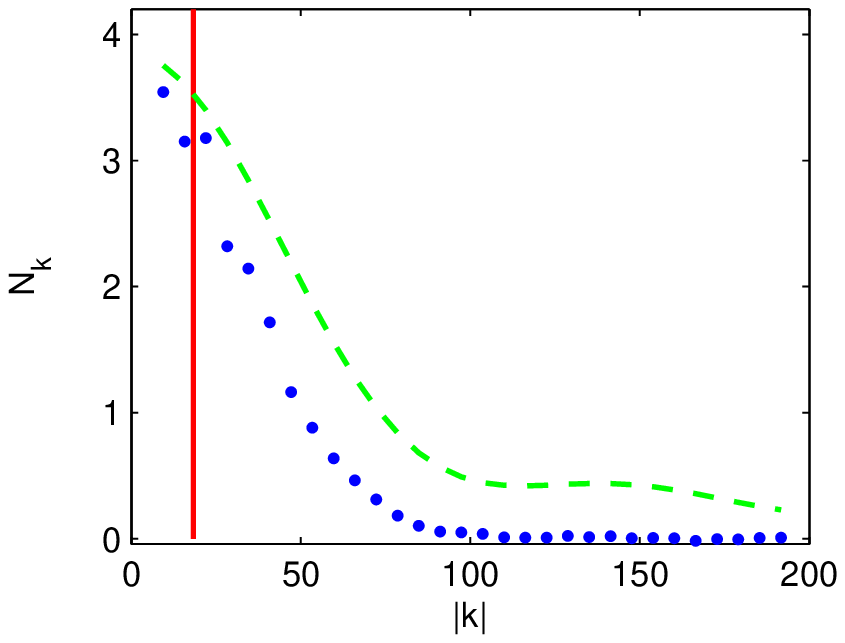}}
\hspace{0mm}
\subfigure[$\,$ $t=1.00 \times t_{f}$ ; $t_{s}=5 \times 10^{-5}$. \label{Fig:timeslice_dsa_t_1.00tf_ts5e-5}]{\includegraphics[width=0.45\textwidth]{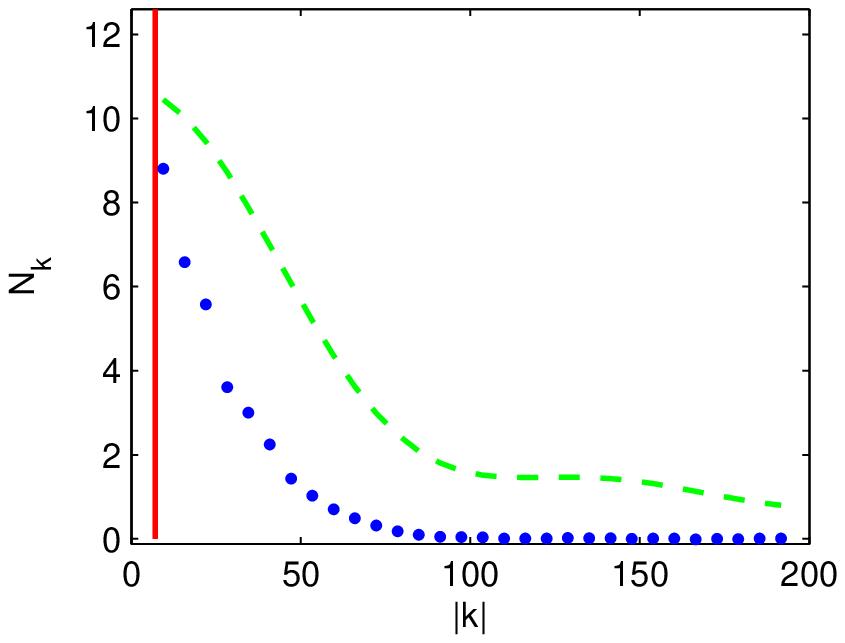}}     }
\caption[Quasi-particle production during rainbow inflation --- I.]{The four figures show time-slices of the  ``quasi-particle'' production $N_{k}$ in a de Sitter like spacetime, for $t_{s}=5\times 10^{-5}$. The blue dots represent the actual data obtained from the simulations for a realistic Bose gas with time-dependent interaction-strength~\cite{Jain:2006ki}. The dashed green line shows the theoretical results for a finite de Sitter calculation obtained in the hydrodynamic limit, as presented in Section~\ref{Sec:Toy.Model.Conventional.Inflation}. The vertical red line indicates the borderline between the nature of the excitations: Everything to the left of the red line is phononic, while everything to right is trans-phononic\,/\,free-particle like.}\label{Fig:timeslice_ts5e-5}.
\end{center}
\end{figure*}
%


%
To apply the method of instantaneous Hamiltonian diagonalization to particle production from a finite de Sitter phase, we assume a continuous function for all $t$, but at the times $t_{0}$ and $t_{f}$ the function is not differentiable:
In the hydrodynamic limit 
\begin{equation}
b_{k}(t) \to b(t)
\end{equation}
we assume
\begin{equation}
b(t) = 1 +[b^{\mathrm{dS}}(t)-1]  \, \Theta_{\mathrm{HS}}(t-t_{0}) - [b^{\mathrm{dS}}(t)-1/X] \, \Theta_{\mathrm{HS}}(t-t_{f}) \,.
\end{equation}

Here $b^{\mathrm{dS}}(t)=\exp[-2H(t-t_{0})]$, such that $b(t_{0})=1$, and $b(t_{f})=1/X$. 
In addition, we define $\omega_{k}(t_{0})=\omega_{k}^{\mathrm{in}}$, and $\omega_{k}(t_{f})=\omega_{k}^{\mathrm{out}}$, such that for $t<t_{0}$ we choose the $\mathrm{in}$-mode functions, given in Eq.~\refb{Eq:Mode.Functions.Minkowski.in}, and similarly for $t>t_{f}$ we choose the $\mathrm{out}$-mode functions, given in Eq.~\refb{Eq:Mode.Functions.Minkowski.out}, for the initial and final flat regions.

Note that for any laboratory set-up the time-scales are finite, but given that this calculation is only valid for modes that are of phononic nature both before \emph{and} after the expansion, we are not free to choose the expansion time arbitrarily. \\

The Bogoliubov transformation is slightly more difficult here, where two single-step processes have to be carried out:
\begin{equation}
\vec{v}_{k}(t) = M(t_{f}) \cdot \vec{v}_{k}^{dS}(t) = M(t_{f}) \cdot M(t_{0}) \cdot \vec{u}_{k}(t) \, . 
\end{equation}

For the mode functions during the de Sitter phase, which are $v_{k}^{\mathrm{dS}}$ and ($v_{k}^{\mathrm{dS}})^{*}$ for $t \in [t_{0},t_{f}]$,  we solve the harmonic oscillator equation with the time-dependent frequency given in Eq.~\refb{Eq:Harmonic.Oscillator.Frequency.Early.Times}. The solution is a linear combination of first order Hankel functions of the first $\mathrm H_{1}^{(1)}$ and second $\mathrm H_{1}^{(2)}$ kind. The normalized mode functions --- with respect to the Wronskian given in Eq.~\refb{Eq:Wronskian} --- are given by
\begin{eqnarray}
\label{Eq:Mode.Functions.deSitter.v}
v_{k}^{\mathrm{dS}} &=& \sqrt{\frac{\pi}{2H}} \, \mathrm H_{1}^{(1)} (R_{k}(t)) \, ,  \\
\label{Eq:Mode.Functions.deSitter.vc}
(v_{k}^{\mathrm{dS}})^{*} &=& \sqrt{\frac{\pi}{2H}} \,\mathrm H_{1}^{(2)} (R_{k}(t))  \, .
 \end{eqnarray}
Note that in this representation the mode functions are indeed a set of complex conjugate functions, since $(\mathrm H^{(1)})^{*}=\mathrm H^{(2)}$; see for example \cite{Abramowitz:1964aa}. The argument of the mode functions is given in terms of $R_{k}(t)$, the ratio between the mode frequency and the Hubble frequency, see Eq.~\refb{Eq:Ratio.deSitter.hydro}. These mode functions are only valid within the hydrodynamic limit, that is for modes that fulfill $\vert k/K \vert \ll \exp(-H \, t)$.
 
To calculate the Bogoliubov coefficients at each step we again apply Eqs.~\refb{Eq.Alpha.General} and \refb{Eq.Beta.General}. 
Below we will briefly outline the necessary steps to calculate the final Bogoliubov coefficients for the whole process.
To begin with we write down the single-step Bogoliubov coefficients, 
\begin{eqnarray}
\label{Eq.Alpha.General.Continous.Normalized}
\alpha_{k}(t) &=&\frac{1}{4i} \left( 2\, W[ u_{k}, v_{k}^{*}]  + v_{k}^{*} u_{k}  (\dot h/h - \dot g/g)  \right), \quad \\
\label{Eq.Beta.General.Continous.Normalized}
\beta_{k}(t) &=&\frac{1}{4i} \left(  2\, W[ v_{k}^{*},u_{k}^{*}]  - v_{k}^{*} u_{k}^{*}  (\dot h/h - \dot g/g)  \right) , \quad
\end{eqnarray}
appropriate to a continuous, but not continuously differentiable scale function at the time $t$.  
At the first step, where we map from plane waves onto Bessel functions, and for our specific scale factor before the step, $g(t)=1$, and after the step, $h(t)=\exp[-2H(t-t_{0})]$, we get
\begin{eqnarray}
\label{Eq.Alpha.General.Continous.Normalized.deSitter.temp1}
\alpha_{k}(t) &=&\frac{1}{2i} \left\{  W[ v_{k}, (v_{k}^{\mathrm{dS}})^{*}]  -H \, v_{k}^{*} v_{k}^{\mathrm{dS}}  \right\}  \, , \\
\label{Eq.Beta.General.Continous.Normalized.deSitter.temp1}
\beta_{k}(t) &=&\frac{1}{2i} \left\{  W[ v_{k}^{*},(v_{k}^{\mathrm{dS}})^{*}]  + H\, v_{k}^{*} (v_{k}^{\mathrm{dS}})^{*}  \right\}  \, . 
\end{eqnarray}
The final coefficients for our specific mode functions are given as
\begin{equation}
\label{Eq.Alpha.deSitter.1step}
\alpha_{k}^{\mathrm{in}}(t_{0}) =
\frac{ \sqrt{\pi \, R_{k}^{\mathrm{in}}}}{-i\, 2}  \e^{i \, \omega^{\mathrm{in}} \,t_{0}}  \left\{\mathrm H_{0}^{(1)}(R_{k}^{\mathrm{in}}) + i \, \mathrm H_{1}^{(1)} (R_{k}^{\mathrm{in}})  \right\} \, ,
\end{equation}
and
\begin{equation}
\label{Eq.Beta.deSitter.1step}
\beta_{k}^{\mathrm{out}}(t_{0}) =\frac{ \sqrt{\pi \, R_{k}^{\mathrm{in}}}}{i\, 2} \e^{i \, \omega^{\mathrm{in}} \,t_{0}}    \left\{ \mathrm H_{0}^{(2)}(R_{k}^{\mathrm{in}}) - i \, \mathrm H_{1}^{(2)} (R_{k}^{\mathrm{in}})  \right\} \, .
\end{equation}
In the last two equations we used $R_{k}^{\mathrm{in}}=R_{k}(t_{0})$ and $\omega_{k}^{\mathrm{in}}=\omega_{k}(t_{0})$.
Instead of calculating the Bogoliubov coefficients for the second step explicitly, we suggest a little shortcut. The transformation at the first step was a mapping from plane waves, represented as $v_{k}$ and $v_{k}^{*}$, onto the de Sitter mode functions $v_{k}^{\mathrm{dS}}$ and $(v_{k}^{\mathrm{dS}})^{*}$: $\vec{v}_{k} = M(t) \vec{v}_{k}^{\mathrm{dS}}$. At the second step the calculation is exactly the opposite, a mapping from the de Sitter mode functions onto plane waves: $\vec{v}_{k}^{\mathrm{dS}} = M(t)^{-1} \vec{v}_{k}$. The inverse of the transition matrix displayed in Eq.~\refb{Eq:M.Transformation.Matrix} is
\begin{equation}
\label{Eq:M.Transformation.Matrix.Inverse}
M^{-1} =
\left(
\begin{array}{cc}
\alpha_{k} & -\beta_{k}^{*} \\ -\beta_{k} & \alpha_{k}^{*}
\end{array}
\right) \, ,
\end{equation}
where we have employed $\vert \alpha_{k}\vert^{2} -\vert \beta_{k}\vert^{2} =1$, since we map between normalized mode functions. Altogether we can formally derive the Bogoliubov coefficients at the second step as:
\begin{eqnarray}
\alpha_{k}^{\mathrm{out}}(t_{f})  &:& \quad 
\alpha_{k}^{\mathrm{in}}(t_{0})
\xrightarrow[ \mathrm{in} \to \mathrm{out}]{t_{0} \to t_{f}}
[\alpha_{k}^{\mathrm{out}}(t_{f})]^{*} \, \\
\beta_{k}^{\mathrm{out}}(t_{f})  &:& \quad 
\beta_{k}^{\mathrm{in}}(t_{0})
\xrightarrow[ \mathrm{in} \to \mathrm{out}]{t_{0} \to t_{f}}
-[\beta_{k}^{\mathrm{out}}(t_{f})] \, ,
\end{eqnarray}
Thus we can write down the Bogoliubov coefficients at the second step, at $t=t_{f}$, without any further calculation being required, as
\begin{equation}
\label{Eq.Alpha.deSitter.2step}
\alpha_{k}^{\mathrm{out}}(t_{f}) =
\frac{ \sqrt{\pi R_{k}^{\mathrm{out}}}}{i\, 2}  \e^{-i  \omega^{\mathrm{out}} t_{f}}  \left\{ \mathrm H_{0}^{(2)}(R_{k}^{\mathrm{out}}) - i \, \mathrm H_{1}^{(2)} (R_{k}^{\mathrm{out}})  \right\} ,
\end{equation}
and
\begin{equation}
\label{Eq.Beta.deSitter.2step}
\beta_{k}^{\mathrm{out}}(t_{f}) =\frac{ \sqrt{\pi R_{k}^{\mathrm{out}}}}{-i\, 2} \e^{-i  \omega^{\mathrm{out}} t_{f}}    \left\{\mathrm H_{0}^{(1)}(R_{k}^{\mathrm{out}}) + i \, \mathrm H_{1}^{(1)} (R_{k}^{\mathrm{out}})  \right\}  .
\end{equation}
Here $R_{k}^{\mathrm{out}}=R_{k}(t_{f})$ and $\omega_{k}^{\mathrm{out}}=\omega_{k}(t_{f})$.

The overall particle production, at later times $t > t_{f}$ can be obtained by simple matrix multiplication, since
\begin{eqnarray}
\label{Eq:M.Transformation.Matrix.deSitter.Overall}
&M_{t>t_{f}} = 
M(t_{0}) \, M(t_{f}) =
\left(
\begin{array}{cc}
\alpha_{k}^{\mathrm{final}} &
\beta_{k}^{\mathrm{final}}\\
(\beta_{k}^{\mathrm{final}})^{*} &
(\alpha_{k}^{\mathrm{final}})^{*} 
\end{array}
\right)  . \; \; \;
\end{eqnarray}
and therefore we get for the final Bogoliubov coefficients
\begin{eqnarray}
\alpha_{k}^{\mathrm{final}} &=&
\alpha_{k}^{\mathrm{in}} \alpha_{k}^{\mathrm{out}}+\beta_{k}^{\mathrm{in}} (\beta_{k}^{\mathrm{out}})^{*} \, ,
\\
\beta_{k}^{\mathrm{final}} &=&
\alpha_{k}^{\mathrm{in}} \beta_{k}^{\mathrm{out}}+\beta_{k}^{\mathrm{in}} (\alpha_{k}^{\mathrm{out}})^{*} \, .
\end{eqnarray}
Note that these coefficients can in some sense be considered as time-dependent Bogoliubov coefficients, where 
\begin{eqnarray}
\alpha_{k}^{\mathrm{final}(t')} &=&
\alpha_{k}^{\mathrm{in}} \alpha_{k}^{\mathrm{out}(t')}+\beta_{k}^{\mathrm{in}} (\beta_{k}^{\mathrm{out}(t')})^{*} \, ,
\\
\beta_{k}^{\mathrm{final}(t')} &=&
\alpha_{k}^{\mathrm{in}} \beta_{k}^{\mathrm{out}(t')}+\beta_{k}^{\mathrm{in}} (\alpha_{k}^{\mathrm{out}(t')})^{*} \, ,
\end{eqnarray}
where we project at any instant of time onto a plane wave basis at a particular $t'$ with the eigen-frequency $\omega_{k}(t')$. But, as we pointed out in Section~\ref{Sec:Meaning.Particles}, as long as the expansion is continuing, the corresponding $\mathrm{out}$-frequency modes do not represent a physically meaningful vacuum state. Nevertheless, for $t>t_{f}$, we stop the expansion and force the Hamiltonian of the system to become static, so that we are able to associate $n_{k}^{\mathrm{final}}=\vert \beta_{k}^{\mathrm{final}}\vert^{2}$ with the ``real'' mode occupation of a mode $k$ per unit volume. \\

A lengthy but straightforward calculation now yields the mode occupation number after stopping the de Sitter like expanding phase. The number of quasi-particles produced depends only on the initial, $R_{k}^{\mathrm{in}}$, and final $R_{k}^{\mathrm{out}}$ frequency ratio:
\begin{eqnarray}
n_{k}^{\mathrm{dS}} = \frac{\pi^{2}}{64}R_{k}^{\mathrm{in}} R_{k}^{\mathrm{out}} \, \times \Big\vert \!\!\!
\nonumber
&-&[ \mathrm H^{(1)}_{1}(R_{k}^{\mathrm{in}}) \, \mathrm H^{(2)}_{1}(R_{k}^{\mathrm{out}}) -\mathrm H^{(2)}_{1}(R_{k}^{\mathrm{in}}) \, \mathrm  H^{(1)}_{1}(R^{\mathrm{out}}) ]^2 \\
\nonumber
&-& [ \mathrm H^{(1)}_{0}(R_{k}^{\mathrm{in}})\, \mathrm H^{(2)}_{0}(R_{k}^{\mathrm{out}})-\mathrm H^{(2)}_{0}(R_{k}^{\mathrm{in}}) \, \mathrm H^{(1)}_{0}(R_{k}^{\mathrm{out}}) ]^2 \\
\nonumber
&-&[ \mathrm H^{(1)}_{0}(R_{k}^{\mathrm{in}}) \, \mathrm H^{(2)}_{1}(R_{k}^{\mathrm{out}})- \mathrm H^{(2)}_{1}(R_{k}^{\mathrm{in}}) \, \mathrm H^{(1)}_{0}(R_{k}^{\mathrm{out}}) ]^2 \\
\nonumber
&-&[ \mathrm H^{(2)}_{0}(R_{k}^{\mathrm{in}}) \, \mathrm H^{(1)}_{1}(R_{k}^{\mathrm{out}})- \mathrm H^{(1)}_{1}(R_{k}^{\mathrm{in}}) \, \mathrm H^{(2)}_{0}(R_{k}^{\mathrm{out}})]^2 \\
\nonumber
&+& 2 \, \mathrm H^{(2)}_{1}(R_{k}^{\mathrm{out}}) \, \mathrm H^{(2)}_{0}(R_{k}^{\mathrm{in}}) \times 
[\mathrm H^{(1)}_{0}(R_{k}^{\mathrm{in}}) \, \mathrm H^{(1)}_{1}(R_{k}^{\mathrm{out}}) - \mathrm H^{(1)}_{1}(R_{k}^{\mathrm{in}}) \, \mathrm H^{(1)}_{0}(R_{k}^{\mathrm{out}})] \\
\nonumber
&+& 2 \, \mathrm H^{(2)}_{1}(R_{k}^{\mathrm{in}}) \, \mathrm H^{(2)}_{0}(R_{k}^{\mathrm{out}}) \times
[ \mathrm H^{(1)}_{1}(R_{k}^{\mathrm{in}}) \, \mathrm H^{(1)}_{0}(R_{k}^{\mathrm{out}})- \mathrm H^{(1)}_{1}(R_{k}^{\mathrm{out}}) \, \mathrm H^{(1)}_{0}(R_{k}^{\mathrm{in}}) ] \Big\vert \, . \\
&\,&
\label{Eq:Mode.Occupation.Finite.deSitter}
\end{eqnarray}

At first sight this formula seems to be rather complicated, but there are two relatively simple and feasible consistency checks. 

For example, we expect no particle production in the limit where the universe has not changed at all. It can easily been seen that for
\begin{equation}
 \lim_{R_{k}^{\mathrm{out}}\to R_{k}^{\mathrm{in}}} n_{k}^{\mathrm{dS}} \to 0 \, ,
\end{equation}
the occupation number indeed goes to zero.

Also interesting are limits resulting form an infinitely slow (adiabatic), that is $H\to 0$, and an infinitely fast (sudden) $H\to \infty$ expansion. In either case we replace the ratios with $R_{k}^{\mathrm{in}}=\omega_{k}^{\mathrm{in}}/H$ and $R_{k}^{\mathrm{out}}=\omega_{k}^{\mathrm{out}}/H$, hold $\omega_{k}^{\mathrm{out}}$ and $\omega_{k}^{\mathrm{out}}$ fixed, while we allow the Hubble parameter $H$ to vary. Such a parameter choice requires $H \, T = \mathrm{constant}$, and therefore we expect the expansion time $T$ to behave inverse proportional to $H$. The leading order, obtained from a Taylor-series-expansion around $H=\infty$,
\begin{equation}
 \lim_{H\to \infty} n_{k}^{\mathrm{dS}} \to  \frac{1}{4} \left\vert \frac{(\omega_{k}^{\mathrm{in}}-\omega_{k}^{\mathrm{out}})^{2}}{\omega_{k}^{\mathrm{in}}\omega_{k}^{\mathrm{out}}} \right\vert + O(1/H^{2}) \, ,
\end{equation}
is as expected --- since $T \to 0$ --- in agreement with the result from the sudden calculation within the hydrodynamic limit, see Eq.~\refb{Eq:nk.sudden}. In contrast, a Taylor-series expansion around $H=0$ yields
\begin{equation}
 \lim_{H\to 0} n_{k}^{\mathrm{dS}} \to 0 \, .
\end{equation}
As expected, within the limit of infinitely slow expansion, we produce zero quasi-particles, we recover adiabatic invariance.
Furthermore if we consider the asymptotic expansion where $1 \ll  R_{k}^{\mathrm{in}} \ll \sqrt{X}$, and employ the asymptotic limits of the Hankel functions~\cite{Abramowitz:1964aa}, we get at linear order:
\begin{equation}
n_{k} \to \frac{1}{2 \pi \, R_{k}^{\mathrm{out}}} = \frac{H}{2 \pi \, \omega_{k}^{\mathrm{out}}} \, .
\end{equation}
Of course, we cannot rely on our calculation in the case of an infinitely long-lasting expansion, since it is based on the validity of the hydrodynamic limit, which is completely inappropriate for $t \gg t_{s}$. However, in our previous numerical simulation of a ``realistic'' Bose gas~\cite{Jain:2006ki} we compared our theoretically obtained result with short-time expansion scenarios, and were able to match them to the phononic part of the particle production spectrum, see Figs.~\ref{Fig:timeslice_ts5e-5} and \ref{Fig:timeslice_ts1e-4}.

In Figure~\ref{Fig:timeslice_ts5e-5}/ Figure~\ref{Fig:timeslice_ts1e-4}, for $t_{s}=5\times 10^{-5}$/ $t_{s}=1\times 10^{-4}$ we see an excellent agreement between our theoretical predictions (dashed green line), and the numerical data (blue dots) for the ``quasi-particle'' production in the phononic regime (left of the vertical red line). At the end of the expansion we see that almost all excitations are trans-phononic, hence the analogy \emph{eventually breaks down}. But, to answer the initial question: Yes, the BEC can be used as an analogue model for cosmological particle production \emph{within certain limits}, as our numerical results clearly affirm.

The calculations we have presented are somewhat tedious, but they are more than worth the effort since there is a fundamental lesson to be learnt from Eq.~\refb{Eq:Mode.Occupation.Finite.deSitter}: It establishes our previous suspicion (see Section~\ref{Sec:Characteristic.Value.Quantum.Process}), that the characteristic value for the particle production is the ratio, $R_{k}(t)$ between the mode frequency, and the Hubble frequency. Comparison with Eq.~\refb{Eq:Mode.Occupation.Finite.deSitter} shows explicitly that the final quantity of particle production --- for truly phononic modes before, during, and after the expansion --- only depends on the initial $R_{k}^{\mathrm{in}}$ and final $R_{k}^{\mathrm{out}}$ frequency ratio. This motivated us to extend the role of the frequency ratio beyond the hydrodynamic limit, and compare our predictions with the data obtained from our simulations, where the non-perturbative corrections are included; see \cite{Jain:2006ki}.

\subsubsection{Qualitative behavior of quantum fluctuations\label{Sec:Qualitative.Behavior.Quantum.Fluctuations}}
%
\begin{figure*}[!h]
\begin{center}
\mbox{
\subfigure[\label{modes_X2000_ts1e-3}]{
\includegraphics[width=0.35\textwidth]{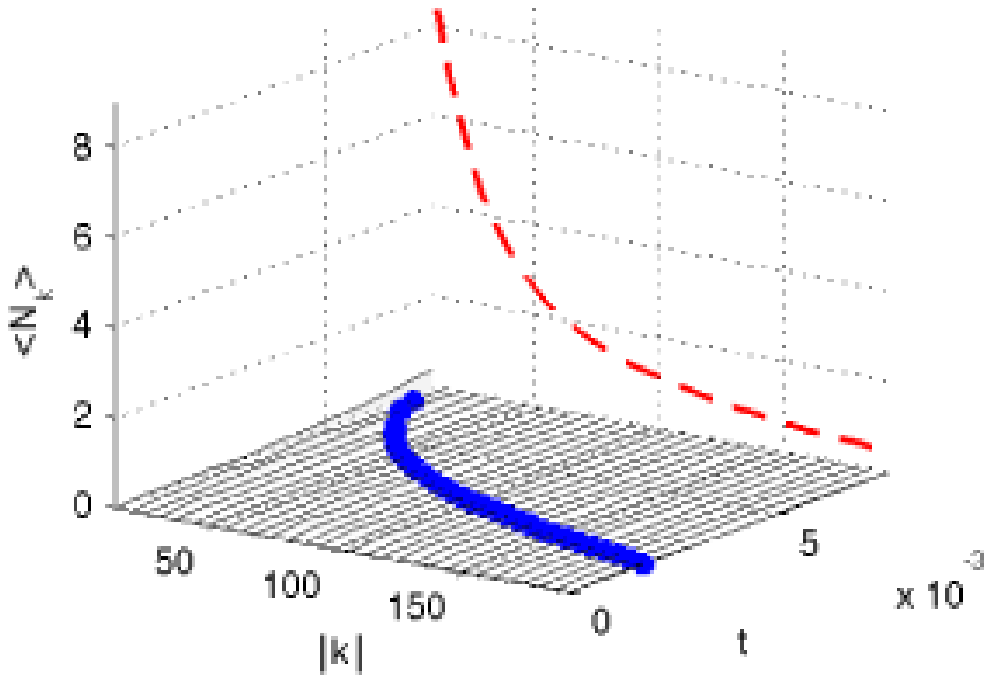}}
\hspace{0mm}
\subfigure[\label{ratio_X2000_ts1e-3}]{
\includegraphics[width=0.32\textwidth]{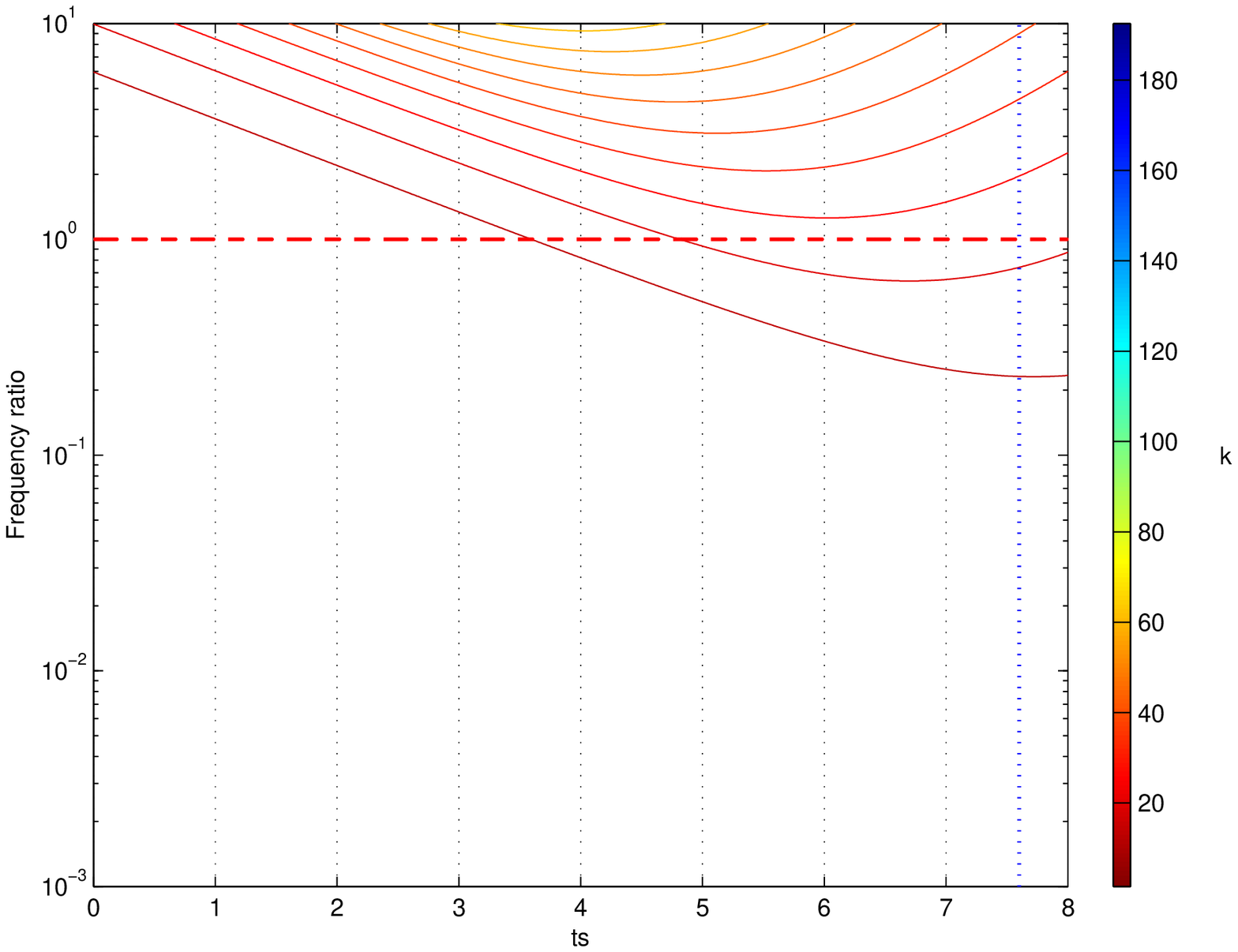}}
}
\mbox{
\subfigure[\label{modes_X2000_ts1e-5}]{
\includegraphics[width=0.35\textwidth]{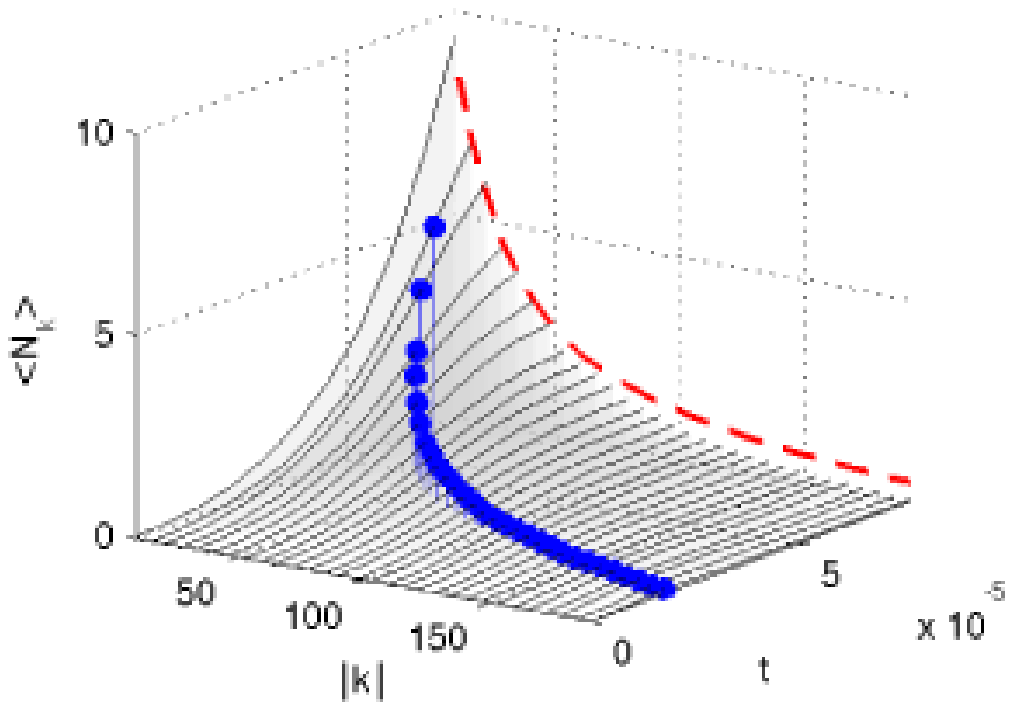}}
\hspace{0mm}
\subfigure[\label{ratio_X2000_ts1e-5}]{
\includegraphics[width=0.32\textwidth]{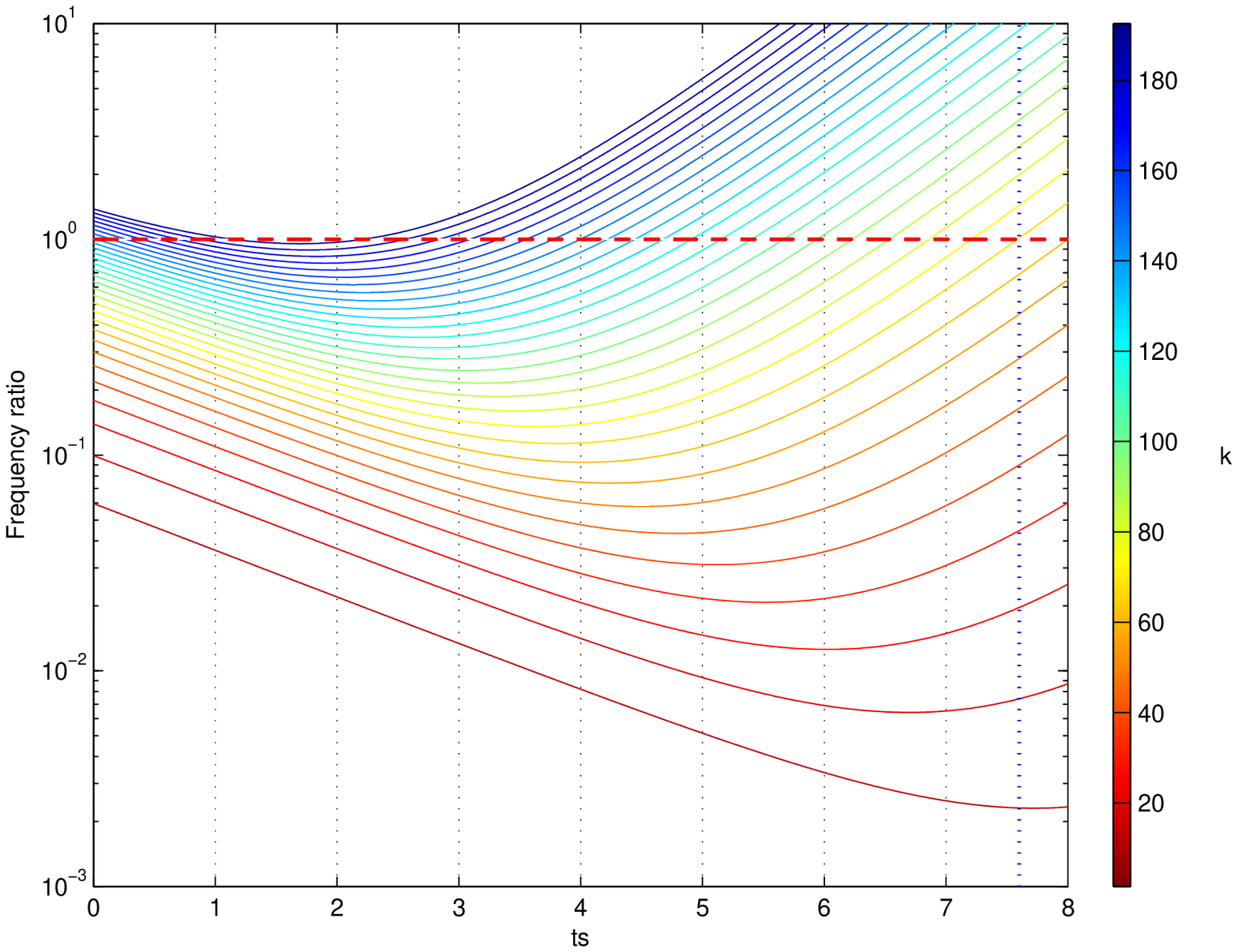}}
}
\caption[Qualitative particle production in Bose gas.]{This figure compares the change in the mode occupation number in each mode on the left side, with frequency ratio $\mathcal{R}_{k}(t)$ of the mode on the right side. The different rows express different scaling times $t_{s}$.   
Each black line on the right hand side indicates the occupation number in the mode $k$ as a function of time. (The red dashed line on the left had side indicates the sudden limit.) On the right hand side we have plotted the frequency ratio $\mathcal{R}_{k}(t)$ for each of those modes with a different color (online only) to indicate the energy difference of the modes; gradually changing from infrared modes (dark red), to ultraviolet modes (dark blue). The horizontal (red dashed) indicates where the frequency ratio $\mathcal{R}_{k}(t)$ is equal to one, while the vertical (blue dashed) line indicates end of the expansion time in our simulations. The blue dots in the figures to the left indicate when the modes cross over from phononic to trans-phononic behavior, or in other words start to decouple from the emergent spacetime. Roughly we can see that the blue dots are located where the frequency ratio has its minimum. The first row Figs.~(a) and (b) correspond to a scaling time of $t_{s}=1\times 10^{-3}$, the second row Figs.~(c) and (d) to $t_{s}=1\times 10^{-5}$. 
}\label{Fig:deSitter.Compare.Nk.Ratio}
\end{center}
\end{figure*}
%

The process of cosmological particle production in an expanding\,/\,collapsing universe can be qualitatively understood by investigating one parameter, the frequency ratio $\mathcal{R}_{k}(t)$ as given in Eq.~\refb{Eq:Ratio.deSitter.eikonal}. 
First, we explain the correlation between the qualitative behavior of the particle production process and this frequency ratio \refb{Eq:Ratio.deSitter.hydro} in the hydrodynamic limit, then transfer these ideas to the emergent rainbow metrics we have introduced in~Section~\ref{Sec:FRW.type.geometries}. \\

\paragraph{Qualitative behavior of particle production in the hydrodynamic limit}
There is a relatively simple way to understand the qualitative behavior of mode functions propagating through an exponentially changing universe, by checking out the effective harmonic oscillator equation for the auxiliary field.
Within the hydrodynamic limit we get
\begin{equation}
\label{Eq:Qualitative.Hydro}
\ddot{\hat \chi}_{k}(t) + \left(\omega_{k}(t)^{2} - H^{2} \right) \hat\chi_{k}(t) = 0 \, ,
\end{equation}
where we used the equation of motion, see Eq.~\refb{Eq:Equation.Motion.Chi}, for the effective time-dependent harmonic oscillator frequency for phononic modes given in Eq.~\refb{Eq:Harmonic.Oscillator.deSitter.Hydrodynamic}. 

Above we have shown that the general solution is a linear combination of first order Hankel functions of the first and second kind, see  Eqn.~\refb{Eq:Mode.Functions.deSitter.v} and \refb{Eq:Mode.Functions.deSitter.vc}. These mode functions are a function of $R_{k}(t)$, and therefore in the limit of $R_{k}\to \infty$ the mode functions approach positive and negative frequency modes, while for $R_{k} \to 0$ the modes stop oscillating, and the modes exhibit exponentially growing and exponentially decaying kinematics.

A simpler way to come to the same answer is to investigate Eq.~\refb{Eq:Qualitative.Hydro} in its obvious limits: These are $\omega_{k}(t) \gg H$, and $\omega_{k}(t) \ll H$, or in terms of the frequency ratio:

\begin{description}
\item[$\mathbf{R_{k}(t) \gg 1}$:] {It is then possible to write down an \emph{approximate} solution for Eq.~\refb{Eq:Qualitative.Hydro},
\begin{eqnarray}
\label{Eq:WKB.Mode.Function.v}
v_{k}^{\mathrm{dS}} &=& \frac{\exp(i \int_{t_{0}}^{t} \omega_{k}(t') \, dt')}{\sqrt{\omega_{k}{t}}} \, , \\
\label{Eq:WKB.Mode.Function.vc}
(v_{k}^{*})^{\mathrm{dS}} &=& \frac{\exp(-i \int_{t_{0}}^{t} \omega_{k}(t') \, dt')}{\sqrt{\omega_{k}{t}}} \, .
\end{eqnarray}
These keep the shape of plane waves, but allow their amplitude and frequency to change as a function of time.  This ansatz is referred to as the WKB approximation, which is valid within the \emph{adiabatic} limit,
when during one oscillation $T=2\pi/ \omega_{k}(t)$ the relative change in the frequency is small (see \cite{Mukhanov:2007aa}),
\begin{equation}
\left\vert \frac{\omega_{k}(t+T)-\omega_{k}(t)}{\omega_{k}(t)} \right\vert \approx  
2\pi \, \left\vert \frac{\dot \omega_{k}}{\omega_{k}^{2}} \right\vert \ll 1 \, .
\end{equation}
For de Sitter spacetimes this equals to
\begin{equation}
\left\vert \frac{\dot \omega_{k}^{\mathrm{dS}}}{(\omega_{k}^{\mathrm{dS}})^{2}} \right\vert 
=   \left\vert \frac{1}{R_{k}(t)}  \right\vert 
\ll 1 \, ,
\end{equation}
the condition that the ratio $R_{k}(t)$ is much larger than one.\\ 

Hence, the particle production process goes down to zero in the infinite past, where the modes oscillations are too fast to ``notice'' the ongoing (relatively) slow change in the universe.}
\item[$\mathbf{R_{k}(t) \ll 1}$:] {Here the differential equation \refb{Eq:Qualitative.Hydro}, reduces to
\begin{equation}
\label{Eq:Qualitative.Hydro.Large.H}
\ddot{\hat \chi}_{k}(t) - H^{2}  \hat\chi_{k}(t) = 0 \, .
\end{equation}
The solutions of this equation are exponentially growing, $\sim \exp(Ht)$, or decaying, $\sim \exp(-Ht)$, mode functions. \\

Therefore, the modes are no longer free to oscillate. Instead they get dragged along with the spacetime fabric.
}
\item[$\mathbf{R_{k}(t)=1}$:]{For any mode $k$ there is a time $t_{\mathrm{crossing}}$ when the frequency ratio is equal to one, and such that:
\begin{eqnarray}
t< t_{\mathrm{crossing}}  \quad  &:&  \quad \Omega_{k}(t)^{2} > 1 \, ,\\
t> t_{\mathrm{crossing}}  \quad  &:&  \quad \Omega_{k}(t)^{2} < 1 \, .
\end{eqnarray}
Therefore it is possible to associate a boundary, the so-called ``\emph{Hubble horizon}'', when $\omega_{k}=H$. For different modes with wavelength $k$ the Hubble horizon occurs at different times. \emph{Subhorizon} modes, \ie, $\omega_{k}>H$ are some sort of more or less distorted plane wave oscillations, while \emph{superhorizon} modes, \ie, $\omega_{k}<H$, approximately satisfy a harmonic oscillator equation with imaginary oscillator frequency, and thus they are exponentially growing or decaying. The ``crossing-time'' $t=t_{\mathrm{crossing}}$ is referred to as the time of ``Horizon crossing''. The \emph{superhorizon} modes are sometimes said to be frozen modes. Following this definition, if we retain the hydrodynamic approximation, then in the infinite future all modes will be frozen out.\\

Please bear in mind that the ``Hubble horizon'' is different from the cosmological horizon, see Section~\ref{Sec:Emergent.Cosmological.Horizons}. One easy way to see this, is to remember that the Hubble horizon is associated with a point in time, which is different for each mode, while the cosmological horizon is the maximum distance for all modes.}
\end{description}

The main lesson for the hydrodynamic limit is that it is not essential to explicitly solve the differential equation. Of course, it is much easier to investigate certain limits employing the resulting mode functions, instead of investigating the limits of the equation of motion. In the hydrodynamic limit both methods are accessible. Beyond this limit, in the eikonal approximation, we suggest a rather different strategy. \\

\paragraph{Qualitative behavior of particle production beyond the hydrodynamic limit:}
\emph{Ex ante} we would like to motivate this section by the remark that while the frequency ratio in the hydrodynamic limit $R_{k}(t)$ is a monotone decreasing function in time, the ratio in the eikonal approximation $\mathcal{R}_{k}(t)$ is not. 
Therefore there is some freedom to obtain different results to the ``conventional'' particle production process. 
We demonstrate the correctness of this assertion by appeal to the numerical simulations reported in \cite{Jain:2006ki}.\\

To obtain a rough estimate on the different qualitative regimes of particle production, we use the experience gained in the hydrodynamic limit, and simply exchange $R_{k}(t) \to \mathcal{R}_{k}(t)$. The eikonal frequency ratio has been introduced in Section~\ref{Sec:Characteristic.Value.Quantum.Process}, see Eq.~\refb{Eq:Ratio.deSitter.eikonal}, as the ratio between the modes (modified) frequency $\omega_{k}(t)=\omega_{0}\sqrt{b_{k}(t)}$, and the (rainbow) Hubble parameter $H_{k}(t)$, see Eqs.~\refb{Eq:Rainbow.Scale.Factor.deSitter} and \refb{Eq:Ratio.deSitter.eikonal}. \\

For early times, when $R_{k}(t)\gg 1$ the hydrodynamic and eikonal ratios are identical, therefore in both cases we are approaching the adiabatic limit, where we expect the particle production process to be negligibly small. \\

As intimated, the overall slope of the eikonal ratio is not a monotone decreasing function, since
\begin{equation}
\label{Eq:Derivative.Ratio.Eikonal}
\dot{\mathcal{R}}_{k} =  \frac{\omega_{0}\,\sqrt{\exp(-2Ht)+(k/K)^{2}}\, [\exp(-2Ht)-2 \, (k/K)^{2}]}{-\exp(-2Ht)} \, ,
\end{equation} 
the time derivative of the eikonal limit changes its sign at 
\begin{equation}
t_{\mathrm{turn}} = \frac{\ln (K^{2}/(2\, k^{2}))}{2H} \, .
\end{equation}
For $t < t_{\mathrm{turn}}$ the slope of the ratio is negative, for $t=t_{\mathrm{turn}}$ the ratio is given by 
\begin{equation}
\mathcal{R}_{k}(t_{\mathrm{turn}})= \frac{3\sqrt{3}}{2} \, \frac{\epsilon_{\mathrm{qp}}}{H} \, k^{2}\, ,
\end{equation}
and for $t>t_{\mathrm{turn}}$ the ratio is positive. Therefore the eikonal ratio has a minimum at $t_{\mathrm{turn}}$, with the maximal particle production around this point. After this point the ratio starts to increase again, and the particle production process will slow down again. \\

To qualitatively describe the particle production process in our specific rainbow spacetime, we suggest the following terminology: 
\begin{description}
\item[$t\to - \infty$:] {At early times almost all modes are ``sub-Hubble-horizon'' modes, and the particle production process is negligible. The modes oscillate with much higher frequencies as their corresponding Hubble frequencies, that is $\mathcal{R}_{k}(t) \gg 1$.}
\item[$t \sim t_{ \mathrm{turn} } $:] {As time goes on the mode frequencies are decreasing, while at the same time the rainbow Hubble frequencies are decreasing as well, see Section~\ref{Sec:What.About.Hubble.Paramter}. Nevertheless, the ratio between them exhibits a minimum at $t_{\mathrm{turn}}$, where the particle production process is expected to be maximal. 
Even if the particle production process is maximal, this does not necessarily imply that the quantity of particle production is noticeable; the modes also need to be ``super-Hubble-horizon'' modes, or in more accurate terminology, we require $\mathcal{R}_{k}(t_{\mathrm{turn}})\ll 1$.}
\item[$t \sim t_{ \mathrm{\mathrm{crossing}} } $:] {If there exists a time $t=t_{\mathrm{crossing}}$, such that $R_{k}(t_{\mathrm{crossing}})\sim 1$, where a mode $k$ is crossing the ``Hubble horizon'', there will be a second time $t=t_{\mathrm{re-entering}}$, where the mode $k$ is re-entering the ``Hubble horizon'', and  $R_{k}(t_{\mathrm{re-entering}})\sim 1$. Altogether, we suggest that it is useful to name the behavior of the modes as follows: The ``freezing of the mode $k$''  in the time period $t_{\mathrm{crossing}} <t< t_{\mathrm{turn}}$, and the ``melting of the mode $k$'' during $t_{\mathrm{turn}}< t < t_{\mathrm{re-entering}}$.}
\end{description}

Note, that we have used $R_{k}(t_{\mathrm{crossing}})\sim 1$ and $R_{k}(t_{\mathrm{re-entering}})\sim 1$, instead of  $R_{k}(t_{\mathrm{crossing}})= 1$ and $R_{k}(t_{\mathrm{re-entering}})= 1$. This is due to the fact that the harmonic oscillator frequency in the eikonal limit, see Eq.~\refb{Eq:Harmonic.Oscillator.Frequency.Eikanol.General}, can be written as
\begin{equation}
\Omega_{k}(t)^{2} = \omega_{k}(t)^{2} - H_{k}(t)^{2} \, [1 - 2\, \exp(2Ht) \, (k/K)^{2}] \, , \;
\end{equation}
and cannot be simplified to $\Omega_{k}(t)^{2} = \omega_{k}(t)^{2} - H_{k}(t)^{2} $.
Here the \emph{exact} times for the ``crossing'' and ``re-entering'' of horizons should be correlated with the sign-change in the harmonic oscillator frequency. Nevertheless, it can be seen that this does not change the qualitative description for the particle process. \\

Another novelty in our qualitative understanding of the particle production process in our FRW rainbow-spacetime, is the connection with the condensed matter point of view:
The minimum of the ratio $\mathcal{R}_{k}(t_{\mathrm{turn}})$ for a particular mode $k$, see Eqs.~\refb{Eq:Derivative.Ratio.Eikonal}, \refb{Eq.Time.Dependent.Effective.Planck.Length}, and \refb{Eq:K}, occurs at
\begin{equation}
\exp(-2Ht) - 2(k/K)^{2}=0 \quad \rightarrow \quad k=  \ell_{\mathrm{Planck}}(t)^{2} / 2 \, . 
\end{equation}
This quantity also appears in the context of conventional condensed matter physics, where it is defined as the crossover between the phonon and free-particle region. This borderline, the inverse of the healing, or coherence length $\xi$~\cite{Pethick:2001aa}, is given by
\begin{equation}
\xi^{-2} = \frac{2\, m \,n_{0} \, U(t)}{\hbar^{2}} = \frac{1}{2} \, \frac{4 m^{2}}{\hbar^{2}} \, \frac{n_{0} \, U(t)}{m} =   \ell_{\mathrm{Planck}}(t)^{2} / 2 \, ,
\end{equation}
which indicates where each mode starts to decouple from the spacetime. In other words, each mode can experience particle production, until it becomes free-particle like. 
Hence, from the BEC side we have a natural understanding of $\mathcal{R}_{k}(t_{\mathrm{turn}})$. \\

To show the qualitative correlation between the modified frequency ratio \refb{Eq:Ratio.deSitter.eikonal} with particle production in our specific rainbow de Sitter spacetime, we have plotted the ratio for several $k$-modes as a function of time, and compared them to number occupation plots, see Figs.~\ref{Fig:deSitter.Compare.Nk.Ratio}. (In Appendix~\ref{Sec:Intro.Rainbow.Inflation} for the two cases we provide larger figures for $t_{s}=1\times 10^{-3}$ and $t_{s}=1\times 10^{-5}$, see Figs.~\ref{Fig:deSitter.Compare.Nk.Ratio.1}-\ref{Fig:deSitter.Compare.Nk.Ratio.4}. In addition we also show two scaling times $t_{s}=1\times 10^{-4}$ and $t_{s}=5\times 10^{-5}$.)

For example, in Figure~\ref{modes_X2000_ts1e-3} we see the ``quasi-particle'' production in a certain mode in the de Sitter region where $t_{s}=1\times 10^{-3} $. The negligibly small population of the modes during, and after, the expansion can be explained by means of the ratio plot to its right, in Figure~\ref{ratio_X2000_ts1e-3}. As pointed out above, the particle production process is large only if $\mathcal{R}_{k} \ll 1$, which is impossible to achieve for such a \emph{relatively} large scaling unit $t_{s}$. We also see, that only the first two modes from the bottom of the infrared scale cross the ``Hubble horizon'', such that $\mathcal{R}_{k} <1$, and they quickly turn around and re-enter the ``Hubble horizon'' after a few e-foldings. Consequently, such an experimental set-up is inappropriate for mimicking cosmological particle production.

On the contrary, Figure~\ref{modes_X2000_ts1e-5} shows a noticeable ``quasi-particle'' production, for the first $100$ modes. Here $t_{s}=1\times 10^{-5}$, and thus the expansion is two magnitudes faster than one we discussed before. The qualitative behavior is roughly in agreement with the mode frequency ratio plotted to the right in Figure~\ref{ratio_X2000_ts1e-5}. The blue dots in the data-figure represent the point when the particular mode becomes free-particle like. The mode occupation process slows down after this point, and as we can read off the two figures, this happens for the fourth-lowest energy modes around $6$ e-foldings. \\

While the emergent spacetime picture was necessary to understand the time-dependent commutator relations --- in terms of the field operator and its conjugate momentum on a time-dependent classical background --- the breakdown of the particle production process during inflation is naturally explained using condensed matter physics, as it is related to the borderline between phononic and trans-phononic particles.

\subsubsection{Emergent cosmological horizons?\label{Sec:Emergent.Cosmological.Horizons}}
The existence of a cosmological horizon in our specific emergent spacetime can be investigated by calculating the maximum distance, $r_{\mathrm{max}}$, travelled by small perturbations initiated at a certain time $t=t_{0}$ and certain point $\vec{r}_{0}$. We can associate a cosmological horizon to each point $(t_{0},\vec{r}_{0})$, if the maximum distance the signal --- here our excitations in the Bose--Einstein condensate --- can travel is finite. This naturally defines a region around the point of emission, and its boundary is the cosmological horizon. In the presence of cosmological horizons two points in spacetime can be causally disconnected if their distance is larger than $r_{\mathrm{max}}$.

In ``conventional'' cosmology cosmological horizons are predicted for an infinitely long-lasting de Sitter universe, but what is the situation in our spacetime emerging from a Bose gas? We would like to address this problem briefly, and show that due to non-perturbative corrections there are, strictly speaking, no emergent cosmological horizons present in the system.

This can easily be shown, given that one only needs to integrate the \emph{group} velocity of a perturbation emitted at $(t_{0},\vec{r}_{0})$,
\begin{equation}
\label{distance_deSitter_max}
r_{\mathrm{max}}
= \lim\limits_{t\rightarrow \infty}  \int \limits_{t_{0}}^{t}  c_{\mathrm{group}} \; dt 
\end{equation}
from $t_{0}$ to the infinite future. In ``conventional'' --- that is Lorentz-invariant spacetimes --- \emph{group} and \emph{phase} velocities are identical.
This is not the case for our specific analogue model, where the \emph{phase},
\begin{equation} 
\label{phase_velocity_deSitter}
c_{\mathrm{phase}}:=\frac{\omega_{k}}{k} = c_{0} \, \sqrt{\exp(-2Ht) + (k/K)^{2}},
\end{equation} 
and the \emph{group} velocity,
\begin{equation}
\label{group_velocity_deSitter}
c_{\mathrm{group}}:=\frac{\partial \omega_{k}}{\partial k}
= \frac{c_{0}^{2} \exp(-2Ht) + 2 (k/K)^{2}}{\sqrt{c_{0}^{2} \exp(-2Ht) + (k/K)^{2}}}
\end{equation}
are different.

Only within the hydrodynamic limit, when $K \to \infty$, are the two velocities equivalent, $c_{\mathrm{group}}(t) \to c_{k}(t)$. Within this limit all individual signals travel with the same speed, and therefore wave packets ``keep their shape'', they show no dispersion.
A signal --- sent at $(t=0,\vec{r}_{0})$ --- propagating forever will only travel a finite distance,
\begin{equation}
\label{distance_deSitter_hydro}
\lim\limits_{K \to \infty} r_{\mathrm{max}}
= \lim\limits_{t' \to \infty}  \int \limits_{t_{0}}^{t'}  \lim\limits_{K \to \infty}  c_{\mathrm{group}} \; dt
= \frac{c_{0}}{H} \, .
\end{equation}
In the past cosmological horizons have been repeatedly mis-interpreted, and we would like to advise the interested reader to carefully read \cite{Rothman:1993aa}.

Returning to our problem of investigating the presence of cosmological horizons in the Bose gas, the maximum distance a signal can travel is given by 
\begin{equation}
\label{distance_deSitter}
r_{\mathrm{max}}
= \lim\limits_{t\rightarrow \infty}  \int \limits_{0}^{t}   c_{\mathrm{group}} \; dt
= \infty \, ,
\end{equation}
and thus is infinite. The emergent spacetime picture completely lacks the concept of (strict) cosmological horizons completely. This result can be generalized, since many super-sonic modifications in the dispersion relations will show similar behavior.

\subsubsection{Long lasting rainbow inflation\label{Sec:Long.Lasting.Rainbow.Inflation}}
%
\begin{figure*}[!htb]
\begin{center}
\mbox{
\subfigure[$\,$ $t_{s}=1\times 10^{-3}$ \label{Fig:deSitter.1e3}]{\includegraphics[width=0.45\textwidth]{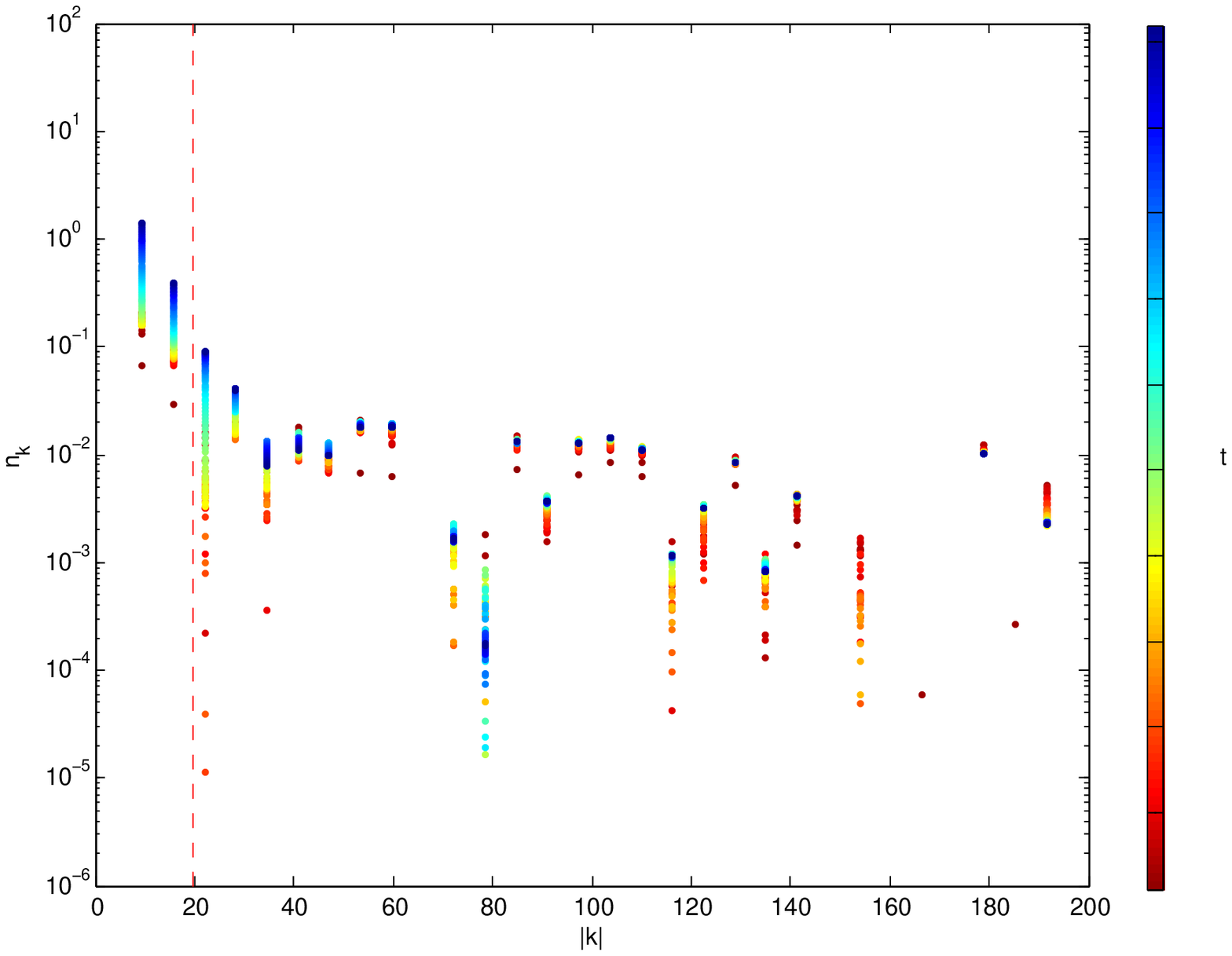}}
\hspace{0mm}
\subfigure[$\,$ $t_{s}=1\times 10^{-4}$ \label{Fig:deSitter.1e4}]{\includegraphics[width=0.45\textwidth]{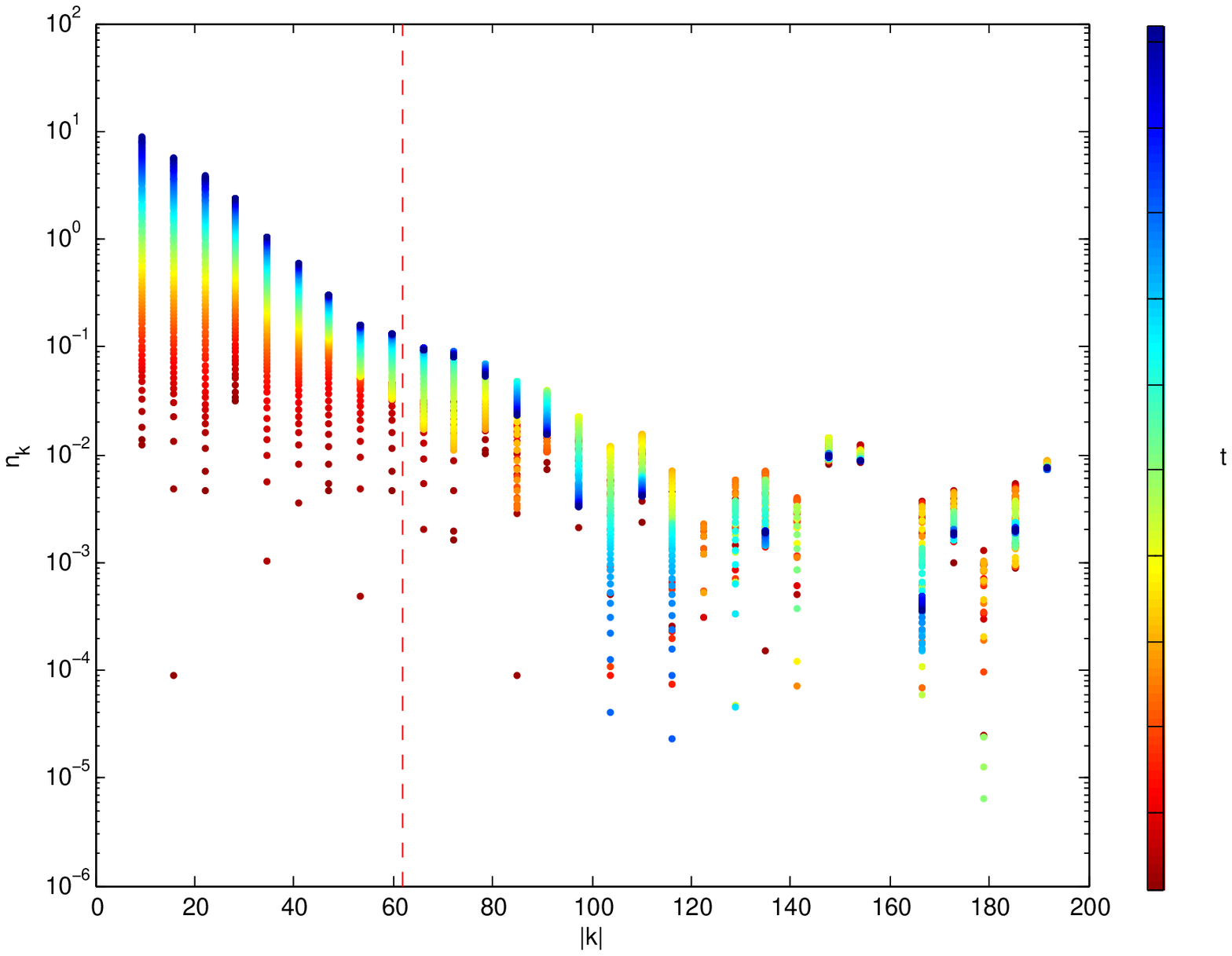}}
}
\mbox{
\subfigure[$\,$ $t_{s}=5\times 10^{-5}$ \label{Fig:deSitter.5e5}]{\includegraphics[width=0.45\textwidth]{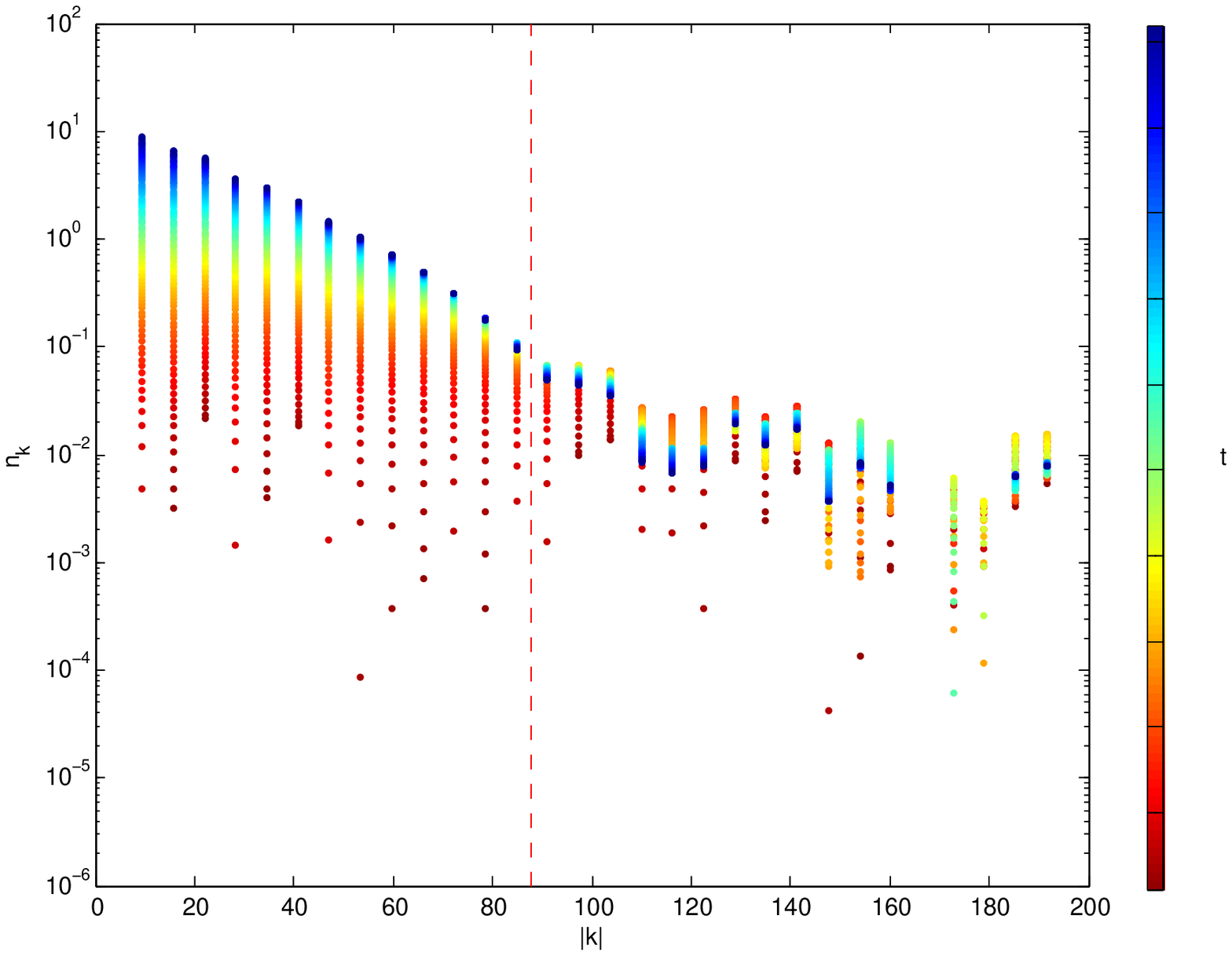}}
\hspace{0mm}
\subfigure[$\,$ $t_{s}=1\times 10^{-5}$ \label{Fig:deSitter.1e5}]{\includegraphics[width=0.45\textwidth]{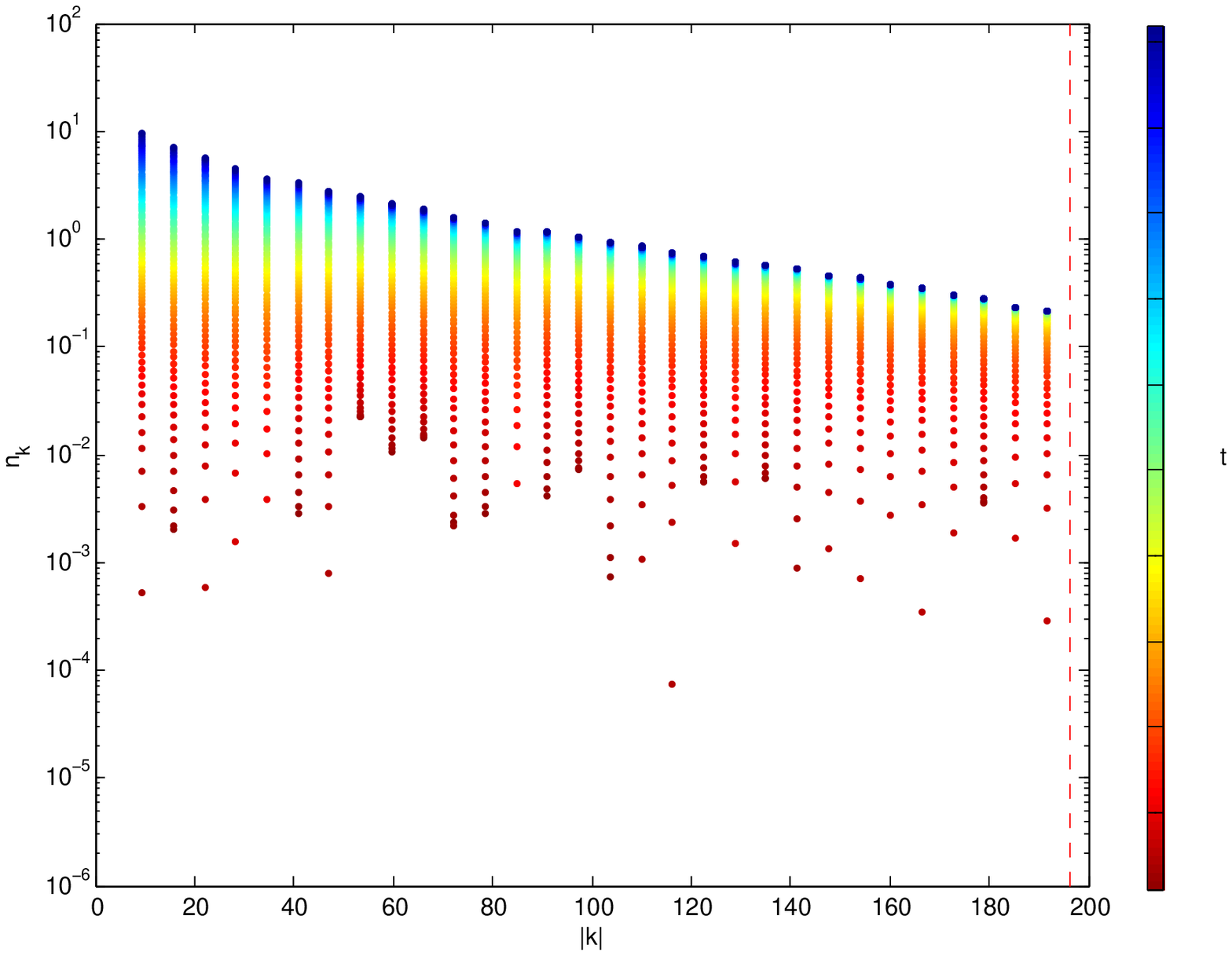}}     }
\caption[Long lasting rainbow inflation.]{Each plot shows the time-evolution of the particle spectrum projected into one plot. As time goes on the data points become darker, changing gradually from white to black. Note that at low momenta --- corresponding to those modes that twice ``cross the Hubble horizon'' by ``freezing'' and ``re-entering'' -- the final quasi-particle density is approximately log-linear, corresponding to $n_{k} \sim C \, \exp(-\sigma \, k)$. Some modes never ``cross the Hubble horizon'', if $\mathcal{R}_{k}(t_{\mathrm{turn}})\geqslant 1$. This can be seen in our plots, where we indicated 
$k_{\mathrm{static}} =  [\sqrt{2} \, 3^{(-3/4)}/ \sqrt{\epsilon_{\mathrm{qp}}}] \cdot \sqrt{H}$ 
in each plot with a vertical (red) dashed line.
\label{Fig:deSitter.Results.Free.Particle.Regime}}.
\end{center}
\end{figure*}
%

The present analogue model does not possess horizons in the cosmological sense, consequently we cannot automatically assume that an infinitely long-lasting de Sitter phase would lead to a Planckian spectrum where the temperature is connected to the surface gravity at the horizon, see~\cite{Gibbons:1977aa}. Thus there is no short-cut to calculate the overall temperature for our everlasting de Sitter rainbow spacetime. \\

At first, it seems to be a highly contrived question to ask for the final spectrum in our emergent rainbow spacetime after an infinitely long-lasting inflationary epoch. However, we demonstrate by means of numerical evidence --- from our computer simulations \cite{Jain:2006ki} --- that a \emph{sufficiently long duration expansion} can be enough to reveal the characteristic final shape of the particle spectrum at the end of inflation. \\

Before we jump to the result, we would like to show that the final Bogoliubov coefficients are indeed time-independent, and that the mode occupation number calculated with the instantaneous Hamiltonian diagonalization method represents ``real'' quasi-particles, due to well-defined $\mathrm{in}$ states in the adiabatic regime, and \emph{also} well-defined $\mathrm{out}$ states in the free-particle regime. The mode functions in our rainbow spacetime interpolate between these two states. \\
%

\paragraph{Pair of coupled harmonic oscillators:}
Given that we are only dealing with one region, we can drop the prefactor in the equation of motion for the auxiliary mode operators $\hat \chi_{k}$:
\begin{equation}
\label{Eq:Equation.Motion.Chi.One.Region}
\ddot{\hat{\chi}}_{k}(t) +\Omega_{k}(t)^{2} \, \hat\chi_{k}(t)  = 0 \, ,
\end{equation}
where
\begin{eqnarray}
t \to + \infty \; &:& \; \ddot{\hat{\chi}}_{k}(t) + (\omega_{k}^{\mathrm{WKB}})^{2}  \, \hat\chi_{k}(t)  = 0 \, , \\
t \to - \infty \; &:& \; \ddot{\hat{\chi}}_{k}(t) + (\omega_{k}^{\mathrm{FP}})^{2} \, \hat\chi_{k}(t)  = 0 \, .
\end{eqnarray}
In the infinite past, when the eikonal scale factor approaches the hydrodynamic (``conventional'') scale factor, the mode functions approach the adiabatic regime. Thus $\omega_{k}^{\mathrm{WKB}}\sim \omega_{k}(t)$, and the mode functions are approximated by Eqs.~\refb{Eq:WKB.Mode.Function.v} and \refb{Eq:WKB.Mode.Function.vc}. These mode functions represent positive and negative frequency modes at every instant of time, in the infinite past they represent the adiabatic vacuum, for instance see \cite{Mukhanov:2007aa,Fulling:1989aa,Birrell:1984aa}.

In the infinite future all excitations behave like free particles, since
\begin{equation}
\omega_{k}^{\mathrm{FP}} = \frac{\omega_{0} \, k}{K} = \frac{\hbar \, k^{2}}{m} \, ,
\end{equation}
represents the kinetic energy of an object with mass $m$; $E=p^{2}/(2m)=\hbar^{2} k^{2}(2m)$. The time-independence is due to the end of the effective expansion in our emergent spacetime. The mode functions are the usual positive and negative frequency modes, as given in Eq.~\refb{Eq:Mode.Functions.Minkowski.out}, where $\omega_{k}^{\mathrm{out}}=\omega_{k}^{\mathrm{FP}}$. \\

%
\begin{figure}[!htb]
\centering
\includegraphics[width=0.60\textwidth]{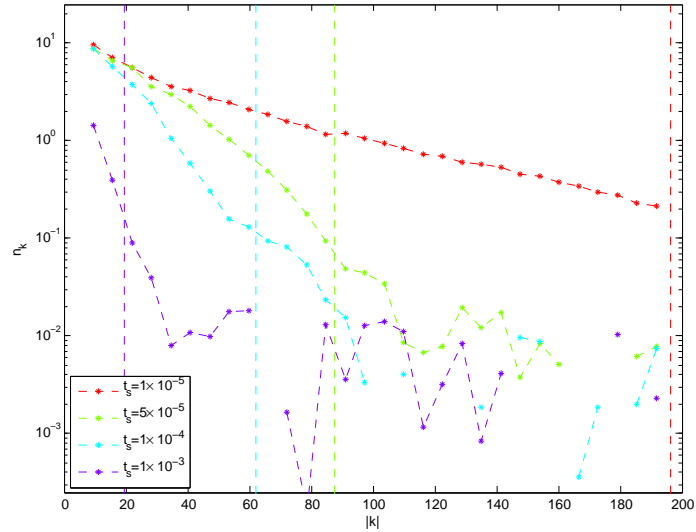}
\caption[Final particle spectra.]{This figure shows the four different particle spectra, belonging to four different scaling times, after a finite rainbow de Sitter phase. (Please note that the ``gaps'' in the quasi-particle spectrum are a result of the logarithmic scale that is incapable of handling (small) negative occupation numbers, a numerical artifact arising in our simulations. For more information about the numerical tools used to obtain these results consult~\cite{Jain:2006ki}.)}
\label{Fig:deSitter.All}
\end{figure}
%

Altogether we are dealing with well-defined \emph{initial and final} vacua states. This kind of problem can be compared with the $\tanh$-expansion, studied for example in~\cite{Fulling:1989aa,Birrell:1984aa}. In this case the Bogoliubov coefficients are --- after an infinitely long-lasting expansion --- time-independent. Please note that this analytic calculation does not by itself give any details 
\emph{when and how} the mode functions change, since the Bogoliubov coefficients for a smooth scale function $b_{k}(t)$ for all $t$ are given by the globally defined time-dependent quantities 
\begin{eqnarray}
\label{Eq.Alpha.General.Continous.Normalized.Smooth}
\alpha_{k}(t) &=&\frac{1}{2i} \, W[ u_{k}, v_{k}^{*}] \, ,  \\
\label{Eq.Beta.General.Continous.Normalized.Smooth}
\beta_{k}(t) &=&\frac{1}{2i} \, W[ v_{k}^{*},u_{k}^{*}]  \, .
\end{eqnarray}
Compare with Eqs.~\refb{Eq.Alpha.General} and \refb{Eq.Beta.General}. The Wronskian of the mode function is time-independent, therefore it can be evaluated at any time. (Note that the general mode functions --- (theoretically) obtained from the eikonal equation of motion --- are not necessarily normalized.) Again, $v_{k}$ and $v_{k}^{*}$ represent the $\mathrm{in}$, and $u_{k}$ and $u_{k}^{*}$ the $\mathrm{out}$ mode functions.  \\

%
\begin{figure}[!htb]
\centering
\includegraphics[width=0.60\textwidth]{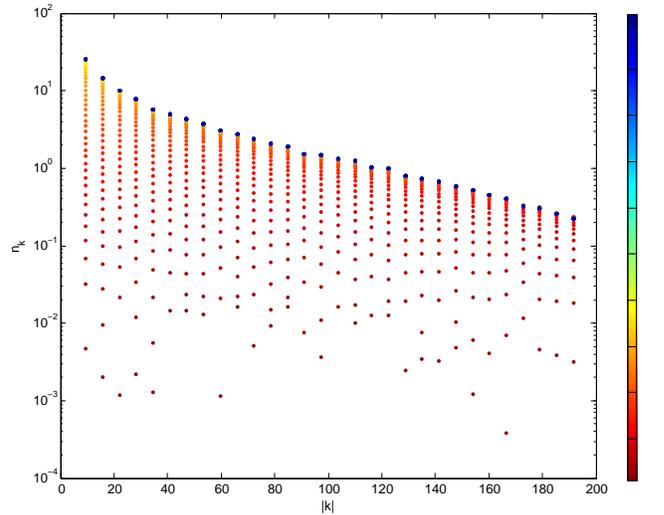}
\caption[Final particle spectra.]{This figure shows the final particle spectra after a finite duration rainbow de Sitter phase for $t_{s}=1\times 10^{-5}$ after all modes have decoupled from the emergent spacetime geometry.}
\label{Fig:deSitter.Longrun}
\end{figure}
%

By means of our numerical results, we see that the final spectrum is, after sufficiently long-lasting expansion, already \emph{rapidly} approaching its final shape, see Figs.~\refb{Fig:deSitter.Results.Free.Particle.Regime}. 
We have plotted the final spectra in one figure, Figure~\ref{Fig:deSitter.All}, so that we easily see the correlation between the slope of the line and the scaling time $t_{s}^{-1}$. 
However the temporal duration in our previous numerical simulation for $t_{s}=1\times 10^{-5}$ has not been sufficiently long. In Figure~\ref{ratio_X2000_ts1e-5} we can see that at the end of the numerical simulation a good fraction of the quantum field modes (\ie, roughly $\vert k \vert < 40$) are frozen. We expect particle production in those modes to contribute significantly to the infrared end of the final spectrum. We repeated the numerical simulation for $t_{s}=1 \times 10^{-5}$ (so that  initially all modes are sub-Hubble horizon modes) with two times the previous duration.
In addition, the initial nonlinearity is now $C = 2 \times 10^7$ instead of $C = 1 \times 10^7$ so that all modes are ``phononic'' at the start of the simulation.
 As shown in Fig.~\refb{Fig:deSitter.Longrun}, at the end of the simulation all modes are trans-phononic, and the particle production process ceases. 
As a result of our numerical simulations the final particle spectrum does not seem to nicely fit a straight line, but it seems conceivable to employ standard line-fitting tools to study the final particle spectrum as a function of $t_{s}$ and $k$. (We are currently investigating this issue.) 

%
\section{Conclusions and outlook \label{Sec:Conclusions.Outlook}}			
%
In this chapter we put the analogy between a parametrically excited Bose--Einstein condensate and cosmological particle production to the test. Knowing that the analogy for mimicking a specific quantum effect in ``conventional'' curved-spacetime quantum-field-theory hinges on the robustness of the effect against model-specific deviations, we derived the ``whole'' model-dependent emergent rainbow spacetime. Similar work on the \emph{acoustic} Hawking effect in subsonic and supersonic (super)-fluids has been carried out in~\cite{Jacobson:1991sd}.\\

There were two main lessons learnt for the analogue model community. First, the specific model we presented --- a uniform gas of atoms with time-dependent atomic interactions --- is in general \emph{not} robust against the non-perturbative ultraviolet corrections. Secondly, we also showed that the analogy is sufficiently good for mimicking some aspects of cosmological particle production for finite changes in the size of the effective universe. We said ``some aspects'', because the analogy only holds for the low-energy part of the spectrum, and therefore the analogy is associated with a certain $k$-range. There are the phononic excitations, bounded by a time-dependent parameter $\vert k \vert <  1/ \ell_{\mathrm{Planck}}(t)$.

Given that we expect significant deviation from the desired quantum effect, one might ask the question
``Is the analogue model we have presented a suitable candidate for laboratory experiments?''
Previously in~\cite{Jain:2006ki}, as well as briefly in the current chapter, we have presented numerical results for cosmological particle production in a ``realistic'' Bose gas. As a matter  of fact, despite many possible sources of difficulties, for example back-reaction-effects\,/\,mode-mixing, the phononic regime shows excellent agreement with the theoretical predictions.\footnote{
In fact, for our numerical simulations the parameters we choose (number of atoms and nonlinearity) were such that we were working in a regime where back-reaction and mode-mixing were negligible or very small effects. But in principle these effects are included by the methodology.}
The Bose--Einstein condensate enables us to prepare and control a quantum field to such an extent, that within a few years time the technology should be able to drive ``inflation'' between two ``natural'' vacua and --- that is the \emph{outstanding} problem~\cite{Schutzhold:2006pv,Weinfurtner:2007aa,Barcelo:2003yk} --- directly measure the resulting spectra. Of course, there are also other models involving a freely expanding condensate cloud, but we suspect similar problems --- a growing ``Planck-length'' and the lack of a (strict) cosmological horizon~\cite{Fedichev:2004on} --- to appear, and would like to stress that those models destroy the condensate during the expansion process. We leave it as an open problem to simulate and check the ``robustness'' of those models. \\

So much for the analogy, but what did we learn from the deviations occurring in our specific emergent spacetime? Are there conclusions to be drawn that are of relevance for the cosmology, or even quantum gravity programme? We leave this to our readers, and merely summarize our experience regarding ``trans-Planckian'' physics in our emergent spacetime:

The emergent spacetime we have presented is an example of emergent Lorentz-symmetry. At the infrared end of the excitation spectrum we exhibit Lorentz invariance, while non-perturbative corrections from the microscopic substructure ``naturally'' break the Lorentz-invariance in the ultraviolet regime~\cite{Liberati:2006sj,Weinfurtner:2006iv,Weinfurtner:2006eq,Weinfurtner:2006eq}. These corrections also alter the spacetime picture from ``ordinary'' spacetimes to the --- more ``unusual'' but conceivable~\cite{Magueijo:2004aa} --- concept of \emph{rainbow} spacetimes. These are momentum-dependent spacetimes, where the $k$-dependence is suppressed in the infrared regime. Further, a time-dependence in the atomic contact potential yields a FRW-type universe for phononic modes, and \emph{rainbow FRW-type spacetimes} for higher-energy modes. The borderline that divides the energy scale  into phononic and trans-phononic intervals, may be viewed as an \emph{analogue Planck-length}. (The physical behavior changes somewhere in between the phononic and free-particle regime, in the same sense that the Planck-scale is expected to exhibit new laws of physics.) The ``Planck-length'' in our model is correlated with the contact potential (scattering length), and thus we are dealing with a time-dependent ``Planck-length''. In our expanding universe the ``Planck-length'' is growing as well~\cite{Visser:2001ix}, such that more and more modes are ``trans-Planckian'' as time goes on. Consequently they gradually decouple from the emergent spacetime picture. They behave more and more like free particles. The \emph{rainbow} scale factor
\begin{equation}
a_{k}(t) = a(t)/ \sqrt{1+k^{2} \, \ell_{\mathrm{Planck}}(t)^{2}} \, ,
\end{equation}
and the \emph{rainbow} Hubble parameter
\begin{equation}
H_{k}(t)= H/ (1 + k^{2} \ell_{\mathrm{Planck}}(t)^{2}) \, ,
\end{equation}
are both momentum-dependent. The growth in the ``Planck-length'' forces the \emph{rainbow} Hubble parameter to approach zero, and therefore the universe gradually --- mode by mode --- effectively stops expanding~\cite{Jacobson:1999aa,Niemeyer:2001aa}. 

This leads to interesting physics for \emph{rainbow} inflation: At early times all modes are ``sub-Hubble-horizon'' modes. As the expansion goes on some --- not all --- modes cross the ``Hubble-horizon'' and become ``superhorizon'' modes, these modes are ``frozen modes'' and get dragged along with the spacetime fabric. Eventually the effective expansion starts to slow down --- due to the growing ``Planck-length'' --- and a process starts that we call ``\emph{the melting of the modes}''. After a while these modes ``re-enter'' the ``Hubble-horizon'' and the particle production process is finished. \\

However useful these results are to the general relativity and cosmology community, we would like to end our conclusion by commenting on the importance of BEC based analogue models to the condensed matter physics community. There are many aspects, \eg, the time-dependent commutation relations, that really seem to require the emergent spacetime picture to fully understand the physics in our parametrically excited condensate. The emergent spacetime and the Bose--Einstein condensate are two aspects of one and the same effect, and one needs to know both points of view to appreciate the full complexity of this novel state of matter.
%
\chapter{Summary \& Conclusions \label{Chap:Summary}}
This thesis has dealt with numerous fundamental issues related to the analogue gravity programme.

We mainly considered linearized fluctuations around some background field whose dynamics is governed by some first-order Lagrangian. In the systems (\ie, fluids and superfluids) we presented in this thesis, we saw that these perturbations experience an effective gravitational field, and that the equation of motion is identical to that for spin-zero scalar fields in an effective curved-spacetime, both at the classical and quantum levels. \\

We gave many examples of the diversity of the emergent geometrical objects. We outlined how to obtain analogue spacetime geometries described by a fully symmetric tensor of arbitrary rank and of Lorentzian\,/\,Riemannian signature. 
For example, the geometry obtained from linearizing around a \emph{single} background field is a rank two symmetric $n \times n$ tensor~$g_{ab}$ --- where $n=d+1$, for $d$ spatial dimensions.
Its signature is not necessarily pseudo-Riemannian. More specifically, we treated effective signature change events, between Riemannian and pseudo-Riemannian\,/\,Lorentzian signature,  in a spacetime emerging from a \emph{single}-component Bose--Einstein condensate.
The rank of the generalized metric tensor depends on the initial background field. We also focused on small perturbations around a two-component scalar field, a Bose--Einstein condensate with atoms in two different hyperfine states. The dominant geometrical object is then a rank four symmetric tensor, a $n \times n \times n \times n$ tensor~$g_{abcd}$, describing a \emph{Finslerian} geometry. 

Interestingly, the notion of \emph{Finslerian} geometry also governs the equation of motion for the quasi-particle excitations around the two-component scalar field. A fine tuning of the condensate parameters is possible, such that the Finslerian structure reduces to either bi-metric, or mono-metric pseudo-Riemannian geometry. The emergent curved-spacetime quantum field theory for the perturbations results in two particles, one with zero mass, and one with non-zero mass. \\

Bose--Einstein condensation is an example of \emph{spontaneous symmetry breaking} as the many Body-Hamiltonian is no longer invariant under $S0(2)$ transformations, $\theta \to \tilde\theta = \theta \, \exp( i \, \alpha)$. Therefore the presence of massive excitations from a Bose--Einstein condensate naively seems to disagree with the Nambu--Goldstone theorem, which predicts massless fields whenever a continuous symmetry is broken.

However, the fundamental Hamiltonian of the two-component system is a functional of 
$\vec\theta = (\theta_{A}, \theta_{B})$. 
In the absence of transitions between the two fields the Hamiltonian exhibits an extra $SO(2)$ symmetry under which $\vec\theta$ transforms as a $2$-component vector. This symmetry is explicitly broken for interacting fields, so that 
$SO(2)_{A} \times SO(2)_{B}  \to  S0(2)_{AB}$. The coupled system is only invariant under simultaneous transformations of the form 
$\theta_{A} \to \tilde\theta_{A}  = \theta_{A} \, \exp(+ i\, \alpha)$, and 
$\theta_{B} \to \tilde\theta_{B} = \theta_{B} \, \exp(- i\, \alpha)$. Thus the spontaneous symmetry breaking during the Bose--Einstein condensation relates to $S0(2)_{AB}$, instead of the individual symmetries. Altogether, linearizing around both fields yields two excitations, where one \emph{has} to be a ``Nambu--Goldstone Boson'' (\ie, a zero-mass excitation), while there are no constraints on the mass of the second quasi-particle. \\

For mimicking gravity, besides the general structure of the emergent spacetime geometry, we also need to know the degrees of freedom of the metric tensor. In the case of a \emph{single} scalar field the acoustic metric (a rank symmetric $n \times n$ two tensor) has $2\, n -1$ non-zero components. But only three of them are functions of time and space: the density, the speed of sound, and the background velocity potential of the fluid\,/\,superfluid. The continuity equation imposes an additional constraint onto the parameters, such that we are left with effectively \emph{two degrees of freedom}. An alternative way to get to the same result is to consider that the Bose--Einstein condensate is a complex-valued field, with only two independent parameters, the density and background phase.

The relatively low number of degrees of freedom, and the conformally flat spatial slice of the acoustic line-element, makes it impossible to mimic all features of general relativity. We have illustrated this by considering the rotating Kerr black hole, where it is not possible to cast the analogue metric into Kerr form. However it is possible to bring the acoustic metric into the form of the equatorial plane of the Kerr black hole, which is sufficient to exhibit almost all of the interesting physics, \eg, apparent horizon and ergoregion. \\

So far we have summarized only the low-energy\,/\,long-distance regime of analogue models for gravity. However, an essential part of this thesis has been based on ``extending'' the analogue model programme towards an emergent spacetime picture --- one that more accurately describes the high-energy\,/\,short-distance part of the excitation spectrum. \\

By definition Lorentz symmetry in the analogue model programme is an emergent symmetry. The Lorentz invariant geometrical object (be it a usual metric or some Finslerian generalization) is a sufficiently good approximation to describe the long-distance behavior for linearized perturbations around the background field. At short distances non-perturbative corrections --- as they are not obtained from any loop calculation --- \emph{not only} modify the dispersion relation,  they also enter into the metric components. In this thesis we showed that the former, the modifications in the dispersion relation, are of interest for issues related to general framework of Lorentz symmetry breaking as a part of quantum gravity phenomenology, while the latter \emph{may} alter the quantum aspects of the analogue model. 

We explicitly calculated the modifications in the dispersion relations for perturbations in two coupled Bose--Einstein condensates. In this part of our calculations we focused on a field-configuration capable of mimicking special relativity in the infrared limit. The emergent quantum field theory yields zero mass \emph{and} non-zero mass quasi-particles, and automatically experiences Lorentz symmetry breaking due to ultraviolet physics. The dispersion relation is of the form
\begin{equation}
\nonumber
\omega_{k}^{2} = \omega_{0}^{2} + \left( 1 + \eta_{2} \right) \; c^{2} \, k^{2} + \eta_{4} \, \left( \frac{\hbar}{M_{\mathrm{LIV}}} \right)^{2} \, k^{4} + \cdots \, .
\end{equation}
Here $\eta_{4}\approx 1$ is of order one for zero mass and non-zero mass particles. While $\eta_{4}$ generically shows no \emph{further} suppression, $\eta_{2}$ is suppressed by the smallness of the ratio of the quasi-particle mass to the scale of the Lorentz violation. In particular the Lorentz symmetry breaking mechanism in our emergent spacetime avoids the naturalness problem that arises in the effective field theory framework.

We also included the non-perturbative corrections for single background fields: In general effective curved spacetimes the ``conventional'' geometry has to be replaced by a momentum-dependent ``rainbow'' geometry. The momentum-dependence in the metric components is suppressed by the analogue ``Planck scale''\,/\,``Lorentz symmetry breaking scale''. The analogue ``Planck scale'', $\ell_{\mathrm{Planck}}=\ell_{\mathrm{Planck}}(t,\mathbf x)$, \emph{may} be a function of time and space. \\

The notion of ``rainbow'' metrics has now supplied us with a more accurate description for the ultraviolet physics, and enabled us to study the robustness of ``cosmological particle production in rainbow metrics'' emerging from a Bose gas. We showed that the analogy is sufficiently good for mimicking parts of cosmological inflation, but in general the particle production process presented in this thesis (\ie, time-dependent modifications of the microscopic interaction-strength) shows significant model-dependent effects in the particle spectrum. In our model sub-Hubble horizon quantum field modes can both cross \emph{and} re-enter the Hubble ``horizon'' \emph{during} the inflationary epoch. 

This behavior of the quantum field modes is due to a time-dependent growing ``Planck length'', and the quantum modes gradually (\ie, mode-by-mode) de-couple from the emergent spacetime picture. The \emph{effective} expansion gradually comes to an end. \\

It is also noteworthy that the commutation relation for the perturbations in terms of the emergent scalar field and its conjugate momentum with respect to the preferred rest-frame --- the laboratory frame --- exhibits an explicit momentum-dependence:
\begin{equation}
\left[ \partial_{t} \hat \theta_{k}, \hat \theta_{k} \right] \sim U  + k^{2} \, .
\end{equation}
Consequently, the size of the quantum fluctuation is growing with the momentum of the quantum field mode. (In a sense, Planck's constant is being generalized in a momentum-dependent manner.) That is, the precision with which you can measure the macroscopic field variable $\langle  \hat \theta_{k} \rangle$ decreases as momentum increases. 

This behavior is related to the fact that as one goes to smaller and smaller distances the emergent spacetime picture has to be replaced by the microscopic substructure, and the notion of the field and its conjugate momentum vanish. In the high-end ultraviolet regime collective excitations are overruled by microscopic physics. \\

To conclude this thesis we summarize \emph{possible} lessons to be learnt for emergent spacetimes. \\  
\textsl{{\bf Emergent spacetime geometries} involve:
\begin{itemize}
\item[$\blacksquare$] A microscopic system of fundamental objects (e.g. strings, atoms or molecules);
\item[$\blacksquare$] a dominant mean field regime,where the microscopic degrees of freedom give way to collective variables;
\item[$\blacksquare$] a geometrical object (e.g. a symmetric tensor) dominating the evolution of linearized classical and quantum excitations around the mean field;
\item[$\blacksquare$] an emergent Lorentz symmetry for the long-distance behavior of the geometrical object;
\end{itemize}
While the geometrical object supplies a measure of ``distance'' for excitations in the mean field limit, the fundamental microscopic objects (and their interactions) are the ones that at the same time determine both the nature of classical and quantum excitations, and the structure of the geometrical object. Independent of the specific microscopic system, we will refer to the resulting general concept as emergent spacetime. The Lorentz symmetry breaking seems to be related to existence of a preferred frame~\cite{Hu:2005wu}.
} 

We do not wish to comment directly on the relevance of emergent spacetimes for the quantum gravity programme, as it is far beyond the scope of this thesis. However we hoped we have convinced the reader of the diversity and utility of the analogue gravity programme, that makes use of interdisciplinary approaches to investigate areas where conventional methods for quantum aspects of gravity seem not to deliver a coherent picture, or show limited progress. \\
\hrule
\begin{flushright}
``The appearance is not supported by any existent different from itself; it has its own being.'' \\
{\bf Being and Nothingness} by \emph{Jean-Paul Sartre}.
\end{flushright}

%
\appendix

\chapter{Publications not included in the thesis}
\newpage
\section[Generating perfect fluid spheres in general relativity]{\centerline{Generating perfect fluid spheres in general relativity} \label{Sec:Perfect.Fluid.Spheres}}

\bigskip
\centerline{\bf Petarpa Boonserm, Matt Visser, and Silke Weinfurtner}
\bigskip
\centerline{Electronic preprint gr-qc/0503007.}
\bigskip
\centerline{Published as {\bf Physical Review D 71 (2005) 124037}}

\vskip 1 cm

\noindent
Ever since Karl Schwarzschild's 1916 discovery of the spacetime geometry describing the interior of a particular idealized general relativistic star --- a static spherically symmetric blob of fluid with position independent density --- the general relativity community has continued to devote considerable time and energy to understanding the general-relativistic static perfect fluid sphere.
Over the last 90 years a tangle of specific perfect fluid spheres has been discovered,
with most of these specific examples seemingly independent from each other.
To bring some order to this chaos, in this article we develop several new transformation theorems that map perfect fluid spheres into perfect fluid spheres.  
These transformation theorems sometimes lead to unexpected connections between previously known perfect fluid spheres, sometimes lead to new previously unknown perfect fluid spheres, and in general can be used to develop a systematic way of classifying the set of all perfect fluid spheres.

\newpage
\section[Solution generating theorems for the TOV equation]{\centerline{Solution generating theorems for the TOV equation} \label{Sec:TOV.Equations}}
\bigskip
\centerline{\bf Petarpa Boonserm, Matt Visser, and Silke Weinfurtner}
\bigskip
\centerline{Electronic preprint gr-qc/0607001.}

\vskip 1 cm

\noindent
 The Tolman--Oppenheimer--Volkov [TOV] equation constrains the
  internal structure of general relativistic static perfect fluid
  spheres. We develop several ``solution generating'' theorems for the
  TOV, whereby any given solution can be ``deformed'' to a new
  solution.  Because the theorems we develop work directly in terms of
  the physical observables --- pressure profile and density profile
  --- it is relatively easy to check the density and pressure profiles
  for physical reasonableness. This work complements our previous
  article [Phys.~Rev.~{\bf D71} (2005) 124307; gr-qc/0503007] wherein
  a similar ``algorithmic'' analysis of the general relativistic
  static perfect fluid sphere was presented in terms of the spacetime
  geometry --- in the present analysis the pressure and density are
  primary and the spacetime geometry is secondary. In particular, our
  ``deformed'' solutions to the TOV equation are conveniently
  parameterized in terms of $\delta \rho_c$ and $\delta p_c$, the
  \emph{shift} in the central density and central pressure.

\newpage
\section[Analogue model of a FRW universe in \emph{realistic} Bose--Einstein condensates]
{\centerline{Analogue model of a FRW universe in Bose--Einstein condensates:} \centerline{Application of the classical field method} \label{Sec:Application.Classical.Field.Methods}}
\bigskip
\centerline{\bf Piyush Jain, Silke Weinfurtner, Matt Visser, and C.~W. Gardiner}
\bigskip
\centerline{Electronic preprint cond-mat/0705.2077}
\bigskip
\centerline{Accepted for publication in {\bf Physical Review A}}

\vskip 1 cm

\noindent
Analogue models of gravity have been motivated by the possibility of investigating phenomena not readily accessible in their cosmological counterparts. 
In this paper, we investigate the analogue of cosmological particle creation in a Friedmann--Robertson--Walker universe by numerically simulating a Bose--Einstein condensate with a time-dependent scattering length. In particular, we
focus on a two-dimensional homogeneous condensate using the classical field method via the truncated Wigner approximation. We show that for various forms of the scaling function the particle production is consistent with the underlying theory in the long wavelength limit. In this context, we further discuss the implications of modified dispersion relations that arise from the microscopic theory of a weakly interacting Bose gas.

\newpage
\section[Effective refractive index tensor for weak field gravity]
{\centerline{Effective refractive index tensor for weak field gravity} \label{Sec:Effective.Refactive.Index}}
\bigskip
\centerline{\bf Petarpa Boonsern, Celine Cattoen, Tristan Faber, Matt Visser, and Silke Weinfurtner}
\bigskip
\centerline{Electronic preprint gr-qc/0411034.}
\bigskip
\centerline{Published as {\bf Class. Quant. Grav. 22 (2006) 1905-1916}}

\vskip 1 cm

\noindent
 Gravitational lensing in a weak but otherwise arbitrary gravitational field can be described in terms of a $3 \times 3$ tensor, the ``effective refractive index''. If the sources generating the gravitational field all have small internal fluxes, stresses, and pressures, then this tensor is automatically isotropic and the ``effective refractive index'' is simply a scalar that can be determined in terms of a classic result involving the Newtonian gravitational potential. In contrast if anisotropic stresses are ever important then the gravitational field acts similarly to an anisotropic crystal. We derive simple formulae for the refractive index tensor, and indicate some situations in which this will be important.

\chapter{Klein--Gordon equation and exterior derivatives\label{Sec:KGE.Exterior.Derivatives}}
%
We use this appendix to enable readers that are somewhat unfamiliar with the notation of exterior differential forms to relate to the alternative representation of the Klein--Gordon equation in terms of exterior differential forms:~\footnote{Note that ``colored'' indices are internal indices.}
\begin{equation}
\label{Eq:KGE.Forms.Differentials}
\ast d(\ast d\hat\phi) = \frac{(-1)^{n}}{\sqrt{-g}}  \, \partial_{\Red{a}}  \left( \sqrt{-g} \, g^{\Red{a}\Blue{b}} \, \partial_{\Blue{b}}\hat\phi  \right)  \, .
\end{equation}
\section{Differential forms and exterior derivatives in brief\label{Sec:Diff.Forms.Exterior.Derivatives}}
A $s$-form is a completely antisymmetric tensor of type $T_{s}^{0}$. For example a $0$-form corresponds to a scalar, a $1$-form to a covector, a $2$-form is a covariant tensor antisymmetric in both indices, and so on.
Exterior derivatives of a tensor produce covariant objects involving derivatives of the tensor, without the notation of connexions (or any parallel transport), such that completely anti-symmetrized partial derivatives of $s$-forms transform as tensors.
The exterior derivative (denoted as $d$) transforms a $s$-form $F$ into a $(s+1)$-form $dF$,
\begin{equation}
\label{Eq:Exterior.Derivative}
\{ dF \}_{a_{1} \, a_{2} \, ... \, a_{s+1}} = (s+1) \, \partial_{[ a_{1}} F_{a_{2} \, a_{3} \, ... \, a_{s+1}]} \,.
\end{equation}
The Levi--Civita tensor is a $n$-form defined by
\begin{equation}
\epsilon_{a_{1} \, a_{2} \, ... \, a_{n}} = \sqrt{-\det(g_{ab})} \, \mathrm{signum}(a_{1} \, a_{2} \, ... \, a_{n})=\sqrt{-g} \, \mathrm{sig} (a_{1} \, a_{2} \, ... \, a_{n}) \, , 
\end{equation}
where the last step is purely cosmetic, to shorten the expressions below. 
We can straightforwardly write down the inverse Levi--Civita symbol as
\begin{equation}
\epsilon^{a_{1} \, a_{2} \, ... \, a_{n}} = \frac{1}{ \sqrt{-\det(g)} } \,  \mathrm{sig}(a_{1} \, a_{2} \, ... \, a_{n}) \, ,
\end{equation}
so that
\begin{equation}
\epsilon^{a \, \Red{c_{1} \, ... \, c_{n-1}}} \, \epsilon_{b \, \Red{c_{1} \, ... \, c_{n-1}}} = (n-1)! \, \delta^{a}{}_{b} \, . 
\end{equation}
Notice that $\mathrm{sig}$ is proportional to an anti-symmetic tensor (form), and hence
\begin{equation}
\mathrm{sig}(a_{1} \, a_{2} \, ... \, a_{n}) = (-1)^{n-1} \, \mathrm{sig}(a_{2} \, ... \, a_{n} \, a_{1}) \, .
\end{equation} 
Finally, we introduce the Hodge star operator (denoted by $\ast$) as the operation that transforms a $s$-form into a $(n-s)$-form,
\begin{equation}
\{ \ast F \}_{a_{1} \, a_{2} \, ... \, a_{n-s}} = \frac{1}{s!} \, \epsilon_{a_{1} \, a_{2} \, ... \, a_{n-s}}{}^{\Red{b_{1} \, b_{2} \, ... b_{s}}} \, \{ F \}_{\Red{b_{1} \, b_{2} \, ... \, b_{s}}} \, ,
\end{equation}
and are now ready to show the equivalence between the index-free version of the Klein--Gordon equation, as pointed out in Eq.~\refb{Eq:KGE.Forms.Differentials}.
\section{Index-free Klein--Gordon equation\label{Sec:Diff.Forms.Exterior.Derivatives.Derivation}}
The field operator $\hat\phi$ is a scalar operator, hence a $0$-form, and applying Eq.~\refb{Eq:Exterior.Derivative} we get,
\begin{equation}
d\hat\phi = \hat \phi_{,a} \, .
\end{equation}
We further apply the Hodge star operator and get,
\begin{equation}
\ast (d\hat\phi) = \ast \hat\phi_{,a} = \epsilon_{a_{1} \, a_{2} \, ... \, a_{n-1}}{}^{\Red{b}}  \, \hat\phi_{,\Red{b}} =\sqrt{-g} \, \mathrm{sig}(a_{1} \, a_{2} \, ... \, {\Red a_{n}}) \, g^{{\Red a_{n}} \, \Blue{b}} \, \hat \phi_{, \Blue{b}} \, .
\end{equation}
Once again we need to apply the exterior derivative,
\begin{eqnarray}
d(\ast d\hat\phi) &=& n \, \partial_{[c} \left( \mathrm{sig}(a_{1} \, a_{2} \, ... \, a_{n-1\,]} \, {\Red a_{n}}) \, g^{{\Red{a_{n}}} \, {\Blue b_{1}}} \, \hat\phi_{,{\Blue b_{1}}} \sqrt{-g} \right) \, , \\
&=& n \, (-1)^{n-1} \; \mathrm{sig}([a_{1} \, a_{2} \, ... \, {\Red a_{n}}) \, \partial_{c]}  \left( g^{{\Red a_{n}} \, {\Blue b_{1}}} \, \hat\phi_{,{\Blue b_{1}}} \sqrt{-g} \right)  \, ,
\end{eqnarray}
and subsequently the Hodge star operator leads us to the final result:
\begin{eqnarray}
\ast d(\ast d\hat \phi) 
&=& \frac{n \, (-1)^{n-1}}{n!} \, \epsilon^{{\Green b_{1} \, b_{2} \, ... \, b_{n-1}} \, {\Cyan b_{n}}} \, \frac{(-1)}{\sqrt{-g}} \, \mathrm{sig}({\Green b_{1} \, b_{2} \, ... \, b_{n-1}} \, {\Red a_{n}}) \, \partial_{\Cyan b_{n}}  \left( \sqrt{-g} \, g^{{\Red a_{n}} \, {\Blue b_{1}}} \, \hat\phi_{,{\Blue b_{1}}} \right) \, , \\
&=& \frac{(-1)^{n}}{\sqrt{-g}} \, \delta^{{\Cyan b_{n}}}{}_{{\Red a_{n}}} \, \partial_{{\Cyan b_{n}}} \, \left( \sqrt{-g} \, g^{{\Red a_{n}} \, {\Blue b_{1}}} \, \hat\phi_{,{\Blue b_{1}}} \right) \, , \\
&=& \frac{(-1)^{n}}{\sqrt{-g}}  \, \partial_{\Red a_{n}} \, \left( \sqrt{-g} \, g^{{\Red a_{n}} \, {\Blue b_{1}}} \, \hat\phi_{,{\Blue b_{1}}} \right) \, .
\end{eqnarray}
Relabeling $a_{n} \to a$ and $b_{1} \to b$ yields Eq.~\refb{Eq:Exterior.Derivative}.

\chapter{Finsler and co--Finsler geometries \label{Sec:Finsler}}  
%
Finsler \index{Finsler!geometry} geometries are sufficiently unusual
that a brief discussion is in order --- especially in view of the fact
that the needs of the physics community are often somewhat at odds
with what the mathematical community might view as the most important
issues.  Below are some elementary results, where we emphasise that
for the time being we are working with ordinary ``Euclidean
signature'' Finsler \index{Finsler!geometry} geometry. For general
references, see~\cite{Cartan:1934aa,Rund:1959aa,Bao:1996aa,Bao:2000aa,Shen:2001aa}.

\section{Basics}

\noindent\emph{Euler theorem:} If $H(z)$ is homogeneous of degree $n$ then
\begin{equation}
z^i\; {\partial H(z)\over\partial z^i} = n \; H(z).
\end{equation}

\noindent\emph{Finsler \index{Finsler!function} function:} Defined on
the ``slit tangent bundle'' $T_{\neq0}(M)$ such that $F:T_{\neq0}(M)\to[0,+\infty)$
where
\begin{equation}
F(x,t): \quad F(x,\lambda  t) = \lambda \; F(x, t),
\end{equation}
and
\begin{equation}
T_{\neq0}(M) = \bigcup_{x\in M} \left[ T_x-\{\vec 0\} \right].
\end{equation}
That is, the Finsler function is a defined only for nonzero tangent
vectors $t \in [T_x-\{\vec 0\}]$, and for any fixed direction is
linear in the size of the vector.

\noindent\emph{Finsler \index{Finsler!distance} distance:} 
\begin{equation}
d_\gamma(x,y) = \int_x^y F(x(\tau),dx/d\tau) \; d\tau; \qquad \tau=\;\hbox{arbitrary parameter}.
\end{equation}

\noindent\emph{Finsler \index{Finsler!metric} metric:} 
\begin{equation}
g_{ij}(x,t) = {1\over2} {\partial^2 [F^2(x,t)]\over\partial t^i\;\partial t^j}.
\end{equation}

\noindent
The first slightly unusual item is the introduction of co--Finsler
\index{co--Finsler~geometry} structure:
\\
\emph{co--Finsler function:} Define a co--Finsler
\index{co--Finsler~geometry} structure on the cotangent bundle by
Legendre transformation of $F^2(x,t)$. That is:
\begin{equation}
G^2(x,p) = t^j(p) \; p_j - F^2(x,t(p))
\end{equation}
where $t(p)$ is defined by the Legendre transformation condition
\begin{equation}
{\partial [F^2]\over\partial t^j}(x,t) = p_j.
\end{equation}
Note
\begin{equation}
{\partial p_j\over\partial t^k} = {\partial [F^2]\over \partial t^j\; \partial t^k} = 2 g_{jk}(x,t),
\end{equation}
which is why we demand the Finsler \index{Finsler!metric} metric be
nonsingular.

\noindent\emph{Lemma:} $G(x,p)$ defined in this way is homogeneous of degree 1.
\\
\emph{Proof:} Note
\begin{equation}
z^i\; {\partial H(z)\over\partial z^i} = n \; H(z)
\end{equation}
implies
\begin{equation}
z^i\; {\partial \over\partial z^i} \left[{\partial^m\over(\partial z)^m} H(z)\right]  
= (n-m) \;\left[{\partial^m\over(\partial z)^m} H(z)\right].
\end{equation}
In particular:
\begin{itemize}
\item $F^2$ is homogeneous of degree 2.
\item $g_{ij}$ is homogeneous of degree 0.
\item $\partial[F^2]/\partial t$ is homogeneous of degree 1.
\item Therefore $p(t)$  is homogeneous of degree 1, 
  \emph{and} $t(p)$ is homogeneous of degree 1.
\item Therefore $t(p) p -F^2(t(p))$ is homogeneous of degree 2.
\item Therefore $G(p)$ is homogeneous of degree 1.
\end{itemize}
Thus from a Finsler \index{Finsler!function} function $F(x,t)$ we can
always construct a co--Finsler \index{co--Finsler~geometry} function
$G(x,p)$ which is homogeneous of degree 1 on the cotangent bundle.

{From} the way the proof is set up it is clearly reversible --- if you
are given a co--Finsler \index{co--Finsler~geometry} function $G(x,p)$
on the cotangent bundle this provides a \emph{natural} way of
extracting the corresponding Finsler function:
\begin{equation}
F^2(x,t) = t \; p(t) - G^2(x,p(t)).
\end{equation}

\section{Connection with the quasi-particle PDE analysis}

{From} the PDE-based analysis we obtain the second-order system of PDEs
\begin{equation}
\partial_a \left( f^{ab}{}_{\A\B}\; \partial_b \theta^B \right)
+ \hbox{lower order terms} = 0.
\label{E:system}
\end{equation}
We are now generalizing in the obvious manner to any arbitrary number
$n$ of interacting BECs, but the analysis is even more general than
that --- it applies to any field-theory normal-mode analysis that
arises from a wide class of Lagrangian based systems~\cite{Barcelo:2001ay,Barcelo:2002dp}.

Going to the eikonal\index{eikonal} approximation this becomes
\begin{equation}
f^{ab}_{\A\B} \; p_a p_b \; \epsilon^B + \hbox{lower-order terms} = 0,
\end{equation}
which leads (neglecting lower order terms for now) to the
Fresnel-like\index{Fresnel~equation} equation
\begin{equation}
\det[ f^{ab}_{\A\B} \; p_a p_b] = 0.
\end{equation}
But by expanding the $n\times n$ determinant ($n$ is the number of
fields, not the dimension of spacetime) we have
\begin{equation}
\det[ f^{ab}_{\A\B} \; p_a p_b]  = Q^{abcd\dots} \; p_a p_b p_c p_d \dots
\end{equation}
where if there are $n$ fields there will be $2n$ factors of $p$. 

Now define
\begin{equation}
Q(x,p) =  Q^{abcd\dots} \; p_a p_b p_c p_d \dots,
\end{equation}
and
\begin{equation}
G(x,p) = \sqrt[2n]{Q(x,p)} = [Q(z,p)]^{1/(2n)},
\end{equation}
then
\begin{itemize}
\item $Q(x,p)$ is homogeneous of degree $2n$.
\item $G(x,p)$ is homogeneous of degree 1, and hence is a co--Finsler
  \index{co--Finsler~geometry} function.
\item We can now Legendre transform $G\to F$, providing a chain
\begin{equation}
Q(x,p) \to  G(x,p) \to F(x,t).
\end{equation}
Can this route be reversed? 
\end{itemize}
\emph{Step 1:} We can always reverse $F(x,t)\to G(x,p)$ by Legendre transformation.

\noindent
\emph{Step 2:} We can always define
\begin{equation}
g^{ab} (x,p) = {1\over2} {\partial\over\partial p_a} {\partial\over\partial p_b} [G(x,p)^2],
\end{equation}
this is homogeneous of degree 0, but is generically not smooth at
$p=0$.

In fact, \emph{if} $g^{ab}(x,p)$ is smooth at $p=0$ then there exits a limit
\begin{equation}
g^{ab}(x,p\to 0)= \Bar{g\,}^{ab}(x),
\end{equation}
but since $g^{ab}(x,p)$ is homogeneous of degree 0 this implies
\begin{equation}
g^{ab}(x,p) = \Bar{g\,}^{ab}(x) \qquad [\forall p],
\end{equation}
and so the geometry simplifies Finsler $\to$ Riemann.

This observation suggests the following definition.

\noindent\emph{Definition:} A co--Finsler \index{co--Finsler~geometry}
function $G(x,p)$ is $2n$-smooth iff the limit
\begin{equation}
{1\over(2n)!} \lim_{p\to0} \left({\partial\over\partial p}\right)^{2n} G(x,p)^{2n} =
\Bar{Q\,}^{abcd\dots}
\end{equation}
exists independent of the direction $p$ in which you approach zero.

\noindent\emph{Lemma:} If $G(x,p)$ is $2n$-smooth then
\begin{equation}
G(x,p)^{2n} = \Bar{Q\,}^{abcd\dots} \; p_a p_b p_c p_d \dots,
\end{equation}
and indeed
\begin{equation}
G(x,p) = \sqrt[2n]{\Bar{Q\,}^{abcd\dots} \; p_a p_b p_c p_d \dots}.
\end{equation}

\noindent
\emph{Proof:} $G^{2n}$ is homogeneous of degree $2n$, so
$(\partial/\partial p)^{2n} G^{2n}$ is homogeneous of degree 0.
Therefore if the limit
\begin{equation}
{1\over(2n)!} \lim_{p\to0} \left({\partial\over\partial p}\right)^{2n} G(x,p)^{2n} =
\Bar{Q\,}^{abcd\dots}
\end{equation}
exists, it follows that
\begin{equation}
{1\over(2n)!}  \left({\partial\over\partial p}\right)^{2n} G(x,p)^{2n} =
\Bar{Q\,}^{abcd\dots} \qquad [\forall p],
\end{equation}
and so the result follows.

\noindent
\emph{Special case $n=1$:} If $G(x,p)$ is 2-smooth then
\begin{equation}
{1\over2}  {\partial^2\over\partial p^a \;\partial p_b } G(x,p)^{2} =
\Bar{Q\,}^{ab} = g^{ab}(x,p),
\end{equation}
and co--Finsler \index{co--Finsler~geometry} $\to$ Riemann.

\noindent 
These observations have a number of implications:
\begin{itemize}
\item For all those co--Finsler \index{co--Finsler~geometry} functions
  that are $2n$ smooth we \emph{can} recover the tensor
  $Q^{abcd\dots}$.
\item Not all co--Finsler \index{co--Finsler~geometry} functions are
  $2n$ smooth, and for those functions we \emph{cannot} extract
  $Q^{abcd\dots}$ in any meaningful way.
\item But those specific co--Finsler \index{co--Finsler~geometry}
  functions that arise from the leading symbol of a 2nd-order system
  of PDEs are naturally $2n$-smooth, and so for the specific
  co--Finsler \index{co--Finsler~geometry} structures we are physically
  interested in
\begin{equation}
Q(x,p) \leftrightarrow G(x,p) \leftrightarrow F(x,t).
\end{equation}
\item Therefore, in the physically interesting case the Finsler
  function $F(x,t)$ encodes all the information present in
  $Q^{abcd\dots}$.
\end{itemize}

\noindent
\emph{Special case $n=2$:} For two fields (appropriate for our 2-BEC
\index{2-BEC~system} system), we can follow the chain
\begin{equation}
\mathbf{f}^{ab} \to Q(x,p) \leftrightarrow G(x,p) \leftrightarrow F(x,t)
\end{equation}
to formally write
\begin{equation}
ds^4 = g_{abcd} \; dx^a dx^b dx^c dx^d,
\end{equation}
or
\begin{equation}
ds = \sqrt[4]{g_{abcd} \; dx^a dx^b dx^c dx^d}.
\end{equation}
This is one of the ``more general'' cases Riemann alludes to in his
inaugural lecture of 1854~\cite{Riemann:1873aa,Clifford:1873aa}.

This discussion makes it clear that the general geometry in our 2-BEC
\index{2-BEC~system} system is a 4-smooth Finsler geometry. It is only
for certain special cases that the Finsler geometry specializes first
to ``multi-metric'' and then to ``mono-metric'' Riemannian geometries.

\section{Lorentzian signature Finsler geometries}

The distinction between Finsler and pseudo--Finsler
\index{pseudo--Finsler~spacetime} geometries has to do with the
distinction between elliptic \index{elliptic~system} and hyperbolic
\index{hyperbolic~system} PDEs. Elliptic \index{elliptic~system} PDEs
lead to ordinary Finsler geometries, hyperbolic PDEs lead to
\emph{pseudo}--Finsler geometries.

Remember that in special relativity we typically define
\begin{equation}
d_\gamma(x,y) = \int_x^y \sqrt{ g_{ab} (dx^a/d\tau) (dx^b/d\tau)} d\tau,
\end{equation}
then
\begin{itemize}
\item $d_\gamma(x,y) \in I\!\!R^+$ for spacelike paths;
\item $d_\gamma(x,y) =0$ for light-like (null) paths;
\item $d_\gamma(x,y) \in I\!\!I^+$ for timelike paths;
\end{itemize}
The point is that even in special relativity\index{special~relativity}
(and by implication in general relativity)\index{general~relativity}
``distances'' do not have to be real numbers. This is why physicists
deal with \emph{pseudo}--Riemannian [Lorentzian] geometries, not
(strictly speaking) Riemannian geometries.

To see how this generalizes in a Finsler situation let us first
consider a co--Finsler \index{co--Finsler~geometry} structure that is
multi-metric, that is:
\begin{equation}
Q(x,p) = \Pi_{i=1}^n (g^{ab}_i p_a p_b),
\end{equation}
where each one of these $n$ factors contains a Lorentzian signature
matrix and so can pass through zero.  Then
\begin{equation}
G(x,p) = \sqrt[2n]{ \Pi_{i=1}^n (g^{ab}_i p_a p_b) },
\end{equation}
and 
\begin{equation}
G(x,p) \in \exp\left({i\pi \ell\over2n}\right) \; I\!\!R^+,
\end{equation}
where
\begin{itemize}
\item $\ell=0 \to G(x,p)\in  I\!\!R^+ \to$ outside all $n$ signal cones;
\item $\ell=n \to G(x,p)\in  I\!\!I^+ \to$ inside all $n$ signal cones.
\end{itemize}
So we can now define
\begin{itemize}
\item Spacelike $\leftrightarrow$ outside all $n$ signal cones $\leftrightarrow$ $G$ real;
\item Null $\leftrightarrow$ on any one of the $n$ signal cones $\leftrightarrow$ $G$ zero;
\item Timelike $\leftrightarrow$ inside all $n$ signal cones $\leftrightarrow$ $G$ imaginary;
\item plus the various ``intermediate'' cases:
\begin{equation}
\hbox{``intermediate''} \leftrightarrow  \hbox{inside $\ell$ of $n$ signal cones} 
\leftrightarrow G\in i^{\ell/n} \times I\!\!R^+.
\end{equation}
\end{itemize}
Now this basic idea survives even if we do not have a multi-metric
theory. The condition $Q(x,p)=0$ defines a polynomial of degree $2n$,
and so defines $n$ nested sheets (possibly crossing in places).
Compare with Courant and Hilbert's discussion of the Monge
cone~\cite{Courant:1989aa}.

That is:
\begin{eqnarray*}
Q(x,p)=0
&\Leftrightarrow&
Q(x,(E,\vec p)) =0;
\\
&\Leftrightarrow& 
\hbox{polynomial of degree $2n$ in $E$ for any fixed $\vec p$;}
\\
&\Leftrightarrow& 
\hbox{in each direction $\exists$ $2n$ roots in $E$;}
\\
&\Leftrightarrow& 
\hbox{corresponds to $n$ [topological] cones.}
\end{eqnarray*}
(These are topological cones, not geometrical cones, and the roots
might happen to be degenerate.)

\noindent\emph{Question:} Should we be worried by the fact that the
co-metric $g^{ab}$ is singular on the signal cone? (In fact on all $n$
of the signal cones.) Not really.  We have
\begin{equation}
G(x,p) = \sqrt[2n]{\Bar{Q\,}^{abcd\dots} \; p_a p_b p_c p_d \dots},
\end{equation}
so
\begin{equation}
g^{ab}(x,p) = {1\over2}  
{\partial^2\over\partial p^a \;\partial p_b } \left( \sqrt[n]{Q(x,p)} \right)
=
{1\over2n} {\partial\over\partial p_b } 
\left\{ Q^{{1\over n}-1} \; Q^{abcd\dots} \; p_b p_c p_d \dots \right\},
\end{equation}
whence
\begin{eqnarray}
g^{ab}(x,p) &=& {1\over2n} 
Q^{{1\over n}-1} \; Q^{abcd\dots} \; p_c p_d \dots 
\\
&&
+ {1\over2n} \left({1\over n}-1\right) Q^{{1\over n}-2} \left[ Q^{acde\dots} \; p_c p_d p_e \dots \right]\; \left[  Q^{bfgh\dots} \; p_f p_g p_h \dots \right],
\nonumber
\end{eqnarray}
which we can write as
\begin{eqnarray}
g^{ab}(x,p) &=& {1\over2n} 
Q^{-(n-1)/n} \; Q^{abcd\dots} \; p_c p_d \dots 
\\
&&
- {1\over2n} {n-1\over n} Q^{-(2n-1)/n} 
\left[ Q^{acde\dots} \; p_c p_d p_e \dots \right]\; \left[  Q^{bfgh\dots} \; p_f p_g p_h \dots \right].
\nonumber
\end{eqnarray}
Yes, this naively looks like it's singular on the signal cone
where $Q(x,p)=0$.  But no, this is not a problem: Consider
\begin{equation}
g^{ab} p_a p_b = {1\over2n}  Q^{-(2n-1)/n} Q - {1\over2n} {n-1\over n} Q^{-(2n-1)/n} Q^2,
\end{equation}
then 
\begin{equation}
  g^{ab} p_a p_b = {1\over2n}\left(1-{n-1\over n}\right) Q^{1/n} = {1\over2n^2} \; Q^{1/n} = 0,
\end{equation}
and this quantity is definitely non-singular.

\section{Summary} 

In short: 
\begin{itemize}
\item pseudo--Finsler \index{pseudo--Finsler~spacetime} functions
  arise naturally from the leading symbol of hyperbolic
  \index{hyperbolic~system} systems of PDEs;
\item pseudo--Finsler \index{pseudo--Finsler~spacetime} geometries
  provide the natural ``geometric'' interpretation of a
  multi-component PDE before fine tuning;
\item In particular the natural geometric interpretation of our 2-BEC
  \index{2-BEC~system} model (before fine tuning) is as a 4-smooth
  pseudo--Finsler \index{pseudo--Finsler~spacetime} geometry.
\end{itemize}

\chapter{Some matrix identities \label{Sec:MatrixIdentities}}     
%
To simplify the flow of argument in the body of the thesis, here we
collect a few basic results on $2\times 2$ matrices that are used in
our analysis.

\section{Determinants}

\noindent{\bf Theorem:} For any two $2\times 2$ matrix
$A$:
\begin{equation}
\label{E:A-det2}
\det(A)=  {1\over2} \left\{ \tr[A]^2-\tr[A^2]\right\} . 
\end{equation}
This is best proved by simply noting
\begin{equation}
\det (A) = \lambda_1 \lambda_2 = 
{1\over2}\left[ (\lambda_1 +\lambda_2)^2 - (\lambda_1^2 +\lambda_2^2) \right]
= {1\over2} \left\{ \tr[A]^2-\tr[A^2]\right\} . 
\end{equation}
If we now define $2\times2$ ``trace reversal'' (in a manner
reminiscent of standard GR) by
\begin{equation}
\Bar{A\,} = A - \tr[A]\;\mathbf{I};  \qquad   \Bar{\Bar{A\,}} = A;
\end{equation}
then this looks even simpler
\begin{equation}
\label{E:A-det23}
\det(A)=  -{1\over2} \tr[A\;\Bar{A\,}] = \det(\Bar{A\,}). 
\end{equation}

\noindent
A simple implication is now:\\
\noindent{\bf Theorem:} For any two $2\times 2$ matrices
$A$ and $B$:
\begin{equation}
\label{E:A-2-matrices2}
\det(A+\lambda \; B) = \det(A) + \lambda\;\left\{ \tr[A] \tr[B] - \tr[  A\;B]
\right\}  + \lambda^2\;\det(B).
\end{equation}
which we can also write as
\begin{equation}
\label{E:A-2-matrices}
\det(A+\lambda \; B) = \det(A) - \lambda\; \tr[  A\;\Bar{B\,}]
+ \lambda^2\;\det(B).
\end{equation}
Note that $\tr[A\;\Bar{B\,}] = \tr[\Bar{A\,}\; B]$.

\section{Hamilton--Cayley theorems}

\noindent{\bf Theorem:} For any two $2\times 2$ matrix
$A$:
\begin{equation}
A^{-1} = \frac{\tr[A] \;\; \mathbf{I} - A }{\det[A]} = -{\Bar{A\,}\over\det{[\Bar{A\,}]}} .
\end{equation}

\noindent{\bf Theorem:} For any two $2\times 2$ matrix
$A$:
\begin{equation}
A^{1/2} = \;\pm \left\{
\frac{A\pm\sqrt{\det A} \;\; \mathbf{I} }{\sqrt{\tr[A]\pm 2  \sqrt{\det A}}}
\right\} .
\end{equation}

\chapter{Junction conditions\label{sec:Junction_conditions}}

For the benefit of readers who may not wish to deal with
differential forms and conjugate momenta as discussed in Chapter~\ref{Chap:Early.Quantum.Cosmology}, we now present a
``low-brow'' calculation of the $\alpha$ and $\beta$ coefficients
--- for the specific physics problem we are interested in --- in
two different coordinate systems.  Both co-ordinate systems lead
to the same physics result, and both give explicit expressions for
the conserved inner product, the modes, how to normalize them, and
of course the junction conditions.

%
\section{Basics}
%
The master PDE that everything starts from is
\begin{equation}
\partial_t \left( - {\hbar\over U} \; \partial_t \theta \right) + {n_0\hbar\over m} \; \nabla^2 \theta = 0
\end{equation}
where $n_0$, $m$ and $\hbar$ are constant, while
\begin{equation}
U(t) = U^{\reg{1}} \;\Theta(t^{\reg{1}\reg{2}}-t) + U^{\reg{2}}
\;\Theta(t-t^{\reg{1}\reg{2}}) \, .
\end{equation}
The most basic forms of the junction conditions are then that the
field operators $ \theta$, and its conjugate momentum $(\hbar/U) \;
\partial_t  \theta$, are continuous at $t^{\reg{1}\reg{2}}$.
If these conditions are not satisfied, then there is no way that
the master PDE can be satisfied at $t^{\reg{1}\reg{2}}$ --- there
would be delta-function contributions on the LHS that would not
cancel against anything on the RHS. These two junction conditions
can be written as
\begin{equation}
[{\theta}]=0;  \qquad\quad  \left[ {\hbar\over U} \; \partial_t
\theta \right] =0.
\end{equation}
Note that we are here working in terms of physical laboratory
time, and will continue to do so until we get to the subsection
where we explain how we  could have equally well done things in
terms of ``canonical time''.
One can read the conserved inner product off by starting directly
from the PDE itself --- noting that this PDE is defined for all
time, and then considering the quantity
\begin{equation}
\label{E:inner-product}
( \theta_a, \theta_b) = \int \left( \theta_a  \; {\hbar\over U} \;
\partial_t  \theta_b -  \theta_b  \; {\hbar\over U} \; \partial_t
\theta_a \right) d^d x \, ,
\end{equation}
where $ \theta_{a}$ and $ \theta_{b}$ are solutions of the master PDE.
Then by Gauss' law (or the fundamental theorem of calculus)
\begin{eqnarray}
&&
( \theta_a, \theta_b)\vert_\mathrm{final} - ( \theta_a, \theta_b)\vert_\mathrm{initial}
\\
&&
\quad \quad = \int \partial_t
\left(  \theta_a \; {\hbar\over U} \; \partial_t  \theta_b
-  \theta_b  \; {\hbar\over U} \; \partial_t  \theta_a \right) d^{d+1} x.
\nonumber
\end{eqnarray}
Now apply Leibnitz' rule, the PDE, and  an integration by parts to
obtain
\begin{equation}
( \theta_a, \theta_b)_\mathrm{final} - ( \theta_a,
\theta_b)_\mathrm{initial} = 0.
\end{equation}
So this is the correct conserved inner product for the
Klein--Gordon-like PDE we are starting with. Note that this
conservation law, the way we have set it up holds for any and all
$\mathrm{initial}$ and $\mathrm{final}$ times, regardless of which
side of the junction they are located.

%
\section{The two non-overlapping  simple regions}
%
In region $\Reg{1}$ ($t<t^{\reg{1}\reg{2}}$) the master PDE
reduces to
\begin{equation}
\partial_t \left( - {\hbar\over U^{\reg{1}}} \; \partial_t  \theta \right) + {n_0\hbar\over m} \; \nabla^2  \theta = 0,
\end{equation}
and can be rearranged to
\begin{equation}
- \partial_t^2  \theta + {n_0\; U^{\reg{1}} \over m} \; \nabla^2
\theta = 0.
\end{equation}
Introducing
\begin{equation}
(c^{\reg{1}})^2 = {n_0\; U^{\reg{1}} \over m} ,
\end{equation}
this becomes
\begin{equation}
- \partial_t^2   \theta + (c^{\reg{1}})^2 \; \nabla^2  \theta = 0,
\end{equation}
with the understanding that this PDE holds only in region
$\Reg{1}$ (\ie, $t<t^{\reg{1}\reg{2}}$).
Similarly, in region $\Reg{2}$ (\ie, $t>t^{\reg{1}\reg{2}}$) the
master PDE reduces to
\begin{equation}
- \partial_t^2  \theta + (c^{\reg{2}})^2 \; \nabla^2  \theta = 0.
\end{equation}
The junction conditions are unchanged, though for convenience we
can write them in terms of $c^2= n_0 \; U/m$ as
\begin{equation}\label{juncinit}
[ \theta]=0;  \qquad\quad  \left[ {1\over c^2} \; \partial_t  \theta
\right] =0.
\end{equation}

In each individual region the field satisfies the usual flat
spacetime Klein-Gordon equation, so solutions are of the form
\begin{equation}
\theta \propto \exp( i [\omega_{k} t - \mathbf{k}  \mathbf{x}]) \, ,
\end{equation}
where $\omega_{k}$ and $\mathbf{k}$ satisfy the dispersion
relation
\begin{equation}
\omega_{k} = c \, k \, ,
\end{equation}
and we have very carefully not yet specified any normalization for
these modes. (Nor is $c$ the same in the two regions, it is either
$c^{\reg{1}}$ or $c^{\reg{2}}$ as appropriate.)

One thing we can say without further calculation is this: In view
of the junction condition $[ \theta]=0$, that is $
\theta(t^{\reg{1}\reg{2}-},\mathbf{x}) =
\theta(t^{\reg{1}\reg{2}+},\mathbf{x})$,  the spatial position
dependence of the solutions to the PDE on the two sides of the
junction must be the same --- this implies that in terms of the
plane waves above we must enforce  $\mathbf{k}$ to be the same in
regions $\Reg{1}$ and $\Reg{2}$, and therefore
\begin{eqnarray}
{\omega^{\reg{1}}_{k}\over c^{\reg{1}}} = k =
{\omega^{\reg{2}}_{k} \over c^{\reg{2}}} \, .
\end{eqnarray}

%
\section{Normalization}
%
Now let us compare the conserved inner product for the PDE we are
physically interested in with the inner product for the naive
Klein--Gordon equation. In particular, if $\mathrm{initial}$ is
before the transition and $\mathrm{final}$ is after the
transition, then in terms of the naive ordinary Klein--Gordon
inner product the conserved inner product for our PDE is
\begin{equation}
 ( \theta_a, \theta_b)\vert_{\mathrm{initial}} = {\hbar\over U^{\reg{1}}} \; ( \theta_a, \theta_b)_\mathrm{naive} \, ;
\end{equation}
\begin{equation}
 ( \theta_a, \theta_b)\vert_{\mathrm{final}} = {\hbar\over U^{\reg{2}}} \; ( \theta_a, \theta_b)_\mathrm{naive} \, ;
\end{equation}
where $( \theta_a, \theta_b)_\mathrm{naive} = \int \left( \theta_a  \;
\partial_t  \theta_b -  \theta_b   \; \partial_t  \theta_a \right) d^d
x$ .

This means that properly normalized modes are
\begin{equation}
\sqrt{U\over\hbar} \; {1\over\sqrt{2\omega_{k}}} \; \exp( i
[\omega_{k} t - \mathbf{k}  \mathbf{x}]),
\end{equation}
which (ignoring a trivial overall constant factor, that does not
change from region $\Reg{1}$ to region $\Reg{2}$) we might as well
write as
\begin{equation}
 {c\over\sqrt{2\omega_{k}}} \; \exp( i [\omega_{k} t - \mathbf{k}  \mathbf{x}]).
\end{equation}
In particular in regions $\Reg{1}$ and $\Reg{2}$ we want to deal
with
\begin{equation}
\mf{1}(t,\mathbf{x}) =
{c^{\reg{1}}\over\sqrt{2\omega^{\reg{1}}_{k}}} \; \exp( i
[\omega^{\reg{1}}_{k} \,t - \mathbf{k} \mathbf{x}]),
\end{equation}
and
\begin{equation}
\mf{2}(t,\mathbf{x}) =
{c^{\reg{2}}\over\sqrt{2\omega^{\reg{2}}_{k}}} \; \exp( i
[\omega^{\reg{2}}_{k}\, t - \mathbf{k}  \mathbf{x}]),
\end{equation}
respectively.

%
\section{Applying the junction conditions}
%
In regions $\Reg{1}$ and $\Reg{2}$ we write the solutions of the
PDE as the real parts of
\begin{eqnarray}
 \theta^{\reg{1}} &=&  A {c^{\reg{1}}\over\sqrt{2\omega^{\reg{1}}_{k}}} \; \exp( i [\omega^{\reg{1}}_{k} t - \mathbf{k}  \mathbf{x}])
\nonumber\\
&& + B  {c^{\reg{1}}_{k}\over\sqrt{2\omega^{\reg{1}}_{k}}} \;
\exp(  i [-\omega^{\reg{1}}_{k} t - \mathbf{k}  \mathbf{x}]),
\end{eqnarray}
and
\begin{eqnarray}
 \theta^{\reg{2}} &=&  C {c^{\reg{2}}\over\sqrt{2\omega^{\reg{2}}_{k}}} \; \exp( i [\omega^{\reg{2}}_{k} t - \mathbf{k}  \mathbf{x}])
\nonumber\\
&& + D  {c^{\reg{2}}\over\sqrt{2\omega^{\reg{2}}_{k}}} \; \exp(  i
[-\omega^{\reg{2}}_{k} t - \mathbf{k} \mathbf{x}]).
\end{eqnarray}
Note that the $\mathbf{x}$ dependence is the same in all four of
these terms and so quietly factors out --- this is why we asserted
that $\mathbf{k}$ had to be the same in both regions.

Applying the first junction condition $[\theta]=0$, at the
transition time $t^{\reg{1}\reg{2}}$, and using the dispersion
relation implies
\begin{eqnarray}
&&\sqrt{c^{\reg{1}}} \left[A  \; \exp( i \omega^{\reg{1}}_{k}
t^{\reg{1}\reg{2}} ) + B \; \exp( - i \omega^{\reg{1}}_{k}
t^{\reg{1}\reg{2}} ) \right]
\nonumber\\
&& = \sqrt{c^{\reg{2}}} \left[ C \; \exp( i \omega^{\reg{2}}_{k}
t^{\reg{1}\reg{2}} ) + D  \; \exp( - i \omega^{\reg{2}}_{k}
t^{\reg{1}\reg{2}} )\right]. \qquad
\end{eqnarray}

The second junction condition, $[\partial_t \theta/c^2]=0$, applied
at the transition time $t^{\reg{1}\reg{2}}$,  implies (after
factoring out all the $c$'s and $\omega$'s, and using the
dispersion relation) that
\begin{eqnarray}
&& {1\over \sqrt{c^{\reg{1}}}} \left[A  \; \exp( i
\omega^{\reg{1}}_{k} t^{\reg{1}\reg{2}} ) - B \; \exp( - i
\omega^{\reg{1}}_{k} t^{\reg{1}\reg{2}} ) \right]
\nonumber\\
&& = {1\over \sqrt{c^{\reg{2}}}}
 \left[ C \; \exp( i \omega^{\reg{2}}_{k} t^{\reg{1}\reg{2}} ) - D  \; \exp( - i \omega^{\reg{2}}_{k} t^{\reg{1}\reg{2}} )\right]. \qquad
\end{eqnarray}

These two junction conditions are enough to completely specify the
transmission matrix of $\alpha$'s and $\beta$'s. Indeed, setting
$A\to 1$, $B\to 0$, $C\to \alpha$, and $D\to \beta$ and solving
for $\alpha$ and $\beta$ we find
\begin{equation}
\alpha = {1\over2} \left( \sqrt{c^{\reg{1}}\over c^{\reg{2}}}  +
\sqrt{c^{\reg{2}}\over c^{\reg{1}}} \right) \exp( -i
[c^{\reg{2}}-c^{\reg{1}}] k t^{\reg{1}\reg{2}});
\end{equation}
\begin{equation}
\beta= {1\over2} \left( \sqrt{c^{\reg{1}}\over c^{\reg{2}}}  -
\sqrt{c^{\reg{2}}\over c^{\reg{1}}} \right) \exp( i
[c^{\reg{2}}+c^{\reg{1}}] k t^{\reg{1}\reg{2}});
\end{equation}
which is exactly the result reported in the body of Chapter~\ref{Chap:Early.Quantum.Cosmology}.

%
\section{Using ``canonical time''}
%
Let us now pick another time coordinate.  Let us call it
``canonical time'' and define it by
\begin{equation}
T = \int {c(t)^2\over (c_*)^2} \; dt =  \int {U(t)\over U_*} \;dt,
\end{equation}
\begin{equation}
dT= {c(t)^2\over (c_*)^2} \; dt =  {U(t)\over U_*} \;dt ,
\end{equation}
where we have introduced a convenient constant reference point
$U_*$ and used this to define a convenient constant reference
speed
\begin{equation}
(c_*)^2 = {n_0 \; U_*\over m}.
\end{equation}

The the master PDE, which was in the original laboratory time
coordinate
\begin{equation}
\partial_t \left( - {\hbar\over U} \; \partial_t \theta \right) + {n_0\hbar\over m} \; \nabla^2 \theta = 0,
\end{equation}
now becomes
\begin{equation}
-\partial_T^2 \theta+ {n_0 U_*^2\over m \; U} \; \nabla^2 \theta = 0.
\end{equation}
Recognizing that ${n_0 U_*^2/(m \; U)} = {(c_*)^4/c^2}$
and defining
\begin{equation}
c_\mathrm{eff}(T) = {(c_*)^2\over c(T)},
\end{equation}
this can be re-written simply in the form of a ``parametrically
excited oscillator''
\begin{equation}
- \partial_T^2\theta + c^2_\mathrm{eff}(T) \; \nabla^2 \theta = 0.
\end{equation}

Assuming an exponential space dependence, separation of variables
yields
\begin{equation}
\theta(t,x) = \tilde\theta(t) \; \exp(-i\mathbf{k}\mathbf{x}),
\end{equation}
so that
\begin{equation}
\partial_T^2\tilde \theta = c^2_\mathrm{eff}(T) \; k^2 \;\tilde \theta.
\end{equation}

The junction conditions, which were originally  in terms of
laboratory time (refer to equation [\ref{juncinit}])
now become, in terms of canonical time, the very simple:
\begin{equation}
[\theta]=0;  \qquad\quad  \left[ \partial_T \theta \right] =0.
\end{equation}
The conserved inner product (\ref{E:inner-product})
can now be re-written
\begin{equation}
(\theta_a,\theta_b) =  {\hbar\over U_* } \;  \int \left( \theta_a
\;\partial_T \theta_b - \theta_b  \; \partial_T \theta_a \right) d^d x,
\end{equation}
that is
\begin{equation}
(\theta_a,\theta_b) =  {\hbar\over U_* } \;
(\theta_a,\theta_b)_\mathrm{naive}.
\end{equation}
So in ``canonical time'' coordinates we have a particularly simple
PDE, elementary junction conditions, and a trivial inner product
--- the only ``tricky'' thing is that we have to use the
``effective'' sound speed $c_\mathrm{eff} = {(c_*)^2/c}$.

Now in region $\Reg{1}$ the original master PDE can be cast into
the form
\begin{equation}
- \partial_T^2\theta + {(c_*)^4\over (c^{\reg{1}})^2} \; \nabla^2
\theta = 0,
\end{equation}
while in region $\Reg{2}$
\begin{equation}
- \partial_T^2\theta + {(c_*)^4\over (c^{\reg{2}})^2} \; \nabla^2
\theta = 0.
\end{equation}

In view of the form of the conserved inner product, the normalized
modes are
\begin{equation}
{1\over \sqrt{2\omega^{\reg{1}}_{k}}} \; \exp( i
[\omega^{\reg{1}}_{k} T - \mathbf{k}  \mathbf{x}]); \qquad
\omega^{\reg{1}}_{k} = {(c_*)^2\over c^{\reg{1}}} \; k;
\end{equation}
\begin{equation}
{1\over \sqrt{2\omega^{\reg{2}}_{k}}} \; \exp( i
[\omega^{\reg{2}}_{k} T - \mathbf{k}  \mathbf{x}]); \qquad
\omega^{\reg{2}}_{k} = {(c_*)^2\over c^{\reg{2}}} \; k.
\end{equation}
Note the ``odd looking'' form of the dispersion relation --- but
this is just because $\omega^{\reg{1}}_{k}$ and
$\omega^{\reg{2}}_{k}$ are not physical frequencies --- they are
``$T$-time frequencies''.

In regions $\Reg{1}$ and $\Reg{2}$ we write the solutions of the
PDE as the real parts of
\begin{eqnarray}
\theta_{\reg{1}} =  A {1\over\sqrt{2\omega^{\reg{1}}_{k}}} \;
\exp( i [\omega^{\reg{1}}_{k} T - \mathbf{k}  \mathbf{x}])
 + B  {1\over\sqrt{2\omega^{\reg{1}}_{k}}} \; \exp(  i
[-\omega^{\reg{1}}_{k} T - \mathbf{k}  \mathbf{x}]);
\end{eqnarray}

\begin{eqnarray}
\theta_{\reg{2}} =  C {1\over\sqrt{2\omega^{\reg{2}}_{k}}} \;
\exp( i [\omega^{\reg{2}}_{k} T - \mathbf{k}  \mathbf{x}])
 + D  {1\over\sqrt{2\omega^{\reg{2}}_{k}}} \; \exp(  i [-\omega^{\reg{2}}_{k} T - \mathbf{k}  \mathbf{x}]).
\end{eqnarray}

The first junction condition $[\theta]=0$, after applying the
dispersion relation is
\begin{eqnarray}
&& \sqrt{c^{\reg{1}}} \left [A \; \exp( i \omega^{\reg{1}}_{k}
T^{\reg{1}\reg{2}} ) + B  \; \exp( - i \omega^{\reg{1}}_{k}
T^{\reg{1}\reg{2}} )\right]
\\
&&
 =
 \sqrt{c^{\reg{2}}} \left[ C \; \exp( i \omega^{\reg{2}}_{k} T^{\reg{1}\reg{2}} ) + D \; \exp( - i \omega^{\reg{2}}_{k} T^{\reg{1}\reg{2}} )\right].
 \nonumber
\end{eqnarray}

The second junction condition $[\partial_T \theta]=0$, after
simplifying using the dispersion relation, now leads to
\begin{eqnarray}
&&{1\over \sqrt{c^{\reg{1}}}} \left [A \; \exp( i
\omega^{\reg{1}}_{k} T^{\reg{1}\reg{2}} ) - B  \; \exp( - i
\omega^{\reg{1}}_{k} T^{\reg{1}\reg{2}} )\right]
\\
&&
 =
{1\over \sqrt{c^{\reg{2}}}} \left[ C \; \exp( i
\omega^{\reg{2}}_{k} T^{\reg{1}\reg{2}} ) - D \; \exp( -
i\omega^{\reg{2}}_{k} T^{\reg{1}\reg{2}} )\right]. \nonumber
\end{eqnarray}
Up to irrelevant phases, these are the same equations as were
derived in laboratory time.

Now setting $A\to 1$, $B\to 0$, $C\to \alpha$, and $D\to \beta$
and the solving for $\alpha$ and $\beta$ we find
\begin{equation}
\alpha = {1\over2} \left( \sqrt{c^{\reg{1}}\over c^{\reg{2}}}  +
\sqrt{c^{\reg{2}}\over c^{\reg{1}}} \right) \exp( i
[\omega^{\reg{1}}_{k}-\omega^{\reg{2}}_{k}] T^{\reg{1}\reg{2}}),
\end{equation}
\begin{equation}
\beta= {1\over2} \left( \sqrt{c^{\reg{1}}\over c^{\reg{2}}}  -
\sqrt{c^{\reg{2}}\over c^{\reg{1}}} \right) \exp( i
[\omega^{\reg{1}}_{k}+\omega^{\reg{2}}_{k}] T^{\reg{1}\reg{2}}).
\end{equation}
The phases on $\alpha$ and $\beta$ are now slightly different,
but this is not new physics --- it has to do with the phases we
picked for our ``normalized modes'' --- these phases  are now
slightly different from the laboratory time calculation. Note the
magnitudes are completely unambiguous:
\begin{equation}
|\alpha| = {1\over2} \left| \sqrt{c^{\reg{1}}\over c^{\reg{2}}}  +
\sqrt{c^{\reg{2}}\over c^{\reg{1}}} \right|;
\end{equation}
\begin{equation}
|\beta|= {1\over2} \left| \sqrt{c^{\reg{1}}\over c^{\reg{2}}}  -
\sqrt{c^{\reg{2}}\over c^{\reg{1}}} \right|.
\end{equation}
In short, this appendix has served to verify that the key
technical parts of the calculation in Chapter~\ref{Chap:Early.Quantum.Cosmology} can mathematically be carried
out in a number of different ways that ultimately lead to the same
physical result.

\chapter{Numerical results for rainbow inflation\label{Sec:Intro.Rainbow.Inflation}}
The graphs in this appendix are bases on numerical work carried out by \emph{Piyush Jain} as part of a close collaborative effort between the current author, Piyush Jain, Professor Crispin Gardiner, and Professor Matt Visser.

\newpage
%
\section{Quasi-particle production during rainbow inflation for $t_{s}=1\times 10^{-4}$}
%
%
\begin{figure*}[!htb]
\begin{center}
\mbox{
\subfigure[$\,$ $t=0.25 \times t_{f}$ ; $t_{s}=1 \times 10^{-4}$. \label{Fig:timeslice_dsa_t_0.25tf_ts1e-4}]{\includegraphics[width=0.45\textwidth]{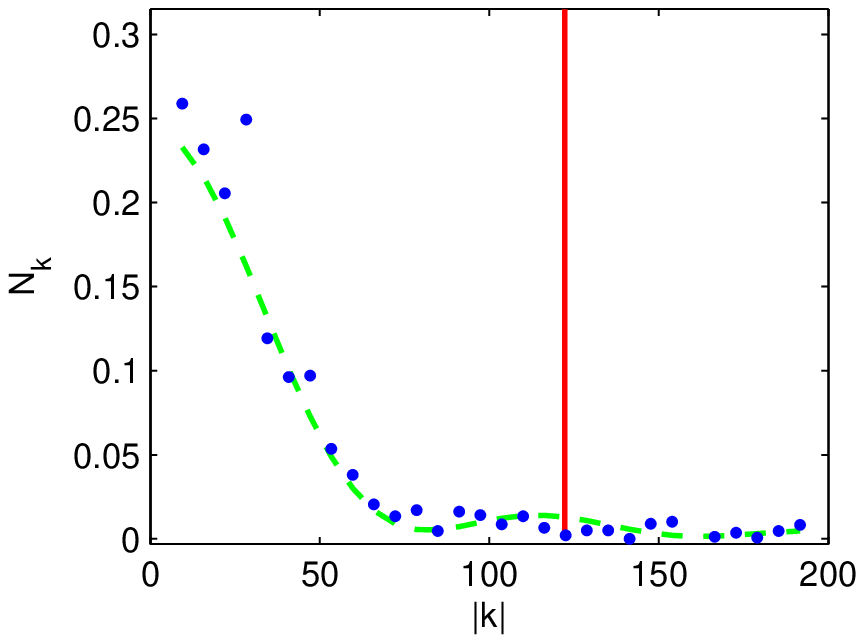}}
\hspace{0mm}
\subfigure[$\,$ $t=0.50 \times t_{f}$ ; $t_{s}=1 \times 10^{-4}$. \label{Fig:timeslice_dsa_t_0.50tf_ts1e-4}]{\includegraphics[width=0.45\textwidth]{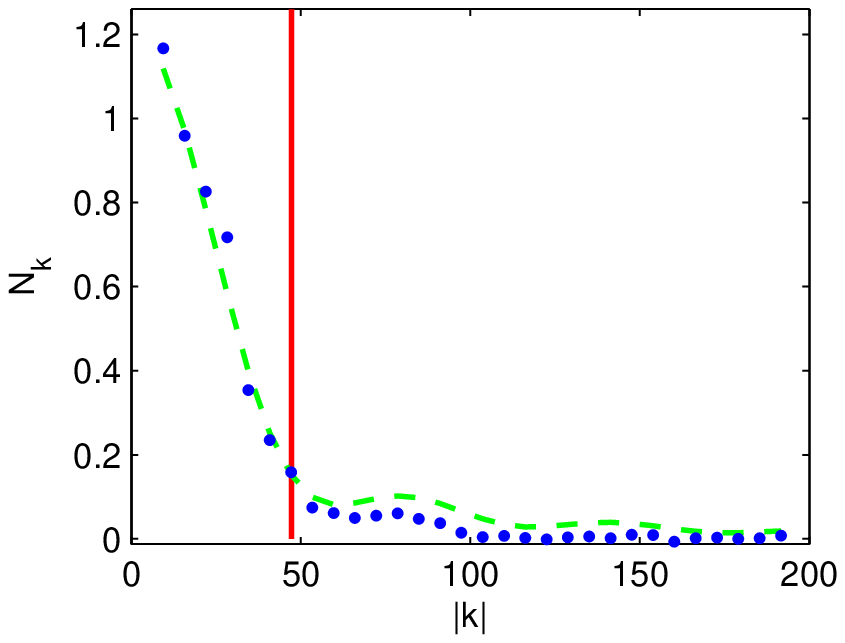}}
}
\mbox{
\subfigure[$\,$ $t=0.75 \times t_{f}$ ; $t_{s}=1 \times 10^{-4}$. \label{Fig:timeslice_dsa_t_0.75tf_ts1e-4}]{\includegraphics[width=0.45\textwidth]{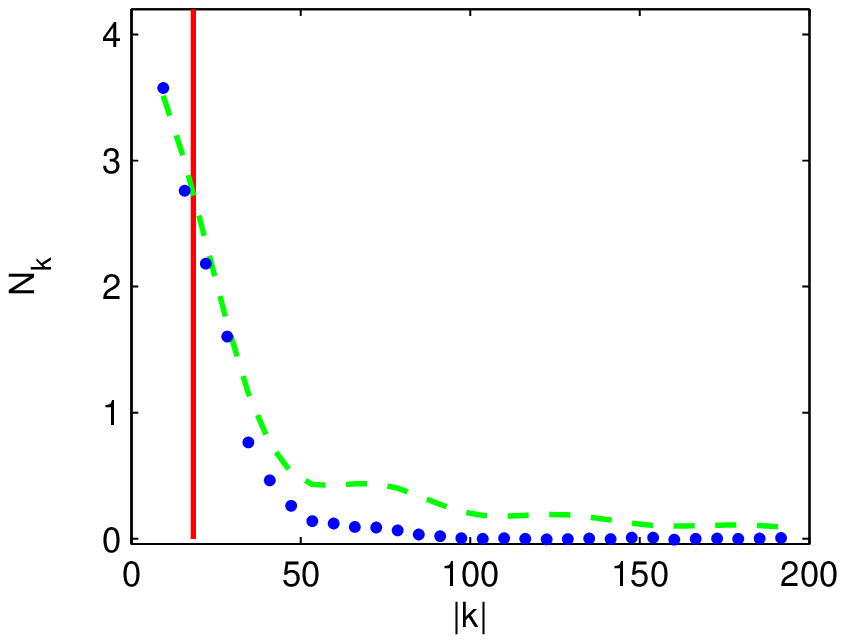}}
\hspace{0mm}
\subfigure[$\,$ $t=1.00 \times t_{f}$ ; $t_{s}=1 \times 10^{-4}$. \label{Fig:timeslice_dsa_t_1.00tf_ts1e-4}]{\includegraphics[width=0.45\textwidth]{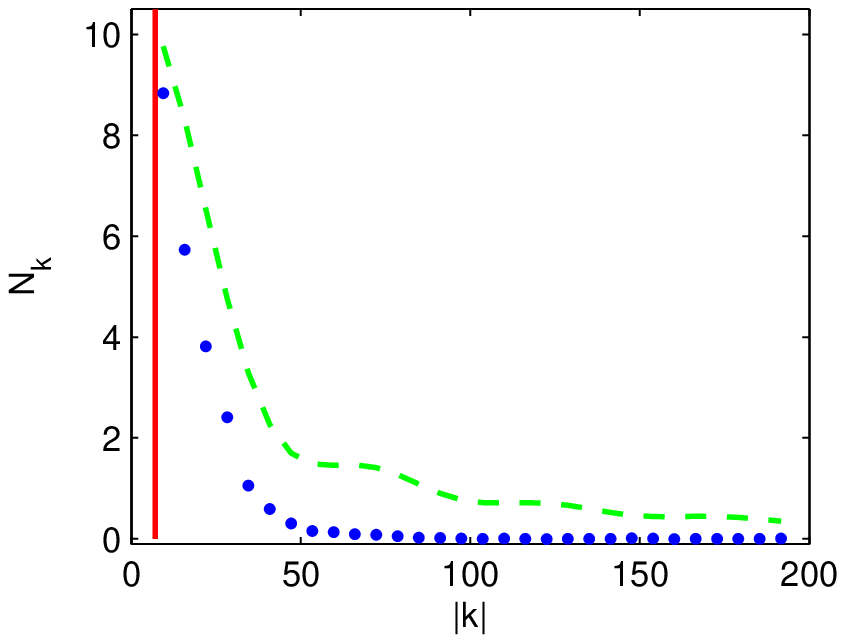}}     }
\caption[Quasi-particle production during rainbow inflation --- II.]{The four figures show time-slices of the  ``quasi-particle'' production $N_{k}$ in a de Sitter like spacetime, for $t_{s}=1\times 10^{-4}$. The blue dots represent the actual data obtained from the simulations for a realistic Bose gas with time-dependent interaction-strength~\cite{Jain:2006ki}. The dashed green line are the theoretical results for a finite de Sitter calculation obtained in the hydrodynamic limit, as presented in Sec.~\refb{Sec:Toy.Model.Conventional.Inflation}. The vertical red line indicates the borderline between the nature of the excitations: Everything to the left of the red line is phononic, while everything to right is trans-phononic/ free-particle like.}\label{Fig:timeslice_ts1e-4}.
\end{center}
\end{figure*}
%

%
\section{Qualitative particle production in Bose gas }
\begin{figure*}[!h]
\begin{center}
\mbox{
\subfigure[\label{modes_X2000_ts1e-3}]{
\includegraphics[width=0.50\textwidth]{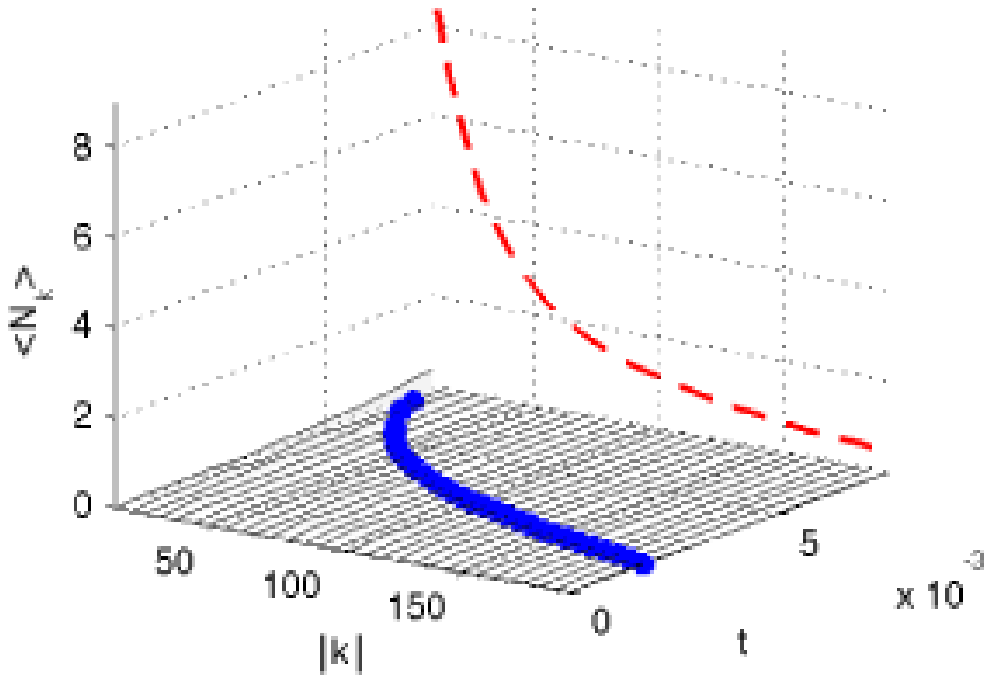}}
\hspace{0mm}
\subfigure[\label{ratio_X2000_ts1e-3}]{
\includegraphics[width=0.45\textwidth]{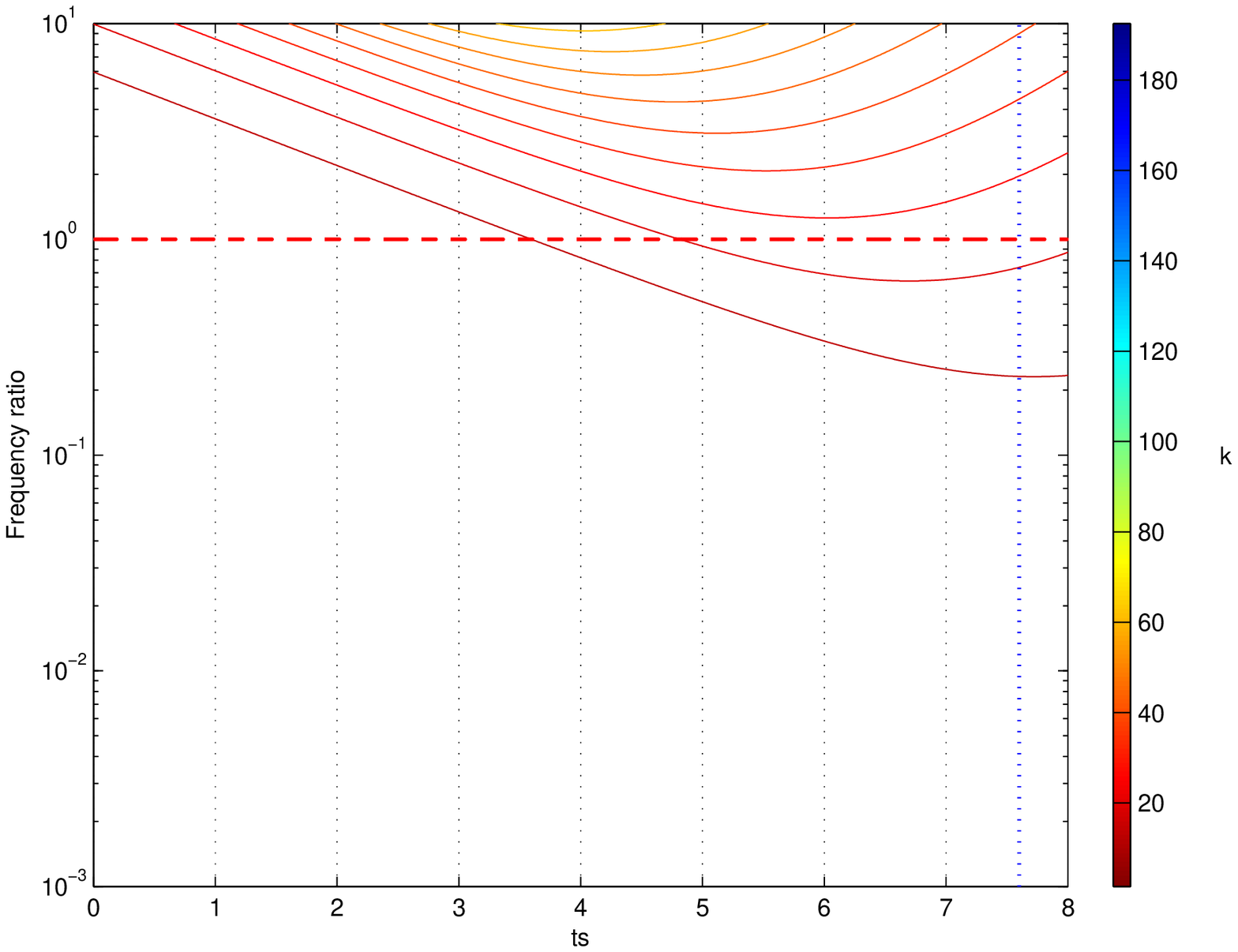}}
}
\caption[Qualitative particle production in Bose gas --- A.]{$t_{s}=1\times 10^{-3}$
}\label{Fig:deSitter.Compare.Nk.Ratio.1}
\end{center}
\end{figure*}

\begin{figure*}[!h]
\begin{center}
\mbox{
\subfigure[\label{modes_X2000_ts1e-4}]{
\includegraphics[width=0.50\textwidth]{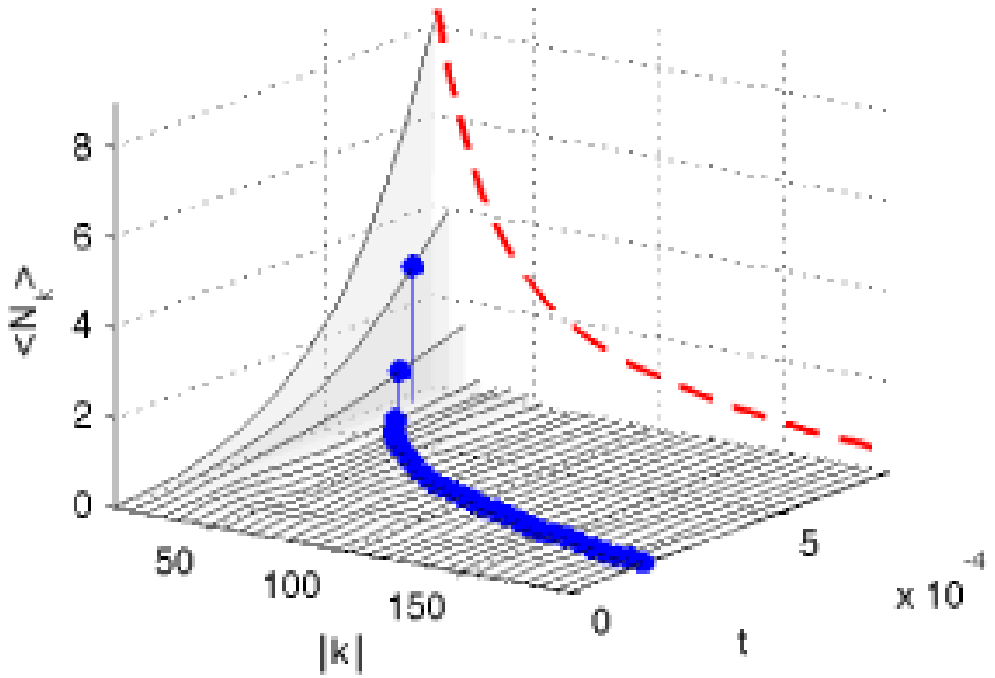}}
\hspace{0mm}
\subfigure[\label{ratio_X2000_ts1e-4}]{
\includegraphics[width=0.45\textwidth]{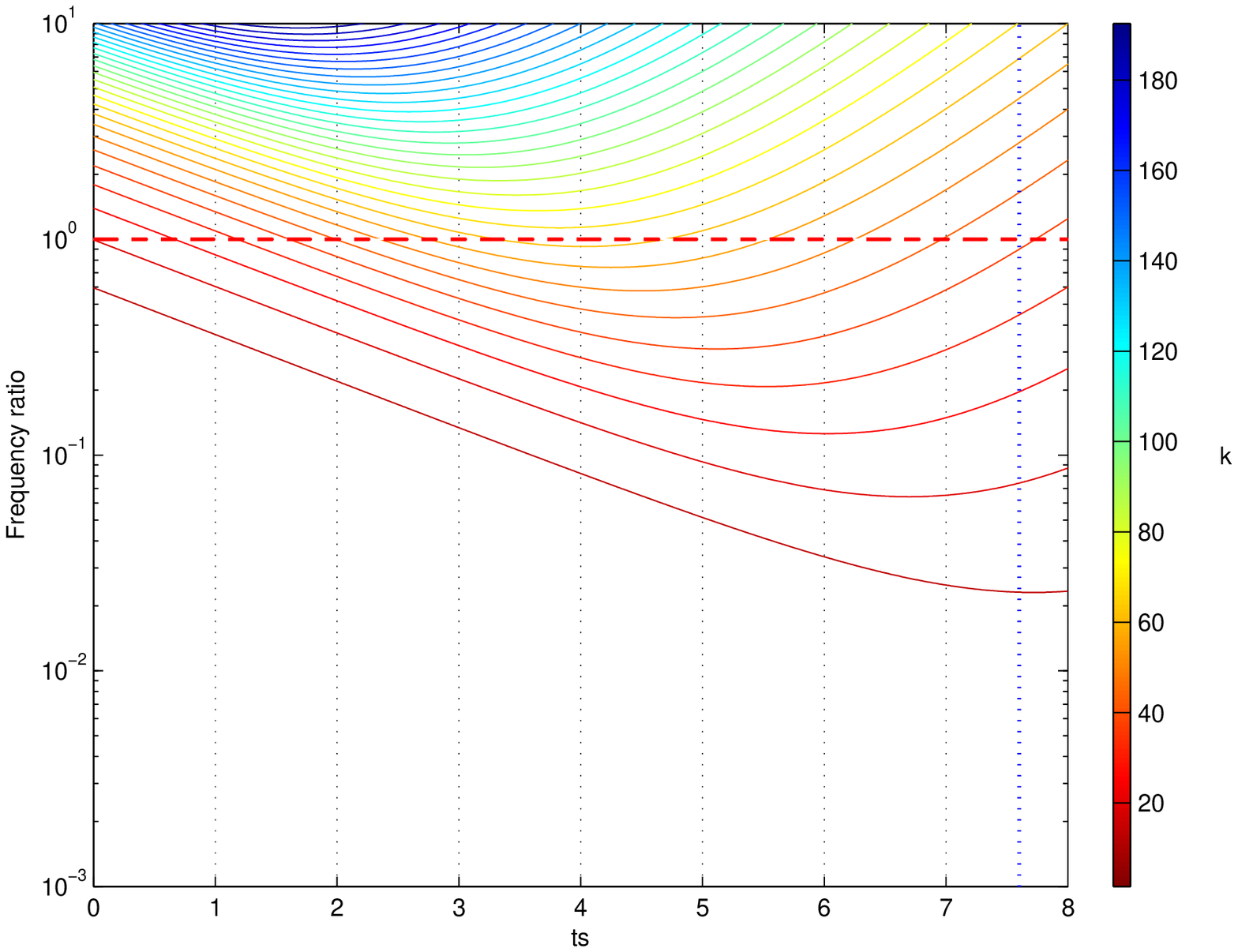}}
}
\caption[Qualitative particle production in Bose gas --- B.]{$t_{s}=1\times 10^{-4}$
}\label{Fig:deSitter.Compare.Nk.Ratio.2}
\end{center}
\end{figure*}

\begin{figure*}[!h]
\begin{center}
\mbox{
\subfigure[\label{modes_X2000_ts5e-5}]{
\includegraphics[width=0.50\textwidth]{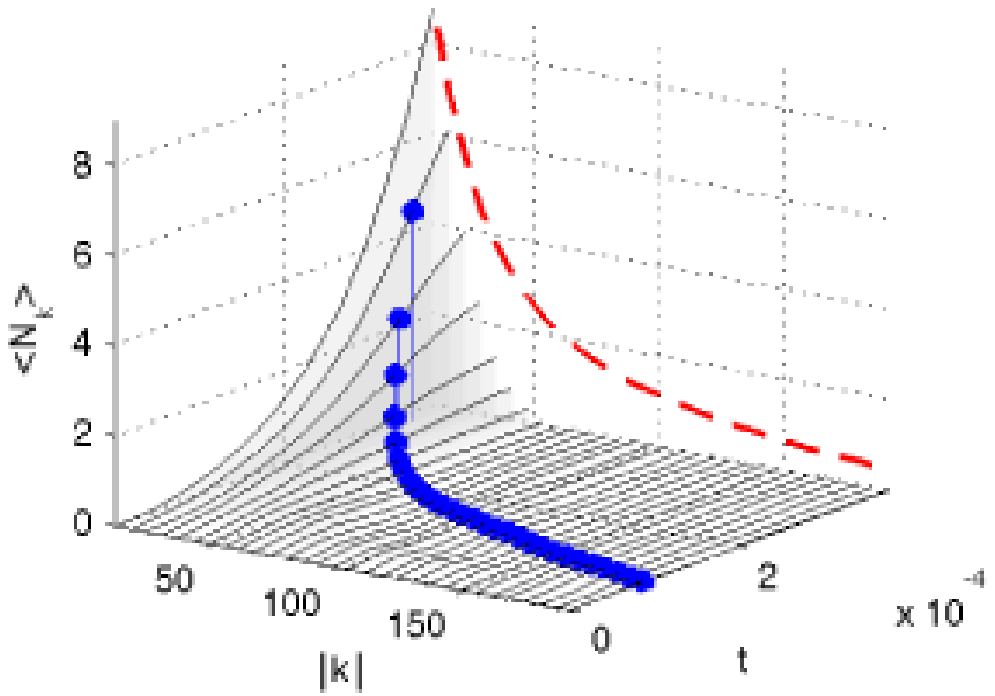}}
\hspace{0mm}
\subfigure[\label{ratio_X2000_ts5e-5}]{
\includegraphics[width=0.45\textwidth]{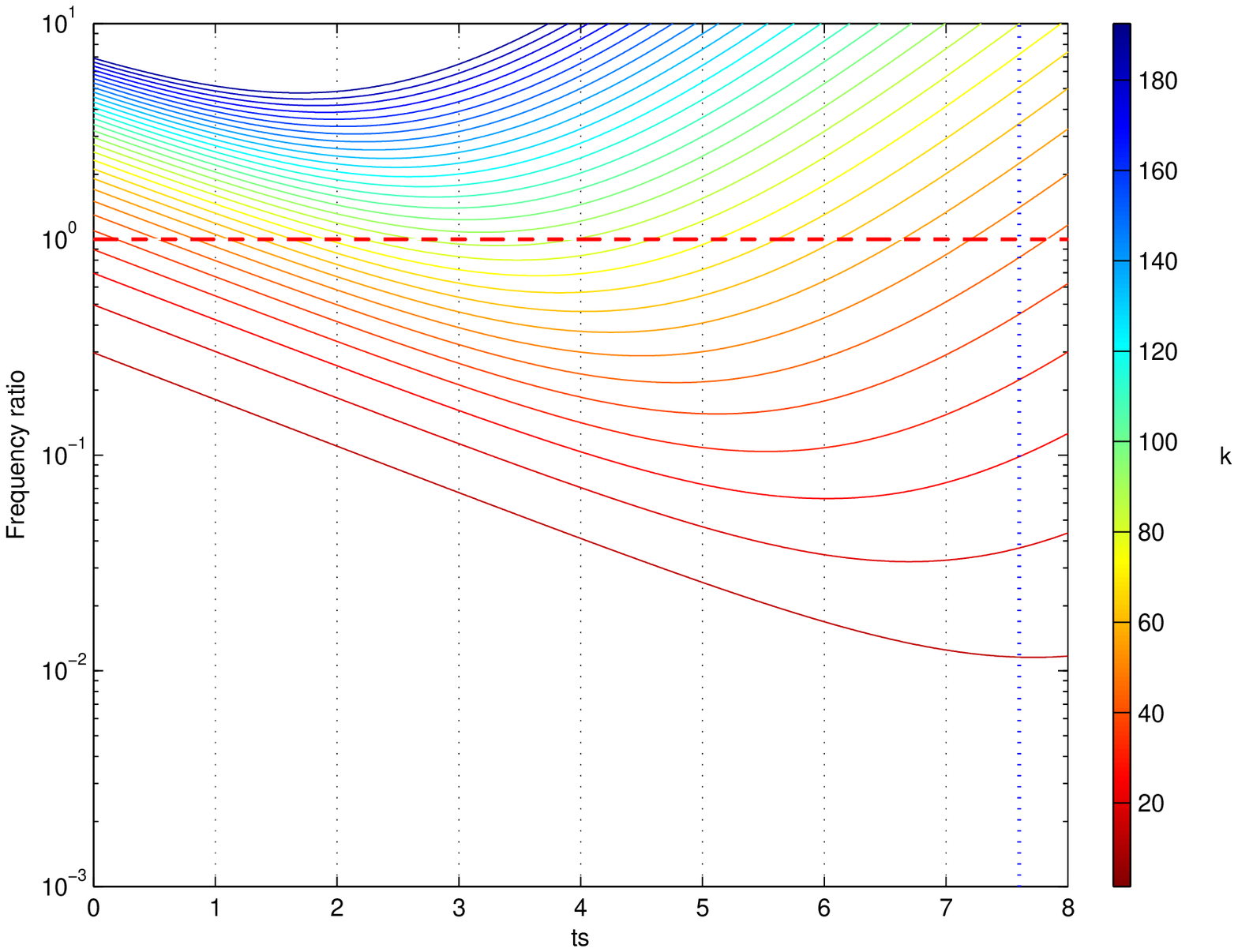}}
}
\caption[Qualitative particle production in Bose gas --- C.]{$t_{s}=5\times 10^{-5}$
}\label{Fig:deSitter.Compare.Nk.Ratio.3}
\end{center}
\end{figure*}

\begin{figure*}[!h]
\begin{center}
\mbox{
\subfigure[\label{modes_X2000_ts1e-5}]{
\includegraphics[width=0.50\textwidth]{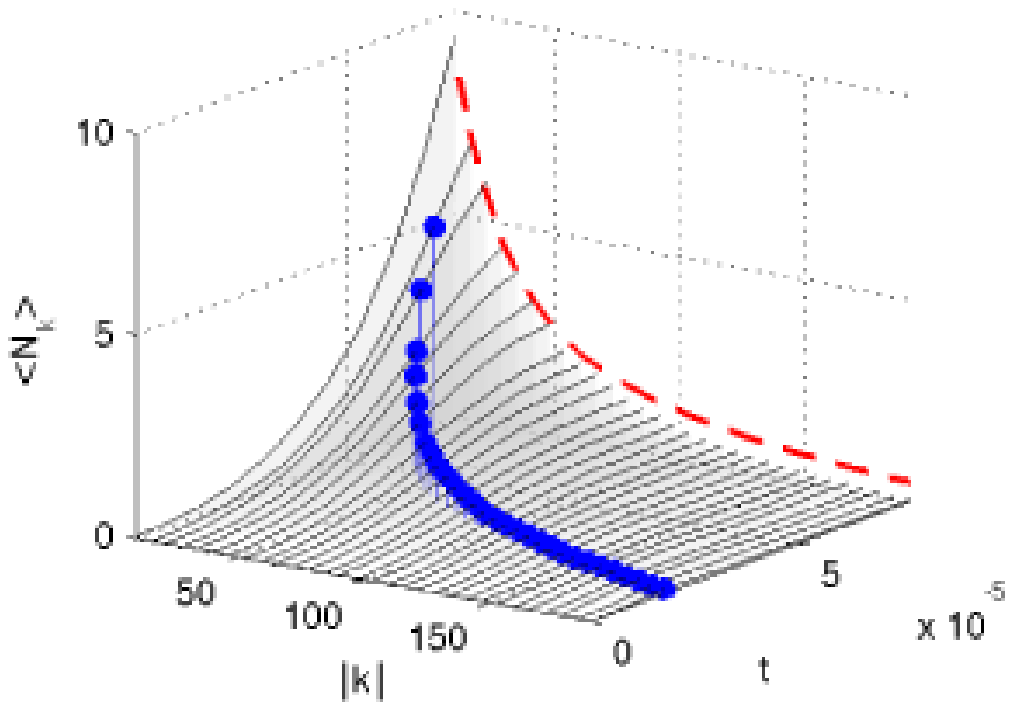}}
\hspace{0mm}
\subfigure[\label{ratio_X2000_ts1e-5}]{
\includegraphics[width=0.45\textwidth]{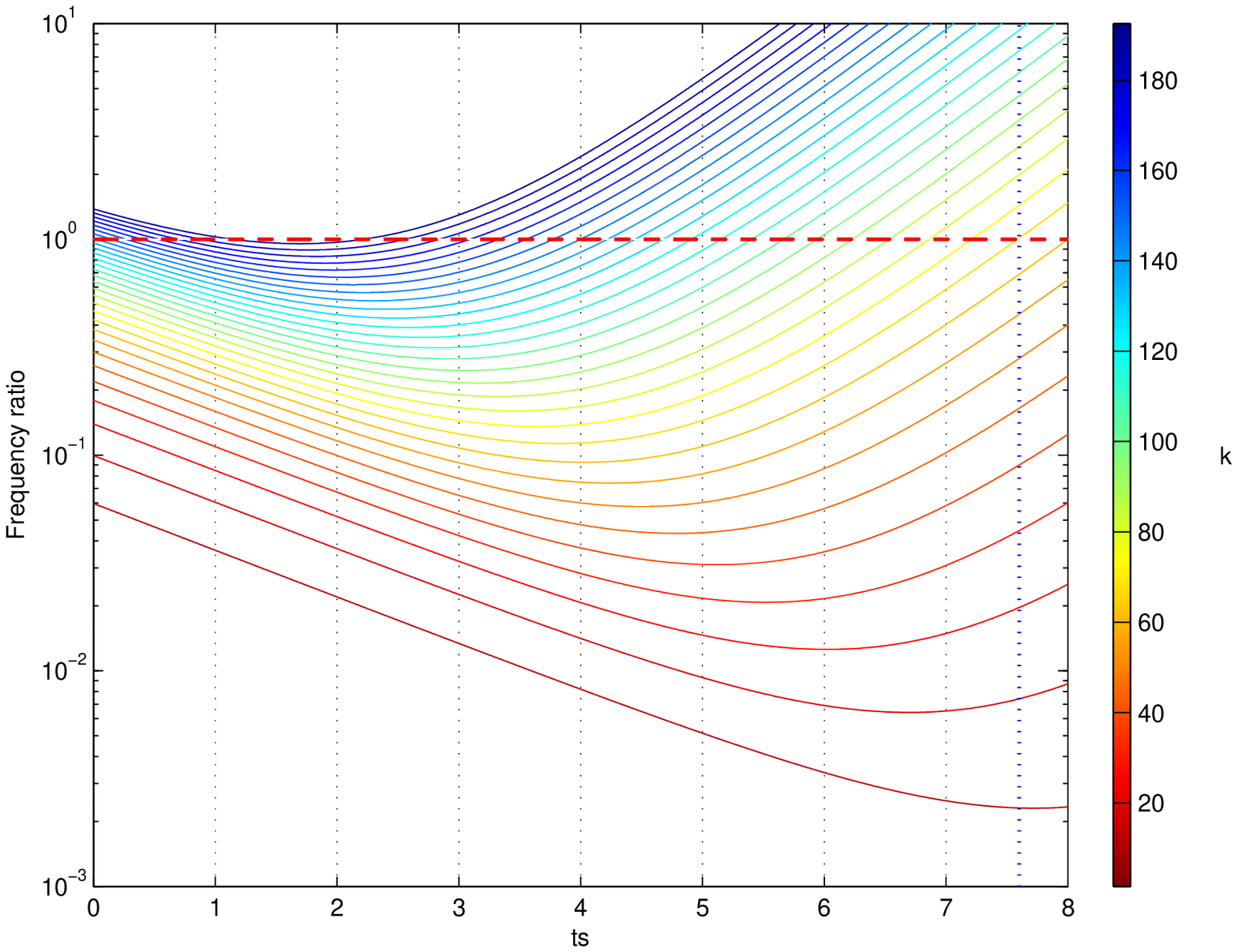}}
}
\caption[Qualitative particle production in Bose gas --- D.]{$t_{s}=5\times 10^{-5}$
}\label{Fig:deSitter.Compare.Nk.Ratio.4}
\end{center}
\end{figure*}

%
Figs.~\ref{Fig:deSitter.Compare.Nk.Ratio.1} - \ref{Fig:deSitter.Compare.Nk.Ratio.4} compare the change in the mode occupation number in each mode on the left side (a), with frequency ratio $\mathcal{R}_{k}(t)$ of the mode on the right side (b). The different rows express different scaling times $t_{s}$.   
Each black line on the right hand side indicates the occupation number in the mode $k$ as a function of time. (The red dashed line on the left had side indicates the sudden limit.) On the right hand side we have plotted the frequency ratio $\mathcal{R}_{k}(t)$ for each of those modes with a different color (online only) to indicate the energy difference of the modes; gradually changing from infrared modes (dark red), to ultraviolet modes (dark blue). The horizontal (red dashed) indicates where the frequency ratio $\mathcal{R}_{k}(t)$ is equal to one, while the vertical (blue dashed) line indicates end of the expansion time in our simulations. The blue dots in the figures to the left indicate when the modes cross over from phononic to trans-phononic behavior, or in other words start to decouple from the emergent spacetime. Roughly we can see that the blue dots are located where the frequency ratio has its minimum.  
%
%

\backmatter
\silkychaptermark{CURRICULUM VITAE}
\chapter{Curriculum Vitae}

\begin{description}
\item[2003-2007] PhD student at Victoria University of Wellington, New Zealand; \\
Project: \emph{Emergent spacetimes}; \\
Supervisor: Prof. M. Visser; \\
	\begin{description}
	\item[2006]
	PhD completion scholarship from the Victoria University of Wellington.
	\item[2005]
	New Zealand Postgraduate Study Abroad Award, from Education New Zealand.
	\item[2003]
	Deutscher Akademischer Austauschdienst (DAAD)
	Gegenstipendium zur \\ wissenschaftlichen Aus- und Fortbildung im Ausland. (Partial stipend 
	for overseas studies.) 
	\item[2003]
	Hartle Prize of the International Society on General Relativity and Gravitation.
	For one of the best student presentations at the 17th international conference on general 
	relativity and gravitation (Dublin, Ireland).
	\end{description}
\item[2002-2003] Diplom thesis at the Max-Plank-Quantumoptics, Garching, Germany; \\
Thesis: \emph{Simulation von Gravitationsobjekten im Bose-Einstein-Kondensat}; \\
Supervisor: Prof. J. I. Cirac; \\
\item[1997-2002] Technical University of Munich, Germany
undergraduate and graduate studies: General relativity, advanced quantum mechanics and theory of condensed matter physics.
\end{description}

\newpage $\,$
\newpage
\silkychaptermark{PUBLICATIONS}
\chapter{Publications}

\subsection*{Publications in Journals \label{Sec:Publications.Journal}}

\begin{itemize}

\item[{\tiny $\blacksquare$}]
{\bf Analogue model of an expanding FRW universe in Bose--Einstein condensates:\\ Application of the classical field method}
 \\{}P.~Jain, S.~Weinfurtner, M.~Visser, and C.~Gardiner
\\{}Accepted for publication in Phys.\ Rev.\ A
  \\{}[arXiv:gr-qc:cond-mat/0705.2077]

\item[{\tiny $\blacksquare$}]
\textbf{ Naturalness in emergent spacetime}
  \\{}S.~Liberati, M.~Visser and S.~Weinfurtner
  \\{}Phys.\ Rev.\ Lett.\  \textbf{ 96}, 151301 (2006)
 \\{}[arXiv:gr-qc/0512139]

\item[{\tiny $\blacksquare$}]
{\bf Analogue quantum gravity phenomenology from a \\ two-component  Bose-Einstein condensate}
  \\{}S.~Liberati, M.~Visser and S.~Weinfurtner
  \\{}Class.\ Quant.\ Grav.\  \textbf{ 23}, 3129 (2006)
   \\{}[arXiv:gr-qc/0510125]

\item[{\tiny $\blacksquare$}]
\textbf{ Massive Klein-Gordon equation from a BEC-based analogue spacetime}
  \\{}M.~Visser and S.~Weinfurtner
  \\{}Phys.\ Rev.\ D \textbf{ 72}, 044020 (2005)
   \\{}[arXiv:gr-qc/0506029]

\item[{\tiny $\blacksquare$}]
\textbf{ Generating perfect fluid spheres in general relativity}
  \\{}P.~Boonserm, M.~Visser and S.~Weinfurtner
  \\{}Phys.\ Rev.\ D \textbf{71}, 124037 (2005)
   \\{}[arXiv:gr-qc/0503007]

\newpage

\item[{\tiny $\blacksquare$}]
\textbf{ Effective refractive index tensor for weak field gravity}
  \\{}P.~Boonserm, C.~Cattoen, T.~Faber, M.~Visser and S.~Weinfurtner
  \\{}Class.\ Quant.\ Grav.\  \textbf{ 22}, 1905 (2005)
   \\{}[arXiv:gr-qc/0411034]

\item[{\tiny $\blacksquare$}]
\textbf{ Vortex geometry for the equatorial slice of the Kerr black hole}
  \\{}M.~Visser and S.~E.~C.~Weinfurtner
  \\{}Class.\ Quant.\ Grav.\  \textbf{ 22}, 2493 (2005)
   \\{}[arXiv:gr-qc/0409014]

\end{itemize}

\subsection*{Articles under review \label{Sec:Publications.Eprint}}
\begin{itemize}

\item[{\tiny $\blacksquare$}]
{\bf Trans-Planckian physics and signature change events in Bose gas hydrodynamics}
  \\{}S.~Weinfurtner, A.~White and M.~Visser
  \\{}[arXiv:gr-qc/0703117]

\item[{\tiny $\blacksquare$}]
{\bf Solution generating theorems for the TOV equation}
  \\{}P.~Boonserm, M.~Visser and S.~Weinfurtner
  \\{}[arXiv:gr-qc/0607001]

\end{itemize}

\subsection*{Conference Proceedings \label{Sec:Publications.Proceedings}}

\begin{itemize}

\item[{\tiny $\blacksquare$}]
{\bf Analogue spacetime based on 2-component Bose-Einstein condensates}
  \\{}S.~Weinfurtner, S.~Liberati and M.~Visser
  \\{}Lecture Notes in Physics \ \textbf{718}, (2006)  
[arXiv:gr-qc/0605121]

\item[{\tiny $\blacksquare$}]
{\bf Solution generating theorems: perfect fluid spheres and the TOV equation}
  \\{}P.~Boonserm, M.~Visser and S.~Weinfurtner
  \\{}[arXiv:gr-qc/0609099]
\\{}{\it To appear in the proceedings of 11th Marcel Grossmann Meeting on Recent Developments in Theoretical and Experimental General Relativity,
Gravitation, and Relativistic Field Theories, Berlin, Germany, 23-29 Jul 2006}

\item[{\tiny $\blacksquare$}]
{\bf Solution generating theorems for perfect fluid spheres}
  \\{}P.~Boonserm, M.~Visser and S.~Weinfurtner
  \\{}[arXiv:gr-qc/0609088]
\\{}{\it To appear in the proceedings of 12th Conference on Recent Developments in Gravity (NEB XII), Nafplio, Greece, 29 Jun - 2 Jul 2006}

\item[{\tiny $\blacksquare$}]
{\bf Modelling Planck-scale Lorentz violation via analogue models}
  \\{}S.~Weinfurtner, S.~Liberati and M.~Visser
  \\{}J.\ Phys.\ Conf.\ Ser.\  {\bf 33}, 373 (2006)
  [arXiv:gr-qc/0512127]

\item[{\tiny $\blacksquare$}]
{\bf Analogue model for quantum gravity phenomenology}
  \\{}S.~Weinfurtner, S.~Liberati and M.~Visser
  \\{}J.\ Phys.\ A  {\bf 39}, 6807 (2006)
  [arXiv:gr-qc/0511105]

\end{itemize}

\newpage 
\addcontentsline{toc}{chapter}{Bibliography}

\newpage
%
%
\end{document}